\documentclass[a4paper,UKenglish,cleveref,autoref,thm-restate]{lipics-v2021}
\usepackage{wjd}
\usepackage{comment}
\usepackage{proof-dashed}
\usepackage{unixode}

\usepackage[dvipsnames]{xcolor}
\definecolor{britishracinggreen}{rgb}{0.0, 0.26, 0.15}
\hypersetup{
    bookmarks=true,         % show bookmarks bar?
    unicode=true,          % non-Latin characters in Acrobat’s bookmarks
    pdftoolbar=true,        % show Acrobat’s toolbar?
    pdfmenubar=true,        % show Acrobat’s menu?
    pdffitwindow=false,     % window fit to page when opened
    pdftitle={Birkhoff's Variety Theorem Formalized in Agda},    % title
    pdfauthor={William DeMeo},     % author
    pdfsubject={},   % subject of the document
    pdfcreator={pdflatex with hyperref},
    pdfproducer={},  % producer of the document
    pdfkeywords= {Agda} {Birkhoff’s HSP theorem} {equational logic}
    {Formalization of mathematics} {Martin-Löf Type Theory} {model theory}
    {universal algebra},
    pdfnewwindow=true,      % links in new window
    colorlinks=true,       % false: boxed links; true: colored links
    linkcolor=blue,          % color of internal links
    citecolor=black,        % color of links to bibliography
    filecolor=black,      % color of file links
    urlcolor=britishracinggreen           % color of external links
}

\usepackage[wjd,links]{agda}
\AgdaNoSpaceAroundCode{}

\newif\ifnonbold % comment this line to restore bold universe symbols

\newif\ifshort
\shorttrue % (to compile short version of the paper)
\shortfalse % (to compile long version of the paper)
\newcommand\seemedium{see~\cite{DeMeo:2021b} for the complete formalization\xspace}
\newcommand\seeshort{see~\cite{DeMeo:2021b}\xspace}

\usepackage{tikz}
\usetikzlibrary{snakes,arrows,shapes}
\usepackage{amsmath}

\crefformat{footnote}{#2\footnotemark[#1]#3}

\ifshort
\title{Birkhoff's Variety Theorem formalized in Agda}
\else
\title{A Machine-checked Proof of Birkhoff's Variety Theorem in Martin-L\"of Type Theory}
\fi

\author{William DeMeo}
       {Thmpr Research}
       {williamdemeo@gmail.com}{https://orcid.org/0000-0003-1832-5690}{supported
         by the CoCoSym project under ERC Consolidator Grant No.~771005.}
\author{Jacques Carette}{McMaster University}{carette@mcmaster.ca}{https://orcid.org/0000-0001-8993-9804}{}
\authorrunning{DeMeo and Carette}

\copyrightfooter


\category{} %optional, e.g. invited paper

\nolinenumbers %uncomment to disable line numbering

\hideLIPIcs  %uncomment to remove references to LIPIcs series (logo, DOI, ...), e.g. when preparing a pre-final version to be uploaded to arXiv or another public repository

\EventEditors{}
\EventNoEds{2}
\EventLongTitle{}
\EventShortTitle{}
\EventAcronym{}
\EventYear{2021}
\EventDate{26 October}
\EventLocation{}
\EventLogo{}
\SeriesVolume{0}
\ArticleNo{0}
\begin{document}

\maketitle

%\newpage %%%%%%%%%%%%%%%%%%%%%%%%%%%%  TOC  %%%%%%%%%%%%%%%%%%%%%%

% \setcounter{tocdepth}{2}
% \tableofcontents

\begin{abstract}
The Agda Universal Algebra Library (\agdaalgebras) is a library of types and programs (theorems and proofs) we developed to formalize the foundations of universal algebra in dependent type theory using the Agda programming language and proof assistant. In this paper we draw on and explain many components of the agda-algebras library, which we extract into a single Agda module in order to present a self-contained formal and constructive proof of Birkhoff's HSP theorem in Martin-L\"of dependent type theory. In the course of our presentation, we highlight some of the more challenging aspects of formalizing the basic definitions and theorems of universal algebra in type theory. Nonetheless, we hope this paper and the agda-algebras library serve as further evidence in support of the claim that dependent type theory and the Agda language, despite the technical demands they place on the user, are accessible to working mathematicians (such as ourselves) who possess sufficient patience and resolve to formally verify their results with a proof assistant. Indeed, the agda-algebras library now includes a substantial collection of definitions, theorems, and proofs from universal algebra, illustrating the expressive power of inductive and dependent types for representing and reasoning about general algebraic and relational structures.

\end{abstract}

\providecommand{\hypertarget}[2]{#2}

%% This is to make the typewriter font inserted by pandoc look more similar to
%% the Agda fonts used in the highlighted code.
\renewcommand\texttt[1]{\textsf{#1}}

\section{Introduction}
The \agdaalgebras library is a repository of types and programs
(theorems and proofs) formalizing the foundations of universal algebra in Martin-Löf
dependent type theory (MLTT) using the \agda programming language.
The library now includes an fairly extensive collection of formal definitions, theorems, and
proofs that codify, in the formal language of type theory, the analogous definitions,
theorems, and proofs of classical, set-theory-based universal algebra and equational
logic. As such, the \agdaalgebras library provides many examples that exhibit the
expressiveness of inductive and dependent types for representing and reasoning about
general algebraic and relational structures in a formal language. The main advantage of
formalizing mathematics in type theory using a proof assistant (like Agda) is that
the software checks the correctness of our proofs by a process known as ``type-checking.''

The first major milestone of the \agdaalgebras project is a formal proof of \emph{Birkhoff's
variety theorem} (also known as the \emph{HSP theorem})~\cite{Birkhoff:1935} in dependent
type theory. Our first formal proof of the theorem, completed in January of 2021,
contained some flaws and there were concerns that the proof was not truly
constructive.\footnote{See the \href{https://github.com/ualib/ualib.github.io/blob/71f173858701398d56224dd79d152c380c0c2b5e/src/lagda/UALib/Birkhoff.lagda}{Birkhoff} module
from the
\href{https://github.com/ualib/ualib.github.io/commit/71f173858701398d56224dd79d152c380c0c2b5e}{15 Jan 2021 commit (71f1738)} of the \href{https://github.com/ualib/ualib.github.io}{ualib/ualib.gitlab.io} repository~\cite{ualib_v1.0.0}.}
We are confident that the version we present here---based on version 2.0.0 of the
\agdaalgebras library---is fully constructive and correct.\footnote{Specifically, see the \href{https://github.com/ualib/agda-algebras/commit/ab859caf814566fe32205f76bd0a4ac1e6007147}{30 Nov 2021 commit (ab859ca)} of the \agdaalgebras library~\cite{ualib_v2.0.0}.}
To the best of our knowledge, ours is the first formulation of the HSP theorem in MLTT, and the
first formal, machine-verified proof of Birkhoff's celebrated 1935 result.

In this paper, we present a self-contained formal proof of the HSP theorem by
extracting into a single Agda module a subset of the \agdaalgebras library, including only
the pieces we need for the proof.  The main body of the paper is generated by a literate
Agda file, available online,\footnote{See
\url{https://github.com/ualib/agda-algebras/blob/master/src/Demos/HSP.lagda}}
that others can type-check, using \agda version 2.6.2 and \agdastdlib version 1.7, to
verify its correctness.
\ifshort

In order to present the proof in under 20 pages, we omit some of the formalities,
but strive to preserve the essential content and character of the development.
We leave out the routine or overly technical components, as well as anything that does not
seem to offer any insight into the central ideas of the proof. For readers wishing to see all the
details, we have posted on the arXiv an unabridged version of the paper, which includes
every line of code of our formal proof of Birkhoff's theorem in a single, self-contained
(apart from a few dozen imports from the \agdastdlib) Agda module~\cite{DeMeo:2021b}.
\else
We include here every line of code of our formal proof of Birkhoff's theorem
in a single, self-contained (apart from a few dozen imports from the \agdastdlib) Agda
module.
%For readers preferring a more concise presentation, an abridged version of this paper is also
%available~\cite{DeMeo:2021c}.
\fi

In the course of this presentation we highlight some of the challenging aspects of
formalizing the basic definitions and theorems of universal algebra in type theory.
%In particular, we touch on some technical issues that arise when attempting
%to constructively prove basic results of our field.  Nonetheless,
One positive contribution of this project is that it lends support to the claim that
dependent type theory and the Agda language, despite the technical demands they place on
the user, are accessible to working mathematicians (such as ourselves) who possess
sufficient patience and resolve to codify their work in type theory in order to formally verify
their results with a proof assistant.

\ifshort\else
Our presentation gives a sobering glimpse of the technical hurdles that must be overcome
to conduct research in mathematics using dependent type theory and the Agda language.
Nonetheless we hope our work does not discourage anyone from investing in these
technologies and we remain committed to the use and promotion of type theory and proof
assistants in general and in our own research. Indeed, we are excited to share the
gratifying outcomes and achievements that resulted from attaining some degree of mastery
of type theory, interactive theorem proving, and the Agda language.
\fi

\section{Preliminaries}

\subsection{Logical foundations}

An Agda program typically begins by setting some language options and by
importing types from existing Agda libraries. The language options are specified
using the {\footnotesize \ak{OPTIONS}} \emph{pragma} which affects the way Agda behaves by controlling
the deduction rules that are available and the logical axioms
that are assumed when the program is type-checked to verify its
correctness. Every Agda program in the \agdaalgebras library, including the
module \DemosHSPlagda described in this paper,\footnote{available at \url{https://github.com/ualib/agda-algebras/blob/master/src/Demos/HSP.lagda}} begins with the line
\begin{code}[inline]%
\>[0]\AgdaSymbol{\{-\#}\AgdaSpace{}%
\AgdaKeyword{OPTIONS}\AgdaSpace{}%
\AgdaPragma{--without-K}\AgdaSpace{}%
\AgdaPragma{--exact-split}\AgdaSpace{}%
\AgdaPragma{--safe}\AgdaSpace{}%
\AgdaSymbol{\#-\}}\<%
\end{code}
. Here are brief descriptions of these options, accompanied by links to related documentation.

\begin{itemize}
\item
\AgdaPragma{without-K} disables \href{https://ncatlab.org/nlab/show/axiom+K+%28type+theory%29}{Streicher's K axiom}.
See the \href{https://agda.readthedocs.io/en/v2.6.1/language/without-k.html}{section on axiom K} in
\ifshort
\cite{agdaref-axiomk}.
\else
the \href{https://agda.readthedocs.io/en/v2.6.1.3/language}{Agda Language Reference Manual}~\cite{agdaref-axiomk}.
\fi

\item
\AgdaPragma{exact-split} makes Agda accept only those definitions that behave like so-called {\it judgmental} equalities.
See the \href{https://agda.readthedocs.io/en/v2.6.1/tools/command-line-options.html#pattern-matching-and-equality}%
{Pattern matching and equality} section of
\ifshort
\cite{agdatools-patternmatching}.
\else
the \href{https://agda.readthedocs.io/en/v2.6.1.3/tools/}{Agda Tools} documentation~\cite{agdatools-patternmatching}.
\fi

\item
\AgdaPragma{safe} ensures that nothing is postulated outright---every non-MLTT axiom has to be an explicit assumption (e.g., an argument to a function or module).
See the \href{https://agda.readthedocs.io/en/v2.6.1/tools/command-line-options.html#cmdoption-safe}{cmdoption-safe} section of~\cite{agdaref-safeagda}.
\end{itemize}

The {\footnotesize \AgdaKeyword{OPTIONS}} pragma is usually followed by the start of a module and a list of
import directives. For example, the present module (\DemosHSPlagda)
\ifshort includes the following imports.
\else
begins as follows.
\fi

\begin{code}[hide]%
\>[0]\AgdaSymbol{\{-\#}\AgdaSpace{}%
\AgdaKeyword{OPTIONS}\AgdaSpace{}%
\AgdaPragma{--without-K}\AgdaSpace{}%
\AgdaPragma{--exact-split}\AgdaSpace{}%
\AgdaPragma{--safe}\AgdaSpace{}%
\AgdaSymbol{\#-\}}\<%
\end{code}
\ifshort\else
\begin{code}%
\>[0]\<%
\\
\>[0]\AgdaComment{--\ Import\ universe\ levels\ and\ Signature\ type\ (described\ below)\ from\ the\ agda-algebras\ library.}\<%
\\
\>[0]\AgdaKeyword{open}\AgdaSpace{}%
\AgdaKeyword{import}\AgdaSpace{}%
\AgdaModule{Algebras.Basic}\AgdaSpace{}%
\AgdaKeyword{using}\AgdaSpace{}%
\AgdaSymbol{(}\AgdaSpace{}%
\AgdaGeneralizable{𝓞}\AgdaSpace{}%
\AgdaSymbol{;}\AgdaSpace{}%
\AgdaGeneralizable{𝓥}\AgdaSpace{}%
\AgdaSymbol{;}\AgdaSpace{}%
\AgdaFunction{Signature}\AgdaSpace{}%
\AgdaSymbol{)}\<%
\\
\\[\AgdaEmptyExtraSkip]%
\>[0]\AgdaKeyword{module}\AgdaSpace{}%
\AgdaModule{Demos.HSP}\AgdaSpace{}%
\AgdaSymbol{\{}\AgdaBound{𝑆}\AgdaSpace{}%
\AgdaSymbol{:}\AgdaSpace{}%
\AgdaFunction{Signature}\AgdaSpace{}%
\AgdaGeneralizable{𝓞}\AgdaSpace{}%
\AgdaGeneralizable{𝓥}\AgdaSymbol{\}}\AgdaSpace{}%
\AgdaKeyword{where}\<%
\end{code}
\fi
\begin{code}%
\>[0]\<%
\\
\>[0]\AgdaComment{--\ Import\ 16\ definitions\ from\ the\ Agda\ Standard\ Library.}\<%
\\
\>[0]\AgdaKeyword{open}\AgdaSpace{}%
\AgdaKeyword{import}%
\>[13]\AgdaModule{Data.Unit.Polymorphic}%
\>[61]\AgdaKeyword{using}\AgdaSpace{}%
\AgdaSymbol{(}\AgdaSpace{}%
\AgdaFunction{⊤}\AgdaSpace{}%
\AgdaSymbol{;}\AgdaSpace{}%
\AgdaFunction{tt}%
\>[99]\AgdaSymbol{)}\<%
\\
\>[0]\AgdaKeyword{open}\AgdaSpace{}%
\AgdaKeyword{import}%
\>[13]\AgdaModule{Function}%
\>[61]\AgdaKeyword{using}\AgdaSpace{}%
\AgdaSymbol{(}\AgdaSpace{}%
\AgdaFunction{id}\AgdaSpace{}%
\AgdaSymbol{;}\AgdaSpace{}%
\AgdaFunction{flip}\AgdaSpace{}%
\AgdaSymbol{;}\AgdaSpace{}%
\AgdaOperator{\AgdaFunction{\AgdaUnderscore{}∘\AgdaUnderscore{}}}%
\>[99]\AgdaSymbol{)}\<%
\\
\>[0]\AgdaKeyword{open}\AgdaSpace{}%
\AgdaKeyword{import}%
\>[13]\AgdaModule{Level}%
\>[61]\AgdaKeyword{using}\AgdaSpace{}%
\AgdaSymbol{(}\AgdaSpace{}%
\AgdaPostulate{Level}%
\>[99]\AgdaSymbol{)}\<%
\\
\>[0]\AgdaKeyword{open}\AgdaSpace{}%
\AgdaKeyword{import}%
\>[13]\AgdaModule{Relation.Binary}%
\>[61]\AgdaKeyword{using}\AgdaSpace{}%
\AgdaSymbol{(}\AgdaSpace{}%
\AgdaFunction{Rel}\AgdaSpace{}%
\AgdaSymbol{;}\AgdaSpace{}%
\AgdaRecord{Setoid}\AgdaSpace{}%
\AgdaSymbol{;}\AgdaSpace{}%
\AgdaRecord{IsEquivalence}%
\>[99]\AgdaSymbol{)}\<%
\\
\>[0]\AgdaKeyword{open}\AgdaSpace{}%
\AgdaKeyword{import}%
\>[13]\AgdaModule{Relation.Binary.Definitions}%
\>[61]\AgdaKeyword{using}\AgdaSpace{}%
\AgdaSymbol{(}\AgdaSpace{}%
\AgdaFunction{Reflexive}\AgdaSpace{}%
\AgdaSymbol{;}\AgdaSpace{}%
\AgdaFunction{Symmetric}%
\>[99]\AgdaSymbol{)}\<%
\\
\>[61]\AgdaKeyword{using}\AgdaSpace{}%
\AgdaSymbol{(}\AgdaSpace{}%
\AgdaFunction{Transitive}\AgdaSpace{}%
\AgdaSymbol{;}\AgdaSpace{}%
\AgdaFunction{Sym}\AgdaSpace{}%
\AgdaSymbol{;}\AgdaSpace{}%
\AgdaFunction{Trans}%
\>[99]\AgdaSymbol{)}\<%
\\
\>[0]\AgdaKeyword{open}\AgdaSpace{}%
\AgdaKeyword{import}%
\>[13]\AgdaModule{Relation.Binary.PropositionalEquality}%
\>[61]\AgdaKeyword{using}\AgdaSpace{}%
\AgdaSymbol{(}\AgdaSpace{}%
\AgdaOperator{\AgdaDatatype{\AgdaUnderscore{}≡\AgdaUnderscore{}}}%
\>[99]\AgdaSymbol{)}\<%
\\
\>[0]\AgdaKeyword{open}\AgdaSpace{}%
\AgdaKeyword{import}%
\>[13]\AgdaModule{Relation.Unary}%
\>[61]\AgdaKeyword{using}\AgdaSpace{}%
\AgdaSymbol{(}\AgdaSpace{}%
\AgdaFunction{Pred}\AgdaSpace{}%
\AgdaSymbol{;}\AgdaSpace{}%
\AgdaOperator{\AgdaFunction{\AgdaUnderscore{}⊆\AgdaUnderscore{}}}\AgdaSpace{}%
\AgdaSymbol{;}\AgdaSpace{}%
\AgdaOperator{\AgdaFunction{\AgdaUnderscore{}∈\AgdaUnderscore{}}}%
\>[99]\AgdaSymbol{)}\<%
\\
\\[\AgdaEmptyExtraSkip]%
\>[0]\AgdaComment{--\ Import\ 23\ definitions\ from\ the\ Agda\ Standard\ Library\ and\ rename\ 12\ of\ them.}\<%
\\
\>[0]\AgdaKeyword{open}\AgdaSpace{}%
\AgdaKeyword{import}%
\>[13]\AgdaModule{Agda.Primitive}%
\>[29]\AgdaKeyword{renaming}\AgdaSpace{}%
\AgdaSymbol{(}\AgdaSpace{}%
\AgdaPrimitive{Set}%
\>[47]\AgdaSymbol{to}\AgdaSpace{}%
\AgdaPrimitive{Type}%
\>[58]\AgdaSymbol{)}%
\>[61]\AgdaKeyword{using}\AgdaSpace{}%
\AgdaSymbol{(}\AgdaSpace{}%
\AgdaOperator{\AgdaPrimitive{\AgdaUnderscore{}⊔\AgdaUnderscore{}}}\AgdaSpace{}%
\AgdaSymbol{;}\AgdaSpace{}%
\AgdaPrimitive{lsuc}%
\>[99]\AgdaSymbol{)}\<%
\\
\>[0]\AgdaKeyword{open}\AgdaSpace{}%
\AgdaKeyword{import}%
\>[13]\AgdaModule{Data.Product}%
\>[29]\AgdaKeyword{renaming}\AgdaSpace{}%
\AgdaSymbol{(}\AgdaSpace{}%
\AgdaField{proj₁}%
\>[47]\AgdaSymbol{to}\AgdaSpace{}%
\AgdaField{fst}%
\>[58]\AgdaSymbol{)}\<%
\\
\>[29]\AgdaKeyword{renaming}\AgdaSpace{}%
\AgdaSymbol{(}\AgdaSpace{}%
\AgdaField{proj₂}%
\>[47]\AgdaSymbol{to}\AgdaSpace{}%
\AgdaField{snd}%
\>[58]\AgdaSymbol{)}%
\>[61]\AgdaKeyword{using}\AgdaSpace{}%
\AgdaSymbol{(}\AgdaSpace{}%
\AgdaOperator{\AgdaFunction{\AgdaUnderscore{}×\AgdaUnderscore{}}}\AgdaSpace{}%
\AgdaSymbol{;}\AgdaSpace{}%
\AgdaOperator{\AgdaInductiveConstructor{\AgdaUnderscore{},\AgdaUnderscore{}}}\AgdaSpace{}%
\AgdaSymbol{;}\AgdaSpace{}%
\AgdaRecord{Σ}\AgdaSpace{}%
\AgdaSymbol{;}\AgdaSpace{}%
\AgdaFunction{Σ-syntax}%
\>[99]\AgdaSymbol{)}\<%
\\
\>[0]\AgdaKeyword{open}\AgdaSpace{}%
\AgdaKeyword{import}%
\>[13]\AgdaModule{Function}%
\>[29]\AgdaKeyword{renaming}\AgdaSpace{}%
\AgdaSymbol{(}\AgdaSpace{}%
\AgdaRecord{Func}%
\>[47]\AgdaSymbol{to}\AgdaSpace{}%
\AgdaRecord{\AgdaUnderscore{}⟶\AgdaUnderscore{}}%
\>[58]\AgdaSymbol{)}%
\>[61]\AgdaKeyword{using}\AgdaSpace{}%
\AgdaSymbol{(}\AgdaSpace{}%
\AgdaRecord{Injection}\AgdaSpace{}%
\AgdaSymbol{;}\AgdaSpace{}%
\AgdaRecord{Surjection}%
\>[99]\AgdaSymbol{)}\<%
\\
\>[0]\AgdaKeyword{open}%
\>[13]\AgdaModule{\AgdaUnderscore{}⟶\AgdaUnderscore{}}%
\>[29]\AgdaKeyword{renaming}\AgdaSpace{}%
\AgdaSymbol{(}\AgdaSpace{}%
\AgdaField{f}%
\>[47]\AgdaSymbol{to}\AgdaSpace{}%
\AgdaField{\AgdaUnderscore{}⟨\$⟩\AgdaUnderscore{}}%
\>[58]\AgdaSymbol{)}%
\>[61]\AgdaKeyword{using}\AgdaSpace{}%
\AgdaSymbol{(}\AgdaSpace{}%
\AgdaField{cong}%
\>[99]\AgdaSymbol{)}\<%
\\
\>[0]\AgdaKeyword{open}%
\>[13]\AgdaModule{Setoid}%
\>[29]\AgdaKeyword{renaming}\AgdaSpace{}%
\AgdaSymbol{(}\AgdaSpace{}%
\AgdaFunction{refl}%
\>[47]\AgdaSymbol{to}\AgdaSpace{}%
\AgdaFunction{reflˢ}%
\>[58]\AgdaSymbol{)}\<%
\\
\>[29]\AgdaKeyword{renaming}\AgdaSpace{}%
\AgdaSymbol{(}\AgdaSpace{}%
\AgdaFunction{sym}%
\>[47]\AgdaSymbol{to}\AgdaSpace{}%
\AgdaFunction{symˢ}%
\>[58]\AgdaSymbol{)}\<%
\\
\>[29]\AgdaKeyword{renaming}\AgdaSpace{}%
\AgdaSymbol{(}\AgdaSpace{}%
\AgdaFunction{trans}%
\>[47]\AgdaSymbol{to}\AgdaSpace{}%
\AgdaFunction{transˢ}%
\>[58]\AgdaSymbol{)}\<%
\\
\>[29]\AgdaKeyword{renaming}\AgdaSpace{}%
\AgdaSymbol{(}\AgdaSpace{}%
\AgdaOperator{\AgdaField{\AgdaUnderscore{}≈\AgdaUnderscore{}}}%
\>[47]\AgdaSymbol{to}\AgdaSpace{}%
\AgdaOperator{\AgdaField{\AgdaUnderscore{}≈ˢ\AgdaUnderscore{}}}%
\>[58]\AgdaSymbol{)}%
\>[61]\AgdaKeyword{using}\AgdaSpace{}%
\AgdaSymbol{(}\AgdaSpace{}%
\AgdaField{Carrier}\AgdaSpace{}%
\AgdaSymbol{;}\AgdaSpace{}%
\AgdaField{isEquivalence}%
\>[99]\AgdaSymbol{)}\<%
\\
\>[0]\AgdaKeyword{open}%
\>[13]\AgdaModule{IsEquivalence}%
\>[29]\AgdaKeyword{renaming}\AgdaSpace{}%
\AgdaSymbol{(}\AgdaSpace{}%
\AgdaField{refl}%
\>[47]\AgdaSymbol{to}\AgdaSpace{}%
\AgdaField{reflᵉ}%
\>[58]\AgdaSymbol{)}\<%
\\
\>[29]\AgdaKeyword{renaming}\AgdaSpace{}%
\AgdaSymbol{(}\AgdaSpace{}%
\AgdaField{sym}%
\>[47]\AgdaSymbol{to}\AgdaSpace{}%
\AgdaField{symᵉ}%
\>[58]\AgdaSymbol{)}\<%
\\
\>[29]\AgdaKeyword{renaming}\AgdaSpace{}%
\AgdaSymbol{(}\AgdaSpace{}%
\AgdaField{trans}%
\>[47]\AgdaSymbol{to}\AgdaSpace{}%
\AgdaField{transᵉ}%
\>[58]\AgdaSymbol{)}%
\>[61]\AgdaKeyword{using}\AgdaSpace{}%
\AgdaSymbol{()}\<%
\\
\>[0]\<%
\end{code}
\ifshort\else
\begin{code}%
\>[0]\AgdaComment{--\ Assign\ handles\ to\ 3\ modules\ of\ the\ Agda\ Standard\ Library.}\<%
\\
\>[0]\AgdaKeyword{import}%
\>[13]\AgdaModule{Function.Definitions}%
\>[52]\AgdaSymbol{as}\AgdaSpace{}%
\AgdaModule{FD}\<%
\\
\>[0]\AgdaKeyword{import}%
\>[13]\AgdaModule{Relation.Binary.PropositionalEquality}%
\>[52]\AgdaSymbol{as}\AgdaSpace{}%
\AgdaModule{≡}\<%
\\
\>[0]\AgdaKeyword{import}%
\>[13]\AgdaModule{Relation.Binary.Reasoning.Setoid}%
\>[52]\AgdaSymbol{as}\AgdaSpace{}%
\AgdaModule{SetoidReasoning}\<%
\\
\\[\AgdaEmptyExtraSkip]%
\>[0]\AgdaKeyword{private}\AgdaSpace{}%
\AgdaKeyword{variable}\<%
\\
\>[0][@{}l@{\AgdaIndent{0}}]%
\>[1]\AgdaGeneralizable{α}\AgdaSpace{}%
\AgdaGeneralizable{ρᵃ}\AgdaSpace{}%
\AgdaGeneralizable{β}\AgdaSpace{}%
\AgdaGeneralizable{ρᵇ}\AgdaSpace{}%
\AgdaGeneralizable{γ}\AgdaSpace{}%
\AgdaGeneralizable{ρᶜ}\AgdaSpace{}%
\AgdaGeneralizable{δ}\AgdaSpace{}%
\AgdaGeneralizable{ρᵈ}\AgdaSpace{}%
\AgdaGeneralizable{ρ}\AgdaSpace{}%
\AgdaGeneralizable{χ}\AgdaSpace{}%
\AgdaGeneralizable{ℓ}\AgdaSpace{}%
\AgdaSymbol{:}\AgdaSpace{}%
\AgdaPostulate{Level}\<%
\\
\>[1]\AgdaGeneralizable{Γ}\AgdaSpace{}%
\AgdaGeneralizable{Δ}\AgdaSpace{}%
\AgdaSymbol{:}\AgdaSpace{}%
\AgdaPrimitive{Type}\AgdaSpace{}%
\AgdaGeneralizable{χ}\<%
\\
\>[1]\AgdaGeneralizable{𝑓}\AgdaSpace{}%
\AgdaSymbol{:}\AgdaSpace{}%
\AgdaField{fst}\AgdaSpace{}%
\AgdaBound{𝑆}\<%
\\
\>[0]\<%
\end{code}
\fi
Note that the above imports include some adjustments to ``standard Agda'' syntax to suit our own taste.
In particular, the following conventions used throughout the \agdaalgebras library and this paper: we use \AgdaPrimitive{Type} in place of \AgdaPrimitive{Set}, the infix long arrow symbol,
\AgdaRecord{\AgdaUnderscore{}⟶\AgdaUnderscore{}}, instead of \AgdaRecord{Func} (the type of ``setoid functions'' discussed in §\ref{setoid-functions} below), and the symbol \aofld{\au{}⟨\$⟩\au{}} in place of \afld{f} (application of the map of a setoid function); we use
\AgdaField{fst} and \AgdaField{snd}, and sometimes \AgdaOperator{\AgdaFunction{∣\AgdaUnderscore{}∣}} and
\AgdaOperator{\AgdaFunction{∥\AgdaUnderscore{}∥}}, to denote the first and second
projections out of the product type
\AgdaOperator{\AgdaFunction{\AgdaUnderscore{}×\AgdaUnderscore{}}}.
\ifshort\else

\begin{code}%
\>[0]\<%
\\
\>[0]\AgdaKeyword{module}\AgdaSpace{}%
\AgdaModule{\AgdaUnderscore{}}\AgdaSpace{}%
\AgdaSymbol{\{}\AgdaBound{A}\AgdaSpace{}%
\AgdaSymbol{:}\AgdaSpace{}%
\AgdaPrimitive{Type}\AgdaSpace{}%
\AgdaGeneralizable{α}\AgdaSpace{}%
\AgdaSymbol{\}\{}\AgdaBound{B}\AgdaSpace{}%
\AgdaSymbol{:}\AgdaSpace{}%
\AgdaBound{A}\AgdaSpace{}%
\AgdaSymbol{→}\AgdaSpace{}%
\AgdaPrimitive{Type}\AgdaSpace{}%
\AgdaGeneralizable{β}\AgdaSymbol{\}}\AgdaSpace{}%
\AgdaKeyword{where}\<%
\\
\>[0][@{}l@{\AgdaIndent{0}}]%
\>[1]\AgdaOperator{\AgdaFunction{∣\AgdaUnderscore{}∣}}\AgdaSpace{}%
\AgdaSymbol{:}\AgdaSpace{}%
\AgdaFunction{Σ[}\AgdaSpace{}%
\AgdaBound{x}\AgdaSpace{}%
\AgdaFunction{∈}\AgdaSpace{}%
\AgdaBound{A}\AgdaSpace{}%
\AgdaFunction{]}\AgdaSpace{}%
\AgdaBound{B}\AgdaSpace{}%
\AgdaBound{x}\AgdaSpace{}%
\AgdaSymbol{→}\AgdaSpace{}%
\AgdaBound{A}\<%
\\
\>[1]\AgdaOperator{\AgdaFunction{∣\AgdaUnderscore{}∣}}\AgdaSpace{}%
\AgdaSymbol{=}\AgdaSpace{}%
\AgdaField{fst}\<%
\\
\>[1]\AgdaOperator{\AgdaFunction{∥\AgdaUnderscore{}∥}}\AgdaSpace{}%
\AgdaSymbol{:}\AgdaSpace{}%
\AgdaSymbol{(}\AgdaBound{z}\AgdaSpace{}%
\AgdaSymbol{:}\AgdaSpace{}%
\AgdaFunction{Σ[}\AgdaSpace{}%
\AgdaBound{a}\AgdaSpace{}%
\AgdaFunction{∈}\AgdaSpace{}%
\AgdaBound{A}\AgdaSpace{}%
\AgdaFunction{]}\AgdaSpace{}%
\AgdaBound{B}\AgdaSpace{}%
\AgdaBound{a}\AgdaSymbol{)}\AgdaSpace{}%
\AgdaSymbol{→}\AgdaSpace{}%
\AgdaBound{B}\AgdaSpace{}%
\AgdaOperator{\AgdaFunction{∣}}\AgdaSpace{}%
\AgdaBound{z}\AgdaSpace{}%
\AgdaOperator{\AgdaFunction{∣}}\<%
\\
\>[1]\AgdaOperator{\AgdaFunction{∥\AgdaUnderscore{}∥}}\AgdaSpace{}%
\AgdaSymbol{=}\AgdaSpace{}%
\AgdaField{snd}\<%
\end{code}
\fi

%% -----------------------------------------------------------------------------
\subsection{Setoids}\label{setoids}
A \defn{setoid} is a pair (\ab A, \af{≈}) where \ab A is a type and \af{≈}
is an equivalence relation on \ab A. Setoids seem to have gotten a bad wrap
in some parts of the interactive theorem proving community because of the extra
overhead they require. However, we feel they are ideally suited to
representing the basic objects of informal mathematics (i.e., sets)
in a constructive, type-theoretic way.

In informal mathematical discourse, a set typically comes equipped with an equivalence
relation manifesting the notion of equality of elements of the set. We
often take this equivalence for granted or view it as self-evident; rarely do we
take pains to define it explicitly. While well-suited to informal
mathematics, this approach is inadequate for formal, machine-checked proofs.

The \agdaalgebras library was first developed without setoids, relying exclusively
on the inductive equality type \ad{\au{}≡\au{}}, defined in \am{Agda.Builtin.Equality},
along with some experimental, domain-specific types for equivalence classes, quotients, etc.
One consequence of this design decision was that the formalization of many
theorems required postulating function extensionality, an axiom that is known to be neither provable
nor refutable in pure Martin-Löf type theory.\footnote{See the section
\href{https://www.cs.bham.ac.uk/~mhe/HoTT-UF-in-Agda-Lecture-Notes/HoTT-UF-Agda.html\#funextfromua}
{\textit{Function extensionality from univalence}} in~\cite{MHE, MHE:2019}.}

In contrast, our current approach using setoids makes the equality relation
of a given type explicit.  A primary motivation for this choice is to avoid the need for
additional axioms and to make it clearer that the formal proofs in the \agdaalgebras
library are fully \emph{constructive} (as defined in~\cite{nlab:constructive_mathematics})
and confined to \emph{Martin-Löf dependent type theory} (as defined in~\cite{nlab:martin-loef_dependent_type_theory}).
In particular, we make no appeals to classical axioms like Choice or Excluded Middle, nor
do we postulate function extensionality at any point in the present work.\footnote{The \defn{function extensionality
axiom} asserts that two point-wise equal functions are equal. There remain some modules in the \agdaalgebras library that
occasionally postulate this axiom, but we don't make use of the axiom here.}
We are confident that the \agdaalgebras library is now fully constructive and free from any hidden
assumptions or inconsistencies that could be used to fool a type-checker.\footnote{As of 26 Nov 2021, the latest version of \agdaalgebras is 2.0.0; see~\cite{ualib_v2.0.0}.}

%% -----------------------------------------------------------------------------
\subsection{Setoid functions}
\label{setoid-functions}
In addition to the \ar{Setoid} type, much of our code employs the
standard library's \ar{Func} type which represents a function from one
setoid to another and packages such a function with a proof (called \afld{cong}) that
the function respects the underlying setoid equalities. As mentioned above, we renamed
\ar{Func} to the more visually appealing infix long arrow symbol,
\AgdaRecord{\AgdaUnderscore{}⟶\AgdaUnderscore{}}, and  throughout the paper we
refer to inhabitants of this type as ``setoid functions.''

\ifshort\else
An example of a setoid function is the identity function from a setoid to itself.
We define it, along with a binary composition operation for setoid functions,
\AgdaOperator{\AgdaFunction{⟨∘⟩}}, as follows.

\begin{code}%
\>[0]\<%
\\
\>[0]\AgdaFunction{𝑖𝑑}\AgdaSpace{}%
\AgdaSymbol{:}\AgdaSpace{}%
\AgdaSymbol{\{}\AgdaBound{A}\AgdaSpace{}%
\AgdaSymbol{:}\AgdaSpace{}%
\AgdaRecord{Setoid}\AgdaSpace{}%
\AgdaGeneralizable{α}\AgdaSpace{}%
\AgdaGeneralizable{ρᵃ}\AgdaSymbol{\}}\AgdaSpace{}%
\AgdaSymbol{→}\AgdaSpace{}%
\AgdaBound{A}\AgdaSpace{}%
\AgdaOperator{\AgdaRecord{⟶}}\AgdaSpace{}%
\AgdaBound{A}\<%
\\
\>[0]\AgdaFunction{𝑖𝑑}\AgdaSpace{}%
\AgdaSymbol{\{}\AgdaBound{A}\AgdaSymbol{\}}\AgdaSpace{}%
\AgdaSymbol{=}\AgdaSpace{}%
\AgdaKeyword{record}\AgdaSpace{}%
\AgdaSymbol{\{}\AgdaSpace{}%
\AgdaField{f}\AgdaSpace{}%
\AgdaSymbol{=}\AgdaSpace{}%
\AgdaFunction{id}\AgdaSpace{}%
\AgdaSymbol{;}\AgdaSpace{}%
\AgdaField{cong}\AgdaSpace{}%
\AgdaSymbol{=}\AgdaSpace{}%
\AgdaFunction{id}\AgdaSpace{}%
\AgdaSymbol{\}}\<%
\\
\\[\AgdaEmptyExtraSkip]%
\>[0]\AgdaOperator{\AgdaFunction{\AgdaUnderscore{}⟨∘⟩\AgdaUnderscore{}}}\AgdaSpace{}%
\AgdaSymbol{:}%
\>[9]\AgdaSymbol{\{}\AgdaBound{A}\AgdaSpace{}%
\AgdaSymbol{:}\AgdaSpace{}%
\AgdaRecord{Setoid}\AgdaSpace{}%
\AgdaGeneralizable{α}\AgdaSpace{}%
\AgdaGeneralizable{ρᵃ}\AgdaSymbol{\}}\AgdaSpace{}%
\AgdaSymbol{\{}\AgdaBound{B}\AgdaSpace{}%
\AgdaSymbol{:}\AgdaSpace{}%
\AgdaRecord{Setoid}\AgdaSpace{}%
\AgdaGeneralizable{β}\AgdaSpace{}%
\AgdaGeneralizable{ρᵇ}\AgdaSymbol{\}}\AgdaSpace{}%
\AgdaSymbol{\{}\AgdaBound{C}\AgdaSpace{}%
\AgdaSymbol{:}\AgdaSpace{}%
\AgdaRecord{Setoid}\AgdaSpace{}%
\AgdaGeneralizable{γ}\AgdaSpace{}%
\AgdaGeneralizable{ρᶜ}\AgdaSymbol{\}}\<%
\\
\>[0][@{}l@{\AgdaIndent{0}}]%
\>[1]\AgdaSymbol{→}%
\>[9]\AgdaBound{B}\AgdaSpace{}%
\AgdaOperator{\AgdaRecord{⟶}}\AgdaSpace{}%
\AgdaBound{C}%
\>[16]\AgdaSymbol{→}%
\>[19]\AgdaBound{A}\AgdaSpace{}%
\AgdaOperator{\AgdaRecord{⟶}}\AgdaSpace{}%
\AgdaBound{B}%
\>[26]\AgdaSymbol{→}%
\>[29]\AgdaBound{A}\AgdaSpace{}%
\AgdaOperator{\AgdaRecord{⟶}}\AgdaSpace{}%
\AgdaBound{C}\<%
\\
\\[\AgdaEmptyExtraSkip]%
\>[0]\AgdaBound{f}\AgdaSpace{}%
\AgdaOperator{\AgdaFunction{⟨∘⟩}}\AgdaSpace{}%
\AgdaBound{g}\AgdaSpace{}%
\AgdaSymbol{=}\AgdaSpace{}%
\AgdaKeyword{record}%
\>[18]\AgdaSymbol{\{}\AgdaSpace{}%
\AgdaField{f}\AgdaSpace{}%
\AgdaSymbol{=}\AgdaSpace{}%
\AgdaSymbol{(}\AgdaOperator{\AgdaField{\AgdaUnderscore{}⟨\$⟩\AgdaUnderscore{}}}\AgdaSpace{}%
\AgdaBound{f}\AgdaSymbol{)}\AgdaSpace{}%
\AgdaOperator{\AgdaFunction{∘}}\AgdaSpace{}%
\AgdaSymbol{(}\AgdaOperator{\AgdaField{\AgdaUnderscore{}⟨\$⟩\AgdaUnderscore{}}}\AgdaSpace{}%
\AgdaBound{g}\AgdaSymbol{)}\<%
\\
\>[18]\AgdaSymbol{;}\AgdaSpace{}%
\AgdaField{cong}\AgdaSpace{}%
\AgdaSymbol{=}\AgdaSpace{}%
\AgdaSymbol{(}\AgdaField{cong}\AgdaSpace{}%
\AgdaBound{f}\AgdaSymbol{)}\AgdaSpace{}%
\AgdaOperator{\AgdaFunction{∘}}\AgdaSpace{}%
\AgdaSymbol{(}\AgdaField{cong}\AgdaSpace{}%
\AgdaBound{g}\AgdaSymbol{)}\AgdaSpace{}%
\AgdaSymbol{\}}\<%
\end{code}
\fi

\paragraph*{Inverses of setoid functions}
We begin by defining an inductive type that represents the \emph{image} of a function.\footnote{cf.~the \ualmodule{Overture.Func.Inverses} module of the \agdaalgebras library.}

\begin{code}%
\>[0]\<%
\\
\>[0]\AgdaKeyword{module}\AgdaSpace{}%
\AgdaModule{\AgdaUnderscore{}}\AgdaSpace{}%
\AgdaSymbol{\{}\AgdaBound{𝑨}\AgdaSpace{}%
\AgdaSymbol{:}\AgdaSpace{}%
\AgdaRecord{Setoid}\AgdaSpace{}%
\AgdaGeneralizable{α}\AgdaSpace{}%
\AgdaGeneralizable{ρᵃ}\AgdaSymbol{\}\{}\AgdaBound{𝑩}\AgdaSpace{}%
\AgdaSymbol{:}\AgdaSpace{}%
\AgdaRecord{Setoid}\AgdaSpace{}%
\AgdaGeneralizable{β}\AgdaSpace{}%
\AgdaGeneralizable{ρᵇ}\AgdaSymbol{\}}\AgdaSpace{}%
\AgdaKeyword{where}\<%
\\
\>[0][@{}l@{\AgdaIndent{0}}]%
\>[1]\AgdaKeyword{open}\AgdaSpace{}%
\AgdaModule{Setoid}\AgdaSpace{}%
\AgdaBound{𝑩}\AgdaSpace{}%
\AgdaKeyword{using}\AgdaSpace{}%
\AgdaSymbol{(}\AgdaSpace{}%
\AgdaOperator{\AgdaField{\AgdaUnderscore{}≈\AgdaUnderscore{}}}\AgdaSpace{}%
\AgdaSymbol{;}\AgdaSpace{}%
\AgdaFunction{sym}\AgdaSpace{}%
\AgdaSymbol{)}\AgdaSpace{}%
\AgdaKeyword{renaming}\AgdaSpace{}%
\AgdaSymbol{(}\AgdaSpace{}%
\AgdaField{Carrier}\AgdaSpace{}%
\AgdaSymbol{to}\AgdaSpace{}%
\AgdaField{B}\AgdaSpace{}%
\AgdaSymbol{)}\<%
\\
\\[\AgdaEmptyExtraSkip]%
\>[1]\AgdaKeyword{data}\AgdaSpace{}%
\AgdaOperator{\AgdaDatatype{Image\AgdaUnderscore{}∋\AgdaUnderscore{}}}\AgdaSpace{}%
\AgdaSymbol{(}\AgdaBound{f}\AgdaSpace{}%
\AgdaSymbol{:}\AgdaSpace{}%
\AgdaBound{𝑨}\AgdaSpace{}%
\AgdaOperator{\AgdaRecord{⟶}}\AgdaSpace{}%
\AgdaBound{𝑩}\AgdaSymbol{)}\AgdaSpace{}%
\AgdaSymbol{:}\AgdaSpace{}%
\AgdaField{B}\AgdaSpace{}%
\AgdaSymbol{→}\AgdaSpace{}%
\AgdaPrimitive{Type}\AgdaSpace{}%
\AgdaSymbol{(}\AgdaBound{α}\AgdaSpace{}%
\AgdaOperator{\AgdaPrimitive{⊔}}\AgdaSpace{}%
\AgdaBound{β}\AgdaSpace{}%
\AgdaOperator{\AgdaPrimitive{⊔}}\AgdaSpace{}%
\AgdaBound{ρᵇ}\AgdaSymbol{)}\AgdaSpace{}%
\AgdaKeyword{where}\<%
\\
\>[1][@{}l@{\AgdaIndent{0}}]%
\>[2]\AgdaInductiveConstructor{eq}\AgdaSpace{}%
\AgdaSymbol{:}\AgdaSpace{}%
\AgdaSymbol{\{}\AgdaBound{b}\AgdaSpace{}%
\AgdaSymbol{:}\AgdaSpace{}%
\AgdaField{B}\AgdaSymbol{\}}\AgdaSpace{}%
\AgdaSymbol{→}\AgdaSpace{}%
\AgdaSymbol{∀}\AgdaSpace{}%
\AgdaBound{a}\AgdaSpace{}%
\AgdaSymbol{→}\AgdaSpace{}%
\AgdaBound{b}\AgdaSpace{}%
\AgdaOperator{\AgdaField{≈}}\AgdaSpace{}%
\AgdaBound{f}\AgdaSpace{}%
\AgdaOperator{\AgdaField{⟨\$⟩}}\AgdaSpace{}%
\AgdaBound{a}\AgdaSpace{}%
\AgdaSymbol{→}\AgdaSpace{}%
\AgdaOperator{\AgdaDatatype{Image}}\AgdaSpace{}%
\AgdaBound{f}\AgdaSpace{}%
\AgdaOperator{\AgdaDatatype{∋}}\AgdaSpace{}%
\AgdaBound{b}\<%
\\
\>[0]\<%
\end{code}
An inhabitant of \aod{Image} \ab f \aod{∋} \ab b is a dependent pair \AgdaPair{a}{p},
where \AgdaTyped{a}{A} and \ab p~\as :~\ab b \af{≈} \ab f~\ab a is a proof that
\ab f maps \ab a to \ab b.  Since the proof that \ab b
belongs to the image of \ab f is always accompanied by a witness \AgdaTyped{a}{A}, we can
actually \emph{compute} a range-restricted right-inverse of \ab f, as follows.
%. For convenience, we define this
%inverse function and give it the name \af{Inv}.

\begin{code}%
\>[0]\<%
\\
\>[0][@{}l@{\AgdaIndent{1}}]%
\>[1]\AgdaFunction{Inv}\AgdaSpace{}%
\AgdaSymbol{:}\AgdaSpace{}%
\AgdaSymbol{(}\AgdaBound{f}\AgdaSpace{}%
\AgdaSymbol{:}\AgdaSpace{}%
\AgdaBound{𝑨}\AgdaSpace{}%
\AgdaOperator{\AgdaRecord{⟶}}\AgdaSpace{}%
\AgdaBound{𝑩}\AgdaSymbol{)\{}\AgdaBound{b}\AgdaSpace{}%
\AgdaSymbol{:}\AgdaSpace{}%
\AgdaField{B}\AgdaSymbol{\}}\AgdaSpace{}%
\AgdaSymbol{→}\AgdaSpace{}%
\AgdaOperator{\AgdaDatatype{Image}}\AgdaSpace{}%
\AgdaBound{f}\AgdaSpace{}%
\AgdaOperator{\AgdaDatatype{∋}}\AgdaSpace{}%
\AgdaBound{b}\AgdaSpace{}%
\AgdaSymbol{→}\AgdaSpace{}%
\AgdaField{Carrier}\AgdaSpace{}%
\AgdaBound{𝑨}\<%
\\
\>[1]\AgdaFunction{Inv}\AgdaSpace{}%
\AgdaSymbol{\AgdaUnderscore{}}\AgdaSpace{}%
\AgdaSymbol{(}\AgdaInductiveConstructor{eq}\AgdaSpace{}%
\AgdaBound{a}\AgdaSpace{}%
\AgdaSymbol{\AgdaUnderscore{})}\AgdaSpace{}%
\AgdaSymbol{=}\AgdaSpace{}%
\AgdaBound{a}\<%
\\
\>[0]\<%
\end{code}
For each \ab b : \afld{B}, given a pair \AgdaPair{a}{p}~\as :~\aod{Image}~\ab f~\aod{∋}~\ab b witnessing the fact that \ab b belongs to the image of \ab f, the function \af{Inv} simply returns the witness \ab a, which is a preimage of \ab b under \ab f.
Let's formally verify that \af{Inv} \ab f is indeed the (range-restricted) right-inverse of \ab f.

\begin{code}%
\>[0]\<%
\\
\>[0][@{}l@{\AgdaIndent{1}}]%
\>[1]\AgdaFunction{InvIsInverseʳ}\AgdaSpace{}%
\AgdaSymbol{:}\AgdaSpace{}%
\AgdaSymbol{\{}\AgdaBound{f}\AgdaSpace{}%
\AgdaSymbol{:}\AgdaSpace{}%
\AgdaBound{𝑨}\AgdaSpace{}%
\AgdaOperator{\AgdaRecord{⟶}}\AgdaSpace{}%
\AgdaBound{𝑩}\AgdaSymbol{\}\{}\AgdaBound{b}\AgdaSpace{}%
\AgdaSymbol{:}\AgdaSpace{}%
\AgdaField{B}\AgdaSymbol{\}(}\AgdaBound{q}\AgdaSpace{}%
\AgdaSymbol{:}\AgdaSpace{}%
\AgdaOperator{\AgdaDatatype{Image}}\AgdaSpace{}%
\AgdaBound{f}\AgdaSpace{}%
\AgdaOperator{\AgdaDatatype{∋}}\AgdaSpace{}%
\AgdaBound{b}\AgdaSymbol{)}\AgdaSpace{}%
\AgdaSymbol{→}\AgdaSpace{}%
\AgdaBound{f}\AgdaSpace{}%
\AgdaOperator{\AgdaField{⟨\$⟩}}\AgdaSpace{}%
\AgdaSymbol{(}\AgdaFunction{Inv}\AgdaSpace{}%
\AgdaBound{f}\AgdaSpace{}%
\AgdaBound{q}\AgdaSymbol{)}\AgdaSpace{}%
\AgdaOperator{\AgdaField{≈}}\AgdaSpace{}%
\AgdaBound{b}\<%
\\
\>[1]\AgdaFunction{InvIsInverseʳ}\AgdaSpace{}%
\AgdaSymbol{(}\AgdaInductiveConstructor{eq}\AgdaSpace{}%
\AgdaSymbol{\AgdaUnderscore{}}\AgdaSpace{}%
\AgdaBound{p}\AgdaSymbol{)}\AgdaSpace{}%
\AgdaSymbol{=}\AgdaSpace{}%
\AgdaFunction{sym}\AgdaSpace{}%
\AgdaBound{p}\<%
\end{code}

\paragraph*{Injective and surjective setoid functions}
If \ab{f} % : \ab{𝑨} \aor{⟶} \ab{𝑩}
is a setoid function from % \ab{𝑨} =
(\ab A, \af{≈ᴬ}) to
% \ab{𝑩} =
(\ab B, \af{≈ᴮ}), then we call \ab f \defn{injective} provided
\as{∀} (\ab{a₀} \ab{a₁} \as : \ab{A}), \ab{f} \aofld{⟨\$⟩} \ab{a₀} \af{≈ᴮ} \ab{f} \aofld{⟨\$⟩} \ab{a₁}
implies \ab{a₀} \af{≈ᴬ} \ab{a₁}; we call \ab{f} \defn{surjective} provided
\as{∀} (\AgdaTyped{b}{B}), \as{∃}~(\AgdaTyped{a}{A}) such that \ab{f} \aofld{⟨\$⟩} \ab{a} \af{≈ᴮ} \ab{b}.
The \agdastdlib represents injective functions on bare types by the
type \af{Injective}, and uses this to define the \af{IsInjective} type to represent
the property of being an injective setoid function. Similarly, the type \af{IsSurjective}
represents the property of being a surjective setoid function. \af{SurjInv} represents the \emph{right-inverse} of a surjective function.
\ifshort %%% BEGIN SHORT VERSION ONLY
 We omit the straightforward definitions and proofs of these types, but \seeshort for details.
\else    %%% END SHORT VERSION ONLY
         %%% BEGIN LONG VERSION ONLY SECTION
 We reproduce the definitions and prove some of their properties
 inside the next submodule where we first set the stage by declaring two
 setoids \ab{𝑨} and \ab{𝑩}, naming their equality relations, and making some
 definitions from the standard library available.

\begin{code}%
\>[0]\<%
\\
\>[0]\AgdaKeyword{module}\AgdaSpace{}%
\AgdaModule{\AgdaUnderscore{}}\AgdaSpace{}%
\AgdaSymbol{\{}\AgdaBound{𝑨}\AgdaSpace{}%
\AgdaSymbol{:}\AgdaSpace{}%
\AgdaRecord{Setoid}\AgdaSpace{}%
\AgdaGeneralizable{α}\AgdaSpace{}%
\AgdaGeneralizable{ρᵃ}\AgdaSymbol{\}\{}\AgdaBound{𝑩}\AgdaSpace{}%
\AgdaSymbol{:}\AgdaSpace{}%
\AgdaRecord{Setoid}\AgdaSpace{}%
\AgdaGeneralizable{β}\AgdaSpace{}%
\AgdaGeneralizable{ρᵇ}\AgdaSymbol{\}}\AgdaSpace{}%
\AgdaKeyword{where}\<%
\\
\>[0][@{}l@{\AgdaIndent{0}}]%
\>[1]\AgdaKeyword{open}\AgdaSpace{}%
\AgdaModule{Setoid}\AgdaSpace{}%
\AgdaBound{𝑨}\AgdaSpace{}%
\AgdaKeyword{using}\AgdaSpace{}%
\AgdaSymbol{()}\AgdaSpace{}%
\AgdaKeyword{renaming}\AgdaSpace{}%
\AgdaSymbol{(}\AgdaSpace{}%
\AgdaOperator{\AgdaField{\AgdaUnderscore{}≈\AgdaUnderscore{}}}\AgdaSpace{}%
\AgdaSymbol{to}\AgdaSpace{}%
\AgdaOperator{\AgdaField{\AgdaUnderscore{}≈ᴬ\AgdaUnderscore{}}}\AgdaSpace{}%
\AgdaSymbol{)}\<%
\\
\>[1]\AgdaKeyword{open}\AgdaSpace{}%
\AgdaModule{Setoid}\AgdaSpace{}%
\AgdaBound{𝑩}\AgdaSpace{}%
\AgdaKeyword{using}\AgdaSpace{}%
\AgdaSymbol{()}\AgdaSpace{}%
\AgdaKeyword{renaming}\AgdaSpace{}%
\AgdaSymbol{(}\AgdaSpace{}%
\AgdaOperator{\AgdaField{\AgdaUnderscore{}≈\AgdaUnderscore{}}}\AgdaSpace{}%
\AgdaSymbol{to}\AgdaSpace{}%
\AgdaOperator{\AgdaField{\AgdaUnderscore{}≈ᴮ\AgdaUnderscore{}}}\AgdaSpace{}%
\AgdaSymbol{)}\<%
\\
\>[1]\AgdaKeyword{open}\AgdaSpace{}%
\AgdaModule{FD}\AgdaSpace{}%
\AgdaOperator{\AgdaFunction{\AgdaUnderscore{}≈ᴬ\AgdaUnderscore{}}}\AgdaSpace{}%
\AgdaOperator{\AgdaField{\AgdaUnderscore{}≈ᴮ\AgdaUnderscore{}}}\<%
\\
\\[\AgdaEmptyExtraSkip]%
\>[1]\AgdaFunction{IsInjective}\AgdaSpace{}%
\AgdaSymbol{:}\AgdaSpace{}%
\AgdaSymbol{(}\AgdaBound{𝑨}\AgdaSpace{}%
\AgdaOperator{\AgdaRecord{⟶}}\AgdaSpace{}%
\AgdaBound{𝑩}\AgdaSymbol{)}\AgdaSpace{}%
\AgdaSymbol{→}%
\>[26]\AgdaPrimitive{Type}\AgdaSpace{}%
\AgdaSymbol{(}\AgdaBound{α}\AgdaSpace{}%
\AgdaOperator{\AgdaPrimitive{⊔}}\AgdaSpace{}%
\AgdaBound{ρᵃ}\AgdaSpace{}%
\AgdaOperator{\AgdaPrimitive{⊔}}\AgdaSpace{}%
\AgdaBound{ρᵇ}\AgdaSymbol{)}\<%
\\
\>[1]\AgdaFunction{IsInjective}\AgdaSpace{}%
\AgdaBound{f}\AgdaSpace{}%
\AgdaSymbol{=}\AgdaSpace{}%
\AgdaFunction{Injective}\AgdaSpace{}%
\AgdaSymbol{(}\AgdaOperator{\AgdaField{\AgdaUnderscore{}⟨\$⟩\AgdaUnderscore{}}}\AgdaSpace{}%
\AgdaBound{f}\AgdaSymbol{)}\<%
\\
\\[\AgdaEmptyExtraSkip]%
\>[1]\AgdaFunction{IsSurjective}\AgdaSpace{}%
\AgdaSymbol{:}\AgdaSpace{}%
\AgdaSymbol{(}\AgdaBound{𝑨}\AgdaSpace{}%
\AgdaOperator{\AgdaRecord{⟶}}\AgdaSpace{}%
\AgdaBound{𝑩}\AgdaSymbol{)}\AgdaSpace{}%
\AgdaSymbol{→}%
\>[27]\AgdaPrimitive{Type}\AgdaSpace{}%
\AgdaSymbol{(}\AgdaBound{α}\AgdaSpace{}%
\AgdaOperator{\AgdaPrimitive{⊔}}\AgdaSpace{}%
\AgdaBound{β}\AgdaSpace{}%
\AgdaOperator{\AgdaPrimitive{⊔}}\AgdaSpace{}%
\AgdaBound{ρᵇ}\AgdaSymbol{)}\<%
\\
\>[1]\AgdaFunction{IsSurjective}\AgdaSpace{}%
\AgdaBound{F}\AgdaSpace{}%
\AgdaSymbol{=}\AgdaSpace{}%
\AgdaSymbol{∀}\AgdaSpace{}%
\AgdaSymbol{\{}\AgdaBound{y}\AgdaSymbol{\}}\AgdaSpace{}%
\AgdaSymbol{→}\AgdaSpace{}%
\AgdaOperator{\AgdaDatatype{Image}}\AgdaSpace{}%
\AgdaBound{F}\AgdaSpace{}%
\AgdaOperator{\AgdaDatatype{∋}}\AgdaSpace{}%
\AgdaBound{y}\<%
\\
\\[\AgdaEmptyExtraSkip]%
\>[1]\AgdaFunction{SurjInv}\AgdaSpace{}%
\AgdaSymbol{:}\AgdaSpace{}%
\AgdaSymbol{(}\AgdaBound{f}\AgdaSpace{}%
\AgdaSymbol{:}\AgdaSpace{}%
\AgdaBound{𝑨}\AgdaSpace{}%
\AgdaOperator{\AgdaRecord{⟶}}\AgdaSpace{}%
\AgdaBound{𝑩}\AgdaSymbol{)}\AgdaSpace{}%
\AgdaSymbol{→}\AgdaSpace{}%
\AgdaFunction{IsSurjective}\AgdaSpace{}%
\AgdaBound{f}\AgdaSpace{}%
\AgdaSymbol{→}\AgdaSpace{}%
\AgdaField{Carrier}\AgdaSpace{}%
\AgdaBound{𝑩}\AgdaSpace{}%
\AgdaSymbol{→}\AgdaSpace{}%
\AgdaField{Carrier}\AgdaSpace{}%
\AgdaBound{𝑨}\<%
\\
\>[1]\AgdaFunction{SurjInv}\AgdaSpace{}%
\AgdaBound{f}\AgdaSpace{}%
\AgdaBound{fonto}\AgdaSpace{}%
\AgdaBound{b}\AgdaSpace{}%
\AgdaSymbol{=}\AgdaSpace{}%
\AgdaFunction{Inv}\AgdaSpace{}%
\AgdaBound{f}\AgdaSpace{}%
\AgdaSymbol{(}\AgdaBound{fonto}\AgdaSpace{}%
\AgdaSymbol{\{}\AgdaBound{b}\AgdaSymbol{\})}\<%
\\
\>[0]\<%
\end{code}

Proving that the composition of injective setoid functions is again injective
is simply a matter of composing the two assumed witnesses to injectivity.
Proving that surjectivity is preserved under composition is only slightly more involved.

\begin{code}%
\>[0]\<%
\\
\>[0]\AgdaKeyword{module}\AgdaSpace{}%
\AgdaModule{\AgdaUnderscore{}}%
\>[10]\AgdaSymbol{\{}\AgdaBound{𝑨}\AgdaSpace{}%
\AgdaSymbol{:}\AgdaSpace{}%
\AgdaRecord{Setoid}\AgdaSpace{}%
\AgdaGeneralizable{α}\AgdaSpace{}%
\AgdaGeneralizable{ρᵃ}\AgdaSymbol{\}\{}\AgdaBound{𝑩}\AgdaSpace{}%
\AgdaSymbol{:}\AgdaSpace{}%
\AgdaRecord{Setoid}\AgdaSpace{}%
\AgdaGeneralizable{β}\AgdaSpace{}%
\AgdaGeneralizable{ρᵇ}\AgdaSymbol{\}\{}\AgdaBound{𝑪}\AgdaSpace{}%
\AgdaSymbol{:}\AgdaSpace{}%
\AgdaRecord{Setoid}\AgdaSpace{}%
\AgdaGeneralizable{γ}\AgdaSpace{}%
\AgdaGeneralizable{ρᶜ}\AgdaSymbol{\}}\<%
\\
\>[10]\AgdaSymbol{(}\AgdaBound{f}\AgdaSpace{}%
\AgdaSymbol{:}\AgdaSpace{}%
\AgdaBound{𝑨}\AgdaSpace{}%
\AgdaOperator{\AgdaRecord{⟶}}\AgdaSpace{}%
\AgdaBound{𝑩}\AgdaSymbol{)(}\AgdaBound{g}\AgdaSpace{}%
\AgdaSymbol{:}\AgdaSpace{}%
\AgdaBound{𝑩}\AgdaSpace{}%
\AgdaOperator{\AgdaRecord{⟶}}\AgdaSpace{}%
\AgdaBound{𝑪}\AgdaSymbol{)}\AgdaSpace{}%
\AgdaKeyword{where}\<%
\\
\\[\AgdaEmptyExtraSkip]%
\>[0][@{}l@{\AgdaIndent{0}}]%
\>[1]\AgdaFunction{∘-IsInjective}\AgdaSpace{}%
\AgdaSymbol{:}\AgdaSpace{}%
\AgdaFunction{IsInjective}\AgdaSpace{}%
\AgdaBound{f}\AgdaSpace{}%
\AgdaSymbol{→}\AgdaSpace{}%
\AgdaFunction{IsInjective}\AgdaSpace{}%
\AgdaBound{g}\AgdaSpace{}%
\AgdaSymbol{→}\AgdaSpace{}%
\AgdaFunction{IsInjective}\AgdaSpace{}%
\AgdaSymbol{(}\AgdaBound{g}\AgdaSpace{}%
\AgdaOperator{\AgdaFunction{⟨∘⟩}}\AgdaSpace{}%
\AgdaBound{f}\AgdaSymbol{)}\<%
\\
\>[1]\AgdaFunction{∘-IsInjective}\AgdaSpace{}%
\AgdaBound{finj}\AgdaSpace{}%
\AgdaBound{ginj}\AgdaSpace{}%
\AgdaSymbol{=}\AgdaSpace{}%
\AgdaBound{finj}\AgdaSpace{}%
\AgdaOperator{\AgdaFunction{∘}}\AgdaSpace{}%
\AgdaBound{ginj}\<%
\\
\\[\AgdaEmptyExtraSkip]%
\>[1]\AgdaFunction{∘-IsSurjective}\AgdaSpace{}%
\AgdaSymbol{:}\AgdaSpace{}%
\AgdaFunction{IsSurjective}\AgdaSpace{}%
\AgdaBound{f}\AgdaSpace{}%
\AgdaSymbol{→}\AgdaSpace{}%
\AgdaFunction{IsSurjective}\AgdaSpace{}%
\AgdaBound{g}\AgdaSpace{}%
\AgdaSymbol{→}\AgdaSpace{}%
\AgdaFunction{IsSurjective}\AgdaSpace{}%
\AgdaSymbol{(}\AgdaBound{g}\AgdaSpace{}%
\AgdaOperator{\AgdaFunction{⟨∘⟩}}\AgdaSpace{}%
\AgdaBound{f}\AgdaSymbol{)}\<%
\\
\>[1]\AgdaFunction{∘-IsSurjective}\AgdaSpace{}%
\AgdaBound{fonto}\AgdaSpace{}%
\AgdaBound{gonto}\AgdaSpace{}%
\AgdaSymbol{\{}\AgdaBound{y}\AgdaSymbol{\}}\AgdaSpace{}%
\AgdaSymbol{=}\AgdaSpace{}%
\AgdaFunction{Goal}\<%
\\
\>[1][@{}l@{\AgdaIndent{0}}]%
\>[2]\AgdaKeyword{where}\<%
\\
\>[2]\AgdaFunction{mp}\AgdaSpace{}%
\AgdaSymbol{:}\AgdaSpace{}%
\AgdaOperator{\AgdaDatatype{Image}}\AgdaSpace{}%
\AgdaBound{g}\AgdaSpace{}%
\AgdaOperator{\AgdaDatatype{∋}}\AgdaSpace{}%
\AgdaBound{y}\AgdaSpace{}%
\AgdaSymbol{→}\AgdaSpace{}%
\AgdaOperator{\AgdaDatatype{Image}}\AgdaSpace{}%
\AgdaBound{g}\AgdaSpace{}%
\AgdaOperator{\AgdaFunction{⟨∘⟩}}\AgdaSpace{}%
\AgdaBound{f}\AgdaSpace{}%
\AgdaOperator{\AgdaDatatype{∋}}\AgdaSpace{}%
\AgdaBound{y}\<%
\\
\>[2]\AgdaFunction{mp}\AgdaSpace{}%
\AgdaSymbol{(}\AgdaInductiveConstructor{eq}\AgdaSpace{}%
\AgdaBound{c}\AgdaSpace{}%
\AgdaBound{p}\AgdaSymbol{)}\AgdaSpace{}%
\AgdaSymbol{=}\AgdaSpace{}%
\AgdaFunction{η}\AgdaSpace{}%
\AgdaBound{fonto}\<%
\\
\>[2][@{}l@{\AgdaIndent{0}}]%
\>[3]\AgdaKeyword{where}\<%
\\
\>[3]\AgdaKeyword{open}\AgdaSpace{}%
\AgdaModule{Setoid}\AgdaSpace{}%
\AgdaBound{𝑪}\AgdaSpace{}%
\AgdaKeyword{using}\AgdaSpace{}%
\AgdaSymbol{(}\AgdaSpace{}%
\AgdaFunction{trans}\AgdaSpace{}%
\AgdaSymbol{)}\<%
\\
\>[3]\AgdaFunction{η}\AgdaSpace{}%
\AgdaSymbol{:}\AgdaSpace{}%
\AgdaOperator{\AgdaDatatype{Image}}\AgdaSpace{}%
\AgdaBound{f}\AgdaSpace{}%
\AgdaOperator{\AgdaDatatype{∋}}\AgdaSpace{}%
\AgdaBound{c}\AgdaSpace{}%
\AgdaSymbol{→}\AgdaSpace{}%
\AgdaOperator{\AgdaDatatype{Image}}\AgdaSpace{}%
\AgdaBound{g}\AgdaSpace{}%
\AgdaOperator{\AgdaFunction{⟨∘⟩}}\AgdaSpace{}%
\AgdaBound{f}\AgdaSpace{}%
\AgdaOperator{\AgdaDatatype{∋}}\AgdaSpace{}%
\AgdaBound{y}\<%
\\
\>[3]\AgdaFunction{η}\AgdaSpace{}%
\AgdaSymbol{(}\AgdaInductiveConstructor{eq}\AgdaSpace{}%
\AgdaBound{a}\AgdaSpace{}%
\AgdaBound{q}\AgdaSymbol{)}\AgdaSpace{}%
\AgdaSymbol{=}\AgdaSpace{}%
\AgdaInductiveConstructor{eq}\AgdaSpace{}%
\AgdaBound{a}\AgdaSpace{}%
\AgdaSymbol{(}\AgdaFunction{trans}\AgdaSpace{}%
\AgdaBound{p}\AgdaSpace{}%
\AgdaSymbol{(}\AgdaField{cong}\AgdaSpace{}%
\AgdaBound{g}\AgdaSpace{}%
\AgdaBound{q}\AgdaSymbol{))}\<%
\\
\\[\AgdaEmptyExtraSkip]%
\>[2]\AgdaFunction{Goal}\AgdaSpace{}%
\AgdaSymbol{:}\AgdaSpace{}%
\AgdaOperator{\AgdaDatatype{Image}}\AgdaSpace{}%
\AgdaBound{g}\AgdaSpace{}%
\AgdaOperator{\AgdaFunction{⟨∘⟩}}\AgdaSpace{}%
\AgdaBound{f}\AgdaSpace{}%
\AgdaOperator{\AgdaDatatype{∋}}\AgdaSpace{}%
\AgdaBound{y}\<%
\\
\>[2]\AgdaFunction{Goal}\AgdaSpace{}%
\AgdaSymbol{=}\AgdaSpace{}%
\AgdaFunction{mp}\AgdaSpace{}%
\AgdaBound{gonto}\<%
\end{code}
\fi      %%% END LONG VERSION ONLY SECTION

\paragraph*{Kernels of setoid functions}
The \defn{kernel} of a function \ab f~\as :~\ab A~\as{→}~\ab B (where \ab A and \ab B are bare types) is defined
informally by \{\AgdaPair{x}{y} \aod{∈} \ab A \aof{×} \ab A \as : \ab f \ab x \as{=} \ab f \ab y \}.
This can be represented in Agda in a number of ways, but for our purposes it is most
convenient to define the kernel as an inhabitant of a (unary) predicate over the square of
the function's domain, as follows.

\begin{code}%
\>[0]\<%
\\
\>[0]\AgdaFunction{kernel}\AgdaSpace{}%
\AgdaSymbol{:}\AgdaSpace{}%
\AgdaSymbol{\{}\AgdaBound{A}\AgdaSpace{}%
\AgdaSymbol{:}\AgdaSpace{}%
\AgdaPrimitive{Type}\AgdaSpace{}%
\AgdaGeneralizable{α}\AgdaSymbol{\}\{}\AgdaBound{B}\AgdaSpace{}%
\AgdaSymbol{:}\AgdaSpace{}%
\AgdaPrimitive{Type}\AgdaSpace{}%
\AgdaGeneralizable{β}\AgdaSymbol{\}}\AgdaSpace{}%
\AgdaSymbol{→}\AgdaSpace{}%
\AgdaFunction{Rel}\AgdaSpace{}%
\AgdaBound{B}\AgdaSpace{}%
\AgdaGeneralizable{ρ}\AgdaSpace{}%
\AgdaSymbol{→}\AgdaSpace{}%
\AgdaSymbol{(}\AgdaBound{A}\AgdaSpace{}%
\AgdaSymbol{→}\AgdaSpace{}%
\AgdaBound{B}\AgdaSymbol{)}\AgdaSpace{}%
\AgdaSymbol{→}\AgdaSpace{}%
\AgdaFunction{Pred}\AgdaSpace{}%
\AgdaSymbol{(}\AgdaBound{A}\AgdaSpace{}%
\AgdaOperator{\AgdaFunction{×}}\AgdaSpace{}%
\AgdaBound{A}\AgdaSymbol{)}\AgdaSpace{}%
\AgdaGeneralizable{ρ}\<%
\\
\>[0]\AgdaFunction{kernel}\AgdaSpace{}%
\AgdaOperator{\AgdaBound{\AgdaUnderscore{}≈\AgdaUnderscore{}}}\AgdaSpace{}%
\AgdaBound{f}\AgdaSpace{}%
\AgdaSymbol{(}\AgdaBound{x}\AgdaSpace{}%
\AgdaOperator{\AgdaInductiveConstructor{,}}\AgdaSpace{}%
\AgdaBound{y}\AgdaSymbol{)}\AgdaSpace{}%
\AgdaSymbol{=}\AgdaSpace{}%
\AgdaBound{f}\AgdaSpace{}%
\AgdaBound{x}\AgdaSpace{}%
\AgdaOperator{\AgdaBound{≈}}\AgdaSpace{}%
\AgdaBound{f}\AgdaSpace{}%
\AgdaBound{y}\<%
\\
\>[0]\<%
\end{code}
The kernel of a \emph{setoid} function \ab f \as : \ab{𝐴} \aor{⟶} \ab{𝐵} is \{\AgdaPair{x}{y} \as{∈} \ab A \aof{×} \ab A \as : \ab f \aofld{⟨\$⟩} \ab x \af{≈} \ab f \aofld{⟨\$⟩} \ab y\},
where \af{\au{}≈\au} denotes equality in \ab{𝐵}. This can be formalized in Agda as follows.

\ifshort\else
\begin{code}%
\>[0]\<%
\\
\>[0]\AgdaKeyword{module}\AgdaSpace{}%
\AgdaModule{\AgdaUnderscore{}}\AgdaSpace{}%
\AgdaSymbol{\{}\AgdaBound{𝐴}\AgdaSpace{}%
\AgdaSymbol{:}\AgdaSpace{}%
\AgdaRecord{Setoid}\AgdaSpace{}%
\AgdaGeneralizable{α}\AgdaSpace{}%
\AgdaGeneralizable{ρᵃ}\AgdaSymbol{\}\{}\AgdaBound{𝐵}\AgdaSpace{}%
\AgdaSymbol{:}\AgdaSpace{}%
\AgdaRecord{Setoid}\AgdaSpace{}%
\AgdaGeneralizable{β}\AgdaSpace{}%
\AgdaGeneralizable{ρᵇ}\AgdaSymbol{\}}\AgdaSpace{}%
\AgdaKeyword{where}\<%
\\
\>[0][@{}l@{\AgdaIndent{0}}]%
\>[1]\AgdaKeyword{open}\AgdaSpace{}%
\AgdaModule{Setoid}\AgdaSpace{}%
\AgdaBound{𝐴}\AgdaSpace{}%
\AgdaKeyword{using}\AgdaSpace{}%
\AgdaSymbol{()}\AgdaSpace{}%
\AgdaKeyword{renaming}\AgdaSpace{}%
\AgdaSymbol{(}\AgdaSpace{}%
\AgdaField{Carrier}\AgdaSpace{}%
\AgdaSymbol{to}\AgdaSpace{}%
\AgdaField{A}\AgdaSpace{}%
\AgdaSymbol{)}\<%
\end{code}
\fi
\begin{code}%
\>[0]\<%
\\
\>[1]\AgdaFunction{ker}\AgdaSpace{}%
\AgdaSymbol{:}\AgdaSpace{}%
\AgdaSymbol{(}\AgdaBound{𝐴}\AgdaSpace{}%
\AgdaOperator{\AgdaRecord{⟶}}\AgdaSpace{}%
\AgdaBound{𝐵}\AgdaSymbol{)}\AgdaSpace{}%
\AgdaSymbol{→}\AgdaSpace{}%
\AgdaFunction{Pred}\AgdaSpace{}%
\AgdaSymbol{(}\AgdaFunction{A}\AgdaSpace{}%
\AgdaOperator{\AgdaFunction{×}}\AgdaSpace{}%
\AgdaFunction{A}\AgdaSymbol{)}\AgdaSpace{}%
\AgdaBound{ρᵇ}\<%
\\
\>[1]\AgdaFunction{ker}\AgdaSpace{}%
\AgdaBound{g}\AgdaSpace{}%
\AgdaSymbol{(}\AgdaBound{x}\AgdaSpace{}%
\AgdaOperator{\AgdaInductiveConstructor{,}}\AgdaSpace{}%
\AgdaBound{y}\AgdaSymbol{)}\AgdaSpace{}%
\AgdaSymbol{=}\AgdaSpace{}%
\AgdaBound{g}\AgdaSpace{}%
\AgdaOperator{\AgdaField{⟨\$⟩}}\AgdaSpace{}%
\AgdaBound{x}\AgdaSpace{}%
\AgdaOperator{\AgdaFunction{≈}}\AgdaSpace{}%
\AgdaBound{g}\AgdaSpace{}%
\AgdaOperator{\AgdaField{⟨\$⟩}}\AgdaSpace{}%
\AgdaBound{y}\AgdaSpace{}%
\AgdaKeyword{where}\AgdaSpace{}%
\AgdaKeyword{open}\AgdaSpace{}%
\AgdaModule{Setoid}\AgdaSpace{}%
\AgdaBound{𝐵}\AgdaSpace{}%
\AgdaKeyword{using}\AgdaSpace{}%
\AgdaSymbol{(}\AgdaSpace{}%
\AgdaOperator{\AgdaField{\AgdaUnderscore{}≈\AgdaUnderscore{}}}\AgdaSpace{}%
\AgdaSymbol{)}\<%
\end{code}

%% -------------------------------------------------------------------------------------

\section{Types for Basic Universal Algebra}
\label{types-for-basic-universal-algebra}
In this section we develop a working vocabulary and formal types for classical,
single-sorted, set-based universal algebra.
We cover a number of important concepts, but we limit ourselves to those
concepts required in our formal proof of Birkhoff's HSP theorem.
In each case, we give a type-theoretic version of the informal definition,
followed by a formal implementation of the definition in MLTT using the Agda language.

\ifshort\else
This section is organized into the following subsections:
§\ref{signatures} defines a general notion of \emph{signature} of a structure and then defines a type that represent signatures;
§\ref{algebras} does the same for \emph{algebraic structures} and \emph{product algebras};
§\ref{homomorphisms} defines \emph{homomorphisms}, \emph{monomorphisms}, and \emph{epimorphisms}, presents types that codify these concepts and formally verifies some of their basic properties;
§§\ref{subalgebras}--\ref{terms} do the same for \emph{subalgebras} and \emph{terms}, respectively.
\fi

%% -----------------------------------------------------------------------------
\subsection{Signatures}
\label{signatures}

\ifshort
An (algebraic) \defn{signature}
\else
In model theory, the \defn{signature} of a structure is a quadruple \ab{𝑆} = (\ab{C},
\ab{F}, \ab{R}, \ab{ρ}) consisting of three (possibly empty) sets \ab{C}, \ab{F}, and
\ab{R}---called \emph{constant}, \emph{function}, and \emph{relation} symbols,
respectively---along with a function \ab{ρ} : \ab{C} \as{+} \ab{F} \as{+} \ab{R}
\as{→} \ab{N} that assigns an \emph{arity} to each symbol. Often, but not always, \ab{N}
is taken to be the set of natural numbers.

As our focus here is universal algebra, we consider the restricted notion of an
\emph{algebraic signature}, that is, a signature for ``purely algebraic'' structures. Such
a signature
\fi
is a pair \ab{𝑆} = \AgdaPair{F}{ρ} where \ab{F} is a collection of
\defn{operation symbols} and \ab{ρ} : \ab{F} \as{→} \ab{N} is an \defn{arity function}
which maps each operation symbol to its arity. Here, \ab{N} denotes the \emph{arity type}.
Heuristically, the arity \ab{ρ} \ab{f} of an operation symbol \ab{f} \as{∈} \ab{F} may be
thought of as the number of arguments that \ab{f} takes as ``input.''

The \agdaalgebras library represents an algebraic signature as an
inhabitant of the following dependent pair type:

\begin{center}

\AgdaFunction{Signature}\AgdaSpace{}%
\AgdaSymbol{:}\AgdaSpace{}%
\AgdaSymbol{(}\AgdaBound{𝓞}\AgdaSpace{}%
\AgdaBound{𝓥}\AgdaSpace{}%
\AgdaSymbol{:}\AgdaSpace{}%
\AgdaPostulate{Level}\AgdaSymbol{)}\AgdaSpace{}%
\AgdaSymbol{→}\AgdaSpace{}%
\AgdaPrimitive{Type}\AgdaSpace{}%
\AgdaSymbol{(}\AgdaPrimitive{lsuc}\AgdaSpace{}%
\AgdaSymbol{(}\AgdaBound{𝓞}\AgdaSpace{}%
\AgdaOperator{\AgdaPrimitive{⊔}}\AgdaSpace{}%
\AgdaBound{𝓥}\AgdaSymbol{))}\\[4pt]
\AgdaFunction{Signature}\AgdaSpace{}%
\AgdaBound{𝓞}\AgdaSpace{}%
\AgdaBound{𝓥}\AgdaSpace{}%
\AgdaSymbol{=}\AgdaSpace{}%
\AgdaFunction{Σ[}\AgdaSpace{}%
\AgdaBound{F}\AgdaSpace{}%
\AgdaFunction{∈}\AgdaSpace{}%
\AgdaPrimitive{Type}\AgdaSpace{}%
\AgdaBound{𝓞}\AgdaSpace{}%
\AgdaFunction{]}\AgdaSpace{}%
\AgdaSymbol{(}\AgdaBound{F}\AgdaSpace{}%
\AgdaSymbol{→}\AgdaSpace{}%
\AgdaPrimitive{Type}\AgdaSpace{}%
\AgdaBound{𝓥}\AgdaSymbol{)}

\end{center}

Using special syntax for the first and second
projections---\AgdaOperator{\AgdaFunction{∣\AgdaUnderscore{}∣}} and
\AgdaOperator{\AgdaFunction{∥\AgdaUnderscore{}∥}} (resp.)---if
\ab{𝑆} \as{:} \af{Signature} \ab{𝓞} \ab{𝓥} is a signature, then
\aof{∣} \ab{𝑆} \aof{∣} denotes the set of operation symbols and \aof{∥} \ab{𝑆} \aof{∥} denotes the arity function.
Thus, if \ab{f} \as{:} \aof{∣} \ab{𝑆} \aof{∣} is an operation symbol in the
signature \ab{𝑆}, then \aof{∥} \ab{𝑆} \aof{∥} \ab{f} is the arity of \ab{f}.

We need to augment the ordinary \af{Signature} type so that it supports algebras over
setoid domains.
\ifshort\else
To do so---following Andreas Abel's lead (cf.~\cite{Abel:2021})---we
define an operator that translates an ordinary signature into a \defn{setoid signature},
that is, a signature over a setoid domain.
\fi
This raises a minor technical issue concerning
the dependent types involved in the definition.
\ifshort\else
Some readers might find the resolution of
this issue instructive, so let's discuss it briefly.
\fi
If we are given two operations \ab{f} and \ab{g}, a tuple \ab{u} \as{:} \aof{∥} \ab{𝑆} \aof{∥} \ab{f} \as{→}
\ab{A} of arguments for \ab{f}, and a tuple \ab{v} \as{:} \aof{∥} \ab{𝑆}
\aof{∥} \ab{g} \as{→} \ab{A} of arguments for \ab{g}, and if we know that \ab f is identically equal to
\ab{g}---that is, \ab{f} \aod{≡} \ab{g} (intensionally)---then we should be able to
check whether \ab u and \ab v are pointwise equal.  Technically, though, \ab{u} and
\ab{v} inhabit different types, so, in order to compare them, we must convince Agda
that \ab u and \ab v inhabit the same type. Of course, this requires an appeal to the
hypothesis \ab f \aod{≡} \ab g, as we see in the definition of \af{EqArgs} below (adapted
from Andreas Abel's development~\cite{Abel:2021}), which neatly resolves this minor
technicality.

\begin{code}%
\>[0]\<%
\\
\>[0]\AgdaFunction{EqArgs}\AgdaSpace{}%
\AgdaSymbol{:}%
\>[10]\AgdaSymbol{\{}\AgdaBound{𝑆}\AgdaSpace{}%
\AgdaSymbol{:}\AgdaSpace{}%
\AgdaFunction{Signature}\AgdaSpace{}%
\AgdaBound{𝓞}\AgdaSpace{}%
\AgdaBound{𝓥}\AgdaSymbol{\}\{}\AgdaBound{ξ}\AgdaSpace{}%
\AgdaSymbol{:}\AgdaSpace{}%
\AgdaRecord{Setoid}\AgdaSpace{}%
\AgdaGeneralizable{α}\AgdaSpace{}%
\AgdaGeneralizable{ρᵃ}\AgdaSymbol{\}}\<%
\\
\>[0][@{}l@{\AgdaIndent{0}}]%
\>[1]\AgdaSymbol{→}%
\>[10]\AgdaSymbol{∀}\AgdaSpace{}%
\AgdaSymbol{\{}\AgdaBound{f}\AgdaSpace{}%
\AgdaBound{g}\AgdaSymbol{\}}\AgdaSpace{}%
\AgdaSymbol{→}\AgdaSpace{}%
\AgdaBound{f}\AgdaSpace{}%
\AgdaOperator{\AgdaDatatype{≡}}\AgdaSpace{}%
\AgdaBound{g}\AgdaSpace{}%
\AgdaSymbol{→}\AgdaSpace{}%
\AgdaSymbol{(}\AgdaOperator{\AgdaFunction{∥}}\AgdaSpace{}%
\AgdaBound{𝑆}\AgdaSpace{}%
\AgdaOperator{\AgdaFunction{∥}}\AgdaSpace{}%
\AgdaBound{f}\AgdaSpace{}%
\AgdaSymbol{→}\AgdaSpace{}%
\AgdaField{Carrier}\AgdaSpace{}%
\AgdaBound{ξ}\AgdaSymbol{)}\AgdaSpace{}%
\AgdaSymbol{→}\AgdaSpace{}%
\AgdaSymbol{(}\AgdaOperator{\AgdaFunction{∥}}\AgdaSpace{}%
\AgdaBound{𝑆}\AgdaSpace{}%
\AgdaOperator{\AgdaFunction{∥}}\AgdaSpace{}%
\AgdaBound{g}\AgdaSpace{}%
\AgdaSymbol{→}\AgdaSpace{}%
\AgdaField{Carrier}\AgdaSpace{}%
\AgdaBound{ξ}\AgdaSymbol{)}\AgdaSpace{}%
\AgdaSymbol{→}\AgdaSpace{}%
\AgdaPrimitive{Type}\AgdaSpace{}%
\AgdaSymbol{(}\AgdaBound{𝓥}\AgdaSpace{}%
\AgdaOperator{\AgdaPrimitive{⊔}}\AgdaSpace{}%
\AgdaGeneralizable{ρᵃ}\AgdaSymbol{)}\<%
\\
\\[\AgdaEmptyExtraSkip]%
\>[0]\AgdaFunction{EqArgs}\AgdaSpace{}%
\AgdaSymbol{\{}\AgdaArgument{ξ}\AgdaSpace{}%
\AgdaSymbol{=}\AgdaSpace{}%
\AgdaBound{ξ}\AgdaSymbol{\}}\AgdaSpace{}%
\AgdaInductiveConstructor{≡.refl}\AgdaSpace{}%
\AgdaBound{u}\AgdaSpace{}%
\AgdaBound{v}\AgdaSpace{}%
\AgdaSymbol{=}\AgdaSpace{}%
\AgdaSymbol{∀}\AgdaSpace{}%
\AgdaBound{i}\AgdaSpace{}%
\AgdaSymbol{→}\AgdaSpace{}%
\AgdaBound{u}\AgdaSpace{}%
\AgdaBound{i}\AgdaSpace{}%
\AgdaOperator{\AgdaFunction{≈}}\AgdaSpace{}%
\AgdaBound{v}\AgdaSpace{}%
\AgdaBound{i}\AgdaSpace{}%
\AgdaKeyword{where}\AgdaSpace{}%
\AgdaKeyword{open}\AgdaSpace{}%
\AgdaModule{Setoid}\AgdaSpace{}%
\AgdaBound{ξ}\AgdaSpace{}%
\AgdaKeyword{using}\AgdaSpace{}%
\AgdaSymbol{(}\AgdaSpace{}%
\AgdaOperator{\AgdaField{\AgdaUnderscore{}≈\AgdaUnderscore{}}}\AgdaSpace{}%
\AgdaSymbol{)}\<%
\\
\>[0]\<%
\end{code}

Finally, we are ready to define an operator which
translates an ordinary (algebraic) signature into a signature of algebras over setoids.
\ifshort\else
We denote this operator by \aof{⟨\AgdaUnderscore{}⟩} and define it as follows.
\fi

\begin{code}%
\>[0]\<%
\\
\>[0]\AgdaOperator{\AgdaFunction{⟨\AgdaUnderscore{}⟩}}\AgdaSpace{}%
\AgdaSymbol{:}\AgdaSpace{}%
\AgdaFunction{Signature}\AgdaSpace{}%
\AgdaBound{𝓞}\AgdaSpace{}%
\AgdaBound{𝓥}\AgdaSpace{}%
\AgdaSymbol{→}\AgdaSpace{}%
\AgdaRecord{Setoid}\AgdaSpace{}%
\AgdaGeneralizable{α}\AgdaSpace{}%
\AgdaGeneralizable{ρᵃ}\AgdaSpace{}%
\AgdaSymbol{→}\AgdaSpace{}%
\AgdaRecord{Setoid}\AgdaSpace{}%
\AgdaSymbol{\AgdaUnderscore{}}\AgdaSpace{}%
\AgdaSymbol{\AgdaUnderscore{}}\<%
\\
\\[\AgdaEmptyExtraSkip]%
\>[0]\AgdaField{Carrier}%
\>[9]\AgdaSymbol{(}\AgdaOperator{\AgdaFunction{⟨}}\AgdaSpace{}%
\AgdaBound{𝑆}\AgdaSpace{}%
\AgdaOperator{\AgdaFunction{⟩}}\AgdaSpace{}%
\AgdaBound{ξ}\AgdaSymbol{)}%
\>[34]\AgdaSymbol{=}\AgdaSpace{}%
\AgdaFunction{Σ[}\AgdaSpace{}%
\AgdaBound{f}\AgdaSpace{}%
\AgdaFunction{∈}\AgdaSpace{}%
\AgdaOperator{\AgdaFunction{∣}}\AgdaSpace{}%
\AgdaBound{𝑆}\AgdaSpace{}%
\AgdaOperator{\AgdaFunction{∣}}\AgdaSpace{}%
\AgdaFunction{]}\AgdaSpace{}%
\AgdaSymbol{(}\AgdaOperator{\AgdaFunction{∥}}\AgdaSpace{}%
\AgdaBound{𝑆}\AgdaSpace{}%
\AgdaOperator{\AgdaFunction{∥}}\AgdaSpace{}%
\AgdaBound{f}\AgdaSpace{}%
\AgdaSymbol{→}\AgdaSpace{}%
\AgdaBound{ξ}\AgdaSpace{}%
\AgdaSymbol{.}\AgdaField{Carrier}\AgdaSymbol{)}\<%
\\
\>[0]\AgdaOperator{\AgdaField{\AgdaUnderscore{}≈ˢ\AgdaUnderscore{}}}%
\>[9]\AgdaSymbol{(}\AgdaOperator{\AgdaFunction{⟨}}\AgdaSpace{}%
\AgdaBound{𝑆}\AgdaSpace{}%
\AgdaOperator{\AgdaFunction{⟩}}\AgdaSpace{}%
\AgdaBound{ξ}\AgdaSymbol{)(}\AgdaBound{f}\AgdaSpace{}%
\AgdaOperator{\AgdaInductiveConstructor{,}}\AgdaSpace{}%
\AgdaBound{u}\AgdaSymbol{)(}\AgdaBound{g}\AgdaSpace{}%
\AgdaOperator{\AgdaInductiveConstructor{,}}\AgdaSpace{}%
\AgdaBound{v}\AgdaSymbol{)}%
\>[34]\AgdaSymbol{=}\AgdaSpace{}%
\AgdaFunction{Σ[}\AgdaSpace{}%
\AgdaBound{eqv}\AgdaSpace{}%
\AgdaFunction{∈}\AgdaSpace{}%
\AgdaBound{f}\AgdaSpace{}%
\AgdaOperator{\AgdaDatatype{≡}}\AgdaSpace{}%
\AgdaBound{g}\AgdaSpace{}%
\AgdaFunction{]}\AgdaSpace{}%
\AgdaFunction{EqArgs}\AgdaSymbol{\{}\AgdaArgument{ξ}\AgdaSpace{}%
\AgdaSymbol{=}\AgdaSpace{}%
\AgdaBound{ξ}\AgdaSymbol{\}}\AgdaSpace{}%
\AgdaBound{eqv}\AgdaSpace{}%
\AgdaBound{u}\AgdaSpace{}%
\AgdaBound{v}\<%
\\
\\[\AgdaEmptyExtraSkip]%
\>[0]\AgdaField{reflᵉ}%
\>[8]\AgdaSymbol{(}\AgdaField{isEquivalence}\AgdaSpace{}%
\AgdaSymbol{(}\AgdaOperator{\AgdaFunction{⟨}}\AgdaSpace{}%
\AgdaBound{𝑆}\AgdaSpace{}%
\AgdaOperator{\AgdaFunction{⟩}}\AgdaSpace{}%
\AgdaBound{ξ}\AgdaSymbol{))}%
\>[60]\AgdaSymbol{=}\AgdaSpace{}%
\AgdaInductiveConstructor{≡.refl}\AgdaSpace{}%
\AgdaOperator{\AgdaInductiveConstructor{,}}\AgdaSpace{}%
\AgdaSymbol{λ}\AgdaSpace{}%
\AgdaBound{i}\AgdaSpace{}%
\AgdaSymbol{→}\AgdaSpace{}%
\AgdaFunction{reflˢ}%
\>[85]\AgdaBound{ξ}\<%
\\
\>[0]\AgdaField{symᵉ}%
\>[8]\AgdaSymbol{(}\AgdaField{isEquivalence}\AgdaSpace{}%
\AgdaSymbol{(}\AgdaOperator{\AgdaFunction{⟨}}\AgdaSpace{}%
\AgdaBound{𝑆}\AgdaSpace{}%
\AgdaOperator{\AgdaFunction{⟩}}\AgdaSpace{}%
\AgdaBound{ξ}\AgdaSymbol{))}\AgdaSpace{}%
\AgdaSymbol{(}\AgdaInductiveConstructor{≡.refl}\AgdaSpace{}%
\AgdaOperator{\AgdaInductiveConstructor{,}}\AgdaSpace{}%
\AgdaBound{g}\AgdaSymbol{)}%
\>[60]\AgdaSymbol{=}\AgdaSpace{}%
\AgdaInductiveConstructor{≡.refl}\AgdaSpace{}%
\AgdaOperator{\AgdaInductiveConstructor{,}}\AgdaSpace{}%
\AgdaSymbol{λ}\AgdaSpace{}%
\AgdaBound{i}\AgdaSpace{}%
\AgdaSymbol{→}\AgdaSpace{}%
\AgdaFunction{symˢ}%
\>[85]\AgdaBound{ξ}\AgdaSpace{}%
\AgdaSymbol{(}\AgdaBound{g}\AgdaSpace{}%
\AgdaBound{i}\AgdaSymbol{)}\<%
\\
\>[0]\AgdaField{transᵉ}%
\>[8]\AgdaSymbol{(}\AgdaField{isEquivalence}\AgdaSpace{}%
\AgdaSymbol{(}\AgdaOperator{\AgdaFunction{⟨}}\AgdaSpace{}%
\AgdaBound{𝑆}\AgdaSpace{}%
\AgdaOperator{\AgdaFunction{⟩}}\AgdaSpace{}%
\AgdaBound{ξ}\AgdaSymbol{))}\AgdaSpace{}%
\AgdaSymbol{(}\AgdaInductiveConstructor{≡.refl}\AgdaSpace{}%
\AgdaOperator{\AgdaInductiveConstructor{,}}\AgdaSpace{}%
\AgdaBound{g}\AgdaSymbol{)(}\AgdaInductiveConstructor{≡.refl}\AgdaSpace{}%
\AgdaOperator{\AgdaInductiveConstructor{,}}\AgdaSpace{}%
\AgdaBound{h}\AgdaSymbol{)}%
\>[60]\AgdaSymbol{=}\AgdaSpace{}%
\AgdaInductiveConstructor{≡.refl}\AgdaSpace{}%
\AgdaOperator{\AgdaInductiveConstructor{,}}\AgdaSpace{}%
\AgdaSymbol{λ}\AgdaSpace{}%
\AgdaBound{i}\AgdaSpace{}%
\AgdaSymbol{→}\AgdaSpace{}%
\AgdaFunction{transˢ}%
\>[85]\AgdaBound{ξ}\AgdaSpace{}%
\AgdaSymbol{(}\AgdaBound{g}\AgdaSpace{}%
\AgdaBound{i}\AgdaSymbol{)}\AgdaSpace{}%
\AgdaSymbol{(}\AgdaBound{h}\AgdaSpace{}%
\AgdaBound{i}\AgdaSymbol{)}\<%
\end{code}

%% -----------------------------------------------------------------------------
\subsection{Algebras}\label{algebras}
Informally, an \defn{algebraic structure in the signature} \ab{𝑆} = (\ab{F}, \ab{ρ}), or
\ab{𝑆}-\defn{algebra}, is denoted by \ab{𝑨} = (\ab{A}, \ab{Fᴬ}) and consists of
\begin{itemize}
\item a \emph{nonempty} set (or type) \ab A, called the \defn{domain} (or \defn{carrier} or
\defn{universe}) of the algebra;
\item a collection \ab{Fᴬ} :=
  \{ \ab{fᴬ} \as{∣} \ab f \as{∈} \ab F, \ab{fᴬ} \as :
    (\ab{ρ} \ab f \as{→} \ab A) \as{→} \ab A \} of \defn{operations} on \ab{A};
\item a (potentially empty) collection of \defn{identities} satisfied by elements and
operations of \ab{𝑨}.
\end{itemize}
The \agdaalgebras library represents algebras as inhabitants of a record type with two
fields:\footnote{We postpone introducing identities until~§\ref{equational-logic}.}
\begin{itemize}
\item \afld{Domain}, representing the domain of the algebra;
\item \afld{Interp}, representing the \emph{interpretation} in the algebra of each
operation symbol in \ab{𝑆}.
\end{itemize}
The \afld{Domain} is a setoid whose \afld{Carrier} denotes the domain of the algebra and
whose equivalence relation denotes equality of elements of the domain.

Here is the definition of the \ar{Algebra} type followed by an explanation of how the
standard library's \ar{Func} type is used to represent the interpretation of operation
symbols in an algebra.

\begin{code}%
\>[0]\<%
\\
\>[0]\AgdaKeyword{record}\AgdaSpace{}%
\AgdaRecord{Algebra}\AgdaSpace{}%
\AgdaBound{α}\AgdaSpace{}%
\AgdaBound{ρ}\AgdaSpace{}%
\AgdaSymbol{:}\AgdaSpace{}%
\AgdaPrimitive{Type}\AgdaSpace{}%
\AgdaSymbol{(}\AgdaBound{𝓞}\AgdaSpace{}%
\AgdaOperator{\AgdaPrimitive{⊔}}\AgdaSpace{}%
\AgdaBound{𝓥}\AgdaSpace{}%
\AgdaOperator{\AgdaPrimitive{⊔}}\AgdaSpace{}%
\AgdaPrimitive{lsuc}\AgdaSpace{}%
\AgdaSymbol{(}\AgdaBound{α}\AgdaSpace{}%
\AgdaOperator{\AgdaPrimitive{⊔}}\AgdaSpace{}%
\AgdaBound{ρ}\AgdaSymbol{))}\AgdaSpace{}%
\AgdaKeyword{where}\<%
\\
\>[0][@{}l@{\AgdaIndent{0}}]%
\>[1]\AgdaKeyword{field}%
\>[8]\AgdaField{Domain}%
\>[16]\AgdaSymbol{:}\AgdaSpace{}%
\AgdaRecord{Setoid}\AgdaSpace{}%
\AgdaBound{α}\AgdaSpace{}%
\AgdaBound{ρ}\<%
\\
\>[8]\AgdaField{Interp}%
\>[16]\AgdaSymbol{:}\AgdaSpace{}%
\AgdaOperator{\AgdaFunction{⟨}}\AgdaSpace{}%
\AgdaBound{𝑆}\AgdaSpace{}%
\AgdaOperator{\AgdaFunction{⟩}}\AgdaSpace{}%
\AgdaField{Domain}\AgdaSpace{}%
\AgdaOperator{\AgdaRecord{⟶}}\AgdaSpace{}%
\AgdaField{Domain}\<%
\\
\>[0]\<%
\end{code}
Recall, we renamed Agda's \ar{Func} type, preferring instead the long-arrow symbol
\AgdaRecord{⟶}, so the \afld{Interp} field has type \ar{Func} (\aof{⟨} \ab{𝑆} \aof{⟩}
\afld{Domain}) \afld{Domain}, a record type with two fields:
\begin{itemize}
\item a function  \ab{f} \as : \afld{Carrier} (\aof{⟨} \ab{𝑆} \aof{⟩} \afld{Domain})
  \as{→} \afld{Carrier} \afld{Domain} representing the operation;
\item a proof \af{cong} \as : \ab f \aof{Preserves \au{}≈₁\au{} \aor{⟶} \au{}≈₂\au{}} that the
operation preserves the relevant setoid equalities.
\end{itemize}
Thus, for each operation symbol in the signature \ab{𝑆}, we have a setoid function
\ab f---with domain a power of \afld{Domain} and codomain \afld{Domain}---along with
a proof that this function respects the setoid equalities.  The latter means that the
operation \ab{f} is accompanied by a proof of the following: ∀ \ab u \ab v in
\afld{Carrier} (\aof{⟨} \ab{𝑆} \aof{⟩} \afld{Domain}), if \ab u \af{≈₁} \ab v, then \ab{f}
\aofld{⟨\$⟩} \ab{u} \af{≈₂} \ab{f} \aofld{⟨\$⟩} \ab{v}.

In the \agdaalgebras library is defined some syntactic sugar that helps to make our
formalizations easier to read and comprehend.
The following are three examples of such syntax that we use below: if \ab{𝑨} is an algebra, then
\begin{itemize}
\item \aof{𝔻[ \ab{𝑨} ]} denotes the setoid \afld{Domain} \ab{𝑨},
\item \aof{𝕌[ \ab{𝑨} ]} is the underlying carrier of the algebra \ab{𝑨}, and
\item \ab f \aof{̂} \ab{𝑨} denotes the interpretation in the algebra \ab{𝑨} of the operation symbol \ab f.
\end{itemize}
\ifshort %%% BEGIN SHORT VERSION ONLY
 We omit the straightforward formal definitions of these types.
\else    %%% END SHORT VERSION ONLY
         %%% BEGIN LONG VERSION ONLY SECTION
\begin{code}%
\>[0]\AgdaKeyword{open}\AgdaSpace{}%
\AgdaModule{Algebra}\<%
\\
\>[0]\AgdaOperator{\AgdaFunction{𝔻[\AgdaUnderscore{}]}}\AgdaSpace{}%
\AgdaSymbol{:}\AgdaSpace{}%
\AgdaRecord{Algebra}\AgdaSpace{}%
\AgdaGeneralizable{α}\AgdaSpace{}%
\AgdaGeneralizable{ρᵃ}\AgdaSpace{}%
\AgdaSymbol{→}%
\>[23]\AgdaRecord{Setoid}\AgdaSpace{}%
\AgdaGeneralizable{α}\AgdaSpace{}%
\AgdaGeneralizable{ρᵃ}\<%
\\
\>[0]\AgdaOperator{\AgdaFunction{𝔻[}}\AgdaSpace{}%
\AgdaBound{𝑨}\AgdaSpace{}%
\AgdaOperator{\AgdaFunction{]}}\AgdaSpace{}%
\AgdaSymbol{=}\AgdaSpace{}%
\AgdaField{Domain}\AgdaSpace{}%
\AgdaBound{𝑨}\<%
\\
\>[0]\AgdaOperator{\AgdaFunction{𝕌[\AgdaUnderscore{}]}}\AgdaSpace{}%
\AgdaSymbol{:}\AgdaSpace{}%
\AgdaRecord{Algebra}\AgdaSpace{}%
\AgdaGeneralizable{α}\AgdaSpace{}%
\AgdaGeneralizable{ρᵃ}\AgdaSpace{}%
\AgdaSymbol{→}%
\>[23]\AgdaPrimitive{Type}\AgdaSpace{}%
\AgdaGeneralizable{α}\<%
\\
\>[0]\AgdaOperator{\AgdaFunction{𝕌[}}\AgdaSpace{}%
\AgdaBound{𝑨}\AgdaSpace{}%
\AgdaOperator{\AgdaFunction{]}}\AgdaSpace{}%
\AgdaSymbol{=}\AgdaSpace{}%
\AgdaField{Carrier}\AgdaSpace{}%
\AgdaSymbol{(}\AgdaField{Domain}\AgdaSpace{}%
\AgdaBound{𝑨}\AgdaSymbol{)}\<%
\\
\>[0]\AgdaOperator{\AgdaFunction{\AgdaUnderscore{}̂\AgdaUnderscore{}}}\AgdaSpace{}%
\AgdaSymbol{:}\AgdaSpace{}%
\AgdaSymbol{(}\AgdaBound{f}\AgdaSpace{}%
\AgdaSymbol{:}\AgdaSpace{}%
\AgdaOperator{\AgdaFunction{∣}}\AgdaSpace{}%
\AgdaBound{𝑆}\AgdaSpace{}%
\AgdaOperator{\AgdaFunction{∣}}\AgdaSymbol{)(}\AgdaBound{𝑨}\AgdaSpace{}%
\AgdaSymbol{:}\AgdaSpace{}%
\AgdaRecord{Algebra}\AgdaSpace{}%
\AgdaGeneralizable{α}\AgdaSpace{}%
\AgdaGeneralizable{ρᵃ}\AgdaSymbol{)}\AgdaSpace{}%
\AgdaSymbol{→}\AgdaSpace{}%
\AgdaSymbol{(}\AgdaOperator{\AgdaFunction{∥}}\AgdaSpace{}%
\AgdaBound{𝑆}\AgdaSpace{}%
\AgdaOperator{\AgdaFunction{∥}}\AgdaSpace{}%
\AgdaBound{f}%
\>[48]\AgdaSymbol{→}%
\>[51]\AgdaOperator{\AgdaFunction{𝕌[}}\AgdaSpace{}%
\AgdaBound{𝑨}\AgdaSpace{}%
\AgdaOperator{\AgdaFunction{]}}\AgdaSymbol{)}\AgdaSpace{}%
\AgdaSymbol{→}\AgdaSpace{}%
\AgdaOperator{\AgdaFunction{𝕌[}}\AgdaSpace{}%
\AgdaBound{𝑨}\AgdaSpace{}%
\AgdaOperator{\AgdaFunction{]}}\<%
\\
\>[0]\AgdaBound{f}\AgdaSpace{}%
\AgdaOperator{\AgdaFunction{̂}}\AgdaSpace{}%
\AgdaBound{𝑨}\AgdaSpace{}%
\AgdaSymbol{=}\AgdaSpace{}%
\AgdaSymbol{λ}\AgdaSpace{}%
\AgdaBound{a}\AgdaSpace{}%
\AgdaSymbol{→}\AgdaSpace{}%
\AgdaSymbol{(}\AgdaField{Interp}\AgdaSpace{}%
\AgdaBound{𝑨}\AgdaSymbol{)}\AgdaSpace{}%
\AgdaOperator{\AgdaField{⟨\$⟩}}\AgdaSpace{}%
\AgdaSymbol{(}\AgdaBound{f}\AgdaSpace{}%
\AgdaOperator{\AgdaInductiveConstructor{,}}\AgdaSpace{}%
\AgdaBound{a}\AgdaSymbol{)}\<%
\end{code}
\fi

%% -----------------------------------------------------------------------------
\paragraph*{Universe levels of algebra types}
The hierarchy of type universes in Agda is structured as follows:
\ap{Type} \ab{ℓ} : \ap{Type} (\ap{lsuc} \ab{ℓ}), \ap{Type} (\ap{lsuc} \ab{ℓ}) : \ap{Type}
(\ap{lsuc} (\ap{lsuc} \ab{ℓ})), …. This means that \ap{Type} \ab{ℓ} has type \ap{Type}
(\ap{lsuc} \ab{ℓ}), etc.  However, this does \emph{not} imply that \ap{Type} \ab{ℓ} :
\ap{Type} (\ap{lsuc} (\ap{lsuc} \ab{ℓ})). In other words, Agda's universe hierarchy is
\emph{noncumulative}.
\ifshort
An
\else
This can be advantageous as it becomes possible to treat universe
levels more generally and precisely. On the other hand, an
\fi
unfortunate side-effect of this noncumulativity is that it can sometimes seem unreasonably
difficult to convince Agda that a program or proof is correct.
\ifshort\else
This aspect of the language was one of the few stumbling
blocks we encountered while learning how to use Agda for formalizing universal algebra in
type theory. Although some may consider this to be one of the least interesting and most
technical aspects of this paper, others might find the presentation more helpful if we
resist the urge to gloss over these technicalities.
\fi
Therefore, it seems worthwhile to explain how we make use
of the general universe lifting and lowering functions, available in the \agdastdlib, to
develop domain-specific tools for dealing with Agda's noncumulative universe hierarchy.

\ifshort\else
Let us be more concrete about what is at issue by considering a typical example. Agda
frequently encounters problems during the type-checking process and responds by printing a
message like the following.
{\color{red}{\small
\begin{verbatim}
  HSP.lagda:498,20-23
  α != 𝓞 ⊔ 𝓥 ⊔ (lsuc α) when checking that... has type...
\end{verbatim}}}
\noindent Here Agda informs us that it encountered universe level \ab{α} on line 498 of
the HSP module, where it was expecting level \ab{𝓞}~\aop{⊔}~\ab{𝓥}~\aop{⊔}~(\ap{lsuc}
\ab{α}). In this case, we tried to use an algebra inhabiting the type \ar{Algebra}
\ab{α} \ab{ρᵃ} whereas Agda expected an inhabitant of the type \ar{Algebra} (\ab{𝓞}
\aop{⊔} \ab{𝓥} \aop{⊔} (\ap{lsuc} \ab{α})) \ab{ρᵃ}.
\fi
To resolve such problems, we use the \AgdaRecord{Lift} record type of the \agdastdlib,
which takes a type inhabiting a particular universe and embeds it into a higher universe.
Specializing the \ar{Lift} type to our domain of interest, the \agdaalgebras library
defines a function called \af{Lift-Alg}%
\ifshort
, whose interface is the following.
\vskip-2mm
\else
.

\begin{code}%
\>[0]\<%
\\
\>[0]\AgdaKeyword{module}\AgdaSpace{}%
\AgdaModule{\AgdaUnderscore{}}\AgdaSpace{}%
\AgdaSymbol{(}\AgdaBound{𝑨}\AgdaSpace{}%
\AgdaSymbol{:}\AgdaSpace{}%
\AgdaRecord{Algebra}\AgdaSpace{}%
\AgdaGeneralizable{α}\AgdaSpace{}%
\AgdaGeneralizable{ρᵃ}\AgdaSymbol{)}\AgdaSpace{}%
\AgdaKeyword{where}\<%
\\
\>[0][@{}l@{\AgdaIndent{0}}]%
\>[1]\AgdaKeyword{open}\AgdaSpace{}%
\AgdaModule{Setoid}\AgdaSpace{}%
\AgdaOperator{\AgdaFunction{𝔻[}}\AgdaSpace{}%
\AgdaBound{𝑨}\AgdaSpace{}%
\AgdaOperator{\AgdaFunction{]}}\AgdaSpace{}%
\AgdaKeyword{using}\AgdaSpace{}%
\AgdaSymbol{(}\AgdaSpace{}%
\AgdaOperator{\AgdaField{\AgdaUnderscore{}≈\AgdaUnderscore{}}}\AgdaSpace{}%
\AgdaSymbol{;}\AgdaSpace{}%
\AgdaFunction{refl}\AgdaSpace{}%
\AgdaSymbol{;}\AgdaSpace{}%
\AgdaFunction{sym}\AgdaSpace{}%
\AgdaSymbol{;}\AgdaSpace{}%
\AgdaFunction{trans}\AgdaSpace{}%
\AgdaSymbol{)}\AgdaSpace{}%
\AgdaSymbol{;}\AgdaSpace{}%
\AgdaKeyword{open}\AgdaSpace{}%
\AgdaModule{Level}\<%
\\
\>[1]\AgdaFunction{Lift-Algˡ}\AgdaSpace{}%
\AgdaSymbol{:}\AgdaSpace{}%
\AgdaSymbol{(}\AgdaBound{ℓ}\AgdaSpace{}%
\AgdaSymbol{:}\AgdaSpace{}%
\AgdaPostulate{Level}\AgdaSymbol{)}\AgdaSpace{}%
\AgdaSymbol{→}\AgdaSpace{}%
\AgdaRecord{Algebra}\AgdaSpace{}%
\AgdaSymbol{(}\AgdaBound{α}\AgdaSpace{}%
\AgdaOperator{\AgdaPrimitive{⊔}}\AgdaSpace{}%
\AgdaBound{ℓ}\AgdaSymbol{)}\AgdaSpace{}%
\AgdaBound{ρᵃ}\<%
\\
\>[1]\AgdaField{Domain}\AgdaSpace{}%
\AgdaSymbol{(}\AgdaFunction{Lift-Algˡ}\AgdaSpace{}%
\AgdaBound{ℓ}\AgdaSymbol{)}\AgdaSpace{}%
\AgdaSymbol{=}\<%
\\
\>[1][@{}l@{\AgdaIndent{0}}]%
\>[2]\AgdaKeyword{record}%
\>[10]\AgdaSymbol{\{}\AgdaSpace{}%
\AgdaField{Carrier}%
\>[27]\AgdaSymbol{=}\AgdaSpace{}%
\AgdaRecord{Lift}\AgdaSpace{}%
\AgdaBound{ℓ}\AgdaSpace{}%
\AgdaOperator{\AgdaFunction{𝕌[}}\AgdaSpace{}%
\AgdaBound{𝑨}\AgdaSpace{}%
\AgdaOperator{\AgdaFunction{]}}\<%
\\
\>[10]\AgdaSymbol{;}\AgdaSpace{}%
\AgdaOperator{\AgdaField{\AgdaUnderscore{}≈\AgdaUnderscore{}}}%
\>[27]\AgdaSymbol{=}\AgdaSpace{}%
\AgdaSymbol{λ}\AgdaSpace{}%
\AgdaBound{x}\AgdaSpace{}%
\AgdaBound{y}\AgdaSpace{}%
\AgdaSymbol{→}\AgdaSpace{}%
\AgdaField{lower}\AgdaSpace{}%
\AgdaBound{x}\AgdaSpace{}%
\AgdaOperator{\AgdaFunction{≈}}\AgdaSpace{}%
\AgdaField{lower}\AgdaSpace{}%
\AgdaBound{y}\<%
\\
\>[10]\AgdaSymbol{;}\AgdaSpace{}%
\AgdaField{isEquivalence}%
\>[27]\AgdaSymbol{=}\AgdaSpace{}%
\AgdaKeyword{record}\AgdaSpace{}%
\AgdaSymbol{\{}\AgdaSpace{}%
\AgdaField{refl}\AgdaSpace{}%
\AgdaSymbol{=}\AgdaSpace{}%
\AgdaFunction{refl}\AgdaSpace{}%
\AgdaSymbol{;}\AgdaSpace{}%
\AgdaField{sym}\AgdaSpace{}%
\AgdaSymbol{=}\AgdaSpace{}%
\AgdaFunction{sym}\AgdaSpace{}%
\AgdaSymbol{;}\AgdaSpace{}%
\AgdaField{trans}\AgdaSpace{}%
\AgdaSymbol{=}\AgdaSpace{}%
\AgdaFunction{trans}\AgdaSpace{}%
\AgdaSymbol{\}\}}\<%
\\
\\[\AgdaEmptyExtraSkip]%
\>[1]\AgdaField{Interp}\AgdaSpace{}%
\AgdaSymbol{(}\AgdaFunction{Lift-Algˡ}\AgdaSpace{}%
\AgdaBound{ℓ}\AgdaSymbol{)}\AgdaSpace{}%
\AgdaOperator{\AgdaField{⟨\$⟩}}\AgdaSpace{}%
\AgdaSymbol{(}\AgdaBound{f}\AgdaSpace{}%
\AgdaOperator{\AgdaInductiveConstructor{,}}\AgdaSpace{}%
\AgdaBound{la}\AgdaSymbol{)}\AgdaSpace{}%
\AgdaSymbol{=}\AgdaSpace{}%
\AgdaInductiveConstructor{lift}\AgdaSpace{}%
\AgdaSymbol{((}\AgdaBound{f}\AgdaSpace{}%
\AgdaOperator{\AgdaFunction{̂}}\AgdaSpace{}%
\AgdaBound{𝑨}\AgdaSymbol{)}\AgdaSpace{}%
\AgdaSymbol{(}\AgdaField{lower}\AgdaSpace{}%
\AgdaOperator{\AgdaFunction{∘}}\AgdaSpace{}%
\AgdaBound{la}\AgdaSymbol{))}\<%
\\
\>[1]\AgdaField{cong}\AgdaSpace{}%
\AgdaSymbol{(}\AgdaField{Interp}\AgdaSpace{}%
\AgdaSymbol{(}\AgdaFunction{Lift-Algˡ}\AgdaSpace{}%
\AgdaBound{ℓ}\AgdaSymbol{))}\AgdaSpace{}%
\AgdaSymbol{(}\AgdaInductiveConstructor{≡.refl}\AgdaSpace{}%
\AgdaOperator{\AgdaInductiveConstructor{,}}\AgdaSpace{}%
\AgdaBound{lab}\AgdaSymbol{)}\AgdaSpace{}%
\AgdaSymbol{=}\AgdaSpace{}%
\AgdaField{cong}\AgdaSpace{}%
\AgdaSymbol{(}\AgdaField{Interp}\AgdaSpace{}%
\AgdaBound{𝑨}\AgdaSymbol{)}\AgdaSpace{}%
\AgdaSymbol{((}\AgdaInductiveConstructor{≡.refl}\AgdaSpace{}%
\AgdaOperator{\AgdaInductiveConstructor{,}}\AgdaSpace{}%
\AgdaBound{lab}\AgdaSymbol{))}\<%
\\
\\[\AgdaEmptyExtraSkip]%
\>[1]\AgdaFunction{Lift-Algʳ}\AgdaSpace{}%
\AgdaSymbol{:}\AgdaSpace{}%
\AgdaSymbol{(}\AgdaBound{ℓ}\AgdaSpace{}%
\AgdaSymbol{:}\AgdaSpace{}%
\AgdaPostulate{Level}\AgdaSymbol{)}\AgdaSpace{}%
\AgdaSymbol{→}\AgdaSpace{}%
\AgdaRecord{Algebra}\AgdaSpace{}%
\AgdaBound{α}\AgdaSpace{}%
\AgdaSymbol{(}\AgdaBound{ρᵃ}\AgdaSpace{}%
\AgdaOperator{\AgdaPrimitive{⊔}}\AgdaSpace{}%
\AgdaBound{ℓ}\AgdaSymbol{)}\<%
\\
\>[1]\AgdaField{Domain}\AgdaSpace{}%
\AgdaSymbol{(}\AgdaFunction{Lift-Algʳ}\AgdaSpace{}%
\AgdaBound{ℓ}\AgdaSymbol{)}\AgdaSpace{}%
\AgdaSymbol{=}\<%
\\
\>[1][@{}l@{\AgdaIndent{0}}]%
\>[2]\AgdaKeyword{record}%
\>[10]\AgdaSymbol{\{}\AgdaSpace{}%
\AgdaField{Carrier}%
\>[27]\AgdaSymbol{=}\AgdaSpace{}%
\AgdaOperator{\AgdaFunction{𝕌[}}\AgdaSpace{}%
\AgdaBound{𝑨}\AgdaSpace{}%
\AgdaOperator{\AgdaFunction{]}}\<%
\\
\>[10]\AgdaSymbol{;}\AgdaSpace{}%
\AgdaOperator{\AgdaField{\AgdaUnderscore{}≈\AgdaUnderscore{}}}%
\>[27]\AgdaSymbol{=}\AgdaSpace{}%
\AgdaSymbol{λ}\AgdaSpace{}%
\AgdaBound{x}\AgdaSpace{}%
\AgdaBound{y}\AgdaSpace{}%
\AgdaSymbol{→}\AgdaSpace{}%
\AgdaRecord{Lift}\AgdaSpace{}%
\AgdaBound{ℓ}\AgdaSpace{}%
\AgdaSymbol{(}\AgdaBound{x}\AgdaSpace{}%
\AgdaOperator{\AgdaFunction{≈}}\AgdaSpace{}%
\AgdaBound{y}\AgdaSymbol{)}\<%
\\
\>[10]\AgdaSymbol{;}\AgdaSpace{}%
\AgdaField{isEquivalence}%
\>[27]\AgdaSymbol{=}\AgdaSpace{}%
\AgdaKeyword{record}%
\>[37]\AgdaSymbol{\{}\AgdaSpace{}%
\AgdaField{refl}%
\>[45]\AgdaSymbol{=}\AgdaSpace{}%
\AgdaInductiveConstructor{lift}\AgdaSpace{}%
\AgdaFunction{refl}\<%
\\
\>[37]\AgdaSymbol{;}\AgdaSpace{}%
\AgdaField{sym}%
\>[45]\AgdaSymbol{=}\AgdaSpace{}%
\AgdaInductiveConstructor{lift}\AgdaSpace{}%
\AgdaOperator{\AgdaFunction{∘}}\AgdaSpace{}%
\AgdaFunction{sym}\AgdaSpace{}%
\AgdaOperator{\AgdaFunction{∘}}\AgdaSpace{}%
\AgdaField{lower}\<%
\\
\>[37]\AgdaSymbol{;}\AgdaSpace{}%
\AgdaField{trans}\AgdaSpace{}%
\AgdaSymbol{=}\AgdaSpace{}%
\AgdaSymbol{λ}\AgdaSpace{}%
\AgdaBound{x}\AgdaSpace{}%
\AgdaBound{y}\AgdaSpace{}%
\AgdaSymbol{→}\AgdaSpace{}%
\AgdaInductiveConstructor{lift}\AgdaSpace{}%
\AgdaSymbol{(}\AgdaFunction{trans}\AgdaSpace{}%
\AgdaSymbol{(}\AgdaField{lower}\AgdaSpace{}%
\AgdaBound{x}\AgdaSymbol{)(}\AgdaField{lower}\AgdaSpace{}%
\AgdaBound{y}\AgdaSymbol{))}\AgdaSpace{}%
\AgdaSymbol{\}\}}\<%
\\
\\[\AgdaEmptyExtraSkip]%
\>[1]\AgdaField{Interp}\AgdaSpace{}%
\AgdaSymbol{(}\AgdaFunction{Lift-Algʳ}\AgdaSpace{}%
\AgdaBound{ℓ}\AgdaSpace{}%
\AgdaSymbol{)}\AgdaSpace{}%
\AgdaOperator{\AgdaField{⟨\$⟩}}\AgdaSpace{}%
\AgdaSymbol{(}\AgdaBound{f}\AgdaSpace{}%
\AgdaOperator{\AgdaInductiveConstructor{,}}\AgdaSpace{}%
\AgdaBound{la}\AgdaSymbol{)}\AgdaSpace{}%
\AgdaSymbol{=}\AgdaSpace{}%
\AgdaSymbol{(}\AgdaBound{f}\AgdaSpace{}%
\AgdaOperator{\AgdaFunction{̂}}\AgdaSpace{}%
\AgdaBound{𝑨}\AgdaSymbol{)}\AgdaSpace{}%
\AgdaBound{la}\<%
\\
\>[1]\AgdaField{cong}\AgdaSpace{}%
\AgdaSymbol{(}\AgdaField{Interp}\AgdaSpace{}%
\AgdaSymbol{(}\AgdaFunction{Lift-Algʳ}\AgdaSpace{}%
\AgdaBound{ℓ}\AgdaSymbol{))(}\AgdaInductiveConstructor{≡.refl}\AgdaSpace{}%
\AgdaOperator{\AgdaInductiveConstructor{,}}\AgdaSpace{}%
\AgdaBound{lab}\AgdaSymbol{)}\AgdaSpace{}%
\AgdaSymbol{=}\AgdaSpace{}%
\AgdaInductiveConstructor{lift}\AgdaSymbol{(}\AgdaField{cong}\AgdaSymbol{(}\AgdaField{Interp}\AgdaSpace{}%
\AgdaBound{𝑨}\AgdaSymbol{)(}\AgdaInductiveConstructor{≡.refl}\AgdaSpace{}%
\AgdaOperator{\AgdaInductiveConstructor{,}}\AgdaSpace{}%
\AgdaSymbol{λ}\AgdaSpace{}%
\AgdaBound{i}\AgdaSpace{}%
\AgdaSymbol{→}\AgdaSpace{}%
\AgdaField{lower}\AgdaSpace{}%
\AgdaSymbol{(}\AgdaBound{lab}\AgdaSpace{}%
\AgdaBound{i}\AgdaSymbol{)))}\<%
\end{code}
\fi

\begin{code}%
\>[0]\<%
\\
\>[0]\AgdaFunction{Lift-Alg}\AgdaSpace{}%
\AgdaSymbol{:}\AgdaSpace{}%
\AgdaSymbol{(}\AgdaBound{𝑨}\AgdaSpace{}%
\AgdaSymbol{:}\AgdaSpace{}%
\AgdaRecord{Algebra}\AgdaSpace{}%
\AgdaGeneralizable{α}\AgdaSpace{}%
\AgdaGeneralizable{ρᵃ}\AgdaSymbol{)(}\AgdaBound{ℓ₀}\AgdaSpace{}%
\AgdaBound{ℓ₁}\AgdaSpace{}%
\AgdaSymbol{:}\AgdaSpace{}%
\AgdaPostulate{Level}\AgdaSymbol{)}\AgdaSpace{}%
\AgdaSymbol{→}\AgdaSpace{}%
\AgdaRecord{Algebra}\AgdaSpace{}%
\AgdaSymbol{(}\AgdaGeneralizable{α}\AgdaSpace{}%
\AgdaOperator{\AgdaPrimitive{⊔}}\AgdaSpace{}%
\AgdaBound{ℓ₀}\AgdaSymbol{)}\AgdaSpace{}%
\AgdaSymbol{(}\AgdaGeneralizable{ρᵃ}\AgdaSpace{}%
\AgdaOperator{\AgdaPrimitive{⊔}}\AgdaSpace{}%
\AgdaBound{ℓ₁}\AgdaSymbol{)}\<%
\end{code}
\ifshort
\vskip2mm
\else
\begin{code}%
\>[0]\AgdaFunction{Lift-Alg}\AgdaSpace{}%
\AgdaBound{𝑨}\AgdaSpace{}%
\AgdaBound{ℓ₀}\AgdaSpace{}%
\AgdaBound{ℓ₁}\AgdaSpace{}%
\AgdaSymbol{=}\AgdaSpace{}%
\AgdaFunction{Lift-Algʳ}\AgdaSpace{}%
\AgdaSymbol{(}\AgdaFunction{Lift-Algˡ}\AgdaSpace{}%
\AgdaBound{𝑨}\AgdaSpace{}%
\AgdaBound{ℓ₀}\AgdaSymbol{)}\AgdaSpace{}%
\AgdaBound{ℓ₁}\<%
\\
\>[0]\<%
\end{code}
\fi
\noindent To see why the \af{Lift-Alg} function is useful, recall that our definition of the algebra
record type uses two universe level parameters corresponding to those of the algebra's
underlying domain setoid.
\ifshort\else
Concretely, an algebra of type \ar{Algebra} \ab{α} \ab{ρᵃ} has a
\afld{Domain} of type \ar{Setoid} \ab{α} \ab{ρᵃ}. This packages a ``carrier set''
(\afld{Carrier}), inhabiting \ap{Type} \ab{α}, with an equality on \afld{Carrier} of type
\af{Rel} \afld{Carrier} \ab{ρᵃ}.
\fi
The \af{Lift-Alg} function takes an algebra---one whose carrier inhabits \ap{Type \ab{α}}
with equality of type \af{Rel} \afld{Carrier} \ab{ρᵃ}---and constructs a new algebra whose
carrier inhabits \ap{Type} (\ab{α} \ap{⊔} \ab{ℓ₀}) with equality of type \af{Rel}
\afld{Carrier} (\ab{ρᵃ} \ap{⊔} \ab{ℓ₁}). This lifting operation would be worthless without
a useful semantic connection between the input and output algebras.
Fortunately, it is easy to prove that \af{Lift-Alg} is an \defn{algebraic invariant},
which is to say that the resulting ``lifted'' algebra has the same algebraic properties as
the original algebra, a fact we will codify later in a type called \af{Lift-≅}.

\paragraph*{Product Algebras}
Here we review the (informal) definition of the \defn{product} of a family of
\ab{𝑆}-algebras and then define a type which formalizes this notion in Agda.
Let \ab{ι} be a universe and \ab I~:~\ap{Type}~\ab{ι} a type (the ``indexing type'').
Then the dependent function type \ab{𝒜}~:~\ab
I~\as{→}~\ab{Algebra}~\ab{α}~\ab{ρᵃ} represents an \defn{indexed family of algebras}.
Denote by \af{⨅}~\ab{𝒜} the \defn{product of algebras} in \ab{𝒜} (or \defn{product
algebra}), by which we mean the algebra whose domain is the Cartesian product \af{Π}~\ab
i~꞉~\ab I~\af{,}~\aof{𝔻[~\ab{𝒜}~\ab i~]} of the domains of the algebras in \ab{𝒜}, and
whose operations are those arising by pointwise interpretation in the obvious way: if
\ab{f} is a \ab J-ary operation symbol and if
\ab a~:~\af{Π}~\ab i~꞉~\ab I~\af{,}~\ab J~\as{→}~\aof{𝔻[~\ab{𝒜}~\ab i~]} is, for each
\ab i~:~\ab I, a \ab J-tuple of elements of the domain \aof{𝔻[~\ab{𝒜}~\ab i~]}, then
we define the interpretation of \ab f in \af{⨅}~\ab{𝒜} by\\[-2mm]

(\ab{f}~\af{̂}~\af{⨅}~\ab{𝒜}) \ab a := \as{λ}~(\ab i~:~\ab I)~\as{→}
(\ab{f}~\af{̂}~\ab{𝒜}~\ab i)(\ab{a}~\ab i).\\[8pt]
In the \agdaalgebras library we define the function \af{⨅} which formalizes this
notion of \defn{product algebra} in MLTT.
\ifshort
Here we just show the interface, but \seeshort for the complete definition of \af{⨅}.

\else
Here is the formal definition.

\fi
\begin{code}%
\>[0]\AgdaKeyword{module}\AgdaSpace{}%
\AgdaModule{\AgdaUnderscore{}}\AgdaSpace{}%
\AgdaSymbol{\{}\AgdaBound{ι}\AgdaSpace{}%
\AgdaSymbol{:}\AgdaSpace{}%
\AgdaPostulate{Level}\AgdaSymbol{\}\{}\AgdaBound{I}\AgdaSpace{}%
\AgdaSymbol{:}\AgdaSpace{}%
\AgdaPrimitive{Type}\AgdaSpace{}%
\AgdaBound{ι}\AgdaSpace{}%
\AgdaSymbol{\}}\AgdaSpace{}%
\AgdaKeyword{where}\<%
\\
\>[0][@{}l@{\AgdaIndent{0}}]%
\>[1]\AgdaFunction{⨅}\AgdaSpace{}%
\AgdaSymbol{:}\AgdaSpace{}%
\AgdaSymbol{(}\AgdaBound{𝒜}\AgdaSpace{}%
\AgdaSymbol{:}\AgdaSpace{}%
\AgdaBound{I}\AgdaSpace{}%
\AgdaSymbol{→}\AgdaSpace{}%
\AgdaRecord{Algebra}\AgdaSpace{}%
\AgdaGeneralizable{α}\AgdaSpace{}%
\AgdaGeneralizable{ρᵃ}\AgdaSymbol{)}\AgdaSpace{}%
\AgdaSymbol{→}\AgdaSpace{}%
\AgdaRecord{Algebra}\AgdaSpace{}%
\AgdaSymbol{(}\AgdaGeneralizable{α}\AgdaSpace{}%
\AgdaOperator{\AgdaPrimitive{⊔}}\AgdaSpace{}%
\AgdaBound{ι}\AgdaSymbol{)}\AgdaSpace{}%
\AgdaSymbol{(}\AgdaGeneralizable{ρᵃ}\AgdaSpace{}%
\AgdaOperator{\AgdaPrimitive{⊔}}\AgdaSpace{}%
\AgdaBound{ι}\AgdaSymbol{)}\<%
\end{code}
\ifshort\else
\begin{code}%
\>[1]\AgdaField{Domain}\AgdaSpace{}%
\AgdaSymbol{(}\AgdaFunction{⨅}\AgdaSpace{}%
\AgdaBound{𝒜}\AgdaSymbol{)}\AgdaSpace{}%
\AgdaSymbol{=}\<%
\\
\>[1][@{}l@{\AgdaIndent{0}}]%
\>[2]\AgdaKeyword{record}%
\>[1112I]\AgdaSymbol{\{}\AgdaSpace{}%
\AgdaField{Carrier}\AgdaSpace{}%
\AgdaSymbol{=}\AgdaSpace{}%
\AgdaSymbol{∀}\AgdaSpace{}%
\AgdaBound{i}\AgdaSpace{}%
\AgdaSymbol{→}\AgdaSpace{}%
\AgdaOperator{\AgdaFunction{𝕌[}}\AgdaSpace{}%
\AgdaBound{𝒜}\AgdaSpace{}%
\AgdaBound{i}\AgdaSpace{}%
\AgdaOperator{\AgdaFunction{]}}\<%
\\
\>[.][@{}l@{}]\<[1112I]%
\>[9]\AgdaSymbol{;}\AgdaSpace{}%
\AgdaOperator{\AgdaField{\AgdaUnderscore{}≈\AgdaUnderscore{}}}\AgdaSpace{}%
\AgdaSymbol{=}\AgdaSpace{}%
\AgdaSymbol{λ}\AgdaSpace{}%
\AgdaBound{a}\AgdaSpace{}%
\AgdaBound{b}\AgdaSpace{}%
\AgdaSymbol{→}\AgdaSpace{}%
\AgdaSymbol{∀}\AgdaSpace{}%
\AgdaBound{i}\AgdaSpace{}%
\AgdaSymbol{→}\AgdaSpace{}%
\AgdaSymbol{(}\AgdaOperator{\AgdaField{\AgdaUnderscore{}≈ˢ\AgdaUnderscore{}}}\AgdaSpace{}%
\AgdaOperator{\AgdaFunction{𝔻[}}\AgdaSpace{}%
\AgdaBound{𝒜}\AgdaSpace{}%
\AgdaBound{i}\AgdaSpace{}%
\AgdaOperator{\AgdaFunction{]}}\AgdaSymbol{)}\AgdaSpace{}%
\AgdaSymbol{(}\AgdaBound{a}\AgdaSpace{}%
\AgdaBound{i}\AgdaSymbol{)(}\AgdaBound{b}\AgdaSpace{}%
\AgdaBound{i}\AgdaSymbol{)}\<%
\\
\>[9]\AgdaSymbol{;}%
\>[1139I]\AgdaField{isEquivalence}\AgdaSpace{}%
\AgdaSymbol{=}\<%
\\
\>[1139I][@{}l@{\AgdaIndent{0}}]%
\>[12]\AgdaKeyword{record}%
\>[20]\AgdaSymbol{\{}\AgdaSpace{}%
\AgdaField{refl}%
\>[29]\AgdaSymbol{=}\AgdaSpace{}%
\AgdaSymbol{λ}\AgdaSpace{}%
\AgdaBound{i}\AgdaSpace{}%
\AgdaSymbol{→}%
\>[42]\AgdaField{reflᵉ}%
\>[50]\AgdaSymbol{(}\AgdaField{isEquivalence}\AgdaSpace{}%
\AgdaOperator{\AgdaFunction{𝔻[}}\AgdaSpace{}%
\AgdaBound{𝒜}\AgdaSpace{}%
\AgdaBound{i}\AgdaSpace{}%
\AgdaOperator{\AgdaFunction{]}}\AgdaSymbol{)}\<%
\\
\>[20]\AgdaSymbol{;}\AgdaSpace{}%
\AgdaField{sym}%
\>[29]\AgdaSymbol{=}\AgdaSpace{}%
\AgdaSymbol{λ}\AgdaSpace{}%
\AgdaBound{x}\AgdaSpace{}%
\AgdaBound{i}\AgdaSpace{}%
\AgdaSymbol{→}%
\>[42]\AgdaField{symᵉ}%
\>[50]\AgdaSymbol{(}\AgdaField{isEquivalence}\AgdaSpace{}%
\AgdaOperator{\AgdaFunction{𝔻[}}\AgdaSpace{}%
\AgdaBound{𝒜}\AgdaSpace{}%
\AgdaBound{i}\AgdaSpace{}%
\AgdaOperator{\AgdaFunction{]}}\AgdaSymbol{)(}\AgdaBound{x}\AgdaSpace{}%
\AgdaBound{i}\AgdaSymbol{)}\<%
\\
\>[20]\AgdaSymbol{;}\AgdaSpace{}%
\AgdaField{trans}%
\>[29]\AgdaSymbol{=}\AgdaSpace{}%
\AgdaSymbol{λ}\AgdaSpace{}%
\AgdaBound{x}\AgdaSpace{}%
\AgdaBound{y}\AgdaSpace{}%
\AgdaBound{i}\AgdaSpace{}%
\AgdaSymbol{→}%
\>[42]\AgdaField{transᵉ}%
\>[50]\AgdaSymbol{(}\AgdaField{isEquivalence}\AgdaSpace{}%
\AgdaOperator{\AgdaFunction{𝔻[}}\AgdaSpace{}%
\AgdaBound{𝒜}\AgdaSpace{}%
\AgdaBound{i}\AgdaSpace{}%
\AgdaOperator{\AgdaFunction{]}}\AgdaSymbol{)(}\AgdaBound{x}\AgdaSpace{}%
\AgdaBound{i}\AgdaSymbol{)(}\AgdaBound{y}\AgdaSpace{}%
\AgdaBound{i}\AgdaSymbol{)}\AgdaSpace{}%
\AgdaSymbol{\}\}}\<%
\\
\>[1]\AgdaField{Interp}\AgdaSpace{}%
\AgdaSymbol{(}\AgdaFunction{⨅}\AgdaSpace{}%
\AgdaBound{𝒜}\AgdaSymbol{)}\AgdaSpace{}%
\AgdaOperator{\AgdaField{⟨\$⟩}}\AgdaSpace{}%
\AgdaSymbol{(}\AgdaBound{f}\AgdaSpace{}%
\AgdaOperator{\AgdaInductiveConstructor{,}}\AgdaSpace{}%
\AgdaBound{a}\AgdaSymbol{)}\AgdaSpace{}%
\AgdaSymbol{=}\AgdaSpace{}%
\AgdaSymbol{λ}\AgdaSpace{}%
\AgdaBound{i}\AgdaSpace{}%
\AgdaSymbol{→}\AgdaSpace{}%
\AgdaSymbol{(}\AgdaBound{f}\AgdaSpace{}%
\AgdaOperator{\AgdaFunction{̂}}\AgdaSpace{}%
\AgdaSymbol{(}\AgdaBound{𝒜}\AgdaSpace{}%
\AgdaBound{i}\AgdaSymbol{))}\AgdaSpace{}%
\AgdaSymbol{(}\AgdaFunction{flip}\AgdaSpace{}%
\AgdaBound{a}\AgdaSpace{}%
\AgdaBound{i}\AgdaSymbol{)}\<%
\\
\>[1]\AgdaField{cong}\AgdaSpace{}%
\AgdaSymbol{(}\AgdaField{Interp}\AgdaSpace{}%
\AgdaSymbol{(}\AgdaFunction{⨅}\AgdaSpace{}%
\AgdaBound{𝒜}\AgdaSymbol{))}\AgdaSpace{}%
\AgdaSymbol{(}\AgdaInductiveConstructor{≡.refl}\AgdaSpace{}%
\AgdaOperator{\AgdaInductiveConstructor{,}}\AgdaSpace{}%
\AgdaBound{f=g}\AgdaSpace{}%
\AgdaSymbol{)}\AgdaSpace{}%
\AgdaSymbol{=}\AgdaSpace{}%
\AgdaSymbol{λ}\AgdaSpace{}%
\AgdaBound{i}\AgdaSpace{}%
\AgdaSymbol{→}\AgdaSpace{}%
\AgdaField{cong}\AgdaSpace{}%
\AgdaSymbol{(}\AgdaField{Interp}\AgdaSpace{}%
\AgdaSymbol{(}\AgdaBound{𝒜}\AgdaSpace{}%
\AgdaBound{i}\AgdaSymbol{))}\AgdaSpace{}%
\AgdaSymbol{(}\AgdaInductiveConstructor{≡.refl}\AgdaSpace{}%
\AgdaOperator{\AgdaInductiveConstructor{,}}\AgdaSpace{}%
\AgdaFunction{flip}\AgdaSpace{}%
\AgdaBound{f=g}\AgdaSpace{}%
\AgdaBound{i}\AgdaSpace{}%
\AgdaSymbol{)}\<%
\end{code}
\fi

%% -------------------------------------------------------------------------------------
\subsection{Homomorphisms}\label{homomorphisms}
Throughout this section, and the rest of the paper unless stated otherwise, \ab{𝑨} and \ab{𝑩}
will denote \ab{𝑆}-algebras inhabiting the types \af{Algebra} \ab{α} \ab{ρᵃ} and
\af{Algebra} \ab{β} \ab{ρᵇ}, respectively.

A \defn{homomorphism} (or ``hom'') from
\ab{𝑨} to \ab{𝑩} is a setoid function \ab{h}~:~\aof{𝔻[~\ab{𝑨}~]} \aor{⟶} \aof{𝔻[~\ab{𝑩}~]}
that is \defn{compatible} with all basic operations; that is, for
every operation symbol \ab{f} : \af{∣~\ab{𝑆}~∣} and all tuples
\ab{a} : \af{∥~\ab{𝑆}~∥}~\ab{f} \as{→} \aof{𝕌[~\ab{𝑨}~]}, we have \ab{h} \aofld{⟨\$⟩}
(\ab{f}~\af{̂}~\ab{𝑨}) \ab{a} \af{≈}
(\ab{f}~\af{̂}~\ab{𝑩}) \ab{h} \aofld{⟨\$⟩} (\ab{a} \au{}).
To formalize this concept in Agda, we first define the type \af{compatible-map-op}
representing the assertion that a given setoid function
\ab{h}~:~\aof{𝔻[~\ab{𝑨}~]} \aor{⟶} \aof{𝔻[~\ab{𝑩}~]} commutes with a given
operation symbol \ab{f}. Then we generalize over operation symbols in the definition
of \af{compatible-map}, the type of compatible maps from (the domain of) \ab{𝐴} to
(the domain of) \ab{𝑩}.

\ifshort\else
\begin{code}%
\>[0]\<%
\\
\>[0]\AgdaKeyword{module}\AgdaSpace{}%
\AgdaModule{\AgdaUnderscore{}}\AgdaSpace{}%
\AgdaSymbol{(}\AgdaBound{𝑨}\AgdaSpace{}%
\AgdaSymbol{:}\AgdaSpace{}%
\AgdaRecord{Algebra}\AgdaSpace{}%
\AgdaGeneralizable{α}\AgdaSpace{}%
\AgdaGeneralizable{ρᵃ}\AgdaSymbol{)(}\AgdaBound{𝑩}\AgdaSpace{}%
\AgdaSymbol{:}\AgdaSpace{}%
\AgdaRecord{Algebra}\AgdaSpace{}%
\AgdaGeneralizable{β}\AgdaSpace{}%
\AgdaGeneralizable{ρᵇ}\AgdaSymbol{)}\AgdaSpace{}%
\AgdaKeyword{where}\<%
\end{code}
\fi
\begin{code}%
\>[0]\<%
\\
\>[0][@{}l@{\AgdaIndent{1}}]%
\>[1]\AgdaFunction{compatible-map-op}\AgdaSpace{}%
\AgdaSymbol{:}\AgdaSpace{}%
\AgdaSymbol{(}\AgdaOperator{\AgdaFunction{𝔻[}}\AgdaSpace{}%
\AgdaBound{𝑨}\AgdaSpace{}%
\AgdaOperator{\AgdaFunction{]}}\AgdaSpace{}%
\AgdaOperator{\AgdaRecord{⟶}}\AgdaSpace{}%
\AgdaOperator{\AgdaFunction{𝔻[}}\AgdaSpace{}%
\AgdaBound{𝑩}\AgdaSpace{}%
\AgdaOperator{\AgdaFunction{]}}\AgdaSymbol{)}\AgdaSpace{}%
\AgdaSymbol{→}\AgdaSpace{}%
\AgdaOperator{\AgdaFunction{∣}}\AgdaSpace{}%
\AgdaBound{𝑆}\AgdaSpace{}%
\AgdaOperator{\AgdaFunction{∣}}\AgdaSpace{}%
\AgdaSymbol{→}\AgdaSpace{}%
\AgdaPrimitive{Type}\AgdaSpace{}%
\AgdaSymbol{\AgdaUnderscore{}}\<%
\\
\>[1]\AgdaFunction{compatible-map-op}\AgdaSpace{}%
\AgdaBound{h}\AgdaSpace{}%
\AgdaBound{f}\AgdaSpace{}%
\AgdaSymbol{=}\AgdaSpace{}%
\AgdaSymbol{∀}\AgdaSpace{}%
\AgdaSymbol{\{}\AgdaBound{a}\AgdaSymbol{\}}\AgdaSpace{}%
\AgdaSymbol{→}\AgdaSpace{}%
\AgdaBound{h}\AgdaSpace{}%
\AgdaOperator{\AgdaField{⟨\$⟩}}\AgdaSpace{}%
\AgdaSymbol{(}\AgdaBound{f}\AgdaSpace{}%
\AgdaOperator{\AgdaFunction{̂}}\AgdaSpace{}%
\AgdaBound{𝑨}\AgdaSymbol{)}\AgdaSpace{}%
\AgdaBound{a}\AgdaSpace{}%
\AgdaOperator{\AgdaFunction{≈}}\AgdaSpace{}%
\AgdaSymbol{(}\AgdaBound{f}\AgdaSpace{}%
\AgdaOperator{\AgdaFunction{̂}}\AgdaSpace{}%
\AgdaBound{𝑩}\AgdaSymbol{)}\AgdaSpace{}%
\AgdaSymbol{λ}\AgdaSpace{}%
\AgdaBound{x}\AgdaSpace{}%
\AgdaSymbol{→}\AgdaSpace{}%
\AgdaBound{h}\AgdaSpace{}%
\AgdaOperator{\AgdaField{⟨\$⟩}}\AgdaSpace{}%
\AgdaSymbol{(}\AgdaBound{a}\AgdaSpace{}%
\AgdaBound{x}\AgdaSymbol{)}\<%
\\
\>[1][@{}l@{\AgdaIndent{0}}]%
\>[2]\AgdaKeyword{where}\AgdaSpace{}%
\AgdaKeyword{open}\AgdaSpace{}%
\AgdaModule{Setoid}\AgdaSpace{}%
\AgdaOperator{\AgdaFunction{𝔻[}}\AgdaSpace{}%
\AgdaBound{𝑩}\AgdaSpace{}%
\AgdaOperator{\AgdaFunction{]}}\AgdaSpace{}%
\AgdaKeyword{using}\AgdaSpace{}%
\AgdaSymbol{(}\AgdaSpace{}%
\AgdaOperator{\AgdaField{\AgdaUnderscore{}≈\AgdaUnderscore{}}}\AgdaSpace{}%
\AgdaSymbol{)}\<%
\\
\>[1]\AgdaFunction{compatible-map}\AgdaSpace{}%
\AgdaSymbol{:}\AgdaSpace{}%
\AgdaSymbol{(}\AgdaOperator{\AgdaFunction{𝔻[}}\AgdaSpace{}%
\AgdaBound{𝑨}\AgdaSpace{}%
\AgdaOperator{\AgdaFunction{]}}\AgdaSpace{}%
\AgdaOperator{\AgdaRecord{⟶}}\AgdaSpace{}%
\AgdaOperator{\AgdaFunction{𝔻[}}\AgdaSpace{}%
\AgdaBound{𝑩}\AgdaSpace{}%
\AgdaOperator{\AgdaFunction{]}}\AgdaSymbol{)}\AgdaSpace{}%
\AgdaSymbol{→}\AgdaSpace{}%
\AgdaPrimitive{Type}\AgdaSpace{}%
\AgdaSymbol{\AgdaUnderscore{}}\<%
\\
\>[1]\AgdaFunction{compatible-map}\AgdaSpace{}%
\AgdaBound{h}\AgdaSpace{}%
\AgdaSymbol{=}\AgdaSpace{}%
\AgdaSymbol{∀}\AgdaSpace{}%
\AgdaSymbol{\{}\AgdaBound{f}\AgdaSymbol{\}}\AgdaSpace{}%
\AgdaSymbol{→}\AgdaSpace{}%
\AgdaFunction{compatible-map-op}\AgdaSpace{}%
\AgdaBound{h}\AgdaSpace{}%
\AgdaBound{f}\<%
\\
\>[0]\<%
\end{code}
Using these we define a record type \ar{IsHom} representing the property of being
a homomorphism, and finally the type \af{hom} of homomorphisms from \ab{𝑨} to \ab{𝐵}.

\begin{code}%
\>[0]\<%
\\
\>[0][@{}l@{\AgdaIndent{1}}]%
\>[1]\AgdaKeyword{record}\AgdaSpace{}%
\AgdaRecord{IsHom}\AgdaSpace{}%
\AgdaSymbol{(}\AgdaBound{h}\AgdaSpace{}%
\AgdaSymbol{:}\AgdaSpace{}%
\AgdaOperator{\AgdaFunction{𝔻[}}\AgdaSpace{}%
\AgdaBound{𝑨}\AgdaSpace{}%
\AgdaOperator{\AgdaFunction{]}}\AgdaSpace{}%
\AgdaOperator{\AgdaRecord{⟶}}\AgdaSpace{}%
\AgdaOperator{\AgdaFunction{𝔻[}}\AgdaSpace{}%
\AgdaBound{𝑩}\AgdaSpace{}%
\AgdaOperator{\AgdaFunction{]}}\AgdaSymbol{)}\AgdaSpace{}%
\AgdaSymbol{:}\AgdaSpace{}%
\AgdaPrimitive{Type}\AgdaSpace{}%
\AgdaSymbol{(}\AgdaBound{𝓞}\AgdaSpace{}%
\AgdaOperator{\AgdaPrimitive{⊔}}\AgdaSpace{}%
\AgdaBound{𝓥}\AgdaSpace{}%
\AgdaOperator{\AgdaPrimitive{⊔}}\AgdaSpace{}%
\AgdaBound{α}\AgdaSpace{}%
\AgdaOperator{\AgdaPrimitive{⊔}}\AgdaSpace{}%
\AgdaBound{ρᵇ}\AgdaSymbol{)}\AgdaSpace{}%
\AgdaKeyword{where}\<%
\\
\>[1][@{}l@{\AgdaIndent{0}}]%
\>[2]\AgdaKeyword{constructor}\AgdaSpace{}%
\AgdaInductiveConstructor{mkhom}\AgdaSpace{}%
\AgdaSymbol{;}\AgdaSpace{}%
\AgdaKeyword{field}\AgdaSpace{}%
\AgdaField{compatible}\AgdaSpace{}%
\AgdaSymbol{:}\AgdaSpace{}%
\AgdaFunction{compatible-map}\AgdaSpace{}%
\AgdaBound{h}\<%
\\
\>[1]\AgdaFunction{hom}\AgdaSpace{}%
\AgdaSymbol{:}\AgdaSpace{}%
\AgdaPrimitive{Type}\AgdaSpace{}%
\AgdaSymbol{\AgdaUnderscore{}}\<%
\\
\>[1]\AgdaFunction{hom}\AgdaSpace{}%
\AgdaSymbol{=}\AgdaSpace{}%
\AgdaRecord{Σ}\AgdaSpace{}%
\AgdaSymbol{(}\AgdaOperator{\AgdaFunction{𝔻[}}\AgdaSpace{}%
\AgdaBound{𝑨}\AgdaSpace{}%
\AgdaOperator{\AgdaFunction{]}}\AgdaSpace{}%
\AgdaOperator{\AgdaRecord{⟶}}\AgdaSpace{}%
\AgdaOperator{\AgdaFunction{𝔻[}}\AgdaSpace{}%
\AgdaBound{𝑩}\AgdaSpace{}%
\AgdaOperator{\AgdaFunction{]}}\AgdaSymbol{)}\AgdaSpace{}%
\AgdaRecord{IsHom}\<%
\\
\>[0]\<%
\end{code}
Thus, an inhabitant of \af{hom} is a pair (\ab h , \ab p) whose first component is
a setoid function from the domain of \ab{𝑨} to that of \ab{𝑩} and whose second component
is \ab p : \ar{IsHom} \ab h, a proof that \ab h is a homomorphism.

A \defn{monomorphism} (resp. \defn{epimorphism}) is an injective (resp. surjective)
homomorphism.  The \agdaalgebras library defines types \ar{IsMon} and \ar{IsEpi} to
represent these properties, as well as \af{mon} and \af{epi}, the types of monomorphisms
and epimorphisms, respectively.
\ifshort %%% BEGIN SHORT VERSION ONLY
\else    %%% BEGIN LONG VERSION ONLY

\begin{code}%
\>[0]\<%
\\
\>[0][@{}l@{\AgdaIndent{1}}]%
\>[1]\AgdaKeyword{record}\AgdaSpace{}%
\AgdaRecord{IsMon}\AgdaSpace{}%
\AgdaSymbol{(}\AgdaBound{h}\AgdaSpace{}%
\AgdaSymbol{:}\AgdaSpace{}%
\AgdaOperator{\AgdaFunction{𝔻[}}\AgdaSpace{}%
\AgdaBound{𝑨}\AgdaSpace{}%
\AgdaOperator{\AgdaFunction{]}}\AgdaSpace{}%
\AgdaOperator{\AgdaRecord{⟶}}\AgdaSpace{}%
\AgdaOperator{\AgdaFunction{𝔻[}}\AgdaSpace{}%
\AgdaBound{𝑩}\AgdaSpace{}%
\AgdaOperator{\AgdaFunction{]}}\AgdaSymbol{)}\AgdaSpace{}%
\AgdaSymbol{:}\AgdaSpace{}%
\AgdaPrimitive{Type}\AgdaSpace{}%
\AgdaSymbol{(}\AgdaBound{𝓞}\AgdaSpace{}%
\AgdaOperator{\AgdaPrimitive{⊔}}\AgdaSpace{}%
\AgdaBound{𝓥}\AgdaSpace{}%
\AgdaOperator{\AgdaPrimitive{⊔}}\AgdaSpace{}%
\AgdaBound{α}\AgdaSpace{}%
\AgdaOperator{\AgdaPrimitive{⊔}}\AgdaSpace{}%
\AgdaBound{ρᵃ}\AgdaSpace{}%
\AgdaOperator{\AgdaPrimitive{⊔}}\AgdaSpace{}%
\AgdaBound{ρᵇ}\AgdaSymbol{)}\AgdaSpace{}%
\AgdaKeyword{where}\<%
\\
\>[1][@{}l@{\AgdaIndent{0}}]%
\>[2]\AgdaKeyword{field}%
\>[9]\AgdaField{isHom}\AgdaSpace{}%
\AgdaSymbol{:}\AgdaSpace{}%
\AgdaRecord{IsHom}\AgdaSpace{}%
\AgdaBound{h}\<%
\\
\>[9]\AgdaField{isInjective}\AgdaSpace{}%
\AgdaSymbol{:}\AgdaSpace{}%
\AgdaFunction{IsInjective}\AgdaSpace{}%
\AgdaBound{h}\<%
\\
\\[\AgdaEmptyExtraSkip]%
\>[2]\AgdaFunction{HomReduct}\AgdaSpace{}%
\AgdaSymbol{:}\AgdaSpace{}%
\AgdaFunction{hom}\<%
\\
\>[2]\AgdaFunction{HomReduct}\AgdaSpace{}%
\AgdaSymbol{=}\AgdaSpace{}%
\AgdaBound{h}\AgdaSpace{}%
\AgdaOperator{\AgdaInductiveConstructor{,}}\AgdaSpace{}%
\AgdaField{isHom}\<%
\\
\\[\AgdaEmptyExtraSkip]%
\>[1]\AgdaFunction{mon}\AgdaSpace{}%
\AgdaSymbol{:}\AgdaSpace{}%
\AgdaPrimitive{Type}\AgdaSpace{}%
\AgdaSymbol{\AgdaUnderscore{}}\<%
\\
\>[1]\AgdaFunction{mon}\AgdaSpace{}%
\AgdaSymbol{=}\AgdaSpace{}%
\AgdaRecord{Σ}\AgdaSpace{}%
\AgdaSymbol{(}\AgdaOperator{\AgdaFunction{𝔻[}}\AgdaSpace{}%
\AgdaBound{𝑨}\AgdaSpace{}%
\AgdaOperator{\AgdaFunction{]}}\AgdaSpace{}%
\AgdaOperator{\AgdaRecord{⟶}}\AgdaSpace{}%
\AgdaOperator{\AgdaFunction{𝔻[}}\AgdaSpace{}%
\AgdaBound{𝑩}\AgdaSpace{}%
\AgdaOperator{\AgdaFunction{]}}\AgdaSymbol{)}\AgdaSpace{}%
\AgdaRecord{IsMon}\<%
\\
\>[0]\<%
\end{code}
As with \af{hom}, the type \af{mon} is a dependent product type; each inhabitant is a pair
consisting of a setoid function, say, \ab h, along with a proof that \ab h is a
monomorphism.

\begin{code}%
\>[0]\<%
\\
\>[0][@{}l@{\AgdaIndent{1}}]%
\>[1]\AgdaKeyword{record}\AgdaSpace{}%
\AgdaRecord{IsEpi}\AgdaSpace{}%
\AgdaSymbol{(}\AgdaBound{h}\AgdaSpace{}%
\AgdaSymbol{:}\AgdaSpace{}%
\AgdaOperator{\AgdaFunction{𝔻[}}\AgdaSpace{}%
\AgdaBound{𝑨}\AgdaSpace{}%
\AgdaOperator{\AgdaFunction{]}}\AgdaSpace{}%
\AgdaOperator{\AgdaRecord{⟶}}\AgdaSpace{}%
\AgdaOperator{\AgdaFunction{𝔻[}}\AgdaSpace{}%
\AgdaBound{𝑩}\AgdaSpace{}%
\AgdaOperator{\AgdaFunction{]}}\AgdaSymbol{)}\AgdaSpace{}%
\AgdaSymbol{:}\AgdaSpace{}%
\AgdaPrimitive{Type}\AgdaSpace{}%
\AgdaSymbol{(}\AgdaBound{𝓞}\AgdaSpace{}%
\AgdaOperator{\AgdaPrimitive{⊔}}\AgdaSpace{}%
\AgdaBound{𝓥}\AgdaSpace{}%
\AgdaOperator{\AgdaPrimitive{⊔}}\AgdaSpace{}%
\AgdaBound{α}\AgdaSpace{}%
\AgdaOperator{\AgdaPrimitive{⊔}}\AgdaSpace{}%
\AgdaBound{β}\AgdaSpace{}%
\AgdaOperator{\AgdaPrimitive{⊔}}\AgdaSpace{}%
\AgdaBound{ρᵇ}\AgdaSymbol{)}\AgdaSpace{}%
\AgdaKeyword{where}\<%
\\
\>[1][@{}l@{\AgdaIndent{0}}]%
\>[2]\AgdaKeyword{field}%
\>[9]\AgdaField{isHom}\AgdaSpace{}%
\AgdaSymbol{:}\AgdaSpace{}%
\AgdaRecord{IsHom}\AgdaSpace{}%
\AgdaBound{h}\<%
\\
\>[9]\AgdaField{isSurjective}\AgdaSpace{}%
\AgdaSymbol{:}\AgdaSpace{}%
\AgdaFunction{IsSurjective}\AgdaSpace{}%
\AgdaBound{h}\<%
\\
\\[\AgdaEmptyExtraSkip]%
\>[2]\AgdaFunction{HomReduct}\AgdaSpace{}%
\AgdaSymbol{:}\AgdaSpace{}%
\AgdaFunction{hom}\<%
\\
\>[2]\AgdaFunction{HomReduct}\AgdaSpace{}%
\AgdaSymbol{=}\AgdaSpace{}%
\AgdaBound{h}\AgdaSpace{}%
\AgdaOperator{\AgdaInductiveConstructor{,}}\AgdaSpace{}%
\AgdaField{isHom}\<%
\\
\\[\AgdaEmptyExtraSkip]%
\>[1]\AgdaFunction{epi}\AgdaSpace{}%
\AgdaSymbol{:}\AgdaSpace{}%
\AgdaPrimitive{Type}\AgdaSpace{}%
\AgdaSymbol{\AgdaUnderscore{}}\<%
\\
\>[1]\AgdaFunction{epi}\AgdaSpace{}%
\AgdaSymbol{=}\AgdaSpace{}%
\AgdaRecord{Σ}\AgdaSpace{}%
\AgdaSymbol{(}\AgdaOperator{\AgdaFunction{𝔻[}}\AgdaSpace{}%
\AgdaBound{𝑨}\AgdaSpace{}%
\AgdaOperator{\AgdaFunction{]}}\AgdaSpace{}%
\AgdaOperator{\AgdaRecord{⟶}}\AgdaSpace{}%
\AgdaOperator{\AgdaFunction{𝔻[}}\AgdaSpace{}%
\AgdaBound{𝑩}\AgdaSpace{}%
\AgdaOperator{\AgdaFunction{]}}\AgdaSymbol{)}\AgdaSpace{}%
\AgdaRecord{IsEpi}\<%
\end{code}

Here are two mere utilities that are useful for translating between types.

\begin{code}%
\>[0]\AgdaKeyword{open}\AgdaSpace{}%
\AgdaModule{IsHom}\AgdaSpace{}%
\AgdaSymbol{;}\AgdaSpace{}%
\AgdaKeyword{open}\AgdaSpace{}%
\AgdaModule{IsMon}\AgdaSpace{}%
\AgdaSymbol{;}\AgdaSpace{}%
\AgdaKeyword{open}\AgdaSpace{}%
\AgdaModule{IsEpi}\<%
\\
\\[\AgdaEmptyExtraSkip]%
\>[0]\AgdaKeyword{module}\AgdaSpace{}%
\AgdaModule{\AgdaUnderscore{}}\AgdaSpace{}%
\AgdaSymbol{(}\AgdaBound{𝑨}\AgdaSpace{}%
\AgdaSymbol{:}\AgdaSpace{}%
\AgdaRecord{Algebra}\AgdaSpace{}%
\AgdaGeneralizable{α}\AgdaSpace{}%
\AgdaGeneralizable{ρᵃ}\AgdaSymbol{)(}\AgdaBound{𝑩}\AgdaSpace{}%
\AgdaSymbol{:}\AgdaSpace{}%
\AgdaRecord{Algebra}\AgdaSpace{}%
\AgdaGeneralizable{β}\AgdaSpace{}%
\AgdaGeneralizable{ρᵇ}\AgdaSymbol{)}\AgdaSpace{}%
\AgdaKeyword{where}\<%
\\
\\[\AgdaEmptyExtraSkip]%
\>[0][@{}l@{\AgdaIndent{0}}]%
\>[1]\AgdaFunction{mon→intohom}\AgdaSpace{}%
\AgdaSymbol{:}\AgdaSpace{}%
\AgdaFunction{mon}\AgdaSpace{}%
\AgdaBound{𝑨}\AgdaSpace{}%
\AgdaBound{𝑩}\AgdaSpace{}%
\AgdaSymbol{→}\AgdaSpace{}%
\AgdaFunction{Σ[}\AgdaSpace{}%
\AgdaBound{h}\AgdaSpace{}%
\AgdaFunction{∈}\AgdaSpace{}%
\AgdaFunction{hom}\AgdaSpace{}%
\AgdaBound{𝑨}\AgdaSpace{}%
\AgdaBound{𝑩}\AgdaSpace{}%
\AgdaFunction{]}\AgdaSpace{}%
\AgdaFunction{IsInjective}\AgdaSpace{}%
\AgdaOperator{\AgdaFunction{∣}}\AgdaSpace{}%
\AgdaBound{h}\AgdaSpace{}%
\AgdaOperator{\AgdaFunction{∣}}\<%
\\
\>[1]\AgdaFunction{mon→intohom}\AgdaSpace{}%
\AgdaSymbol{(}\AgdaBound{hh}\AgdaSpace{}%
\AgdaOperator{\AgdaInductiveConstructor{,}}\AgdaSpace{}%
\AgdaBound{hhM}\AgdaSymbol{)}\AgdaSpace{}%
\AgdaSymbol{=}\AgdaSpace{}%
\AgdaSymbol{(}\AgdaBound{hh}\AgdaSpace{}%
\AgdaOperator{\AgdaInductiveConstructor{,}}\AgdaSpace{}%
\AgdaField{isHom}\AgdaSpace{}%
\AgdaBound{hhM}\AgdaSymbol{)}\AgdaSpace{}%
\AgdaOperator{\AgdaInductiveConstructor{,}}\AgdaSpace{}%
\AgdaField{isInjective}\AgdaSpace{}%
\AgdaBound{hhM}\<%
\\
\\[\AgdaEmptyExtraSkip]%
\>[1]\AgdaFunction{epi→ontohom}\AgdaSpace{}%
\AgdaSymbol{:}\AgdaSpace{}%
\AgdaFunction{epi}\AgdaSpace{}%
\AgdaBound{𝑨}\AgdaSpace{}%
\AgdaBound{𝑩}\AgdaSpace{}%
\AgdaSymbol{→}\AgdaSpace{}%
\AgdaFunction{Σ[}\AgdaSpace{}%
\AgdaBound{h}\AgdaSpace{}%
\AgdaFunction{∈}\AgdaSpace{}%
\AgdaFunction{hom}\AgdaSpace{}%
\AgdaBound{𝑨}\AgdaSpace{}%
\AgdaBound{𝑩}\AgdaSpace{}%
\AgdaFunction{]}\AgdaSpace{}%
\AgdaFunction{IsSurjective}\AgdaSpace{}%
\AgdaOperator{\AgdaFunction{∣}}\AgdaSpace{}%
\AgdaBound{h}\AgdaSpace{}%
\AgdaOperator{\AgdaFunction{∣}}\<%
\\
\>[1]\AgdaFunction{epi→ontohom}\AgdaSpace{}%
\AgdaSymbol{(}\AgdaBound{hh}\AgdaSpace{}%
\AgdaOperator{\AgdaInductiveConstructor{,}}\AgdaSpace{}%
\AgdaBound{hhE}\AgdaSymbol{)}\AgdaSpace{}%
\AgdaSymbol{=}\AgdaSpace{}%
\AgdaSymbol{(}\AgdaBound{hh}\AgdaSpace{}%
\AgdaOperator{\AgdaInductiveConstructor{,}}\AgdaSpace{}%
\AgdaField{isHom}\AgdaSpace{}%
\AgdaBound{hhE}\AgdaSymbol{)}\AgdaSpace{}%
\AgdaOperator{\AgdaInductiveConstructor{,}}\AgdaSpace{}%
\AgdaField{isSurjective}\AgdaSpace{}%
\AgdaBound{hhE}\<%
\end{code}

\paragraph*{Composition of homomorphisms}
\fi      %%% END LONG VERSION ONLY SECTION
The composition of homomorphisms is again a homomorphism, and similarly for epimorphisms (and monomorphisms).
\ifshort
The proofs of these facts are straightforward so we omit them, but give them names,
\af{∘-hom} and \af{∘-epi}, so we can refer to them below.
\else

\begin{code}%
\>[0]\<%
\\
\>[0]\AgdaKeyword{module}\AgdaSpace{}%
\AgdaModule{\AgdaUnderscore{}}%
\>[10]\AgdaSymbol{\{}\AgdaBound{𝑨}\AgdaSpace{}%
\AgdaSymbol{:}\AgdaSpace{}%
\AgdaRecord{Algebra}\AgdaSpace{}%
\AgdaGeneralizable{α}\AgdaSpace{}%
\AgdaGeneralizable{ρᵃ}\AgdaSymbol{\}}\AgdaSpace{}%
\AgdaSymbol{\{}\AgdaBound{𝑩}\AgdaSpace{}%
\AgdaSymbol{:}\AgdaSpace{}%
\AgdaRecord{Algebra}\AgdaSpace{}%
\AgdaGeneralizable{β}\AgdaSpace{}%
\AgdaGeneralizable{ρᵇ}\AgdaSymbol{\}}\AgdaSpace{}%
\AgdaSymbol{\{}\AgdaBound{𝑪}\AgdaSpace{}%
\AgdaSymbol{:}\AgdaSpace{}%
\AgdaRecord{Algebra}\AgdaSpace{}%
\AgdaGeneralizable{γ}\AgdaSpace{}%
\AgdaGeneralizable{ρᶜ}\AgdaSymbol{\}}\<%
\\
\>[10]\AgdaSymbol{\{}\AgdaBound{g}\AgdaSpace{}%
\AgdaSymbol{:}\AgdaSpace{}%
\AgdaOperator{\AgdaFunction{𝔻[}}\AgdaSpace{}%
\AgdaBound{𝑨}\AgdaSpace{}%
\AgdaOperator{\AgdaFunction{]}}\AgdaSpace{}%
\AgdaOperator{\AgdaRecord{⟶}}\AgdaSpace{}%
\AgdaOperator{\AgdaFunction{𝔻[}}\AgdaSpace{}%
\AgdaBound{𝑩}\AgdaSpace{}%
\AgdaOperator{\AgdaFunction{]}}\AgdaSymbol{\}\{}\AgdaBound{h}\AgdaSpace{}%
\AgdaSymbol{:}\AgdaSpace{}%
\AgdaOperator{\AgdaFunction{𝔻[}}\AgdaSpace{}%
\AgdaBound{𝑩}\AgdaSpace{}%
\AgdaOperator{\AgdaFunction{]}}\AgdaSpace{}%
\AgdaOperator{\AgdaRecord{⟶}}\AgdaSpace{}%
\AgdaOperator{\AgdaFunction{𝔻[}}\AgdaSpace{}%
\AgdaBound{𝑪}\AgdaSpace{}%
\AgdaOperator{\AgdaFunction{]}}\AgdaSymbol{\}}\AgdaSpace{}%
\AgdaKeyword{where}\<%
\\
\\[\AgdaEmptyExtraSkip]%
\>[0][@{}l@{\AgdaIndent{0}}]%
\>[2]\AgdaKeyword{open}\AgdaSpace{}%
\AgdaModule{Setoid}\AgdaSpace{}%
\AgdaOperator{\AgdaFunction{𝔻[}}\AgdaSpace{}%
\AgdaBound{𝑪}\AgdaSpace{}%
\AgdaOperator{\AgdaFunction{]}}\AgdaSpace{}%
\AgdaKeyword{using}\AgdaSpace{}%
\AgdaSymbol{(}\AgdaSpace{}%
\AgdaFunction{trans}\AgdaSpace{}%
\AgdaSymbol{)}\<%
\\
\\[\AgdaEmptyExtraSkip]%
\>[2]\AgdaFunction{∘-is-hom}\AgdaSpace{}%
\AgdaSymbol{:}\AgdaSpace{}%
\AgdaRecord{IsHom}\AgdaSpace{}%
\AgdaBound{𝑨}\AgdaSpace{}%
\AgdaBound{𝑩}\AgdaSpace{}%
\AgdaBound{g}\AgdaSpace{}%
\AgdaSymbol{→}\AgdaSpace{}%
\AgdaRecord{IsHom}\AgdaSpace{}%
\AgdaBound{𝑩}\AgdaSpace{}%
\AgdaBound{𝑪}\AgdaSpace{}%
\AgdaBound{h}\AgdaSpace{}%
\AgdaSymbol{→}\AgdaSpace{}%
\AgdaRecord{IsHom}\AgdaSpace{}%
\AgdaBound{𝑨}\AgdaSpace{}%
\AgdaBound{𝑪}\AgdaSpace{}%
\AgdaSymbol{(}\AgdaBound{h}\AgdaSpace{}%
\AgdaOperator{\AgdaFunction{⟨∘⟩}}\AgdaSpace{}%
\AgdaBound{g}\AgdaSymbol{)}\<%
\\
\>[2]\AgdaFunction{∘-is-hom}\AgdaSpace{}%
\AgdaBound{ghom}\AgdaSpace{}%
\AgdaBound{hhom}\AgdaSpace{}%
\AgdaSymbol{=}\AgdaSpace{}%
\AgdaInductiveConstructor{mkhom}\AgdaSpace{}%
\AgdaFunction{c}\<%
\\
\>[2][@{}l@{\AgdaIndent{0}}]%
\>[3]\AgdaKeyword{where}\<%
\\
\>[3]\AgdaFunction{c}\AgdaSpace{}%
\AgdaSymbol{:}\AgdaSpace{}%
\AgdaFunction{compatible-map}\AgdaSpace{}%
\AgdaBound{𝑨}\AgdaSpace{}%
\AgdaBound{𝑪}\AgdaSpace{}%
\AgdaSymbol{(}\AgdaBound{h}\AgdaSpace{}%
\AgdaOperator{\AgdaFunction{⟨∘⟩}}\AgdaSpace{}%
\AgdaBound{g}\AgdaSymbol{)}\<%
\\
\>[3]\AgdaFunction{c}\AgdaSpace{}%
\AgdaSymbol{=}\AgdaSpace{}%
\AgdaFunction{trans}\AgdaSpace{}%
\AgdaSymbol{(}\AgdaField{cong}\AgdaSpace{}%
\AgdaBound{h}\AgdaSpace{}%
\AgdaSymbol{(}\AgdaField{compatible}\AgdaSpace{}%
\AgdaBound{ghom}\AgdaSymbol{))}\AgdaSpace{}%
\AgdaSymbol{(}\AgdaField{compatible}\AgdaSpace{}%
\AgdaBound{hhom}\AgdaSymbol{)}\<%
\\
\\[\AgdaEmptyExtraSkip]%
\>[2]\AgdaFunction{∘-is-epi}\AgdaSpace{}%
\AgdaSymbol{:}\AgdaSpace{}%
\AgdaRecord{IsEpi}\AgdaSpace{}%
\AgdaBound{𝑨}\AgdaSpace{}%
\AgdaBound{𝑩}\AgdaSpace{}%
\AgdaBound{g}\AgdaSpace{}%
\AgdaSymbol{→}\AgdaSpace{}%
\AgdaRecord{IsEpi}\AgdaSpace{}%
\AgdaBound{𝑩}\AgdaSpace{}%
\AgdaBound{𝑪}\AgdaSpace{}%
\AgdaBound{h}\AgdaSpace{}%
\AgdaSymbol{→}\AgdaSpace{}%
\AgdaRecord{IsEpi}\AgdaSpace{}%
\AgdaBound{𝑨}\AgdaSpace{}%
\AgdaBound{𝑪}\AgdaSpace{}%
\AgdaSymbol{(}\AgdaBound{h}\AgdaSpace{}%
\AgdaOperator{\AgdaFunction{⟨∘⟩}}\AgdaSpace{}%
\AgdaBound{g}\AgdaSymbol{)}\<%
\\
\>[2]\AgdaFunction{∘-is-epi}\AgdaSpace{}%
\AgdaBound{gE}\AgdaSpace{}%
\AgdaBound{hE}\AgdaSpace{}%
\AgdaSymbol{=}\AgdaSpace{}%
\AgdaKeyword{record}%
\>[27]\AgdaSymbol{\{}\AgdaSpace{}%
\AgdaField{isHom}\AgdaSpace{}%
\AgdaSymbol{=}\AgdaSpace{}%
\AgdaFunction{∘-is-hom}\AgdaSpace{}%
\AgdaSymbol{(}\AgdaField{isHom}\AgdaSpace{}%
\AgdaBound{gE}\AgdaSymbol{)}\AgdaSpace{}%
\AgdaSymbol{(}\AgdaField{isHom}\AgdaSpace{}%
\AgdaBound{hE}\AgdaSymbol{)}\<%
\\
\>[27]\AgdaSymbol{;}\AgdaSpace{}%
\AgdaField{isSurjective}\AgdaSpace{}%
\AgdaSymbol{=}\AgdaSpace{}%
\AgdaFunction{∘-IsSurjective}\AgdaSpace{}%
\AgdaBound{g}\AgdaSpace{}%
\AgdaBound{h}\AgdaSpace{}%
\AgdaSymbol{(}\AgdaField{isSurjective}\AgdaSpace{}%
\AgdaBound{gE}\AgdaSymbol{)}\AgdaSpace{}%
\AgdaSymbol{(}\AgdaField{isSurjective}\AgdaSpace{}%
\AgdaBound{hE}\AgdaSymbol{)}\AgdaSpace{}%
\AgdaSymbol{\}}\<%
\\
\\[\AgdaEmptyExtraSkip]%
\>[0]\AgdaKeyword{module}\AgdaSpace{}%
\AgdaModule{\AgdaUnderscore{}}\AgdaSpace{}%
\AgdaSymbol{\{}\AgdaBound{𝑨}\AgdaSpace{}%
\AgdaSymbol{:}\AgdaSpace{}%
\AgdaRecord{Algebra}\AgdaSpace{}%
\AgdaGeneralizable{α}\AgdaSpace{}%
\AgdaGeneralizable{ρᵃ}\AgdaSymbol{\}}\AgdaSpace{}%
\AgdaSymbol{\{}\AgdaBound{𝑩}\AgdaSpace{}%
\AgdaSymbol{:}\AgdaSpace{}%
\AgdaRecord{Algebra}\AgdaSpace{}%
\AgdaGeneralizable{β}\AgdaSpace{}%
\AgdaGeneralizable{ρᵇ}\AgdaSymbol{\}}\AgdaSpace{}%
\AgdaSymbol{\{}\AgdaBound{𝑪}\AgdaSpace{}%
\AgdaSymbol{:}\AgdaSpace{}%
\AgdaRecord{Algebra}\AgdaSpace{}%
\AgdaGeneralizable{γ}\AgdaSpace{}%
\AgdaGeneralizable{ρᶜ}\AgdaSymbol{\}}\AgdaSpace{}%
\AgdaKeyword{where}\<%
\\
\\[\AgdaEmptyExtraSkip]%
\>[0][@{}l@{\AgdaIndent{0}}]%
\>[2]\AgdaFunction{∘-hom}\AgdaSpace{}%
\AgdaSymbol{:}\AgdaSpace{}%
\AgdaFunction{hom}\AgdaSpace{}%
\AgdaBound{𝑨}\AgdaSpace{}%
\AgdaBound{𝑩}\AgdaSpace{}%
\AgdaSymbol{→}\AgdaSpace{}%
\AgdaFunction{hom}\AgdaSpace{}%
\AgdaBound{𝑩}\AgdaSpace{}%
\AgdaBound{𝑪}%
\>[29]\AgdaSymbol{→}\AgdaSpace{}%
\AgdaFunction{hom}\AgdaSpace{}%
\AgdaBound{𝑨}\AgdaSpace{}%
\AgdaBound{𝑪}\<%
\\
\>[2]\AgdaFunction{∘-hom}\AgdaSpace{}%
\AgdaSymbol{(}\AgdaBound{h}\AgdaSpace{}%
\AgdaOperator{\AgdaInductiveConstructor{,}}\AgdaSpace{}%
\AgdaBound{hhom}\AgdaSymbol{)}\AgdaSpace{}%
\AgdaSymbol{(}\AgdaBound{g}\AgdaSpace{}%
\AgdaOperator{\AgdaInductiveConstructor{,}}\AgdaSpace{}%
\AgdaBound{ghom}\AgdaSymbol{)}\AgdaSpace{}%
\AgdaSymbol{=}\AgdaSpace{}%
\AgdaSymbol{(}\AgdaBound{g}\AgdaSpace{}%
\AgdaOperator{\AgdaFunction{⟨∘⟩}}\AgdaSpace{}%
\AgdaBound{h}\AgdaSymbol{)}\AgdaSpace{}%
\AgdaOperator{\AgdaInductiveConstructor{,}}\AgdaSpace{}%
\AgdaFunction{∘-is-hom}\AgdaSpace{}%
\AgdaBound{hhom}\AgdaSpace{}%
\AgdaBound{ghom}\<%
\\
\\[\AgdaEmptyExtraSkip]%
\>[2]\AgdaFunction{∘-epi}\AgdaSpace{}%
\AgdaSymbol{:}\AgdaSpace{}%
\AgdaFunction{epi}\AgdaSpace{}%
\AgdaBound{𝑨}\AgdaSpace{}%
\AgdaBound{𝑩}\AgdaSpace{}%
\AgdaSymbol{→}\AgdaSpace{}%
\AgdaFunction{epi}\AgdaSpace{}%
\AgdaBound{𝑩}\AgdaSpace{}%
\AgdaBound{𝑪}%
\>[29]\AgdaSymbol{→}\AgdaSpace{}%
\AgdaFunction{epi}\AgdaSpace{}%
\AgdaBound{𝑨}\AgdaSpace{}%
\AgdaBound{𝑪}\<%
\\
\>[2]\AgdaFunction{∘-epi}\AgdaSpace{}%
\AgdaSymbol{(}\AgdaBound{h}\AgdaSpace{}%
\AgdaOperator{\AgdaInductiveConstructor{,}}\AgdaSpace{}%
\AgdaBound{hepi}\AgdaSymbol{)}\AgdaSpace{}%
\AgdaSymbol{(}\AgdaBound{g}\AgdaSpace{}%
\AgdaOperator{\AgdaInductiveConstructor{,}}\AgdaSpace{}%
\AgdaBound{gepi}\AgdaSymbol{)}\AgdaSpace{}%
\AgdaSymbol{=}\AgdaSpace{}%
\AgdaSymbol{(}\AgdaBound{g}\AgdaSpace{}%
\AgdaOperator{\AgdaFunction{⟨∘⟩}}\AgdaSpace{}%
\AgdaBound{h}\AgdaSymbol{)}\AgdaSpace{}%
\AgdaOperator{\AgdaInductiveConstructor{,}}\AgdaSpace{}%
\AgdaFunction{∘-is-epi}\AgdaSpace{}%
\AgdaBound{hepi}\AgdaSpace{}%
\AgdaBound{gepi}\<%
\end{code}

\paragraph*{Universe lifting of homomorphisms}
Here we define the identity homomorphism for setoid algebras. Then we prove that the
operations of lifting and lowering of a setoid algebra are homomorphisms.

\begin{code}%
\>[0]\<%
\\
\>[0]\AgdaFunction{𝒾𝒹}\AgdaSpace{}%
\AgdaSymbol{:}\AgdaSpace{}%
\AgdaSymbol{\{}\AgdaBound{𝑨}\AgdaSpace{}%
\AgdaSymbol{:}\AgdaSpace{}%
\AgdaRecord{Algebra}\AgdaSpace{}%
\AgdaGeneralizable{α}\AgdaSpace{}%
\AgdaGeneralizable{ρᵃ}\AgdaSymbol{\}}\AgdaSpace{}%
\AgdaSymbol{→}\AgdaSpace{}%
\AgdaFunction{hom}\AgdaSpace{}%
\AgdaBound{𝑨}\AgdaSpace{}%
\AgdaBound{𝑨}\<%
\\
\>[0]\AgdaFunction{𝒾𝒹}\AgdaSpace{}%
\AgdaSymbol{\{}\AgdaArgument{𝑨}\AgdaSpace{}%
\AgdaSymbol{=}\AgdaSpace{}%
\AgdaBound{𝑨}\AgdaSymbol{\}}\AgdaSpace{}%
\AgdaSymbol{=}\AgdaSpace{}%
\AgdaFunction{𝑖𝑑}\AgdaSpace{}%
\AgdaOperator{\AgdaInductiveConstructor{,}}\AgdaSpace{}%
\AgdaInductiveConstructor{mkhom}\AgdaSpace{}%
\AgdaSymbol{(}\AgdaFunction{reflexive}\AgdaSpace{}%
\AgdaInductiveConstructor{≡.refl}\AgdaSymbol{)}\AgdaSpace{}%
\AgdaKeyword{where}\AgdaSpace{}%
\AgdaKeyword{open}\AgdaSpace{}%
\AgdaModule{Setoid}\AgdaSpace{}%
\AgdaSymbol{(}\AgdaSpace{}%
\AgdaField{Domain}\AgdaSpace{}%
\AgdaBound{𝑨}\AgdaSpace{}%
\AgdaSymbol{)}\AgdaSpace{}%
\AgdaKeyword{using}\AgdaSpace{}%
\AgdaSymbol{(}\AgdaSpace{}%
\AgdaFunction{reflexive}\AgdaSpace{}%
\AgdaSymbol{)}\<%
\\
\\[\AgdaEmptyExtraSkip]%
\>[0]\AgdaKeyword{module}\AgdaSpace{}%
\AgdaModule{\AgdaUnderscore{}}\AgdaSpace{}%
\AgdaSymbol{\{}\AgdaBound{𝑨}\AgdaSpace{}%
\AgdaSymbol{:}\AgdaSpace{}%
\AgdaRecord{Algebra}\AgdaSpace{}%
\AgdaGeneralizable{α}\AgdaSpace{}%
\AgdaGeneralizable{ρᵃ}\AgdaSymbol{\}\{}\AgdaBound{ℓ}\AgdaSpace{}%
\AgdaSymbol{:}\AgdaSpace{}%
\AgdaPostulate{Level}\AgdaSymbol{\}}\AgdaSpace{}%
\AgdaKeyword{where}\<%
\\
\>[0][@{}l@{\AgdaIndent{0}}]%
\>[1]\AgdaKeyword{open}\AgdaSpace{}%
\AgdaModule{Setoid}\AgdaSpace{}%
\AgdaOperator{\AgdaFunction{𝔻[}}\AgdaSpace{}%
\AgdaBound{𝑨}\AgdaSpace{}%
\AgdaOperator{\AgdaFunction{]}}%
\>[33]\AgdaKeyword{using}\AgdaSpace{}%
\AgdaSymbol{(}\AgdaSpace{}%
\AgdaFunction{reflexive}\AgdaSpace{}%
\AgdaSymbol{)}%
\>[54]\AgdaKeyword{renaming}\AgdaSpace{}%
\AgdaSymbol{(}\AgdaSpace{}%
\AgdaOperator{\AgdaField{\AgdaUnderscore{}≈\AgdaUnderscore{}}}\AgdaSpace{}%
\AgdaSymbol{to}\AgdaSpace{}%
\AgdaOperator{\AgdaField{\AgdaUnderscore{}≈₁\AgdaUnderscore{}}}\AgdaSpace{}%
\AgdaSymbol{;}\AgdaSpace{}%
\AgdaFunction{refl}\AgdaSpace{}%
\AgdaSymbol{to}\AgdaSpace{}%
\AgdaFunction{refl₁}\AgdaSpace{}%
\AgdaSymbol{)}\<%
\\
\>[1]\AgdaKeyword{open}\AgdaSpace{}%
\AgdaModule{Setoid}\AgdaSpace{}%
\AgdaOperator{\AgdaFunction{𝔻[}}\AgdaSpace{}%
\AgdaFunction{Lift-Algˡ}\AgdaSpace{}%
\AgdaBound{𝑨}\AgdaSpace{}%
\AgdaBound{ℓ}\AgdaSpace{}%
\AgdaOperator{\AgdaFunction{]}}%
\>[33]\AgdaKeyword{using}\AgdaSpace{}%
\AgdaSymbol{()}%
\>[54]\AgdaKeyword{renaming}\AgdaSpace{}%
\AgdaSymbol{(}\AgdaSpace{}%
\AgdaOperator{\AgdaField{\AgdaUnderscore{}≈\AgdaUnderscore{}}}\AgdaSpace{}%
\AgdaSymbol{to}\AgdaSpace{}%
\AgdaOperator{\AgdaField{\AgdaUnderscore{}≈ˡ\AgdaUnderscore{}}}\AgdaSpace{}%
\AgdaSymbol{;}\AgdaSpace{}%
\AgdaFunction{refl}\AgdaSpace{}%
\AgdaSymbol{to}\AgdaSpace{}%
\AgdaFunction{reflˡ}\AgdaSymbol{)}\<%
\\
\>[1]\AgdaKeyword{open}\AgdaSpace{}%
\AgdaModule{Setoid}\AgdaSpace{}%
\AgdaOperator{\AgdaFunction{𝔻[}}\AgdaSpace{}%
\AgdaFunction{Lift-Algʳ}\AgdaSpace{}%
\AgdaBound{𝑨}\AgdaSpace{}%
\AgdaBound{ℓ}\AgdaSpace{}%
\AgdaOperator{\AgdaFunction{]}}%
\>[33]\AgdaKeyword{using}\AgdaSpace{}%
\AgdaSymbol{()}%
\>[54]\AgdaKeyword{renaming}\AgdaSpace{}%
\AgdaSymbol{(}\AgdaSpace{}%
\AgdaOperator{\AgdaField{\AgdaUnderscore{}≈\AgdaUnderscore{}}}\AgdaSpace{}%
\AgdaSymbol{to}\AgdaSpace{}%
\AgdaOperator{\AgdaField{\AgdaUnderscore{}≈ʳ\AgdaUnderscore{}}}\AgdaSpace{}%
\AgdaSymbol{;}\AgdaSpace{}%
\AgdaFunction{refl}\AgdaSpace{}%
\AgdaSymbol{to}\AgdaSpace{}%
\AgdaFunction{reflʳ}\AgdaSymbol{)}\<%
\\
\>[1]\AgdaKeyword{open}\AgdaSpace{}%
\AgdaModule{Level}\<%
\\
\\[\AgdaEmptyExtraSkip]%
\>[1]\AgdaFunction{ToLiftˡ}\AgdaSpace{}%
\AgdaSymbol{:}\AgdaSpace{}%
\AgdaFunction{hom}\AgdaSpace{}%
\AgdaBound{𝑨}\AgdaSpace{}%
\AgdaSymbol{(}\AgdaFunction{Lift-Algˡ}\AgdaSpace{}%
\AgdaBound{𝑨}\AgdaSpace{}%
\AgdaBound{ℓ}\AgdaSymbol{)}\<%
\\
\>[1]\AgdaFunction{ToLiftˡ}\AgdaSpace{}%
\AgdaSymbol{=}\AgdaSpace{}%
\AgdaKeyword{record}\AgdaSpace{}%
\AgdaSymbol{\{}\AgdaSpace{}%
\AgdaField{f}\AgdaSpace{}%
\AgdaSymbol{=}\AgdaSpace{}%
\AgdaInductiveConstructor{lift}\AgdaSpace{}%
\AgdaSymbol{;}\AgdaSpace{}%
\AgdaField{cong}\AgdaSpace{}%
\AgdaSymbol{=}\AgdaSpace{}%
\AgdaFunction{id}\AgdaSpace{}%
\AgdaSymbol{\}}\AgdaSpace{}%
\AgdaOperator{\AgdaInductiveConstructor{,}}\AgdaSpace{}%
\AgdaInductiveConstructor{mkhom}\AgdaSpace{}%
\AgdaSymbol{(}\AgdaFunction{reflexive}\AgdaSpace{}%
\AgdaInductiveConstructor{≡.refl}\AgdaSymbol{)}\<%
\\
\\[\AgdaEmptyExtraSkip]%
\>[1]\AgdaFunction{FromLiftˡ}\AgdaSpace{}%
\AgdaSymbol{:}\AgdaSpace{}%
\AgdaFunction{hom}\AgdaSpace{}%
\AgdaSymbol{(}\AgdaFunction{Lift-Algˡ}\AgdaSpace{}%
\AgdaBound{𝑨}\AgdaSpace{}%
\AgdaBound{ℓ}\AgdaSymbol{)}\AgdaSpace{}%
\AgdaBound{𝑨}\<%
\\
\>[1]\AgdaFunction{FromLiftˡ}\AgdaSpace{}%
\AgdaSymbol{=}\AgdaSpace{}%
\AgdaKeyword{record}\AgdaSpace{}%
\AgdaSymbol{\{}\AgdaSpace{}%
\AgdaField{f}\AgdaSpace{}%
\AgdaSymbol{=}\AgdaSpace{}%
\AgdaField{lower}\AgdaSpace{}%
\AgdaSymbol{;}\AgdaSpace{}%
\AgdaField{cong}\AgdaSpace{}%
\AgdaSymbol{=}\AgdaSpace{}%
\AgdaFunction{id}\AgdaSpace{}%
\AgdaSymbol{\}}\AgdaSpace{}%
\AgdaOperator{\AgdaInductiveConstructor{,}}\AgdaSpace{}%
\AgdaInductiveConstructor{mkhom}\AgdaSpace{}%
\AgdaFunction{reflˡ}\<%
\\
\\[\AgdaEmptyExtraSkip]%
\>[1]\AgdaFunction{ToFromLiftˡ}\AgdaSpace{}%
\AgdaSymbol{:}\AgdaSpace{}%
\AgdaSymbol{∀}\AgdaSpace{}%
\AgdaBound{b}\AgdaSpace{}%
\AgdaSymbol{→}%
\>[22]\AgdaOperator{\AgdaFunction{∣}}\AgdaSpace{}%
\AgdaFunction{ToLiftˡ}\AgdaSpace{}%
\AgdaOperator{\AgdaFunction{∣}}\AgdaSpace{}%
\AgdaOperator{\AgdaField{⟨\$⟩}}\AgdaSpace{}%
\AgdaSymbol{(}\AgdaOperator{\AgdaFunction{∣}}\AgdaSpace{}%
\AgdaFunction{FromLiftˡ}\AgdaSpace{}%
\AgdaOperator{\AgdaFunction{∣}}\AgdaSpace{}%
\AgdaOperator{\AgdaField{⟨\$⟩}}\AgdaSpace{}%
\AgdaBound{b}\AgdaSymbol{)}\AgdaSpace{}%
\AgdaOperator{\AgdaFunction{≈ˡ}}\AgdaSpace{}%
\AgdaBound{b}\<%
\\
\>[1]\AgdaFunction{ToFromLiftˡ}\AgdaSpace{}%
\AgdaBound{b}\AgdaSpace{}%
\AgdaSymbol{=}\AgdaSpace{}%
\AgdaFunction{refl₁}\<%
\\
\\[\AgdaEmptyExtraSkip]%
\>[1]\AgdaFunction{FromToLiftˡ}\AgdaSpace{}%
\AgdaSymbol{:}\AgdaSpace{}%
\AgdaSymbol{∀}\AgdaSpace{}%
\AgdaBound{a}\AgdaSpace{}%
\AgdaSymbol{→}\AgdaSpace{}%
\AgdaOperator{\AgdaFunction{∣}}\AgdaSpace{}%
\AgdaFunction{FromLiftˡ}\AgdaSpace{}%
\AgdaOperator{\AgdaFunction{∣}}\AgdaSpace{}%
\AgdaOperator{\AgdaField{⟨\$⟩}}\AgdaSpace{}%
\AgdaSymbol{(}\AgdaOperator{\AgdaFunction{∣}}\AgdaSpace{}%
\AgdaFunction{ToLiftˡ}\AgdaSpace{}%
\AgdaOperator{\AgdaFunction{∣}}\AgdaSpace{}%
\AgdaOperator{\AgdaField{⟨\$⟩}}\AgdaSpace{}%
\AgdaBound{a}\AgdaSymbol{)}\AgdaSpace{}%
\AgdaOperator{\AgdaFunction{≈₁}}\AgdaSpace{}%
\AgdaBound{a}\<%
\\
\>[1]\AgdaFunction{FromToLiftˡ}\AgdaSpace{}%
\AgdaBound{a}\AgdaSpace{}%
\AgdaSymbol{=}\AgdaSpace{}%
\AgdaFunction{refl₁}\<%
\\
\\[\AgdaEmptyExtraSkip]%
\>[1]\AgdaFunction{ToLiftʳ}\AgdaSpace{}%
\AgdaSymbol{:}\AgdaSpace{}%
\AgdaFunction{hom}\AgdaSpace{}%
\AgdaBound{𝑨}\AgdaSpace{}%
\AgdaSymbol{(}\AgdaFunction{Lift-Algʳ}\AgdaSpace{}%
\AgdaBound{𝑨}\AgdaSpace{}%
\AgdaBound{ℓ}\AgdaSymbol{)}\<%
\\
\>[1]\AgdaFunction{ToLiftʳ}\AgdaSpace{}%
\AgdaSymbol{=}\AgdaSpace{}%
\AgdaKeyword{record}\AgdaSpace{}%
\AgdaSymbol{\{}\AgdaSpace{}%
\AgdaField{f}\AgdaSpace{}%
\AgdaSymbol{=}\AgdaSpace{}%
\AgdaFunction{id}\AgdaSpace{}%
\AgdaSymbol{;}\AgdaSpace{}%
\AgdaField{cong}\AgdaSpace{}%
\AgdaSymbol{=}\AgdaSpace{}%
\AgdaInductiveConstructor{lift}\AgdaSpace{}%
\AgdaSymbol{\}}\AgdaSpace{}%
\AgdaOperator{\AgdaInductiveConstructor{,}}\AgdaSpace{}%
\AgdaInductiveConstructor{mkhom}\AgdaSpace{}%
\AgdaSymbol{(}\AgdaInductiveConstructor{lift}\AgdaSpace{}%
\AgdaSymbol{(}\AgdaFunction{reflexive}\AgdaSpace{}%
\AgdaInductiveConstructor{≡.refl}\AgdaSymbol{))}\<%
\\
\\[\AgdaEmptyExtraSkip]%
\>[1]\AgdaFunction{FromLiftʳ}\AgdaSpace{}%
\AgdaSymbol{:}\AgdaSpace{}%
\AgdaFunction{hom}\AgdaSpace{}%
\AgdaSymbol{(}\AgdaFunction{Lift-Algʳ}\AgdaSpace{}%
\AgdaBound{𝑨}\AgdaSpace{}%
\AgdaBound{ℓ}\AgdaSymbol{)}\AgdaSpace{}%
\AgdaBound{𝑨}\<%
\\
\>[1]\AgdaFunction{FromLiftʳ}\AgdaSpace{}%
\AgdaSymbol{=}\AgdaSpace{}%
\AgdaKeyword{record}\AgdaSpace{}%
\AgdaSymbol{\{}\AgdaSpace{}%
\AgdaField{f}\AgdaSpace{}%
\AgdaSymbol{=}\AgdaSpace{}%
\AgdaFunction{id}\AgdaSpace{}%
\AgdaSymbol{;}\AgdaSpace{}%
\AgdaField{cong}\AgdaSpace{}%
\AgdaSymbol{=}\AgdaSpace{}%
\AgdaField{lower}\AgdaSpace{}%
\AgdaSymbol{\}}\AgdaSpace{}%
\AgdaOperator{\AgdaInductiveConstructor{,}}\AgdaSpace{}%
\AgdaInductiveConstructor{mkhom}\AgdaSpace{}%
\AgdaFunction{reflˡ}\<%
\\
\\[\AgdaEmptyExtraSkip]%
\>[1]\AgdaFunction{ToFromLiftʳ}\AgdaSpace{}%
\AgdaSymbol{:}\AgdaSpace{}%
\AgdaSymbol{∀}\AgdaSpace{}%
\AgdaBound{b}\AgdaSpace{}%
\AgdaSymbol{→}\AgdaSpace{}%
\AgdaOperator{\AgdaFunction{∣}}\AgdaSpace{}%
\AgdaFunction{ToLiftʳ}\AgdaSpace{}%
\AgdaOperator{\AgdaFunction{∣}}\AgdaSpace{}%
\AgdaOperator{\AgdaField{⟨\$⟩}}\AgdaSpace{}%
\AgdaSymbol{(}\AgdaOperator{\AgdaFunction{∣}}\AgdaSpace{}%
\AgdaFunction{FromLiftʳ}\AgdaSpace{}%
\AgdaOperator{\AgdaFunction{∣}}\AgdaSpace{}%
\AgdaOperator{\AgdaField{⟨\$⟩}}\AgdaSpace{}%
\AgdaBound{b}\AgdaSymbol{)}\AgdaSpace{}%
\AgdaOperator{\AgdaFunction{≈ʳ}}\AgdaSpace{}%
\AgdaBound{b}\<%
\\
\>[1]\AgdaFunction{ToFromLiftʳ}\AgdaSpace{}%
\AgdaBound{b}\AgdaSpace{}%
\AgdaSymbol{=}\AgdaSpace{}%
\AgdaInductiveConstructor{lift}\AgdaSpace{}%
\AgdaFunction{refl₁}\<%
\\
\\[\AgdaEmptyExtraSkip]%
\>[1]\AgdaFunction{FromToLiftʳ}\AgdaSpace{}%
\AgdaSymbol{:}\AgdaSpace{}%
\AgdaSymbol{∀}\AgdaSpace{}%
\AgdaBound{a}\AgdaSpace{}%
\AgdaSymbol{→}\AgdaSpace{}%
\AgdaOperator{\AgdaFunction{∣}}\AgdaSpace{}%
\AgdaFunction{FromLiftʳ}\AgdaSpace{}%
\AgdaOperator{\AgdaFunction{∣}}\AgdaSpace{}%
\AgdaOperator{\AgdaField{⟨\$⟩}}\AgdaSpace{}%
\AgdaSymbol{(}\AgdaOperator{\AgdaFunction{∣}}\AgdaSpace{}%
\AgdaFunction{ToLiftʳ}\AgdaSpace{}%
\AgdaOperator{\AgdaFunction{∣}}\AgdaSpace{}%
\AgdaOperator{\AgdaField{⟨\$⟩}}\AgdaSpace{}%
\AgdaBound{a}\AgdaSymbol{)}\AgdaSpace{}%
\AgdaOperator{\AgdaFunction{≈₁}}\AgdaSpace{}%
\AgdaBound{a}\<%
\\
\>[1]\AgdaFunction{FromToLiftʳ}\AgdaSpace{}%
\AgdaBound{a}\AgdaSpace{}%
\AgdaSymbol{=}\AgdaSpace{}%
\AgdaFunction{refl₁}\<%
\\
\\[\AgdaEmptyExtraSkip]%
\\[\AgdaEmptyExtraSkip]%
\>[0]\AgdaKeyword{module}\AgdaSpace{}%
\AgdaModule{\AgdaUnderscore{}}\AgdaSpace{}%
\AgdaSymbol{\{}\AgdaBound{𝑨}\AgdaSpace{}%
\AgdaSymbol{:}\AgdaSpace{}%
\AgdaRecord{Algebra}\AgdaSpace{}%
\AgdaGeneralizable{α}\AgdaSpace{}%
\AgdaGeneralizable{ρᵃ}\AgdaSymbol{\}\{}\AgdaBound{ℓ}\AgdaSpace{}%
\AgdaBound{r}\AgdaSpace{}%
\AgdaSymbol{:}\AgdaSpace{}%
\AgdaPostulate{Level}\AgdaSymbol{\}}\AgdaSpace{}%
\AgdaKeyword{where}\<%
\\
\>[0][@{}l@{\AgdaIndent{0}}]%
\>[1]\AgdaKeyword{open}%
\>[7]\AgdaModule{Setoid}\AgdaSpace{}%
\AgdaOperator{\AgdaFunction{𝔻[}}\AgdaSpace{}%
\AgdaBound{𝑨}\AgdaSpace{}%
\AgdaOperator{\AgdaFunction{]}}%
\>[35]\AgdaKeyword{using}\AgdaSpace{}%
\AgdaSymbol{(}\AgdaSpace{}%
\AgdaFunction{refl}\AgdaSpace{}%
\AgdaSymbol{)}\<%
\\
\>[1]\AgdaKeyword{open}%
\>[7]\AgdaModule{Setoid}\AgdaSpace{}%
\AgdaOperator{\AgdaFunction{𝔻[}}\AgdaSpace{}%
\AgdaFunction{Lift-Alg}\AgdaSpace{}%
\AgdaBound{𝑨}\AgdaSpace{}%
\AgdaBound{ℓ}\AgdaSpace{}%
\AgdaBound{r}\AgdaSpace{}%
\AgdaOperator{\AgdaFunction{]}}%
\>[35]\AgdaKeyword{using}\AgdaSpace{}%
\AgdaSymbol{(}\AgdaSpace{}%
\AgdaOperator{\AgdaField{\AgdaUnderscore{}≈\AgdaUnderscore{}}}\AgdaSpace{}%
\AgdaSymbol{)}\<%
\\
\>[1]\AgdaKeyword{open}%
\>[7]\AgdaModule{Level}\<%
\\
\\[\AgdaEmptyExtraSkip]%
\>[1]\AgdaFunction{ToLift}\AgdaSpace{}%
\AgdaSymbol{:}\AgdaSpace{}%
\AgdaFunction{hom}\AgdaSpace{}%
\AgdaBound{𝑨}\AgdaSpace{}%
\AgdaSymbol{(}\AgdaFunction{Lift-Alg}\AgdaSpace{}%
\AgdaBound{𝑨}\AgdaSpace{}%
\AgdaBound{ℓ}\AgdaSpace{}%
\AgdaBound{r}\AgdaSymbol{)}\<%
\\
\>[1]\AgdaFunction{ToLift}\AgdaSpace{}%
\AgdaSymbol{=}\AgdaSpace{}%
\AgdaFunction{∘-hom}\AgdaSpace{}%
\AgdaFunction{ToLiftˡ}\AgdaSpace{}%
\AgdaFunction{ToLiftʳ}\<%
\\
\\[\AgdaEmptyExtraSkip]%
\>[1]\AgdaFunction{FromLift}\AgdaSpace{}%
\AgdaSymbol{:}\AgdaSpace{}%
\AgdaFunction{hom}\AgdaSpace{}%
\AgdaSymbol{(}\AgdaFunction{Lift-Alg}\AgdaSpace{}%
\AgdaBound{𝑨}\AgdaSpace{}%
\AgdaBound{ℓ}\AgdaSpace{}%
\AgdaBound{r}\AgdaSymbol{)}\AgdaSpace{}%
\AgdaBound{𝑨}\<%
\\
\>[1]\AgdaFunction{FromLift}\AgdaSpace{}%
\AgdaSymbol{=}\AgdaSpace{}%
\AgdaFunction{∘-hom}\AgdaSpace{}%
\AgdaFunction{FromLiftʳ}\AgdaSpace{}%
\AgdaFunction{FromLiftˡ}\<%
\\
\\[\AgdaEmptyExtraSkip]%
\>[1]\AgdaFunction{ToFromLift}\AgdaSpace{}%
\AgdaSymbol{:}\AgdaSpace{}%
\AgdaSymbol{∀}\AgdaSpace{}%
\AgdaBound{b}\AgdaSpace{}%
\AgdaSymbol{→}\AgdaSpace{}%
\AgdaOperator{\AgdaFunction{∣}}\AgdaSpace{}%
\AgdaFunction{ToLift}\AgdaSpace{}%
\AgdaOperator{\AgdaFunction{∣}}\AgdaSpace{}%
\AgdaOperator{\AgdaField{⟨\$⟩}}\AgdaSpace{}%
\AgdaSymbol{(}\AgdaOperator{\AgdaFunction{∣}}\AgdaSpace{}%
\AgdaFunction{FromLift}\AgdaSpace{}%
\AgdaOperator{\AgdaFunction{∣}}\AgdaSpace{}%
\AgdaOperator{\AgdaField{⟨\$⟩}}\AgdaSpace{}%
\AgdaBound{b}\AgdaSymbol{)}\AgdaSpace{}%
\AgdaOperator{\AgdaFunction{≈}}\AgdaSpace{}%
\AgdaBound{b}\<%
\\
\>[1]\AgdaFunction{ToFromLift}\AgdaSpace{}%
\AgdaBound{b}\AgdaSpace{}%
\AgdaSymbol{=}\AgdaSpace{}%
\AgdaInductiveConstructor{lift}\AgdaSpace{}%
\AgdaFunction{refl}\<%
\\
\\[\AgdaEmptyExtraSkip]%
\>[1]\AgdaFunction{ToLift-epi}\AgdaSpace{}%
\AgdaSymbol{:}\AgdaSpace{}%
\AgdaFunction{epi}\AgdaSpace{}%
\AgdaBound{𝑨}\AgdaSpace{}%
\AgdaSymbol{(}\AgdaFunction{Lift-Alg}\AgdaSpace{}%
\AgdaBound{𝑨}\AgdaSpace{}%
\AgdaBound{ℓ}\AgdaSpace{}%
\AgdaBound{r}\AgdaSymbol{)}\<%
\\
\>[1]\AgdaFunction{ToLift-epi}\AgdaSpace{}%
\AgdaSymbol{=}\AgdaSpace{}%
\AgdaOperator{\AgdaFunction{∣}}\AgdaSpace{}%
\AgdaFunction{ToLift}\AgdaSpace{}%
\AgdaOperator{\AgdaFunction{∣}}\AgdaSpace{}%
\AgdaOperator{\AgdaInductiveConstructor{,}}%
\>[28]\AgdaKeyword{record}\AgdaSpace{}%
\AgdaSymbol{\{}\AgdaSpace{}%
\AgdaField{isHom}\AgdaSpace{}%
\AgdaSymbol{=}\AgdaSpace{}%
\AgdaOperator{\AgdaFunction{∥}}\AgdaSpace{}%
\AgdaFunction{ToLift}\AgdaSpace{}%
\AgdaOperator{\AgdaFunction{∥}}\<%
\\
\>[28]\AgdaSymbol{;}\AgdaSpace{}%
\AgdaField{isSurjective}\AgdaSpace{}%
\AgdaSymbol{=}\AgdaSpace{}%
\AgdaSymbol{λ}\AgdaSpace{}%
\AgdaSymbol{\{}\AgdaBound{y}\AgdaSymbol{\}}\AgdaSpace{}%
\AgdaSymbol{→}\AgdaSpace{}%
\AgdaInductiveConstructor{eq}\AgdaSpace{}%
\AgdaSymbol{(}\AgdaOperator{\AgdaFunction{∣}}\AgdaSpace{}%
\AgdaFunction{FromLift}\AgdaSpace{}%
\AgdaOperator{\AgdaFunction{∣}}\AgdaSpace{}%
\AgdaOperator{\AgdaField{⟨\$⟩}}\AgdaSpace{}%
\AgdaBound{y}\AgdaSymbol{)}\AgdaSpace{}%
\AgdaSymbol{(}\AgdaFunction{ToFromLift}\AgdaSpace{}%
\AgdaBound{y}\AgdaSymbol{)}\AgdaSpace{}%
\AgdaSymbol{\}}\<%
\end{code}

\paragraph*{Homomorphisms of product algebras}
Suppose we have an algebra \ab{𝑨}, a type \ab I : \ap{Type} \ab{𝓘}, and a family \ab{ℬ} :
\ab I \as{→} \ar{Algebra} \ab{β} \ab{ρᵇ} of algebras.
We sometimes refer to the inhabitants of \ab{I} as \emph{indices}, and call \ab{ℬ} an
\defn{indexed family of algebras}. If in addition we have a family \ab{𝒽} : (\ab i : \ab
I) → \af{hom} \ab{𝑨} (\ab{ℬ} \ab i) of homomorphisms, then we can construct a homomorphism
from \ab{𝑨} to the product \af{⨅} \ab{ℬ} in the natural way.  We codify the latter in
dependent type theory as follows.

\begin{code}%
\>[0]\<%
\\
\>[0]\AgdaKeyword{module}\AgdaSpace{}%
\AgdaModule{\AgdaUnderscore{}}\AgdaSpace{}%
\AgdaSymbol{\{}\AgdaBound{ι}\AgdaSpace{}%
\AgdaSymbol{:}\AgdaSpace{}%
\AgdaPostulate{Level}\AgdaSymbol{\}\{}\AgdaBound{I}\AgdaSpace{}%
\AgdaSymbol{:}\AgdaSpace{}%
\AgdaPrimitive{Type}\AgdaSpace{}%
\AgdaBound{ι}\AgdaSymbol{\}\{}\AgdaBound{𝑨}\AgdaSpace{}%
\AgdaSymbol{:}\AgdaSpace{}%
\AgdaRecord{Algebra}\AgdaSpace{}%
\AgdaGeneralizable{α}\AgdaSpace{}%
\AgdaGeneralizable{ρᵃ}\AgdaSymbol{\}(}\AgdaBound{ℬ}\AgdaSpace{}%
\AgdaSymbol{:}\AgdaSpace{}%
\AgdaBound{I}\AgdaSpace{}%
\AgdaSymbol{→}\AgdaSpace{}%
\AgdaRecord{Algebra}\AgdaSpace{}%
\AgdaGeneralizable{β}\AgdaSpace{}%
\AgdaGeneralizable{ρᵇ}\AgdaSymbol{)}%
\>[74]\AgdaKeyword{where}\<%
\\
\>[0][@{}l@{\AgdaIndent{0}}]%
\>[1]\AgdaFunction{⨅-hom-co}\AgdaSpace{}%
\AgdaSymbol{:}\AgdaSpace{}%
\AgdaSymbol{(∀(}\AgdaBound{i}\AgdaSpace{}%
\AgdaSymbol{:}\AgdaSpace{}%
\AgdaBound{I}\AgdaSymbol{)}\AgdaSpace{}%
\AgdaSymbol{→}\AgdaSpace{}%
\AgdaFunction{hom}\AgdaSpace{}%
\AgdaBound{𝑨}\AgdaSpace{}%
\AgdaSymbol{(}\AgdaBound{ℬ}\AgdaSpace{}%
\AgdaBound{i}\AgdaSymbol{))}\AgdaSpace{}%
\AgdaSymbol{→}\AgdaSpace{}%
\AgdaFunction{hom}\AgdaSpace{}%
\AgdaBound{𝑨}\AgdaSpace{}%
\AgdaSymbol{(}\AgdaFunction{⨅}\AgdaSpace{}%
\AgdaBound{ℬ}\AgdaSymbol{)}\<%
\\
\>[1]\AgdaFunction{⨅-hom-co}\AgdaSpace{}%
\AgdaBound{𝒽}\AgdaSpace{}%
\AgdaSymbol{=}\AgdaSpace{}%
\AgdaFunction{h}\AgdaSpace{}%
\AgdaOperator{\AgdaInductiveConstructor{,}}\AgdaSpace{}%
\AgdaFunction{hhom}\<%
\\
\>[1][@{}l@{\AgdaIndent{0}}]%
\>[2]\AgdaKeyword{where}\<%
\\
\>[2]\AgdaFunction{h}\AgdaSpace{}%
\AgdaSymbol{:}\AgdaSpace{}%
\AgdaOperator{\AgdaFunction{𝔻[}}\AgdaSpace{}%
\AgdaBound{𝑨}\AgdaSpace{}%
\AgdaOperator{\AgdaFunction{]}}\AgdaSpace{}%
\AgdaOperator{\AgdaRecord{⟶}}\AgdaSpace{}%
\AgdaOperator{\AgdaFunction{𝔻[}}\AgdaSpace{}%
\AgdaFunction{⨅}\AgdaSpace{}%
\AgdaBound{ℬ}\AgdaSpace{}%
\AgdaOperator{\AgdaFunction{]}}\<%
\\
\>[2]\AgdaFunction{h}\AgdaSpace{}%
\AgdaOperator{\AgdaField{⟨\$⟩}}\AgdaSpace{}%
\AgdaBound{a}\AgdaSpace{}%
\AgdaSymbol{=}\AgdaSpace{}%
\AgdaSymbol{λ}\AgdaSpace{}%
\AgdaBound{i}\AgdaSpace{}%
\AgdaSymbol{→}\AgdaSpace{}%
\AgdaOperator{\AgdaFunction{∣}}\AgdaSpace{}%
\AgdaBound{𝒽}\AgdaSpace{}%
\AgdaBound{i}\AgdaSpace{}%
\AgdaOperator{\AgdaFunction{∣}}\AgdaSpace{}%
\AgdaOperator{\AgdaField{⟨\$⟩}}\AgdaSpace{}%
\AgdaBound{a}\<%
\\
\>[2]\AgdaField{cong}\AgdaSpace{}%
\AgdaFunction{h}\AgdaSpace{}%
\AgdaBound{xy}\AgdaSpace{}%
\AgdaBound{i}\AgdaSpace{}%
\AgdaSymbol{=}\AgdaSpace{}%
\AgdaField{cong}\AgdaSpace{}%
\AgdaOperator{\AgdaFunction{∣}}\AgdaSpace{}%
\AgdaBound{𝒽}\AgdaSpace{}%
\AgdaBound{i}\AgdaSpace{}%
\AgdaOperator{\AgdaFunction{∣}}\AgdaSpace{}%
\AgdaBound{xy}\<%
\\
\>[2]\AgdaFunction{hhom}\AgdaSpace{}%
\AgdaSymbol{:}\AgdaSpace{}%
\AgdaRecord{IsHom}\AgdaSpace{}%
\AgdaBound{𝑨}\AgdaSpace{}%
\AgdaSymbol{(}\AgdaFunction{⨅}\AgdaSpace{}%
\AgdaBound{ℬ}\AgdaSymbol{)}\AgdaSpace{}%
\AgdaFunction{h}\<%
\\
\>[2]\AgdaField{compatible}\AgdaSpace{}%
\AgdaFunction{hhom}\AgdaSpace{}%
\AgdaSymbol{=}\AgdaSpace{}%
\AgdaSymbol{λ}\AgdaSpace{}%
\AgdaBound{i}\AgdaSpace{}%
\AgdaSymbol{→}\AgdaSpace{}%
\AgdaField{compatible}\AgdaSpace{}%
\AgdaOperator{\AgdaFunction{∥}}\AgdaSpace{}%
\AgdaBound{𝒽}\AgdaSpace{}%
\AgdaBound{i}\AgdaSpace{}%
\AgdaOperator{\AgdaFunction{∥}}\<%
\end{code}

\paragraph*{Factorization of homomorphisms}
\fi      %%% END LONG VERSION ONLY SECTION
Another basic fact about homomorphisms that we formalize in the \agdaalgebras library
(as the type \af{HomFactor}) is the following factorization theorem: if \ab g : \af{hom}
\ab{𝑨} \ab{𝑩}, \ab h : \af{hom} \ab{𝑨} \ab{𝑪}, \ab h is surjective, and \af{ker} \ab h
\aof{⊆} \af{ker} \ab g, then there exists \ab{φ} : \af{hom} \ab{𝑪} \ab{𝑩} such that \ab g
= \ab{φ} \aof{∘} \ab h.
\ifshort\else

\begin{code}%
\>[0]\<%
\\
\>[0]\AgdaKeyword{module}\AgdaSpace{}%
\AgdaModule{\AgdaUnderscore{}}%
\>[2098I]\AgdaSymbol{\{}\AgdaBound{𝑨}\AgdaSpace{}%
\AgdaSymbol{:}\AgdaSpace{}%
\AgdaRecord{Algebra}\AgdaSpace{}%
\AgdaGeneralizable{α}\AgdaSpace{}%
\AgdaGeneralizable{ρᵃ}\AgdaSymbol{\}(}\AgdaBound{𝑩}\AgdaSpace{}%
\AgdaSymbol{:}\AgdaSpace{}%
\AgdaRecord{Algebra}\AgdaSpace{}%
\AgdaGeneralizable{β}\AgdaSpace{}%
\AgdaGeneralizable{ρᵇ}\AgdaSymbol{)\{}\AgdaBound{𝑪}\AgdaSpace{}%
\AgdaSymbol{:}\AgdaSpace{}%
\AgdaRecord{Algebra}\AgdaSpace{}%
\AgdaGeneralizable{γ}\AgdaSpace{}%
\AgdaGeneralizable{ρᶜ}\AgdaSymbol{\}}\<%
\\
\>[.][@{}l@{}]\<[2098I]%
\>[9]\AgdaSymbol{(}\AgdaBound{gh}\AgdaSpace{}%
\AgdaSymbol{:}\AgdaSpace{}%
\AgdaFunction{hom}\AgdaSpace{}%
\AgdaBound{𝑨}\AgdaSpace{}%
\AgdaBound{𝑩}\AgdaSymbol{)(}\AgdaBound{hh}\AgdaSpace{}%
\AgdaSymbol{:}\AgdaSpace{}%
\AgdaFunction{hom}\AgdaSpace{}%
\AgdaBound{𝑨}\AgdaSpace{}%
\AgdaBound{𝑪}\AgdaSymbol{)}\AgdaSpace{}%
\AgdaKeyword{where}\<%
\\
\>[0][@{}l@{\AgdaIndent{0}}]%
\>[1]\AgdaKeyword{open}\AgdaSpace{}%
\AgdaModule{Setoid}\AgdaSpace{}%
\AgdaOperator{\AgdaFunction{𝔻[}}\AgdaSpace{}%
\AgdaBound{𝑩}\AgdaSpace{}%
\AgdaOperator{\AgdaFunction{]}}\AgdaSpace{}%
\AgdaKeyword{using}\AgdaSpace{}%
\AgdaSymbol{()}\AgdaSpace{}%
\AgdaKeyword{renaming}\AgdaSpace{}%
\AgdaSymbol{(}\AgdaSpace{}%
\AgdaOperator{\AgdaField{\AgdaUnderscore{}≈\AgdaUnderscore{}}}\AgdaSpace{}%
\AgdaSymbol{to}\AgdaSpace{}%
\AgdaOperator{\AgdaField{\AgdaUnderscore{}≈₂\AgdaUnderscore{}}}\AgdaSpace{}%
\AgdaSymbol{)}\<%
\\
\>[1]\AgdaKeyword{open}\AgdaSpace{}%
\AgdaModule{Setoid}\AgdaSpace{}%
\AgdaOperator{\AgdaFunction{𝔻[}}\AgdaSpace{}%
\AgdaBound{𝑪}\AgdaSpace{}%
\AgdaOperator{\AgdaFunction{]}}\AgdaSpace{}%
\AgdaKeyword{using}\AgdaSpace{}%
\AgdaSymbol{()}\AgdaSpace{}%
\AgdaKeyword{renaming}\AgdaSpace{}%
\AgdaSymbol{(}\AgdaSpace{}%
\AgdaOperator{\AgdaField{\AgdaUnderscore{}≈\AgdaUnderscore{}}}\AgdaSpace{}%
\AgdaSymbol{to}\AgdaSpace{}%
\AgdaOperator{\AgdaField{\AgdaUnderscore{}≈₃\AgdaUnderscore{}}}\AgdaSpace{}%
\AgdaSymbol{)}\<%
\\
\>[1]\AgdaKeyword{private}\AgdaSpace{}%
\AgdaFunction{gfunc}\AgdaSpace{}%
\AgdaSymbol{=}\AgdaSpace{}%
\AgdaOperator{\AgdaFunction{∣}}\AgdaSpace{}%
\AgdaBound{gh}\AgdaSpace{}%
\AgdaOperator{\AgdaFunction{∣}}\AgdaSpace{}%
\AgdaSymbol{;}\AgdaSpace{}%
\AgdaFunction{g}\AgdaSpace{}%
\AgdaSymbol{=}\AgdaSpace{}%
\AgdaOperator{\AgdaField{\AgdaUnderscore{}⟨\$⟩\AgdaUnderscore{}}}\AgdaSpace{}%
\AgdaFunction{gfunc}\AgdaSpace{}%
\AgdaSymbol{;}\AgdaSpace{}%
\AgdaFunction{hfunc}\AgdaSpace{}%
\AgdaSymbol{=}\AgdaSpace{}%
\AgdaOperator{\AgdaFunction{∣}}\AgdaSpace{}%
\AgdaBound{hh}\AgdaSpace{}%
\AgdaOperator{\AgdaFunction{∣}}\AgdaSpace{}%
\AgdaSymbol{;}\AgdaSpace{}%
\AgdaFunction{h}\AgdaSpace{}%
\AgdaSymbol{=}\AgdaSpace{}%
\AgdaOperator{\AgdaField{\AgdaUnderscore{}⟨\$⟩\AgdaUnderscore{}}}\AgdaSpace{}%
\AgdaFunction{hfunc}\<%
\\
\\[\AgdaEmptyExtraSkip]%
\>[1]\AgdaFunction{HomFactor}\AgdaSpace{}%
\AgdaSymbol{:}%
\>[14]\AgdaFunction{kernel}\AgdaSpace{}%
\AgdaOperator{\AgdaFunction{\AgdaUnderscore{}≈₃\AgdaUnderscore{}}}\AgdaSpace{}%
\AgdaFunction{h}\AgdaSpace{}%
\AgdaOperator{\AgdaFunction{⊆}}\AgdaSpace{}%
\AgdaFunction{kernel}\AgdaSpace{}%
\AgdaOperator{\AgdaFunction{\AgdaUnderscore{}≈₂\AgdaUnderscore{}}}\AgdaSpace{}%
\AgdaFunction{g}\<%
\\
\>[1][@{}l@{\AgdaIndent{0}}]%
\>[2]\AgdaSymbol{→}%
\>[14]\AgdaFunction{IsSurjective}\AgdaSpace{}%
\AgdaFunction{hfunc}\<%
\\
\>[2]\AgdaSymbol{→}%
\>[14]\AgdaFunction{Σ[}\AgdaSpace{}%
\AgdaBound{φ}\AgdaSpace{}%
\AgdaFunction{∈}\AgdaSpace{}%
\AgdaFunction{hom}\AgdaSpace{}%
\AgdaBound{𝑪}\AgdaSpace{}%
\AgdaBound{𝑩}\AgdaSpace{}%
\AgdaFunction{]}\AgdaSpace{}%
\AgdaSymbol{∀}\AgdaSpace{}%
\AgdaBound{a}\AgdaSpace{}%
\AgdaSymbol{→}\AgdaSpace{}%
\AgdaFunction{g}\AgdaSpace{}%
\AgdaBound{a}\AgdaSpace{}%
\AgdaOperator{\AgdaFunction{≈₂}}\AgdaSpace{}%
\AgdaOperator{\AgdaFunction{∣}}\AgdaSpace{}%
\AgdaBound{φ}\AgdaSpace{}%
\AgdaOperator{\AgdaFunction{∣}}\AgdaSpace{}%
\AgdaOperator{\AgdaField{⟨\$⟩}}\AgdaSpace{}%
\AgdaFunction{h}\AgdaSpace{}%
\AgdaBound{a}\<%
\\
\>[1]\AgdaFunction{HomFactor}\AgdaSpace{}%
\AgdaBound{Khg}\AgdaSpace{}%
\AgdaBound{hE}\AgdaSpace{}%
\AgdaSymbol{=}\AgdaSpace{}%
\AgdaSymbol{(}\AgdaFunction{φmap}\AgdaSpace{}%
\AgdaOperator{\AgdaInductiveConstructor{,}}\AgdaSpace{}%
\AgdaFunction{φhom}\AgdaSymbol{)}\AgdaSpace{}%
\AgdaOperator{\AgdaInductiveConstructor{,}}\AgdaSpace{}%
\AgdaFunction{gφh}\<%
\\
\>[1][@{}l@{\AgdaIndent{0}}]%
\>[2]\AgdaKeyword{where}\<%
\\
\>[2]\AgdaFunction{kerpres}\AgdaSpace{}%
\AgdaSymbol{:}\AgdaSpace{}%
\AgdaSymbol{∀}\AgdaSpace{}%
\AgdaBound{a₀}\AgdaSpace{}%
\AgdaBound{a₁}\AgdaSpace{}%
\AgdaSymbol{→}\AgdaSpace{}%
\AgdaFunction{h}\AgdaSpace{}%
\AgdaBound{a₀}\AgdaSpace{}%
\AgdaOperator{\AgdaFunction{≈₃}}\AgdaSpace{}%
\AgdaFunction{h}\AgdaSpace{}%
\AgdaBound{a₁}\AgdaSpace{}%
\AgdaSymbol{→}\AgdaSpace{}%
\AgdaFunction{g}\AgdaSpace{}%
\AgdaBound{a₀}\AgdaSpace{}%
\AgdaOperator{\AgdaFunction{≈₂}}\AgdaSpace{}%
\AgdaFunction{g}\AgdaSpace{}%
\AgdaBound{a₁}\<%
\\
\>[2]\AgdaFunction{kerpres}\AgdaSpace{}%
\AgdaBound{a₀}\AgdaSpace{}%
\AgdaBound{a₁}\AgdaSpace{}%
\AgdaBound{hyp}\AgdaSpace{}%
\AgdaSymbol{=}\AgdaSpace{}%
\AgdaBound{Khg}\AgdaSpace{}%
\AgdaBound{hyp}\<%
\\
\\[\AgdaEmptyExtraSkip]%
\>[2]\AgdaFunction{h⁻¹}\AgdaSpace{}%
\AgdaSymbol{:}\AgdaSpace{}%
\AgdaOperator{\AgdaFunction{𝕌[}}\AgdaSpace{}%
\AgdaBound{𝑪}\AgdaSpace{}%
\AgdaOperator{\AgdaFunction{]}}\AgdaSpace{}%
\AgdaSymbol{→}\AgdaSpace{}%
\AgdaOperator{\AgdaFunction{𝕌[}}\AgdaSpace{}%
\AgdaBound{𝑨}\AgdaSpace{}%
\AgdaOperator{\AgdaFunction{]}}\<%
\\
\>[2]\AgdaFunction{h⁻¹}\AgdaSpace{}%
\AgdaSymbol{=}\AgdaSpace{}%
\AgdaFunction{SurjInv}\AgdaSpace{}%
\AgdaFunction{hfunc}\AgdaSpace{}%
\AgdaBound{hE}\<%
\\
\\[\AgdaEmptyExtraSkip]%
\>[2]\AgdaFunction{η}\AgdaSpace{}%
\AgdaSymbol{:}\AgdaSpace{}%
\AgdaSymbol{∀}\AgdaSpace{}%
\AgdaSymbol{\{}\AgdaBound{c}\AgdaSymbol{\}}\AgdaSpace{}%
\AgdaSymbol{→}\AgdaSpace{}%
\AgdaFunction{h}\AgdaSpace{}%
\AgdaSymbol{(}\AgdaFunction{h⁻¹}\AgdaSpace{}%
\AgdaBound{c}\AgdaSymbol{)}\AgdaSpace{}%
\AgdaOperator{\AgdaFunction{≈₃}}\AgdaSpace{}%
\AgdaBound{c}\<%
\\
\>[2]\AgdaFunction{η}\AgdaSpace{}%
\AgdaSymbol{=}\AgdaSpace{}%
\AgdaFunction{InvIsInverseʳ}\AgdaSpace{}%
\AgdaBound{hE}\<%
\\
\\[\AgdaEmptyExtraSkip]%
\>[2]\AgdaKeyword{open}\AgdaSpace{}%
\AgdaModule{Setoid}\AgdaSpace{}%
\AgdaOperator{\AgdaFunction{𝔻[}}\AgdaSpace{}%
\AgdaBound{𝑪}\AgdaSpace{}%
\AgdaOperator{\AgdaFunction{]}}\AgdaSpace{}%
\AgdaKeyword{using}\AgdaSpace{}%
\AgdaSymbol{(}\AgdaSpace{}%
\AgdaFunction{sym}\AgdaSpace{}%
\AgdaSymbol{;}\AgdaSpace{}%
\AgdaFunction{trans}\AgdaSpace{}%
\AgdaSymbol{)}\<%
\\
\>[2]\AgdaFunction{ζ}\AgdaSpace{}%
\AgdaSymbol{:}\AgdaSpace{}%
\AgdaSymbol{∀\{}\AgdaBound{x}\AgdaSpace{}%
\AgdaBound{y}\AgdaSymbol{\}}\AgdaSpace{}%
\AgdaSymbol{→}\AgdaSpace{}%
\AgdaBound{x}\AgdaSpace{}%
\AgdaOperator{\AgdaFunction{≈₃}}\AgdaSpace{}%
\AgdaBound{y}\AgdaSpace{}%
\AgdaSymbol{→}\AgdaSpace{}%
\AgdaFunction{h}\AgdaSpace{}%
\AgdaSymbol{(}\AgdaFunction{h⁻¹}\AgdaSpace{}%
\AgdaBound{x}\AgdaSymbol{)}\AgdaSpace{}%
\AgdaOperator{\AgdaFunction{≈₃}}\AgdaSpace{}%
\AgdaFunction{h}\AgdaSpace{}%
\AgdaSymbol{(}\AgdaFunction{h⁻¹}\AgdaSpace{}%
\AgdaBound{y}\AgdaSymbol{)}\<%
\\
\>[2]\AgdaFunction{ζ}\AgdaSpace{}%
\AgdaBound{xy}\AgdaSpace{}%
\AgdaSymbol{=}\AgdaSpace{}%
\AgdaFunction{trans}\AgdaSpace{}%
\AgdaFunction{η}\AgdaSpace{}%
\AgdaSymbol{(}\AgdaFunction{trans}\AgdaSpace{}%
\AgdaBound{xy}\AgdaSpace{}%
\AgdaSymbol{(}\AgdaFunction{sym}\AgdaSpace{}%
\AgdaFunction{η}\AgdaSymbol{))}\<%
\\
\\[\AgdaEmptyExtraSkip]%
\>[2]\AgdaFunction{φmap}\AgdaSpace{}%
\AgdaSymbol{:}\AgdaSpace{}%
\AgdaOperator{\AgdaFunction{𝔻[}}\AgdaSpace{}%
\AgdaBound{𝑪}\AgdaSpace{}%
\AgdaOperator{\AgdaFunction{]}}\AgdaSpace{}%
\AgdaOperator{\AgdaRecord{⟶}}\AgdaSpace{}%
\AgdaOperator{\AgdaFunction{𝔻[}}\AgdaSpace{}%
\AgdaBound{𝑩}\AgdaSpace{}%
\AgdaOperator{\AgdaFunction{]}}\<%
\\
\>[2]\AgdaOperator{\AgdaField{\AgdaUnderscore{}⟨\$⟩\AgdaUnderscore{}}}\AgdaSpace{}%
\AgdaFunction{φmap}\AgdaSpace{}%
\AgdaSymbol{=}\AgdaSpace{}%
\AgdaFunction{g}\AgdaSpace{}%
\AgdaOperator{\AgdaFunction{∘}}\AgdaSpace{}%
\AgdaFunction{h⁻¹}\<%
\\
\>[2]\AgdaField{cong}\AgdaSpace{}%
\AgdaFunction{φmap}\AgdaSpace{}%
\AgdaSymbol{=}\AgdaSpace{}%
\AgdaBound{Khg}\AgdaSpace{}%
\AgdaOperator{\AgdaFunction{∘}}\AgdaSpace{}%
\AgdaFunction{ζ}\<%
\\
\\[\AgdaEmptyExtraSkip]%
\>[2]\AgdaKeyword{open}\AgdaSpace{}%
\AgdaModule{\AgdaUnderscore{}⟶\AgdaUnderscore{}}\AgdaSpace{}%
\AgdaFunction{φmap}\AgdaSpace{}%
\AgdaKeyword{using}\AgdaSpace{}%
\AgdaSymbol{()}\AgdaSpace{}%
\AgdaKeyword{renaming}\AgdaSpace{}%
\AgdaSymbol{(}\AgdaField{cong}\AgdaSpace{}%
\AgdaSymbol{to}\AgdaSpace{}%
\AgdaField{φcong}\AgdaSymbol{)}\<%
\\
\\[\AgdaEmptyExtraSkip]%
\>[2]\AgdaFunction{gφh}\AgdaSpace{}%
\AgdaSymbol{:}\AgdaSpace{}%
\AgdaSymbol{(}\AgdaBound{a}\AgdaSpace{}%
\AgdaSymbol{:}\AgdaSpace{}%
\AgdaOperator{\AgdaFunction{𝕌[}}\AgdaSpace{}%
\AgdaBound{𝑨}\AgdaSpace{}%
\AgdaOperator{\AgdaFunction{]}}\AgdaSymbol{)}\AgdaSpace{}%
\AgdaSymbol{→}\AgdaSpace{}%
\AgdaFunction{g}\AgdaSpace{}%
\AgdaBound{a}\AgdaSpace{}%
\AgdaOperator{\AgdaFunction{≈₂}}\AgdaSpace{}%
\AgdaFunction{φmap}\AgdaSpace{}%
\AgdaOperator{\AgdaField{⟨\$⟩}}\AgdaSpace{}%
\AgdaFunction{h}\AgdaSpace{}%
\AgdaBound{a}\<%
\\
\>[2]\AgdaFunction{gφh}\AgdaSpace{}%
\AgdaBound{a}\AgdaSpace{}%
\AgdaSymbol{=}\AgdaSpace{}%
\AgdaBound{Khg}\AgdaSpace{}%
\AgdaSymbol{(}\AgdaFunction{sym}\AgdaSpace{}%
\AgdaFunction{η}\AgdaSymbol{)}\<%
\\
\\[\AgdaEmptyExtraSkip]%
\>[2]\AgdaFunction{φcomp}\AgdaSpace{}%
\AgdaSymbol{:}\AgdaSpace{}%
\AgdaFunction{compatible-map}\AgdaSpace{}%
\AgdaBound{𝑪}\AgdaSpace{}%
\AgdaBound{𝑩}\AgdaSpace{}%
\AgdaFunction{φmap}\<%
\\
\>[2]\AgdaFunction{φcomp}\AgdaSpace{}%
\AgdaSymbol{\{}\AgdaBound{f}\AgdaSymbol{\}\{}\AgdaBound{c}\AgdaSymbol{\}}\AgdaSpace{}%
\AgdaSymbol{=}\<%
\\
\>[2][@{}l@{\AgdaIndent{0}}]%
\>[3]\AgdaOperator{\AgdaFunction{begin}}\<%
\\
\>[3][@{}l@{\AgdaIndent{0}}]%
\>[4]\AgdaFunction{φmap}\AgdaSpace{}%
\AgdaOperator{\AgdaField{⟨\$⟩}}%
\>[14]\AgdaSymbol{(}\AgdaBound{f}\AgdaSpace{}%
\AgdaOperator{\AgdaFunction{̂}}\AgdaSpace{}%
\AgdaBound{𝑪}\AgdaSymbol{)}%
\>[40]\AgdaBound{c}%
\>[48]\AgdaFunction{≈˘⟨}%
\>[53]\AgdaFunction{φcong}\AgdaSpace{}%
\AgdaSymbol{(}\AgdaField{cong}\AgdaSpace{}%
\AgdaSymbol{(}\AgdaField{Interp}\AgdaSpace{}%
\AgdaBound{𝑪}\AgdaSymbol{)}\AgdaSpace{}%
\AgdaSymbol{(}\AgdaInductiveConstructor{≡.refl}\AgdaSpace{}%
\AgdaOperator{\AgdaInductiveConstructor{,}}\AgdaSpace{}%
\AgdaSymbol{λ}\AgdaSpace{}%
\AgdaBound{\AgdaUnderscore{}}\AgdaSpace{}%
\AgdaSymbol{→}\AgdaSpace{}%
\AgdaFunction{η}\AgdaSymbol{))}%
\>[97]\AgdaFunction{⟩}\<%
\\
\>[4]\AgdaFunction{g}\AgdaSymbol{(}\AgdaFunction{h⁻¹}\AgdaSymbol{(}%
\>[14]\AgdaSymbol{(}\AgdaBound{f}\AgdaSpace{}%
\AgdaOperator{\AgdaFunction{̂}}\AgdaSpace{}%
\AgdaBound{𝑪}\AgdaSymbol{)}%
\>[23]\AgdaSymbol{(}\AgdaFunction{h}\AgdaSpace{}%
\AgdaOperator{\AgdaFunction{∘}}%
\>[31]\AgdaFunction{h⁻¹}%
\>[36]\AgdaOperator{\AgdaFunction{∘}}%
\>[39]\AgdaBound{c}%
\>[42]\AgdaSymbol{)))}%
\>[48]\AgdaFunction{≈˘⟨}%
\>[53]\AgdaFunction{φcong}\AgdaSpace{}%
\AgdaSymbol{(}\AgdaField{compatible}\AgdaSpace{}%
\AgdaOperator{\AgdaFunction{∥}}\AgdaSpace{}%
\AgdaBound{hh}\AgdaSpace{}%
\AgdaOperator{\AgdaFunction{∥}}\AgdaSymbol{)}%
\>[97]\AgdaFunction{⟩}\<%
\\
\>[4]\AgdaFunction{g}\AgdaSymbol{(}\AgdaFunction{h⁻¹}\AgdaSymbol{(}\AgdaFunction{h}\AgdaSymbol{(}%
\>[14]\AgdaSymbol{(}\AgdaBound{f}\AgdaSpace{}%
\AgdaOperator{\AgdaFunction{̂}}\AgdaSpace{}%
\AgdaBound{𝑨}\AgdaSymbol{)}%
\>[23]\AgdaSymbol{(}%
\>[31]\AgdaFunction{h⁻¹}%
\>[36]\AgdaOperator{\AgdaFunction{∘}}%
\>[39]\AgdaBound{c}%
\>[42]\AgdaSymbol{))))}%
\>[48]\AgdaFunction{≈˘⟨}%
\>[53]\AgdaFunction{gφh}\AgdaSpace{}%
\AgdaSymbol{((}\AgdaBound{f}\AgdaSpace{}%
\AgdaOperator{\AgdaFunction{̂}}\AgdaSpace{}%
\AgdaBound{𝑨}\AgdaSymbol{)(}\AgdaFunction{h⁻¹}\AgdaSpace{}%
\AgdaOperator{\AgdaFunction{∘}}\AgdaSpace{}%
\AgdaBound{c}\AgdaSymbol{))}%
\>[97]\AgdaFunction{⟩}\<%
\\
\>[4]\AgdaFunction{g}\AgdaSymbol{(}%
\>[14]\AgdaSymbol{(}\AgdaBound{f}\AgdaSpace{}%
\AgdaOperator{\AgdaFunction{̂}}\AgdaSpace{}%
\AgdaBound{𝑨}\AgdaSymbol{)}%
\>[23]\AgdaSymbol{(}%
\>[31]\AgdaFunction{h⁻¹}%
\>[36]\AgdaOperator{\AgdaFunction{∘}}%
\>[39]\AgdaBound{c}%
\>[42]\AgdaSymbol{))}%
\>[48]\AgdaFunction{≈⟨}%
\>[53]\AgdaField{compatible}\AgdaSpace{}%
\AgdaOperator{\AgdaFunction{∥}}\AgdaSpace{}%
\AgdaBound{gh}\AgdaSpace{}%
\AgdaOperator{\AgdaFunction{∥}}%
\>[97]\AgdaFunction{⟩}\<%
\\
\>[14]\AgdaSymbol{(}\AgdaBound{f}\AgdaSpace{}%
\AgdaOperator{\AgdaFunction{̂}}\AgdaSpace{}%
\AgdaBound{𝑩}\AgdaSymbol{)}%
\>[23]\AgdaSymbol{(}\AgdaFunction{g}\AgdaSpace{}%
\AgdaOperator{\AgdaFunction{∘}}\AgdaSpace{}%
\AgdaSymbol{(}%
\>[31]\AgdaFunction{h⁻¹}%
\>[36]\AgdaOperator{\AgdaFunction{∘}}%
\>[39]\AgdaBound{c}%
\>[42]\AgdaSymbol{))}%
\>[48]\AgdaOperator{\AgdaFunction{∎}}\AgdaSpace{}%
\AgdaKeyword{where}\AgdaSpace{}%
\AgdaKeyword{open}\AgdaSpace{}%
\AgdaModule{SetoidReasoning}\AgdaSpace{}%
\AgdaOperator{\AgdaFunction{𝔻[}}\AgdaSpace{}%
\AgdaBound{𝑩}\AgdaSpace{}%
\AgdaOperator{\AgdaFunction{]}}\<%
\\
\\[\AgdaEmptyExtraSkip]%
\>[2]\AgdaFunction{φhom}\AgdaSpace{}%
\AgdaSymbol{:}\AgdaSpace{}%
\AgdaRecord{IsHom}\AgdaSpace{}%
\AgdaBound{𝑪}\AgdaSpace{}%
\AgdaBound{𝑩}\AgdaSpace{}%
\AgdaFunction{φmap}\<%
\\
\>[2]\AgdaField{compatible}\AgdaSpace{}%
\AgdaFunction{φhom}\AgdaSpace{}%
\AgdaSymbol{=}\AgdaSpace{}%
\AgdaFunction{φcomp}\<%
\end{code}
\paragraph*{Isomorphisms}
\fi      %%% END LONG VERSION ONLY SECTION

Two structures are \defn{isomorphic} provided there are homomorphisms from each to the
other that compose to the identity. In the \agdaalgebras library we codify this notion as
well as some of its obvious consequences, as a record type called \ar{\au{}≅\au{}}.
\ifshort
Here we display only the essential part of the defition, but \seemedium.
\else
Note that the definition, shown below, includes a proof of the fact that the maps \afld{to} and
\afld{from} are bijective, which makes this fact more accessible.

\begin{code}%
\>[0]\<%
\\
\>[0]\AgdaKeyword{module}\AgdaSpace{}%
\AgdaModule{\AgdaUnderscore{}}\AgdaSpace{}%
\AgdaSymbol{(}\AgdaBound{𝑨}\AgdaSpace{}%
\AgdaSymbol{:}\AgdaSpace{}%
\AgdaRecord{Algebra}\AgdaSpace{}%
\AgdaGeneralizable{α}\AgdaSpace{}%
\AgdaGeneralizable{ρᵃ}\AgdaSymbol{)}\AgdaSpace{}%
\AgdaSymbol{(}\AgdaBound{𝑩}\AgdaSpace{}%
\AgdaSymbol{:}\AgdaSpace{}%
\AgdaRecord{Algebra}\AgdaSpace{}%
\AgdaGeneralizable{β}\AgdaSpace{}%
\AgdaGeneralizable{ρᵇ}\AgdaSymbol{)}\AgdaSpace{}%
\AgdaKeyword{where}\<%
\\
\>[0][@{}l@{\AgdaIndent{0}}]%
\>[1]\AgdaKeyword{open}\AgdaSpace{}%
\AgdaModule{Setoid}\AgdaSpace{}%
\AgdaOperator{\AgdaFunction{𝔻[}}\AgdaSpace{}%
\AgdaBound{𝑨}\AgdaSpace{}%
\AgdaOperator{\AgdaFunction{]}}\AgdaSpace{}%
\AgdaKeyword{using}\AgdaSpace{}%
\AgdaSymbol{()}\AgdaSpace{}%
\AgdaKeyword{renaming}\AgdaSpace{}%
\AgdaSymbol{(}\AgdaSpace{}%
\AgdaOperator{\AgdaField{\AgdaUnderscore{}≈\AgdaUnderscore{}}}\AgdaSpace{}%
\AgdaSymbol{to}\AgdaSpace{}%
\AgdaOperator{\AgdaField{\AgdaUnderscore{}≈ᴬ\AgdaUnderscore{}}}\AgdaSpace{}%
\AgdaSymbol{)}\<%
\\
\>[1]\AgdaKeyword{open}\AgdaSpace{}%
\AgdaModule{Setoid}\AgdaSpace{}%
\AgdaOperator{\AgdaFunction{𝔻[}}\AgdaSpace{}%
\AgdaBound{𝑩}\AgdaSpace{}%
\AgdaOperator{\AgdaFunction{]}}\AgdaSpace{}%
\AgdaKeyword{using}\AgdaSpace{}%
\AgdaSymbol{()}\AgdaSpace{}%
\AgdaKeyword{renaming}\AgdaSpace{}%
\AgdaSymbol{(}\AgdaSpace{}%
\AgdaOperator{\AgdaField{\AgdaUnderscore{}≈\AgdaUnderscore{}}}\AgdaSpace{}%
\AgdaSymbol{to}\AgdaSpace{}%
\AgdaOperator{\AgdaField{\AgdaUnderscore{}≈ᴮ\AgdaUnderscore{}}}\AgdaSpace{}%
\AgdaSymbol{)}\<%
\end{code}
\fi
\begin{code}%
\>[0]\<%
\\
\>[1]\AgdaKeyword{record}\AgdaSpace{}%
\AgdaOperator{\AgdaRecord{\AgdaUnderscore{}≅\AgdaUnderscore{}}}\AgdaSpace{}%
\AgdaSymbol{:}\AgdaSpace{}%
\AgdaPrimitive{Type}\AgdaSpace{}%
\AgdaSymbol{(}\AgdaBound{𝓞}\AgdaSpace{}%
\AgdaOperator{\AgdaPrimitive{⊔}}\AgdaSpace{}%
\AgdaBound{𝓥}\AgdaSpace{}%
\AgdaOperator{\AgdaPrimitive{⊔}}\AgdaSpace{}%
\AgdaBound{α}\AgdaSpace{}%
\AgdaOperator{\AgdaPrimitive{⊔}}\AgdaSpace{}%
\AgdaBound{ρᵃ}\AgdaSpace{}%
\AgdaOperator{\AgdaPrimitive{⊔}}\AgdaSpace{}%
\AgdaBound{β}\AgdaSpace{}%
\AgdaOperator{\AgdaPrimitive{⊔}}\AgdaSpace{}%
\AgdaBound{ρᵇ}\AgdaSpace{}%
\AgdaSymbol{)}\AgdaSpace{}%
\AgdaKeyword{where}\<%
\\
\>[1][@{}l@{\AgdaIndent{0}}]%
\>[2]\AgdaKeyword{constructor}%
\>[15]\AgdaInductiveConstructor{mkiso}\<%
\\
\>[2]\AgdaKeyword{field}%
\>[15]\AgdaField{to}\AgdaSpace{}%
\AgdaSymbol{:}\AgdaSpace{}%
\AgdaFunction{hom}\AgdaSpace{}%
\AgdaBound{𝑨}\AgdaSpace{}%
\AgdaBound{𝑩}\<%
\\
\>[15]\AgdaField{from}\AgdaSpace{}%
\AgdaSymbol{:}\AgdaSpace{}%
\AgdaFunction{hom}\AgdaSpace{}%
\AgdaBound{𝑩}\AgdaSpace{}%
\AgdaBound{𝑨}\<%
\\
\>[15]\AgdaField{to∼from}\AgdaSpace{}%
\AgdaSymbol{:}\AgdaSpace{}%
\AgdaSymbol{∀}\AgdaSpace{}%
\AgdaBound{b}\AgdaSpace{}%
\AgdaSymbol{→}\AgdaSpace{}%
\AgdaOperator{\AgdaFunction{∣}}\AgdaSpace{}%
\AgdaField{to}\AgdaSpace{}%
\AgdaOperator{\AgdaFunction{∣}}%
\>[41]\AgdaOperator{\AgdaField{⟨\$⟩}}\AgdaSpace{}%
\AgdaSymbol{(}\AgdaOperator{\AgdaFunction{∣}}\AgdaSpace{}%
\AgdaField{from}\AgdaSpace{}%
\AgdaOperator{\AgdaFunction{∣}}%
\>[56]\AgdaOperator{\AgdaField{⟨\$⟩}}\AgdaSpace{}%
\AgdaBound{b}\AgdaSymbol{)}%
\>[64]\AgdaOperator{\AgdaFunction{≈ᴮ}}\AgdaSpace{}%
\AgdaBound{b}\<%
\\
\>[15]\AgdaField{from∼to}\AgdaSpace{}%
\AgdaSymbol{:}\AgdaSpace{}%
\AgdaSymbol{∀}\AgdaSpace{}%
\AgdaBound{a}\AgdaSpace{}%
\AgdaSymbol{→}\AgdaSpace{}%
\AgdaOperator{\AgdaFunction{∣}}\AgdaSpace{}%
\AgdaField{from}\AgdaSpace{}%
\AgdaOperator{\AgdaFunction{∣}}%
\>[41]\AgdaOperator{\AgdaField{⟨\$⟩}}\AgdaSpace{}%
\AgdaSymbol{(}\AgdaOperator{\AgdaFunction{∣}}\AgdaSpace{}%
\AgdaField{to}\AgdaSpace{}%
\AgdaOperator{\AgdaFunction{∣}}%
\>[56]\AgdaOperator{\AgdaField{⟨\$⟩}}\AgdaSpace{}%
\AgdaBound{a}\AgdaSymbol{)}%
\>[64]\AgdaOperator{\AgdaFunction{≈ᴬ}}\AgdaSpace{}%
\AgdaBound{a}\<%
\end{code}
\ifshort\else    %%% BEGIN LONG VERSION ONLY
\begin{code}%
\>[0]\<%
\\
\>[2]\AgdaFunction{toIsSurjective}\AgdaSpace{}%
\AgdaSymbol{:}\AgdaSpace{}%
\AgdaFunction{IsSurjective}\AgdaSpace{}%
\AgdaOperator{\AgdaFunction{∣}}\AgdaSpace{}%
\AgdaField{to}\AgdaSpace{}%
\AgdaOperator{\AgdaFunction{∣}}\<%
\\
\>[2]\AgdaFunction{toIsSurjective}\AgdaSpace{}%
\AgdaSymbol{\{}\AgdaBound{y}\AgdaSymbol{\}}\AgdaSpace{}%
\AgdaSymbol{=}\AgdaSpace{}%
\AgdaInductiveConstructor{eq}\AgdaSpace{}%
\AgdaSymbol{(}\AgdaOperator{\AgdaFunction{∣}}\AgdaSpace{}%
\AgdaField{from}\AgdaSpace{}%
\AgdaOperator{\AgdaFunction{∣}}\AgdaSpace{}%
\AgdaOperator{\AgdaField{⟨\$⟩}}\AgdaSpace{}%
\AgdaBound{y}\AgdaSymbol{)}\AgdaSpace{}%
\AgdaSymbol{(}\AgdaFunction{sym}\AgdaSpace{}%
\AgdaSymbol{(}\AgdaField{to∼from}\AgdaSpace{}%
\AgdaBound{y}\AgdaSymbol{))}\<%
\\
\>[2][@{}l@{\AgdaIndent{0}}]%
\>[3]\AgdaKeyword{where}\AgdaSpace{}%
\AgdaKeyword{open}\AgdaSpace{}%
\AgdaModule{Setoid}\AgdaSpace{}%
\AgdaOperator{\AgdaFunction{𝔻[}}\AgdaSpace{}%
\AgdaBound{𝑩}\AgdaSpace{}%
\AgdaOperator{\AgdaFunction{]}}\AgdaSpace{}%
\AgdaKeyword{using}\AgdaSpace{}%
\AgdaSymbol{(}\AgdaSpace{}%
\AgdaFunction{sym}\AgdaSpace{}%
\AgdaSymbol{)}\<%
\\
\\[\AgdaEmptyExtraSkip]%
\>[2]\AgdaFunction{toIsInjective}\AgdaSpace{}%
\AgdaSymbol{:}\AgdaSpace{}%
\AgdaFunction{IsInjective}\AgdaSpace{}%
\AgdaOperator{\AgdaFunction{∣}}\AgdaSpace{}%
\AgdaField{to}\AgdaSpace{}%
\AgdaOperator{\AgdaFunction{∣}}\<%
\\
\>[2]\AgdaFunction{toIsInjective}\AgdaSpace{}%
\AgdaSymbol{\{}\AgdaBound{x}\AgdaSymbol{\}\{}\AgdaBound{y}\AgdaSymbol{\}}\AgdaSpace{}%
\AgdaBound{xy}\AgdaSpace{}%
\AgdaSymbol{=}\AgdaSpace{}%
\AgdaFunction{trans}\AgdaSpace{}%
\AgdaSymbol{(}\AgdaFunction{sym}\AgdaSpace{}%
\AgdaSymbol{(}\AgdaField{from∼to}\AgdaSpace{}%
\AgdaBound{x}\AgdaSymbol{))}\AgdaSpace{}%
\AgdaSymbol{(}\AgdaFunction{trans}\AgdaSpace{}%
\AgdaFunction{ξ}\AgdaSpace{}%
\AgdaSymbol{(}\AgdaField{from∼to}\AgdaSpace{}%
\AgdaBound{y}\AgdaSymbol{))}\<%
\\
\>[2][@{}l@{\AgdaIndent{0}}]%
\>[3]\AgdaKeyword{where}\<%
\\
\>[3]\AgdaKeyword{open}\AgdaSpace{}%
\AgdaModule{Setoid}\AgdaSpace{}%
\AgdaOperator{\AgdaFunction{𝔻[}}\AgdaSpace{}%
\AgdaBound{𝑨}\AgdaSpace{}%
\AgdaOperator{\AgdaFunction{]}}\AgdaSpace{}%
\AgdaKeyword{using}\AgdaSpace{}%
\AgdaSymbol{(}\AgdaSpace{}%
\AgdaFunction{sym}\AgdaSpace{}%
\AgdaSymbol{;}\AgdaSpace{}%
\AgdaFunction{trans}\AgdaSpace{}%
\AgdaSymbol{)}\<%
\\
\>[3]\AgdaFunction{ξ}\AgdaSpace{}%
\AgdaSymbol{:}\AgdaSpace{}%
\AgdaOperator{\AgdaFunction{∣}}\AgdaSpace{}%
\AgdaField{from}\AgdaSpace{}%
\AgdaOperator{\AgdaFunction{∣}}\AgdaSpace{}%
\AgdaOperator{\AgdaField{⟨\$⟩}}\AgdaSpace{}%
\AgdaSymbol{(}\AgdaOperator{\AgdaFunction{∣}}\AgdaSpace{}%
\AgdaField{to}\AgdaSpace{}%
\AgdaOperator{\AgdaFunction{∣}}\AgdaSpace{}%
\AgdaOperator{\AgdaField{⟨\$⟩}}\AgdaSpace{}%
\AgdaBound{x}\AgdaSymbol{)}\AgdaSpace{}%
\AgdaOperator{\AgdaFunction{≈ᴬ}}\AgdaSpace{}%
\AgdaOperator{\AgdaFunction{∣}}\AgdaSpace{}%
\AgdaField{from}\AgdaSpace{}%
\AgdaOperator{\AgdaFunction{∣}}\AgdaSpace{}%
\AgdaOperator{\AgdaField{⟨\$⟩}}\AgdaSpace{}%
\AgdaSymbol{(}\AgdaOperator{\AgdaFunction{∣}}\AgdaSpace{}%
\AgdaField{to}\AgdaSpace{}%
\AgdaOperator{\AgdaFunction{∣}}\AgdaSpace{}%
\AgdaOperator{\AgdaField{⟨\$⟩}}\AgdaSpace{}%
\AgdaBound{y}\AgdaSymbol{)}\<%
\\
\>[3]\AgdaFunction{ξ}\AgdaSpace{}%
\AgdaSymbol{=}\AgdaSpace{}%
\AgdaField{cong}\AgdaSpace{}%
\AgdaOperator{\AgdaFunction{∣}}\AgdaSpace{}%
\AgdaField{from}\AgdaSpace{}%
\AgdaOperator{\AgdaFunction{∣}}\AgdaSpace{}%
\AgdaBound{xy}\<%
\\
\\[\AgdaEmptyExtraSkip]%
\>[2]\AgdaFunction{fromIsSurjective}\AgdaSpace{}%
\AgdaSymbol{:}\AgdaSpace{}%
\AgdaFunction{IsSurjective}\AgdaSpace{}%
\AgdaOperator{\AgdaFunction{∣}}\AgdaSpace{}%
\AgdaField{from}\AgdaSpace{}%
\AgdaOperator{\AgdaFunction{∣}}\<%
\\
\>[2]\AgdaFunction{fromIsSurjective}\AgdaSpace{}%
\AgdaSymbol{\{}\AgdaBound{x}\AgdaSymbol{\}}\AgdaSpace{}%
\AgdaSymbol{=}\AgdaSpace{}%
\AgdaInductiveConstructor{eq}\AgdaSpace{}%
\AgdaSymbol{(}\AgdaOperator{\AgdaFunction{∣}}\AgdaSpace{}%
\AgdaField{to}\AgdaSpace{}%
\AgdaOperator{\AgdaFunction{∣}}\AgdaSpace{}%
\AgdaOperator{\AgdaField{⟨\$⟩}}\AgdaSpace{}%
\AgdaBound{x}\AgdaSymbol{)}\AgdaSpace{}%
\AgdaSymbol{(}\AgdaFunction{sym}\AgdaSpace{}%
\AgdaSymbol{(}\AgdaField{from∼to}\AgdaSpace{}%
\AgdaBound{x}\AgdaSymbol{))}\<%
\\
\>[2][@{}l@{\AgdaIndent{0}}]%
\>[3]\AgdaKeyword{where}\AgdaSpace{}%
\AgdaKeyword{open}\AgdaSpace{}%
\AgdaModule{Setoid}\AgdaSpace{}%
\AgdaOperator{\AgdaFunction{𝔻[}}\AgdaSpace{}%
\AgdaBound{𝑨}\AgdaSpace{}%
\AgdaOperator{\AgdaFunction{]}}\AgdaSpace{}%
\AgdaKeyword{using}\AgdaSpace{}%
\AgdaSymbol{(}\AgdaSpace{}%
\AgdaFunction{sym}\AgdaSpace{}%
\AgdaSymbol{)}\<%
\\
\\[\AgdaEmptyExtraSkip]%
\>[2]\AgdaFunction{fromIsInjective}\AgdaSpace{}%
\AgdaSymbol{:}\AgdaSpace{}%
\AgdaFunction{IsInjective}\AgdaSpace{}%
\AgdaOperator{\AgdaFunction{∣}}\AgdaSpace{}%
\AgdaField{from}\AgdaSpace{}%
\AgdaOperator{\AgdaFunction{∣}}\<%
\\
\>[2]\AgdaFunction{fromIsInjective}\AgdaSpace{}%
\AgdaSymbol{\{}\AgdaBound{x}\AgdaSymbol{\}\{}\AgdaBound{y}\AgdaSymbol{\}}\AgdaSpace{}%
\AgdaBound{xy}\AgdaSpace{}%
\AgdaSymbol{=}\AgdaSpace{}%
\AgdaFunction{trans}\AgdaSpace{}%
\AgdaSymbol{(}\AgdaFunction{sym}\AgdaSpace{}%
\AgdaSymbol{(}\AgdaField{to∼from}\AgdaSpace{}%
\AgdaBound{x}\AgdaSymbol{))}\AgdaSpace{}%
\AgdaSymbol{(}\AgdaFunction{trans}\AgdaSpace{}%
\AgdaFunction{ξ}\AgdaSpace{}%
\AgdaSymbol{(}\AgdaField{to∼from}\AgdaSpace{}%
\AgdaBound{y}\AgdaSymbol{))}\<%
\\
\>[2][@{}l@{\AgdaIndent{0}}]%
\>[3]\AgdaKeyword{where}\<%
\\
\>[3]\AgdaKeyword{open}\AgdaSpace{}%
\AgdaModule{Setoid}\AgdaSpace{}%
\AgdaOperator{\AgdaFunction{𝔻[}}\AgdaSpace{}%
\AgdaBound{𝑩}\AgdaSpace{}%
\AgdaOperator{\AgdaFunction{]}}\AgdaSpace{}%
\AgdaKeyword{using}\AgdaSpace{}%
\AgdaSymbol{(}\AgdaSpace{}%
\AgdaFunction{sym}\AgdaSpace{}%
\AgdaSymbol{;}\AgdaSpace{}%
\AgdaFunction{trans}\AgdaSpace{}%
\AgdaSymbol{)}\<%
\\
\>[3]\AgdaFunction{ξ}\AgdaSpace{}%
\AgdaSymbol{:}\AgdaSpace{}%
\AgdaOperator{\AgdaFunction{∣}}\AgdaSpace{}%
\AgdaField{to}\AgdaSpace{}%
\AgdaOperator{\AgdaFunction{∣}}\AgdaSpace{}%
\AgdaOperator{\AgdaField{⟨\$⟩}}\AgdaSpace{}%
\AgdaSymbol{(}\AgdaOperator{\AgdaFunction{∣}}\AgdaSpace{}%
\AgdaField{from}\AgdaSpace{}%
\AgdaOperator{\AgdaFunction{∣}}\AgdaSpace{}%
\AgdaOperator{\AgdaField{⟨\$⟩}}\AgdaSpace{}%
\AgdaBound{x}\AgdaSymbol{)}\AgdaSpace{}%
\AgdaOperator{\AgdaFunction{≈ᴮ}}\AgdaSpace{}%
\AgdaOperator{\AgdaFunction{∣}}\AgdaSpace{}%
\AgdaField{to}\AgdaSpace{}%
\AgdaOperator{\AgdaFunction{∣}}\AgdaSpace{}%
\AgdaOperator{\AgdaField{⟨\$⟩}}\AgdaSpace{}%
\AgdaSymbol{(}\AgdaOperator{\AgdaFunction{∣}}\AgdaSpace{}%
\AgdaField{from}\AgdaSpace{}%
\AgdaOperator{\AgdaFunction{∣}}\AgdaSpace{}%
\AgdaOperator{\AgdaField{⟨\$⟩}}\AgdaSpace{}%
\AgdaBound{y}\AgdaSymbol{)}\<%
\\
\>[3]\AgdaFunction{ξ}\AgdaSpace{}%
\AgdaSymbol{=}\AgdaSpace{}%
\AgdaField{cong}\AgdaSpace{}%
\AgdaOperator{\AgdaFunction{∣}}\AgdaSpace{}%
\AgdaField{to}\AgdaSpace{}%
\AgdaOperator{\AgdaFunction{∣}}\AgdaSpace{}%
\AgdaBound{xy}\<%
\\
\\[\AgdaEmptyExtraSkip]%
\>[0]\AgdaKeyword{open}\AgdaSpace{}%
\AgdaOperator{\AgdaModule{\AgdaUnderscore{}≅\AgdaUnderscore{}}}\<%
\\
\>[0]\<%
\end{code}

It is easy to prove that \ar{\au{}≅\au{}} is an equivalence relation, as follows.

\begin{code}%
\>[0]\<%
\\
\>[0]\AgdaFunction{≅-refl}\AgdaSpace{}%
\AgdaSymbol{:}\AgdaSpace{}%
\AgdaFunction{Reflexive}\AgdaSpace{}%
\AgdaSymbol{(}\AgdaOperator{\AgdaRecord{\AgdaUnderscore{}≅\AgdaUnderscore{}}}\AgdaSpace{}%
\AgdaSymbol{\{}\AgdaGeneralizable{α}\AgdaSymbol{\}\{}\AgdaGeneralizable{ρᵃ}\AgdaSymbol{\})}\<%
\\
\>[0]\AgdaFunction{≅-refl}\AgdaSpace{}%
\AgdaSymbol{\{}\AgdaBound{α}\AgdaSymbol{\}\{}\AgdaBound{ρᵃ}\AgdaSymbol{\}\{}\AgdaBound{𝑨}\AgdaSymbol{\}}\AgdaSpace{}%
\AgdaSymbol{=}\AgdaSpace{}%
\AgdaInductiveConstructor{mkiso}\AgdaSpace{}%
\AgdaFunction{𝒾𝒹}\AgdaSpace{}%
\AgdaFunction{𝒾𝒹}\AgdaSpace{}%
\AgdaSymbol{(λ}\AgdaSpace{}%
\AgdaBound{b}\AgdaSpace{}%
\AgdaSymbol{→}\AgdaSpace{}%
\AgdaFunction{refl}\AgdaSymbol{)}\AgdaSpace{}%
\AgdaSymbol{λ}\AgdaSpace{}%
\AgdaBound{a}\AgdaSpace{}%
\AgdaSymbol{→}\AgdaSpace{}%
\AgdaFunction{refl}\AgdaSpace{}%
\AgdaKeyword{where}\AgdaSpace{}%
\AgdaKeyword{open}\AgdaSpace{}%
\AgdaModule{Setoid}\AgdaSpace{}%
\AgdaOperator{\AgdaFunction{𝔻[}}\AgdaSpace{}%
\AgdaBound{𝑨}\AgdaSpace{}%
\AgdaOperator{\AgdaFunction{]}}\AgdaSpace{}%
\AgdaKeyword{using}\AgdaSpace{}%
\AgdaSymbol{(}\AgdaSpace{}%
\AgdaFunction{refl}\AgdaSpace{}%
\AgdaSymbol{)}\<%
\\
\>[0]\AgdaFunction{≅-sym}\AgdaSpace{}%
\AgdaSymbol{:}\AgdaSpace{}%
\AgdaFunction{Sym}\AgdaSpace{}%
\AgdaSymbol{(}\AgdaOperator{\AgdaRecord{\AgdaUnderscore{}≅\AgdaUnderscore{}}}\AgdaSymbol{\{}\AgdaGeneralizable{β}\AgdaSymbol{\}\{}\AgdaGeneralizable{ρᵇ}\AgdaSymbol{\})}\AgdaSpace{}%
\AgdaSymbol{(}\AgdaOperator{\AgdaRecord{\AgdaUnderscore{}≅\AgdaUnderscore{}}}\AgdaSymbol{\{}\AgdaGeneralizable{α}\AgdaSymbol{\}\{}\AgdaGeneralizable{ρᵃ}\AgdaSymbol{\})}\<%
\\
\>[0]\AgdaFunction{≅-sym}\AgdaSpace{}%
\AgdaBound{φ}\AgdaSpace{}%
\AgdaSymbol{=}\AgdaSpace{}%
\AgdaInductiveConstructor{mkiso}\AgdaSpace{}%
\AgdaSymbol{(}\AgdaField{from}\AgdaSpace{}%
\AgdaBound{φ}\AgdaSymbol{)}\AgdaSpace{}%
\AgdaSymbol{(}\AgdaField{to}\AgdaSpace{}%
\AgdaBound{φ}\AgdaSymbol{)}\AgdaSpace{}%
\AgdaSymbol{(}\AgdaField{from∼to}\AgdaSpace{}%
\AgdaBound{φ}\AgdaSymbol{)}\AgdaSpace{}%
\AgdaSymbol{(}\AgdaField{to∼from}\AgdaSpace{}%
\AgdaBound{φ}\AgdaSymbol{)}\<%
\\
\\[\AgdaEmptyExtraSkip]%
\>[0]\AgdaFunction{≅-trans}\AgdaSpace{}%
\AgdaSymbol{:}\AgdaSpace{}%
\AgdaFunction{Trans}\AgdaSpace{}%
\AgdaSymbol{(}\AgdaOperator{\AgdaRecord{\AgdaUnderscore{}≅\AgdaUnderscore{}}}\AgdaSpace{}%
\AgdaSymbol{\{}\AgdaGeneralizable{α}\AgdaSymbol{\}\{}\AgdaGeneralizable{ρᵃ}\AgdaSymbol{\})}\AgdaSpace{}%
\AgdaSymbol{(}\AgdaOperator{\AgdaRecord{\AgdaUnderscore{}≅\AgdaUnderscore{}}}\AgdaSymbol{\{}\AgdaGeneralizable{β}\AgdaSymbol{\}\{}\AgdaGeneralizable{ρᵇ}\AgdaSymbol{\})}\AgdaSpace{}%
\AgdaSymbol{(}\AgdaOperator{\AgdaRecord{\AgdaUnderscore{}≅\AgdaUnderscore{}}}\AgdaSymbol{\{}\AgdaGeneralizable{α}\AgdaSymbol{\}\{}\AgdaGeneralizable{ρᵃ}\AgdaSymbol{\}\{}\AgdaGeneralizable{γ}\AgdaSymbol{\}\{}\AgdaGeneralizable{ρᶜ}\AgdaSymbol{\})}\<%
\\
\>[0]\AgdaFunction{≅-trans}\AgdaSpace{}%
\AgdaSymbol{\{}\AgdaArgument{ρᶜ}\AgdaSpace{}%
\AgdaSymbol{=}\AgdaSpace{}%
\AgdaBound{ρᶜ}\AgdaSymbol{\}\{}\AgdaBound{𝑨}\AgdaSymbol{\}\{}\AgdaBound{𝑩}\AgdaSymbol{\}\{}\AgdaBound{𝑪}\AgdaSymbol{\}}\AgdaSpace{}%
\AgdaBound{ab}\AgdaSpace{}%
\AgdaBound{bc}\AgdaSpace{}%
\AgdaSymbol{=}\AgdaSpace{}%
\AgdaInductiveConstructor{mkiso}\AgdaSpace{}%
\AgdaFunction{f}\AgdaSpace{}%
\AgdaFunction{g}\AgdaSpace{}%
\AgdaFunction{τ}\AgdaSpace{}%
\AgdaFunction{ν}\<%
\\
\>[0][@{}l@{\AgdaIndent{0}}]%
\>[1]\AgdaKeyword{where}\<%
\\
\>[1][@{}l@{\AgdaIndent{0}}]%
\>[2]\AgdaFunction{f}\AgdaSpace{}%
\AgdaSymbol{:}\AgdaSpace{}%
\AgdaFunction{hom}\AgdaSpace{}%
\AgdaBound{𝑨}\AgdaSpace{}%
\AgdaBound{𝑪}%
\>[29]\AgdaSymbol{;}%
\>[32]\AgdaFunction{g}\AgdaSpace{}%
\AgdaSymbol{:}\AgdaSpace{}%
\AgdaFunction{hom}\AgdaSpace{}%
\AgdaBound{𝑪}\AgdaSpace{}%
\AgdaBound{𝑨}\<%
\\
\>[2]\AgdaFunction{f}\AgdaSpace{}%
\AgdaSymbol{=}\AgdaSpace{}%
\AgdaFunction{∘-hom}\AgdaSpace{}%
\AgdaSymbol{(}\AgdaField{to}\AgdaSpace{}%
\AgdaBound{ab}\AgdaSymbol{)}\AgdaSpace{}%
\AgdaSymbol{(}\AgdaField{to}\AgdaSpace{}%
\AgdaBound{bc}\AgdaSymbol{)}%
\>[29]\AgdaSymbol{;}%
\>[32]\AgdaFunction{g}\AgdaSpace{}%
\AgdaSymbol{=}\AgdaSpace{}%
\AgdaFunction{∘-hom}\AgdaSpace{}%
\AgdaSymbol{(}\AgdaField{from}\AgdaSpace{}%
\AgdaBound{bc}\AgdaSymbol{)}\AgdaSpace{}%
\AgdaSymbol{(}\AgdaField{from}\AgdaSpace{}%
\AgdaBound{ab}\AgdaSymbol{)}\<%
\\
\\[\AgdaEmptyExtraSkip]%
\>[2]\AgdaKeyword{open}\AgdaSpace{}%
\AgdaModule{Setoid}\AgdaSpace{}%
\AgdaOperator{\AgdaFunction{𝔻[}}\AgdaSpace{}%
\AgdaBound{𝑨}\AgdaSpace{}%
\AgdaOperator{\AgdaFunction{]}}\AgdaSpace{}%
\AgdaKeyword{using}\AgdaSpace{}%
\AgdaSymbol{(}\AgdaSpace{}%
\AgdaOperator{\AgdaField{\AgdaUnderscore{}≈\AgdaUnderscore{}}}\AgdaSpace{}%
\AgdaSymbol{;}\AgdaSpace{}%
\AgdaFunction{trans}\AgdaSpace{}%
\AgdaSymbol{)}\<%
\\
\>[2]\AgdaKeyword{open}\AgdaSpace{}%
\AgdaModule{Setoid}\AgdaSpace{}%
\AgdaOperator{\AgdaFunction{𝔻[}}\AgdaSpace{}%
\AgdaBound{𝑪}\AgdaSpace{}%
\AgdaOperator{\AgdaFunction{]}}\AgdaSpace{}%
\AgdaKeyword{using}\AgdaSpace{}%
\AgdaSymbol{()}\AgdaSpace{}%
\AgdaKeyword{renaming}\AgdaSpace{}%
\AgdaSymbol{(}\AgdaSpace{}%
\AgdaOperator{\AgdaField{\AgdaUnderscore{}≈\AgdaUnderscore{}}}\AgdaSpace{}%
\AgdaSymbol{to}\AgdaSpace{}%
\AgdaOperator{\AgdaField{\AgdaUnderscore{}≈ᶜ\AgdaUnderscore{}}}\AgdaSpace{}%
\AgdaSymbol{;}\AgdaSpace{}%
\AgdaFunction{trans}\AgdaSpace{}%
\AgdaSymbol{to}\AgdaSpace{}%
\AgdaFunction{transᶜ}\AgdaSpace{}%
\AgdaSymbol{)}\<%
\\
\\[\AgdaEmptyExtraSkip]%
\>[2]\AgdaFunction{τ}\AgdaSpace{}%
\AgdaSymbol{:}\AgdaSpace{}%
\AgdaSymbol{∀}\AgdaSpace{}%
\AgdaBound{b}\AgdaSpace{}%
\AgdaSymbol{→}\AgdaSpace{}%
\AgdaOperator{\AgdaFunction{∣}}\AgdaSpace{}%
\AgdaFunction{f}\AgdaSpace{}%
\AgdaOperator{\AgdaFunction{∣}}\AgdaSpace{}%
\AgdaOperator{\AgdaField{⟨\$⟩}}\AgdaSpace{}%
\AgdaSymbol{(}\AgdaOperator{\AgdaFunction{∣}}\AgdaSpace{}%
\AgdaFunction{g}\AgdaSpace{}%
\AgdaOperator{\AgdaFunction{∣}}\AgdaSpace{}%
\AgdaOperator{\AgdaField{⟨\$⟩}}\AgdaSpace{}%
\AgdaBound{b}\AgdaSymbol{)}\AgdaSpace{}%
\AgdaOperator{\AgdaFunction{≈ᶜ}}\AgdaSpace{}%
\AgdaBound{b}\<%
\\
\>[2]\AgdaFunction{τ}\AgdaSpace{}%
\AgdaBound{b}\AgdaSpace{}%
\AgdaSymbol{=}\AgdaSpace{}%
\AgdaFunction{transᶜ}\AgdaSpace{}%
\AgdaSymbol{(}\AgdaField{cong}\AgdaSpace{}%
\AgdaOperator{\AgdaFunction{∣}}\AgdaSpace{}%
\AgdaField{to}\AgdaSpace{}%
\AgdaBound{bc}\AgdaSpace{}%
\AgdaOperator{\AgdaFunction{∣}}\AgdaSpace{}%
\AgdaSymbol{(}\AgdaField{to∼from}\AgdaSpace{}%
\AgdaBound{ab}\AgdaSpace{}%
\AgdaSymbol{(}\AgdaOperator{\AgdaFunction{∣}}\AgdaSpace{}%
\AgdaField{from}\AgdaSpace{}%
\AgdaBound{bc}\AgdaSpace{}%
\AgdaOperator{\AgdaFunction{∣}}\AgdaSpace{}%
\AgdaOperator{\AgdaField{⟨\$⟩}}\AgdaSpace{}%
\AgdaBound{b}\AgdaSymbol{)))}\AgdaSpace{}%
\AgdaSymbol{(}\AgdaField{to∼from}\AgdaSpace{}%
\AgdaBound{bc}\AgdaSpace{}%
\AgdaBound{b}\AgdaSymbol{)}\<%
\\
\\[\AgdaEmptyExtraSkip]%
\>[2]\AgdaFunction{ν}\AgdaSpace{}%
\AgdaSymbol{:}\AgdaSpace{}%
\AgdaSymbol{∀}\AgdaSpace{}%
\AgdaBound{a}\AgdaSpace{}%
\AgdaSymbol{→}\AgdaSpace{}%
\AgdaOperator{\AgdaFunction{∣}}\AgdaSpace{}%
\AgdaFunction{g}\AgdaSpace{}%
\AgdaOperator{\AgdaFunction{∣}}\AgdaSpace{}%
\AgdaOperator{\AgdaField{⟨\$⟩}}\AgdaSpace{}%
\AgdaSymbol{(}\AgdaOperator{\AgdaFunction{∣}}\AgdaSpace{}%
\AgdaFunction{f}\AgdaSpace{}%
\AgdaOperator{\AgdaFunction{∣}}\AgdaSpace{}%
\AgdaOperator{\AgdaField{⟨\$⟩}}\AgdaSpace{}%
\AgdaBound{a}\AgdaSymbol{)}\AgdaSpace{}%
\AgdaOperator{\AgdaFunction{≈}}\AgdaSpace{}%
\AgdaBound{a}\<%
\\
\>[2]\AgdaFunction{ν}\AgdaSpace{}%
\AgdaBound{a}\AgdaSpace{}%
\AgdaSymbol{=}\AgdaSpace{}%
\AgdaFunction{trans}\AgdaSpace{}%
\AgdaSymbol{(}\AgdaField{cong}\AgdaSpace{}%
\AgdaOperator{\AgdaFunction{∣}}\AgdaSpace{}%
\AgdaField{from}\AgdaSpace{}%
\AgdaBound{ab}\AgdaSpace{}%
\AgdaOperator{\AgdaFunction{∣}}\AgdaSpace{}%
\AgdaSymbol{(}\AgdaField{from∼to}\AgdaSpace{}%
\AgdaBound{bc}\AgdaSpace{}%
\AgdaSymbol{(}\AgdaOperator{\AgdaFunction{∣}}\AgdaSpace{}%
\AgdaField{to}\AgdaSpace{}%
\AgdaBound{ab}\AgdaSpace{}%
\AgdaOperator{\AgdaFunction{∣}}\AgdaSpace{}%
\AgdaOperator{\AgdaField{⟨\$⟩}}\AgdaSpace{}%
\AgdaBound{a}\AgdaSymbol{)))}\AgdaSpace{}%
\AgdaSymbol{(}\AgdaField{from∼to}\AgdaSpace{}%
\AgdaBound{ab}\AgdaSpace{}%
\AgdaBound{a}\AgdaSymbol{)}\<%
\end{code}
\fi

\paragraph*{Lift-Alg is an algebraic invariant}
The \af{Lift-Alg} operation neatly resolves the technical problem arising from the
noncumulativity of Agda's universe hierarchy. It does so without changing the algebraic
semantics because isomorphism classes of algebras are closed under \af{Lift-Alg}.
\ifshort
The \agdaalgebras library formalizes this fact by proving the following typing judgment.

\else

\begin{code}%
\>[0]\<%
\\
\>[0]\AgdaKeyword{module}\AgdaSpace{}%
\AgdaModule{\AgdaUnderscore{}}\AgdaSpace{}%
\AgdaSymbol{\{}\AgdaBound{𝑨}\AgdaSpace{}%
\AgdaSymbol{:}\AgdaSpace{}%
\AgdaRecord{Algebra}\AgdaSpace{}%
\AgdaGeneralizable{α}\AgdaSpace{}%
\AgdaGeneralizable{ρᵃ}\AgdaSymbol{\}\{}\AgdaBound{ℓ}\AgdaSpace{}%
\AgdaSymbol{:}\AgdaSpace{}%
\AgdaPostulate{Level}\AgdaSymbol{\}}\AgdaSpace{}%
\AgdaKeyword{where}\<%
\\
\>[0][@{}l@{\AgdaIndent{0}}]%
\>[1]\AgdaFunction{Lift-≅ˡ}\AgdaSpace{}%
\AgdaSymbol{:}\AgdaSpace{}%
\AgdaBound{𝑨}\AgdaSpace{}%
\AgdaOperator{\AgdaRecord{≅}}\AgdaSpace{}%
\AgdaSymbol{(}\AgdaFunction{Lift-Algˡ}\AgdaSpace{}%
\AgdaBound{𝑨}\AgdaSpace{}%
\AgdaBound{ℓ}\AgdaSymbol{)}\<%
\\
\>[1]\AgdaFunction{Lift-≅ˡ}\AgdaSpace{}%
\AgdaSymbol{=}\AgdaSpace{}%
\AgdaInductiveConstructor{mkiso}\AgdaSpace{}%
\AgdaFunction{ToLiftˡ}\AgdaSpace{}%
\AgdaFunction{FromLiftˡ}\AgdaSpace{}%
\AgdaSymbol{(}\AgdaFunction{ToFromLiftˡ}\AgdaSymbol{\{}\AgdaArgument{𝑨}\AgdaSpace{}%
\AgdaSymbol{=}\AgdaSpace{}%
\AgdaBound{𝑨}\AgdaSymbol{\})}\AgdaSpace{}%
\AgdaSymbol{(}\AgdaFunction{FromToLiftˡ}\AgdaSymbol{\{}\AgdaArgument{𝑨}\AgdaSpace{}%
\AgdaSymbol{=}\AgdaSpace{}%
\AgdaBound{𝑨}\AgdaSymbol{\}\{}\AgdaBound{ℓ}\AgdaSymbol{\})}\<%
\\
\>[1]\AgdaFunction{Lift-≅ʳ}\AgdaSpace{}%
\AgdaSymbol{:}\AgdaSpace{}%
\AgdaBound{𝑨}\AgdaSpace{}%
\AgdaOperator{\AgdaRecord{≅}}\AgdaSpace{}%
\AgdaSymbol{(}\AgdaFunction{Lift-Algʳ}\AgdaSpace{}%
\AgdaBound{𝑨}\AgdaSpace{}%
\AgdaBound{ℓ}\AgdaSymbol{)}\<%
\\
\>[1]\AgdaFunction{Lift-≅ʳ}\AgdaSpace{}%
\AgdaSymbol{=}\AgdaSpace{}%
\AgdaInductiveConstructor{mkiso}\AgdaSpace{}%
\AgdaFunction{ToLiftʳ}\AgdaSpace{}%
\AgdaFunction{FromLiftʳ}\AgdaSpace{}%
\AgdaSymbol{(}\AgdaFunction{ToFromLiftʳ}\AgdaSymbol{\{}\AgdaArgument{𝑨}\AgdaSpace{}%
\AgdaSymbol{=}\AgdaSpace{}%
\AgdaBound{𝑨}\AgdaSymbol{\})}\AgdaSpace{}%
\AgdaSymbol{(}\AgdaFunction{FromToLiftʳ}\AgdaSymbol{\{}\AgdaArgument{𝑨}\AgdaSpace{}%
\AgdaSymbol{=}\AgdaSpace{}%
\AgdaBound{𝑨}\AgdaSymbol{\}\{}\AgdaBound{ℓ}\AgdaSymbol{\})}\<%
\end{code}
\fi
\begin{code}%
\>[0]\<%
\\
\>[0]\AgdaFunction{Lift-≅}\AgdaSpace{}%
\AgdaSymbol{:}\AgdaSpace{}%
\AgdaSymbol{\{}\AgdaBound{𝑨}\AgdaSpace{}%
\AgdaSymbol{:}\AgdaSpace{}%
\AgdaRecord{Algebra}\AgdaSpace{}%
\AgdaGeneralizable{α}\AgdaSpace{}%
\AgdaGeneralizable{ρᵃ}\AgdaSymbol{\}\{}\AgdaBound{ℓ}\AgdaSpace{}%
\AgdaBound{ρ}\AgdaSpace{}%
\AgdaSymbol{:}\AgdaSpace{}%
\AgdaPostulate{Level}\AgdaSymbol{\}}\AgdaSpace{}%
\AgdaSymbol{→}\AgdaSpace{}%
\AgdaBound{𝑨}\AgdaSpace{}%
\AgdaOperator{\AgdaRecord{≅}}\AgdaSpace{}%
\AgdaSymbol{(}\AgdaFunction{Lift-Alg}\AgdaSpace{}%
\AgdaBound{𝑨}\AgdaSpace{}%
\AgdaBound{ℓ}\AgdaSpace{}%
\AgdaBound{ρ}\AgdaSymbol{)}\<%
\end{code}
\ifshort\else
\begin{code}%
\>[0]\AgdaFunction{Lift-≅}\AgdaSpace{}%
\AgdaSymbol{=}\AgdaSpace{}%
\AgdaFunction{≅-trans}\AgdaSpace{}%
\AgdaFunction{Lift-≅ˡ}\AgdaSpace{}%
\AgdaFunction{Lift-≅ʳ}\<%
\end{code}
\fi

\paragraph*{Homomorphic images}
Here we describe what we have found to be the most useful
way to represent the class of \emph{homomorphic images} of an algebra in MLTT. For future
reference, we also record the fact that an algebra is its own homomorphic
image. (Here and in \agdaalgebras we use the shorthand \af{ov}~\ab{α} := \ab{𝒪}
\ap{⊔} \ab{𝒱} \ap{⊔} \ab{α}, for any level \ab{α}.)

\ifshort\else
\begin{code}%
\>[0]\<%
\\
\>[0]\AgdaFunction{ov}\AgdaSpace{}%
\AgdaSymbol{:}\AgdaSpace{}%
\AgdaPostulate{Level}\AgdaSpace{}%
\AgdaSymbol{→}\AgdaSpace{}%
\AgdaPostulate{Level}\<%
\\
\>[0]\AgdaFunction{ov}\AgdaSpace{}%
\AgdaBound{α}\AgdaSpace{}%
\AgdaSymbol{=}\AgdaSpace{}%
\AgdaBound{𝓞}\AgdaSpace{}%
\AgdaOperator{\AgdaPrimitive{⊔}}\AgdaSpace{}%
\AgdaBound{𝓥}\AgdaSpace{}%
\AgdaOperator{\AgdaPrimitive{⊔}}\AgdaSpace{}%
\AgdaPrimitive{lsuc}\AgdaSpace{}%
\AgdaBound{α}\<%
\end{code}
\fi
\begin{code}%
\>[0]\<%
\\
\>[0]\AgdaOperator{\AgdaFunction{\AgdaUnderscore{}IsHomImageOf\AgdaUnderscore{}}}\AgdaSpace{}%
\AgdaSymbol{:}\AgdaSpace{}%
\AgdaSymbol{(}\AgdaBound{𝑩}\AgdaSpace{}%
\AgdaSymbol{:}\AgdaSpace{}%
\AgdaRecord{Algebra}\AgdaSpace{}%
\AgdaGeneralizable{β}\AgdaSpace{}%
\AgdaGeneralizable{ρᵇ}\AgdaSymbol{)(}\AgdaBound{𝑨}\AgdaSpace{}%
\AgdaSymbol{:}\AgdaSpace{}%
\AgdaRecord{Algebra}\AgdaSpace{}%
\AgdaGeneralizable{α}\AgdaSpace{}%
\AgdaGeneralizable{ρᵃ}\AgdaSymbol{)}\AgdaSpace{}%
\AgdaSymbol{→}\AgdaSpace{}%
\AgdaPrimitive{Type}\AgdaSpace{}%
\AgdaSymbol{\AgdaUnderscore{}}\<%
\\
\>[0]\AgdaBound{𝑩}\AgdaSpace{}%
\AgdaOperator{\AgdaFunction{IsHomImageOf}}\AgdaSpace{}%
\AgdaBound{𝑨}\AgdaSpace{}%
\AgdaSymbol{=}\AgdaSpace{}%
\AgdaFunction{Σ[}\AgdaSpace{}%
\AgdaBound{φ}\AgdaSpace{}%
\AgdaFunction{∈}\AgdaSpace{}%
\AgdaFunction{hom}\AgdaSpace{}%
\AgdaBound{𝑨}\AgdaSpace{}%
\AgdaBound{𝑩}\AgdaSpace{}%
\AgdaFunction{]}\AgdaSpace{}%
\AgdaFunction{IsSurjective}\AgdaSpace{}%
\AgdaOperator{\AgdaFunction{∣}}\AgdaSpace{}%
\AgdaBound{φ}\AgdaSpace{}%
\AgdaOperator{\AgdaFunction{∣}}\<%
\\
\\[\AgdaEmptyExtraSkip]%
\>[0]\AgdaFunction{HomImages}\AgdaSpace{}%
\AgdaSymbol{:}\AgdaSpace{}%
\AgdaRecord{Algebra}\AgdaSpace{}%
\AgdaGeneralizable{α}\AgdaSpace{}%
\AgdaGeneralizable{ρᵃ}\AgdaSpace{}%
\AgdaSymbol{→}\AgdaSpace{}%
\AgdaPrimitive{Type}\AgdaSpace{}%
\AgdaSymbol{(}\AgdaGeneralizable{α}\AgdaSpace{}%
\AgdaOperator{\AgdaPrimitive{⊔}}\AgdaSpace{}%
\AgdaGeneralizable{ρᵃ}\AgdaSpace{}%
\AgdaOperator{\AgdaPrimitive{⊔}}\AgdaSpace{}%
\AgdaFunction{ov}\AgdaSpace{}%
\AgdaSymbol{(}\AgdaGeneralizable{β}\AgdaSpace{}%
\AgdaOperator{\AgdaPrimitive{⊔}}\AgdaSpace{}%
\AgdaGeneralizable{ρᵇ}\AgdaSymbol{))}\<%
\\
\>[0]\AgdaFunction{HomImages}\AgdaSpace{}%
\AgdaSymbol{\{}\AgdaArgument{β}\AgdaSpace{}%
\AgdaSymbol{=}\AgdaSpace{}%
\AgdaBound{β}\AgdaSymbol{\}\{}\AgdaArgument{ρᵇ}\AgdaSpace{}%
\AgdaSymbol{=}\AgdaSpace{}%
\AgdaBound{ρᵇ}\AgdaSymbol{\}}\AgdaSpace{}%
\AgdaBound{𝑨}\AgdaSpace{}%
\AgdaSymbol{=}\AgdaSpace{}%
\AgdaFunction{Σ[}\AgdaSpace{}%
\AgdaBound{𝑩}\AgdaSpace{}%
\AgdaFunction{∈}\AgdaSpace{}%
\AgdaRecord{Algebra}\AgdaSpace{}%
\AgdaBound{β}\AgdaSpace{}%
\AgdaBound{ρᵇ}\AgdaSpace{}%
\AgdaFunction{]}\AgdaSpace{}%
\AgdaBound{𝑩}\AgdaSpace{}%
\AgdaOperator{\AgdaFunction{IsHomImageOf}}\AgdaSpace{}%
\AgdaBound{𝑨}\<%
\\
\\[\AgdaEmptyExtraSkip]%
\>[0]\AgdaFunction{IdHomImage}\AgdaSpace{}%
\AgdaSymbol{:}\AgdaSpace{}%
\AgdaSymbol{\{}\AgdaBound{𝑨}\AgdaSpace{}%
\AgdaSymbol{:}\AgdaSpace{}%
\AgdaRecord{Algebra}\AgdaSpace{}%
\AgdaGeneralizable{α}\AgdaSpace{}%
\AgdaGeneralizable{ρᵃ}\AgdaSymbol{\}}\AgdaSpace{}%
\AgdaSymbol{→}\AgdaSpace{}%
\AgdaBound{𝑨}\AgdaSpace{}%
\AgdaOperator{\AgdaFunction{IsHomImageOf}}\AgdaSpace{}%
\AgdaBound{𝑨}\<%
\\
\>[0]\AgdaFunction{IdHomImage}\AgdaSpace{}%
\AgdaSymbol{\{}\AgdaArgument{α}\AgdaSpace{}%
\AgdaSymbol{=}\AgdaSpace{}%
\AgdaBound{α}\AgdaSymbol{\}\{}\AgdaArgument{𝑨}\AgdaSpace{}%
\AgdaSymbol{=}\AgdaSpace{}%
\AgdaBound{𝑨}\AgdaSymbol{\}}\AgdaSpace{}%
\AgdaSymbol{=}\AgdaSpace{}%
\AgdaFunction{𝒾𝒹}\AgdaSpace{}%
\AgdaOperator{\AgdaInductiveConstructor{,}}\AgdaSpace{}%
\AgdaSymbol{λ}\AgdaSpace{}%
\AgdaSymbol{\{}\AgdaBound{y}\AgdaSymbol{\}}\AgdaSpace{}%
\AgdaSymbol{→}\AgdaSpace{}%
\AgdaInductiveConstructor{Image\AgdaUnderscore{}∋\AgdaUnderscore{}.eq}\AgdaSpace{}%
\AgdaBound{y}\AgdaSpace{}%
\AgdaFunction{refl}\<%
\\
\>[0][@{}l@{\AgdaIndent{0}}]%
\>[1]\AgdaKeyword{where}\AgdaSpace{}%
\AgdaKeyword{open}\AgdaSpace{}%
\AgdaModule{Setoid}\AgdaSpace{}%
\AgdaOperator{\AgdaFunction{𝔻[}}\AgdaSpace{}%
\AgdaBound{𝑨}\AgdaSpace{}%
\AgdaOperator{\AgdaFunction{]}}\AgdaSpace{}%
\AgdaKeyword{using}\AgdaSpace{}%
\AgdaSymbol{(}\AgdaSpace{}%
\AgdaFunction{refl}\AgdaSpace{}%
\AgdaSymbol{)}\<%
\end{code}
\ifshort\else    %%% BEGIN LONG VERSION ONLY

\medskip

\noindent These types should be self-explanatory, but just to be sure, we pause
to describe the semantics of the Sigma type appearing in the definition of \af{HomImages}.
If \ab{𝑨} : \af{Algebra} \ab{α} \ab{ρᵃ} is an \ab{𝑆}-algebra, then \af{HomImages} \ab{𝑨}
denotes the type of pairs (\ab{𝑩} \aic{,} \ab p) such that \ab{𝑩} : \ar{Algebra} \ab{β} \ab{ρᵇ}
and \ab p is a proof that there exists a homomorphism from \ab{𝑨} onto \ab{𝑩}.
\fi      %%% END LONG VERSION ONLY SECTION

%% -------------------------------------------------------------------------------------
\subsection{Subalgebras}
\label{subalgebras}
Given \ab{𝑆}-algebras \ab{𝑨} and \ab{𝑩}, we say that \ab{𝑨} is a \defn{subalgebra} of
\ab{𝑨} and write \ab{𝑨}~\aof{≤}~\ab{𝑩} just in case \ab{𝑨} can be \emph{homomorphically
embedded} in \ab{𝑩}; in other terms, \ab{𝑨}~\aof{≤}~\ab{𝑩} iff there exists an injective
homomorphism from \ab{𝑨} to \ab{𝑩}. The following definition codifies the \defn{binary
subalgebra relation}, \aof{\au{}≤\au{}}, on the class of \ab{𝑆}-algebras.

\begin{code}%
\>[0]\<%
\\
\>[0]\AgdaOperator{\AgdaFunction{\AgdaUnderscore{}≤\AgdaUnderscore{}}}\AgdaSpace{}%
\AgdaSymbol{:}\AgdaSpace{}%
\AgdaRecord{Algebra}\AgdaSpace{}%
\AgdaGeneralizable{α}\AgdaSpace{}%
\AgdaGeneralizable{ρᵃ}\AgdaSpace{}%
\AgdaSymbol{→}\AgdaSpace{}%
\AgdaRecord{Algebra}\AgdaSpace{}%
\AgdaGeneralizable{β}\AgdaSpace{}%
\AgdaGeneralizable{ρᵇ}\AgdaSpace{}%
\AgdaSymbol{→}\AgdaSpace{}%
\AgdaPrimitive{Type}\AgdaSpace{}%
\AgdaSymbol{\AgdaUnderscore{}}\<%
\\
\>[0]\AgdaBound{𝑨}\AgdaSpace{}%
\AgdaOperator{\AgdaFunction{≤}}\AgdaSpace{}%
\AgdaBound{𝑩}\AgdaSpace{}%
\AgdaSymbol{=}\AgdaSpace{}%
\AgdaFunction{Σ[}\AgdaSpace{}%
\AgdaBound{h}\AgdaSpace{}%
\AgdaFunction{∈}\AgdaSpace{}%
\AgdaFunction{hom}\AgdaSpace{}%
\AgdaBound{𝑨}\AgdaSpace{}%
\AgdaBound{𝑩}\AgdaSpace{}%
\AgdaFunction{]}\AgdaSpace{}%
\AgdaFunction{IsInjective}\AgdaSpace{}%
\AgdaOperator{\AgdaFunction{∣}}\AgdaSpace{}%
\AgdaBound{h}\AgdaSpace{}%
\AgdaOperator{\AgdaFunction{∣}}\<%
\\
\>[0]\<%
\end{code}
Obviously the subalgebra relation is reflexive by the identity monomorphism; it is also
transitive since composition of monomorphisms is a monomorphism.
\ifshort
Here we merely give the formal statements of these assertions, omitting the easy proofs,
but \seeshort for details.
\else\fi

\begin{code}%
\>[0]\<%
\\
\>[0]\AgdaFunction{≤-reflexive}%
\>[14]\AgdaSymbol{:}%
\>[17]\AgdaSymbol{\{}\AgdaBound{𝑨}\AgdaSpace{}%
\AgdaSymbol{:}\AgdaSpace{}%
\AgdaRecord{Algebra}\AgdaSpace{}%
\AgdaGeneralizable{α}\AgdaSpace{}%
\AgdaGeneralizable{ρᵃ}\AgdaSymbol{\}}\AgdaSpace{}%
\AgdaSymbol{→}\AgdaSpace{}%
\AgdaBound{𝑨}\AgdaSpace{}%
\AgdaOperator{\AgdaFunction{≤}}\AgdaSpace{}%
\AgdaBound{𝑨}\<%
\end{code}
\ifshort
\vskip2mm
\else
\begin{code}%
\>[0]\AgdaFunction{≤-reflexive}\AgdaSpace{}%
\AgdaSymbol{\{}\AgdaArgument{𝑨}\AgdaSpace{}%
\AgdaSymbol{=}\AgdaSpace{}%
\AgdaBound{𝑨}\AgdaSymbol{\}}\AgdaSpace{}%
\AgdaSymbol{=}\AgdaSpace{}%
\AgdaFunction{𝒾𝒹}\AgdaSpace{}%
\AgdaOperator{\AgdaInductiveConstructor{,}}\AgdaSpace{}%
\AgdaFunction{id}\<%
\end{code}
\fi
\begin{code}%
\>[0]\AgdaFunction{≤-transitive}%
\>[14]\AgdaSymbol{:}%
\>[17]\AgdaSymbol{\{}\AgdaBound{𝑨}\AgdaSpace{}%
\AgdaSymbol{:}\AgdaSpace{}%
\AgdaRecord{Algebra}\AgdaSpace{}%
\AgdaGeneralizable{α}\AgdaSpace{}%
\AgdaGeneralizable{ρᵃ}\AgdaSymbol{\}\{}\AgdaBound{𝑩}\AgdaSpace{}%
\AgdaSymbol{:}\AgdaSpace{}%
\AgdaRecord{Algebra}\AgdaSpace{}%
\AgdaGeneralizable{β}\AgdaSpace{}%
\AgdaGeneralizable{ρᵇ}\AgdaSymbol{\}\{}\AgdaBound{𝑪}\AgdaSpace{}%
\AgdaSymbol{:}\AgdaSpace{}%
\AgdaRecord{Algebra}\AgdaSpace{}%
\AgdaGeneralizable{γ}\AgdaSpace{}%
\AgdaGeneralizable{ρᶜ}\AgdaSymbol{\}}\<%
\\
\>[0][@{}l@{\AgdaIndent{0}}]%
\>[1]\AgdaSymbol{→}%
\>[17]\AgdaBound{𝑨}\AgdaSpace{}%
\AgdaOperator{\AgdaFunction{≤}}\AgdaSpace{}%
\AgdaBound{𝑩}\AgdaSpace{}%
\AgdaSymbol{→}\AgdaSpace{}%
\AgdaBound{𝑩}\AgdaSpace{}%
\AgdaOperator{\AgdaFunction{≤}}\AgdaSpace{}%
\AgdaBound{𝑪}\AgdaSpace{}%
\AgdaSymbol{→}\AgdaSpace{}%
\AgdaBound{𝑨}\AgdaSpace{}%
\AgdaOperator{\AgdaFunction{≤}}\AgdaSpace{}%
\AgdaBound{𝑪}\<%
\end{code}

\ifshort
\vskip2mm
\else
\begin{code}%
\>[0]\AgdaFunction{≤-transitive}\AgdaSpace{}%
\AgdaSymbol{(}\AgdaSpace{}%
\AgdaBound{f}\AgdaSpace{}%
\AgdaOperator{\AgdaInductiveConstructor{,}}\AgdaSpace{}%
\AgdaBound{finj}\AgdaSpace{}%
\AgdaSymbol{)}\AgdaSpace{}%
\AgdaSymbol{(}\AgdaSpace{}%
\AgdaBound{g}\AgdaSpace{}%
\AgdaOperator{\AgdaInductiveConstructor{,}}\AgdaSpace{}%
\AgdaBound{ginj}\AgdaSpace{}%
\AgdaSymbol{)}\AgdaSpace{}%
\AgdaSymbol{=}\AgdaSpace{}%
\AgdaSymbol{(}\AgdaFunction{∘-hom}\AgdaSpace{}%
\AgdaBound{f}\AgdaSpace{}%
\AgdaBound{g}\AgdaSpace{}%
\AgdaSymbol{)}\AgdaSpace{}%
\AgdaOperator{\AgdaInductiveConstructor{,}}\AgdaSpace{}%
\AgdaFunction{∘-IsInjective}\AgdaSpace{}%
\AgdaOperator{\AgdaFunction{∣}}\AgdaSpace{}%
\AgdaBound{f}\AgdaSpace{}%
\AgdaOperator{\AgdaFunction{∣}}\AgdaSpace{}%
\AgdaOperator{\AgdaFunction{∣}}\AgdaSpace{}%
\AgdaBound{g}\AgdaSpace{}%
\AgdaOperator{\AgdaFunction{∣}}\AgdaSpace{}%
\AgdaBound{finj}\AgdaSpace{}%
\AgdaBound{ginj}\<%
\end{code}
\fi
\noindent If
\ab{𝒜} : \ab I → \af{Algebra} \ab{α} \ab{ρᵃ},
\ab{ℬ} : \ab I → \af{Algebra} \ab{β} \ab{ρᵇ} (families of \ab{𝑆}-algebras) and
\ab{ℬ} \ab i \af{≤} \ab{𝒜} \ab i for all \ab i~:~\ab I, then \af{⨅} \ab{ℬ} is a subalgebra
of \af{⨅} \ab{𝒜}.
\ifshort
Here is how we express this fact in Agda.
\else
\begin{code}%
\>[0]\AgdaKeyword{module}\AgdaSpace{}%
\AgdaModule{\AgdaUnderscore{}}\AgdaSpace{}%
\AgdaSymbol{\{}\AgdaBound{ι}\AgdaSpace{}%
\AgdaSymbol{:}\AgdaSpace{}%
\AgdaPostulate{Level}\AgdaSymbol{\}}\AgdaSpace{}%
\AgdaSymbol{\{}\AgdaBound{I}\AgdaSpace{}%
\AgdaSymbol{:}\AgdaSpace{}%
\AgdaPrimitive{Type}\AgdaSpace{}%
\AgdaBound{ι}\AgdaSymbol{\}\{}\AgdaBound{𝒜}\AgdaSpace{}%
\AgdaSymbol{:}\AgdaSpace{}%
\AgdaBound{I}\AgdaSpace{}%
\AgdaSymbol{→}\AgdaSpace{}%
\AgdaRecord{Algebra}\AgdaSpace{}%
\AgdaGeneralizable{α}\AgdaSpace{}%
\AgdaGeneralizable{ρᵃ}\AgdaSymbol{\}\{}\AgdaBound{ℬ}\AgdaSpace{}%
\AgdaSymbol{:}\AgdaSpace{}%
\AgdaBound{I}\AgdaSpace{}%
\AgdaSymbol{→}\AgdaSpace{}%
\AgdaRecord{Algebra}\AgdaSpace{}%
\AgdaGeneralizable{β}\AgdaSpace{}%
\AgdaGeneralizable{ρᵇ}\AgdaSymbol{\}}\AgdaSpace{}%
\AgdaKeyword{where}\<%
\end{code}
\fi
\begin{code}%
\>[0]\<%
\\
\>[0][@{}l@{\AgdaIndent{1}}]%
\>[1]\AgdaFunction{⨅-≤}\AgdaSpace{}%
\AgdaSymbol{:}\AgdaSpace{}%
\AgdaSymbol{(∀}\AgdaSpace{}%
\AgdaBound{i}\AgdaSpace{}%
\AgdaSymbol{→}\AgdaSpace{}%
\AgdaBound{ℬ}\AgdaSpace{}%
\AgdaBound{i}\AgdaSpace{}%
\AgdaOperator{\AgdaFunction{≤}}\AgdaSpace{}%
\AgdaBound{𝒜}\AgdaSpace{}%
\AgdaBound{i}\AgdaSymbol{)}\AgdaSpace{}%
\AgdaSymbol{→}\AgdaSpace{}%
\AgdaFunction{⨅}\AgdaSpace{}%
\AgdaBound{ℬ}\AgdaSpace{}%
\AgdaOperator{\AgdaFunction{≤}}\AgdaSpace{}%
\AgdaFunction{⨅}\AgdaSpace{}%
\AgdaBound{𝒜}\<%
\end{code}
\ifshort
\vskip2mm
\else
\begin{code}%
\>[1]\AgdaFunction{⨅-≤}\AgdaSpace{}%
\AgdaBound{B≤A}\AgdaSpace{}%
\AgdaSymbol{=}\AgdaSpace{}%
\AgdaSymbol{(}\AgdaFunction{hfunc}\AgdaSpace{}%
\AgdaOperator{\AgdaInductiveConstructor{,}}\AgdaSpace{}%
\AgdaFunction{hhom}\AgdaSymbol{)}\AgdaSpace{}%
\AgdaOperator{\AgdaInductiveConstructor{,}}\AgdaSpace{}%
\AgdaFunction{hM}\<%
\\
\>[1][@{}l@{\AgdaIndent{0}}]%
\>[2]\AgdaKeyword{where}\<%
\\
\>[2]\AgdaFunction{hi}\AgdaSpace{}%
\AgdaSymbol{:}\AgdaSpace{}%
\AgdaSymbol{∀}\AgdaSpace{}%
\AgdaBound{i}\AgdaSpace{}%
\AgdaSymbol{→}\AgdaSpace{}%
\AgdaFunction{hom}\AgdaSpace{}%
\AgdaSymbol{(}\AgdaBound{ℬ}\AgdaSpace{}%
\AgdaBound{i}\AgdaSymbol{)}\AgdaSpace{}%
\AgdaSymbol{(}\AgdaBound{𝒜}\AgdaSpace{}%
\AgdaBound{i}\AgdaSymbol{)}\<%
\\
\>[2]\AgdaFunction{hi}\AgdaSpace{}%
\AgdaBound{i}\AgdaSpace{}%
\AgdaSymbol{=}\AgdaSpace{}%
\AgdaOperator{\AgdaFunction{∣}}\AgdaSpace{}%
\AgdaBound{B≤A}\AgdaSpace{}%
\AgdaBound{i}\AgdaSpace{}%
\AgdaOperator{\AgdaFunction{∣}}\<%
\\
\>[2]\AgdaFunction{hfunc}\AgdaSpace{}%
\AgdaSymbol{:}\AgdaSpace{}%
\AgdaOperator{\AgdaFunction{𝔻[}}\AgdaSpace{}%
\AgdaFunction{⨅}\AgdaSpace{}%
\AgdaBound{ℬ}\AgdaSpace{}%
\AgdaOperator{\AgdaFunction{]}}\AgdaSpace{}%
\AgdaOperator{\AgdaRecord{⟶}}\AgdaSpace{}%
\AgdaOperator{\AgdaFunction{𝔻[}}\AgdaSpace{}%
\AgdaFunction{⨅}\AgdaSpace{}%
\AgdaBound{𝒜}\AgdaSpace{}%
\AgdaOperator{\AgdaFunction{]}}\<%
\\
\>[2]\AgdaSymbol{(}\AgdaFunction{hfunc}\AgdaSpace{}%
\AgdaOperator{\AgdaField{⟨\$⟩}}\AgdaSpace{}%
\AgdaBound{x}\AgdaSymbol{)}\AgdaSpace{}%
\AgdaBound{i}\AgdaSpace{}%
\AgdaSymbol{=}\AgdaSpace{}%
\AgdaOperator{\AgdaFunction{∣}}\AgdaSpace{}%
\AgdaFunction{hi}\AgdaSpace{}%
\AgdaBound{i}\AgdaSpace{}%
\AgdaOperator{\AgdaFunction{∣}}\AgdaSpace{}%
\AgdaOperator{\AgdaField{⟨\$⟩}}\AgdaSpace{}%
\AgdaBound{x}\AgdaSpace{}%
\AgdaBound{i}\<%
\\
\>[2]\AgdaField{cong}\AgdaSpace{}%
\AgdaFunction{hfunc}\AgdaSpace{}%
\AgdaSymbol{=}\AgdaSpace{}%
\AgdaSymbol{λ}\AgdaSpace{}%
\AgdaBound{xy}\AgdaSpace{}%
\AgdaBound{i}\AgdaSpace{}%
\AgdaSymbol{→}\AgdaSpace{}%
\AgdaField{cong}\AgdaSpace{}%
\AgdaOperator{\AgdaFunction{∣}}\AgdaSpace{}%
\AgdaFunction{hi}\AgdaSpace{}%
\AgdaBound{i}\AgdaSpace{}%
\AgdaOperator{\AgdaFunction{∣}}\AgdaSpace{}%
\AgdaSymbol{(}\AgdaBound{xy}\AgdaSpace{}%
\AgdaBound{i}\AgdaSymbol{)}\<%
\\
\>[2]\AgdaFunction{hhom}\AgdaSpace{}%
\AgdaSymbol{:}\AgdaSpace{}%
\AgdaRecord{IsHom}\AgdaSpace{}%
\AgdaSymbol{(}\AgdaFunction{⨅}\AgdaSpace{}%
\AgdaBound{ℬ}\AgdaSymbol{)}\AgdaSpace{}%
\AgdaSymbol{(}\AgdaFunction{⨅}\AgdaSpace{}%
\AgdaBound{𝒜}\AgdaSymbol{)}\AgdaSpace{}%
\AgdaFunction{hfunc}\<%
\\
\>[2]\AgdaField{compatible}\AgdaSpace{}%
\AgdaFunction{hhom}\AgdaSpace{}%
\AgdaSymbol{=}\AgdaSpace{}%
\AgdaSymbol{λ}\AgdaSpace{}%
\AgdaBound{i}\AgdaSpace{}%
\AgdaSymbol{→}\AgdaSpace{}%
\AgdaField{compatible}\AgdaSpace{}%
\AgdaOperator{\AgdaFunction{∥}}\AgdaSpace{}%
\AgdaFunction{hi}\AgdaSpace{}%
\AgdaBound{i}\AgdaSpace{}%
\AgdaOperator{\AgdaFunction{∥}}\<%
\\
\>[2]\AgdaFunction{hM}\AgdaSpace{}%
\AgdaSymbol{:}\AgdaSpace{}%
\AgdaFunction{IsInjective}\AgdaSpace{}%
\AgdaFunction{hfunc}\<%
\\
\>[2]\AgdaFunction{hM}\AgdaSpace{}%
\AgdaSymbol{=}\AgdaSpace{}%
\AgdaSymbol{λ}\AgdaSpace{}%
\AgdaBound{xy}\AgdaSpace{}%
\AgdaBound{i}\AgdaSpace{}%
\AgdaSymbol{→}\AgdaSpace{}%
\AgdaOperator{\AgdaFunction{∥}}\AgdaSpace{}%
\AgdaBound{B≤A}\AgdaSpace{}%
\AgdaBound{i}\AgdaSpace{}%
\AgdaOperator{\AgdaFunction{∥}}\AgdaSpace{}%
\AgdaSymbol{(}\AgdaBound{xy}\AgdaSpace{}%
\AgdaBound{i}\AgdaSymbol{)}\<%
\\
\>[0]\<%
\end{code}
\fi

We conclude this brief subsection on subalgebras
\ifshort
by mentioning the function \af{mon→≤}, which we apply once below; it merely converts a monomorphism into a pair in \aof{≤}.
\else
with two easy facts
that will be useful later. The first merely converts a monomorphism into a pair in the subalgebra relation
while the second is an algebraic invariance property of \aof{≤}.

\begin{code}%
\>[0]\<%
\\
\>[0]\AgdaFunction{mon→≤}%
\>[11]\AgdaSymbol{:}%
\>[14]\AgdaSymbol{\{}\AgdaBound{𝑨}\AgdaSpace{}%
\AgdaSymbol{:}\AgdaSpace{}%
\AgdaRecord{Algebra}\AgdaSpace{}%
\AgdaGeneralizable{α}\AgdaSpace{}%
\AgdaGeneralizable{ρᵃ}\AgdaSymbol{\}\{}\AgdaBound{𝑩}\AgdaSpace{}%
\AgdaSymbol{:}\AgdaSpace{}%
\AgdaRecord{Algebra}\AgdaSpace{}%
\AgdaGeneralizable{β}\AgdaSpace{}%
\AgdaGeneralizable{ρᵇ}\AgdaSymbol{\}}\AgdaSpace{}%
\AgdaSymbol{→}\AgdaSpace{}%
\AgdaFunction{mon}\AgdaSpace{}%
\AgdaBound{𝑨}\AgdaSpace{}%
\AgdaBound{𝑩}\AgdaSpace{}%
\AgdaSymbol{→}\AgdaSpace{}%
\AgdaBound{𝑨}\AgdaSpace{}%
\AgdaOperator{\AgdaFunction{≤}}\AgdaSpace{}%
\AgdaBound{𝑩}\<%
\\
\>[0]\AgdaFunction{mon→≤}\AgdaSpace{}%
\AgdaSymbol{\{}\AgdaArgument{𝑨}\AgdaSpace{}%
\AgdaSymbol{=}\AgdaSpace{}%
\AgdaBound{𝑨}\AgdaSymbol{\}\{}\AgdaBound{𝑩}\AgdaSymbol{\}}\AgdaSpace{}%
\AgdaBound{x}\AgdaSpace{}%
\AgdaSymbol{=}\AgdaSpace{}%
\AgdaFunction{mon→intohom}\AgdaSpace{}%
\AgdaBound{𝑨}\AgdaSpace{}%
\AgdaBound{𝑩}\AgdaSpace{}%
\AgdaBound{x}\<%
\\
\\[\AgdaEmptyExtraSkip]%
\>[0]\AgdaFunction{≅-trans-≤}%
\>[11]\AgdaSymbol{:}%
\>[14]\AgdaSymbol{\{}\AgdaBound{𝑨}\AgdaSpace{}%
\AgdaSymbol{:}\AgdaSpace{}%
\AgdaRecord{Algebra}\AgdaSpace{}%
\AgdaGeneralizable{α}\AgdaSpace{}%
\AgdaGeneralizable{ρᵃ}\AgdaSymbol{\}\{}\AgdaBound{𝑩}\AgdaSpace{}%
\AgdaSymbol{:}\AgdaSpace{}%
\AgdaRecord{Algebra}\AgdaSpace{}%
\AgdaGeneralizable{β}\AgdaSpace{}%
\AgdaGeneralizable{ρᵇ}\AgdaSymbol{\}\{}\AgdaBound{𝑪}\AgdaSpace{}%
\AgdaSymbol{:}\AgdaSpace{}%
\AgdaRecord{Algebra}\AgdaSpace{}%
\AgdaGeneralizable{γ}\AgdaSpace{}%
\AgdaGeneralizable{ρᶜ}\AgdaSymbol{\}}\<%
\\
\>[0][@{}l@{\AgdaIndent{0}}]%
\>[1]\AgdaSymbol{→}%
\>[14]\AgdaBound{𝑨}\AgdaSpace{}%
\AgdaOperator{\AgdaRecord{≅}}\AgdaSpace{}%
\AgdaBound{𝑩}\AgdaSpace{}%
\AgdaSymbol{→}\AgdaSpace{}%
\AgdaBound{𝑩}\AgdaSpace{}%
\AgdaOperator{\AgdaFunction{≤}}\AgdaSpace{}%
\AgdaBound{𝑪}\AgdaSpace{}%
\AgdaSymbol{→}\AgdaSpace{}%
\AgdaBound{𝑨}\AgdaSpace{}%
\AgdaOperator{\AgdaFunction{≤}}\AgdaSpace{}%
\AgdaBound{𝑪}\<%
\\
\>[0]\AgdaFunction{≅-trans-≤}\AgdaSpace{}%
\AgdaBound{A≅B}\AgdaSpace{}%
\AgdaSymbol{(}\AgdaBound{h}\AgdaSpace{}%
\AgdaOperator{\AgdaInductiveConstructor{,}}\AgdaSpace{}%
\AgdaBound{hinj}\AgdaSymbol{)}\AgdaSpace{}%
\AgdaSymbol{=}\AgdaSpace{}%
\AgdaSymbol{(}\AgdaFunction{∘-hom}\AgdaSpace{}%
\AgdaSymbol{(}\AgdaField{to}\AgdaSpace{}%
\AgdaBound{A≅B}\AgdaSymbol{)}\AgdaSpace{}%
\AgdaBound{h}\AgdaSymbol{)}\AgdaSpace{}%
\AgdaOperator{\AgdaInductiveConstructor{,}}\AgdaSpace{}%
\AgdaSymbol{(}\AgdaFunction{∘-IsInjective}\AgdaSpace{}%
\AgdaOperator{\AgdaFunction{∣}}\AgdaSpace{}%
\AgdaField{to}\AgdaSpace{}%
\AgdaBound{A≅B}\AgdaSpace{}%
\AgdaOperator{\AgdaFunction{∣}}\AgdaSpace{}%
\AgdaOperator{\AgdaFunction{∣}}\AgdaSpace{}%
\AgdaBound{h}\AgdaSpace{}%
\AgdaOperator{\AgdaFunction{∣}}\AgdaSpace{}%
\AgdaSymbol{(}\AgdaFunction{toIsInjective}\AgdaSpace{}%
\AgdaBound{A≅B}\AgdaSymbol{)}\AgdaSpace{}%
\AgdaBound{hinj}\AgdaSymbol{)}\<%
\end{code}
\fi

%% -------------------------------------------------------------------------------------

\subsection{Terms}
\label{terms}
Fix a signature \ab{𝑆} and let \ab X denote an arbitrary nonempty collection of variable
symbols. Such a collection of variable symbols is called a \defn{context}.
Assume the symbols in \ab X are distinct from the operation symbols of
\ab{𝑆}, that is \ab X \aof{∩} \aof{∣} \ab{𝑆} \aof{∣} = ∅.
A \defn{word} in the language of \ab{𝑆} is a finite sequence of members of \ab X \aof{∪}
\aof{∣~\ab{𝑆}~∣}. We denote the concatenation of such sequences by simple juxtaposition.
Let \ab{S₀} denote the set of nullary operation symbols of \ab{𝑆}. We define by induction
on \ab n the sets \ab{𝑇ₙ} of \emph{words} over \ab X \aof{∪} \aof{∣~\ab{𝑆}~∣} as
follows (cf.~\cite[Def. 4.19]{Bergman:2012}): \ab{𝑇₀} := \ab X \aof{∪} \ab{S₀} and
\ab{𝑇ₙ₊₁} := \ab{𝑇ₙ} \aof{∪} \ab{𝒯ₙ}, where \ab{𝒯ₙ} is the collection of all \ab f \ab t
such that \ab f : \aof{∣~\ab{𝑆}~∣} and \ab t : \aof{∥~\ab{𝑆}~∥} \ab f \as{→}
\ab{𝑇ₙ}.
\ifshort\else
(Recall, \aof{∥~\ab{𝑆}~∥} \ab f is the arity of the operation symbol \ab f.)
\fi
An \ab{𝑆}-\defn{term} is a term in the language of \ab{𝑆} and the collection of all
\ab{𝑆}-\defn{terms} in the context \ab X is given by \Term{X} := \aof{⋃ₙ} \ab{𝑇ₙ}.

As even its informal definition of \Term{X} is recursive, it should come as no surprise
that the semantics of terms can be usefully represented in type theory as an inductive
type. Indeed, here is such a representation.

\begin{code}%
\>[0]\<%
\\
\>[0]\AgdaKeyword{data}\AgdaSpace{}%
\AgdaDatatype{Term}\AgdaSpace{}%
\AgdaSymbol{(}\AgdaBound{X}\AgdaSpace{}%
\AgdaSymbol{:}\AgdaSpace{}%
\AgdaPrimitive{Type}\AgdaSpace{}%
\AgdaGeneralizable{χ}\AgdaSpace{}%
\AgdaSymbol{)}\AgdaSpace{}%
\AgdaSymbol{:}\AgdaSpace{}%
\AgdaPrimitive{Type}\AgdaSpace{}%
\AgdaSymbol{(}\AgdaFunction{ov}\AgdaSpace{}%
\AgdaBound{χ}\AgdaSymbol{)}%
\>[39]\AgdaKeyword{where}\<%
\\
\>[0][@{}l@{\AgdaIndent{0}}]%
\>[1]\AgdaInductiveConstructor{ℊ}\AgdaSpace{}%
\AgdaSymbol{:}\AgdaSpace{}%
\AgdaBound{X}\AgdaSpace{}%
\AgdaSymbol{→}\AgdaSpace{}%
\AgdaDatatype{Term}\AgdaSpace{}%
\AgdaBound{X}\<%
\\
\>[1]\AgdaInductiveConstructor{node}\AgdaSpace{}%
\AgdaSymbol{:}\AgdaSpace{}%
\AgdaSymbol{(}\AgdaBound{f}\AgdaSpace{}%
\AgdaSymbol{:}\AgdaSpace{}%
\AgdaOperator{\AgdaFunction{∣}}\AgdaSpace{}%
\AgdaBound{𝑆}\AgdaSpace{}%
\AgdaOperator{\AgdaFunction{∣}}\AgdaSymbol{)(}\AgdaBound{t}\AgdaSpace{}%
\AgdaSymbol{:}\AgdaSpace{}%
\AgdaOperator{\AgdaFunction{∥}}\AgdaSpace{}%
\AgdaBound{𝑆}\AgdaSpace{}%
\AgdaOperator{\AgdaFunction{∥}}\AgdaSpace{}%
\AgdaBound{f}\AgdaSpace{}%
\AgdaSymbol{→}\AgdaSpace{}%
\AgdaDatatype{Term}\AgdaSpace{}%
\AgdaBound{X}\AgdaSymbol{)}\AgdaSpace{}%
\AgdaSymbol{→}\AgdaSpace{}%
\AgdaDatatype{Term}\AgdaSpace{}%
\AgdaBound{X}\<%
\\
\>[0]\<%
\end{code}
This basic inductive type represents each term as a tree with an operation symbol at each
\aic{node} and a variable symbol at each leaf \aic{ℊ}%
\ifshort
.
\else
; hence the constructor names
(\aic{ℊ} for ``generator'' and \aic{node} for ``node'').
\fi

\paragraph*{The term algebra}
We enrich the \ad{Term} type with
an inductive type \ad{\au{}≃\au{}} representing equality of terms, then we roll up
into a setoid the types \ad{Term} and \ad{\au{}≃\au{}} along with a proof that
\ad{\au{}≃\au{}} is an equivalence relation. Ultimately we use this setoid of \ab{𝑆}-terms
as the domain of an algebra, called the \emph{term algebra in the signature} \ab{𝑆}.
Here is the equality type on terms.

\ifshort\else
\begin{code}%
\>[0]\<%
\\
\>[0]\AgdaKeyword{module}\AgdaSpace{}%
\AgdaModule{\AgdaUnderscore{}}\AgdaSpace{}%
\AgdaSymbol{\{}\AgdaBound{X}\AgdaSpace{}%
\AgdaSymbol{:}\AgdaSpace{}%
\AgdaPrimitive{Type}\AgdaSpace{}%
\AgdaGeneralizable{χ}\AgdaSpace{}%
\AgdaSymbol{\}}\AgdaSpace{}%
\AgdaKeyword{where}\<%
\end{code}
\fi
\begin{code}%
\>[0]\<%
\\
\>[0][@{}l@{\AgdaIndent{1}}]%
\>[1]\AgdaKeyword{data}\AgdaSpace{}%
\AgdaOperator{\AgdaDatatype{\AgdaUnderscore{}≃\AgdaUnderscore{}}}\AgdaSpace{}%
\AgdaSymbol{:}\AgdaSpace{}%
\AgdaDatatype{Term}\AgdaSpace{}%
\AgdaBound{X}\AgdaSpace{}%
\AgdaSymbol{→}\AgdaSpace{}%
\AgdaDatatype{Term}\AgdaSpace{}%
\AgdaBound{X}\AgdaSpace{}%
\AgdaSymbol{→}\AgdaSpace{}%
\AgdaPrimitive{Type}\AgdaSpace{}%
\AgdaSymbol{(}\AgdaFunction{ov}\AgdaSpace{}%
\AgdaBound{χ}\AgdaSymbol{)}\AgdaSpace{}%
\AgdaKeyword{where}\<%
\\
\>[1][@{}l@{\AgdaIndent{0}}]%
\>[2]\AgdaInductiveConstructor{rfl}\AgdaSpace{}%
\AgdaSymbol{:}\AgdaSpace{}%
\AgdaSymbol{\{}\AgdaBound{x}\AgdaSpace{}%
\AgdaBound{y}\AgdaSpace{}%
\AgdaSymbol{:}\AgdaSpace{}%
\AgdaBound{X}\AgdaSymbol{\}}\AgdaSpace{}%
\AgdaSymbol{→}\AgdaSpace{}%
\AgdaBound{x}\AgdaSpace{}%
\AgdaOperator{\AgdaDatatype{≡}}\AgdaSpace{}%
\AgdaBound{y}\AgdaSpace{}%
\AgdaSymbol{→}\AgdaSpace{}%
\AgdaSymbol{(}\AgdaInductiveConstructor{ℊ}\AgdaSpace{}%
\AgdaBound{x}\AgdaSymbol{)}\AgdaSpace{}%
\AgdaOperator{\AgdaDatatype{≃}}\AgdaSpace{}%
\AgdaSymbol{(}\AgdaInductiveConstructor{ℊ}\AgdaSpace{}%
\AgdaBound{y}\AgdaSymbol{)}\<%
\\
\>[2]\AgdaInductiveConstructor{gnl}\AgdaSpace{}%
\AgdaSymbol{:}\AgdaSpace{}%
\AgdaSymbol{∀}\AgdaSpace{}%
\AgdaSymbol{\{}\AgdaBound{f}\AgdaSymbol{\}\{}\AgdaBound{s}\AgdaSpace{}%
\AgdaBound{t}\AgdaSpace{}%
\AgdaSymbol{:}\AgdaSpace{}%
\AgdaOperator{\AgdaFunction{∥}}\AgdaSpace{}%
\AgdaBound{𝑆}\AgdaSpace{}%
\AgdaOperator{\AgdaFunction{∥}}\AgdaSpace{}%
\AgdaBound{f}\AgdaSpace{}%
\AgdaSymbol{→}\AgdaSpace{}%
\AgdaDatatype{Term}\AgdaSpace{}%
\AgdaBound{X}\AgdaSymbol{\}}\AgdaSpace{}%
\AgdaSymbol{→}\AgdaSpace{}%
\AgdaSymbol{(∀}\AgdaSpace{}%
\AgdaBound{i}\AgdaSpace{}%
\AgdaSymbol{→}\AgdaSpace{}%
\AgdaSymbol{(}\AgdaBound{s}\AgdaSpace{}%
\AgdaBound{i}\AgdaSymbol{)}\AgdaSpace{}%
\AgdaOperator{\AgdaDatatype{≃}}\AgdaSpace{}%
\AgdaSymbol{(}\AgdaBound{t}\AgdaSpace{}%
\AgdaBound{i}\AgdaSymbol{))}\AgdaSpace{}%
\AgdaSymbol{→}\AgdaSpace{}%
\AgdaSymbol{(}\AgdaInductiveConstructor{node}\AgdaSpace{}%
\AgdaBound{f}\AgdaSpace{}%
\AgdaBound{s}\AgdaSymbol{)}\AgdaSpace{}%
\AgdaOperator{\AgdaDatatype{≃}}\AgdaSpace{}%
\AgdaSymbol{(}\AgdaInductiveConstructor{node}\AgdaSpace{}%
\AgdaBound{f}\AgdaSpace{}%
\AgdaBound{t}\AgdaSymbol{)}\<%
\\
\>[0]\<%
\end{code}
It's easy to show that this is an equivalence relation on terms%
\ifshort
; the proof is called \af{≃-isEquiv} in the \agdaalgebras library.
\else
, as follows.

\begin{code}%
\>[0]\<%
\\
\>[0][@{}l@{\AgdaIndent{1}}]%
\>[1]\AgdaFunction{≃-isRefl}%
\>[12]\AgdaSymbol{:}\AgdaSpace{}%
\AgdaFunction{Reflexive}%
\>[29]\AgdaOperator{\AgdaDatatype{\AgdaUnderscore{}≃\AgdaUnderscore{}}}\<%
\\
\>[1]\AgdaFunction{≃-isRefl}\AgdaSpace{}%
\AgdaSymbol{\{}\AgdaInductiveConstructor{ℊ}\AgdaSpace{}%
\AgdaSymbol{\AgdaUnderscore{}\}}\AgdaSpace{}%
\AgdaSymbol{=}\AgdaSpace{}%
\AgdaInductiveConstructor{rfl}\AgdaSpace{}%
\AgdaInductiveConstructor{≡.refl}\<%
\\
\>[1]\AgdaFunction{≃-isRefl}\AgdaSpace{}%
\AgdaSymbol{\{}\AgdaInductiveConstructor{node}\AgdaSpace{}%
\AgdaSymbol{\AgdaUnderscore{}}\AgdaSpace{}%
\AgdaSymbol{\AgdaUnderscore{}\}}\AgdaSpace{}%
\AgdaSymbol{=}\AgdaSpace{}%
\AgdaInductiveConstructor{gnl}\AgdaSpace{}%
\AgdaSymbol{(λ}\AgdaSpace{}%
\AgdaBound{\AgdaUnderscore{}}\AgdaSpace{}%
\AgdaSymbol{→}\AgdaSpace{}%
\AgdaFunction{≃-isRefl}\AgdaSymbol{)}\<%
\\
\\[\AgdaEmptyExtraSkip]%
\>[1]\AgdaFunction{≃-isSym}%
\>[12]\AgdaSymbol{:}\AgdaSpace{}%
\AgdaFunction{Symmetric}%
\>[29]\AgdaOperator{\AgdaDatatype{\AgdaUnderscore{}≃\AgdaUnderscore{}}}\<%
\\
\>[1]\AgdaFunction{≃-isSym}\AgdaSpace{}%
\AgdaSymbol{(}\AgdaInductiveConstructor{rfl}\AgdaSpace{}%
\AgdaBound{x}\AgdaSymbol{)}\AgdaSpace{}%
\AgdaSymbol{=}\AgdaSpace{}%
\AgdaInductiveConstructor{rfl}\AgdaSpace{}%
\AgdaSymbol{(}\AgdaFunction{≡.sym}\AgdaSpace{}%
\AgdaBound{x}\AgdaSymbol{)}\<%
\\
\>[1]\AgdaFunction{≃-isSym}\AgdaSpace{}%
\AgdaSymbol{(}\AgdaInductiveConstructor{gnl}\AgdaSpace{}%
\AgdaBound{x}\AgdaSymbol{)}\AgdaSpace{}%
\AgdaSymbol{=}\AgdaSpace{}%
\AgdaInductiveConstructor{gnl}\AgdaSpace{}%
\AgdaSymbol{(λ}\AgdaSpace{}%
\AgdaBound{i}\AgdaSpace{}%
\AgdaSymbol{→}\AgdaSpace{}%
\AgdaFunction{≃-isSym}\AgdaSpace{}%
\AgdaSymbol{(}\AgdaBound{x}\AgdaSpace{}%
\AgdaBound{i}\AgdaSymbol{))}\<%
\\
\\[\AgdaEmptyExtraSkip]%
\>[1]\AgdaFunction{≃-isTrans}%
\>[12]\AgdaSymbol{:}\AgdaSpace{}%
\AgdaFunction{Transitive}%
\>[29]\AgdaOperator{\AgdaDatatype{\AgdaUnderscore{}≃\AgdaUnderscore{}}}\<%
\\
\>[1]\AgdaFunction{≃-isTrans}\AgdaSpace{}%
\AgdaSymbol{(}\AgdaInductiveConstructor{rfl}\AgdaSpace{}%
\AgdaBound{x}\AgdaSymbol{)}\AgdaSpace{}%
\AgdaSymbol{(}\AgdaInductiveConstructor{rfl}\AgdaSpace{}%
\AgdaBound{y}\AgdaSymbol{)}\AgdaSpace{}%
\AgdaSymbol{=}\AgdaSpace{}%
\AgdaInductiveConstructor{rfl}\AgdaSpace{}%
\AgdaSymbol{(}\AgdaFunction{≡.trans}\AgdaSpace{}%
\AgdaBound{x}\AgdaSpace{}%
\AgdaBound{y}\AgdaSymbol{)}\<%
\\
\>[1]\AgdaFunction{≃-isTrans}\AgdaSpace{}%
\AgdaSymbol{(}\AgdaInductiveConstructor{gnl}\AgdaSpace{}%
\AgdaBound{x}\AgdaSymbol{)}\AgdaSpace{}%
\AgdaSymbol{(}\AgdaInductiveConstructor{gnl}\AgdaSpace{}%
\AgdaBound{y}\AgdaSymbol{)}\AgdaSpace{}%
\AgdaSymbol{=}\AgdaSpace{}%
\AgdaInductiveConstructor{gnl}\AgdaSpace{}%
\AgdaSymbol{(λ}\AgdaSpace{}%
\AgdaBound{i}\AgdaSpace{}%
\AgdaSymbol{→}\AgdaSpace{}%
\AgdaFunction{≃-isTrans}\AgdaSpace{}%
\AgdaSymbol{(}\AgdaBound{x}\AgdaSpace{}%
\AgdaBound{i}\AgdaSymbol{)}\AgdaSpace{}%
\AgdaSymbol{(}\AgdaBound{y}\AgdaSpace{}%
\AgdaBound{i}\AgdaSymbol{))}\<%
\\
\\[\AgdaEmptyExtraSkip]%
\>[1]\AgdaFunction{≃-isEquiv}%
\>[12]\AgdaSymbol{:}\AgdaSpace{}%
\AgdaRecord{IsEquivalence}%
\>[29]\AgdaOperator{\AgdaDatatype{\AgdaUnderscore{}≃\AgdaUnderscore{}}}\<%
\\
\>[1]\AgdaFunction{≃-isEquiv}\AgdaSpace{}%
\AgdaSymbol{=}\AgdaSpace{}%
\AgdaKeyword{record}\AgdaSpace{}%
\AgdaSymbol{\{}\AgdaSpace{}%
\AgdaField{refl}\AgdaSpace{}%
\AgdaSymbol{=}\AgdaSpace{}%
\AgdaFunction{≃-isRefl}\AgdaSpace{}%
\AgdaSymbol{;}\AgdaSpace{}%
\AgdaField{sym}\AgdaSpace{}%
\AgdaSymbol{=}\AgdaSpace{}%
\AgdaFunction{≃-isSym}\AgdaSpace{}%
\AgdaSymbol{;}\AgdaSpace{}%
\AgdaField{trans}\AgdaSpace{}%
\AgdaSymbol{=}\AgdaSpace{}%
\AgdaFunction{≃-isTrans}\AgdaSpace{}%
\AgdaSymbol{\}}\<%
\end{code}
\fi

We now define, for a given signature \ab{𝑆} and context \ab X,
%if the type \Term{X} is nonempty (equivalently, if \ab X or
%\aof{∣~\ab{𝑆}~∣} is nonempty), then
the algebraic structure \T{X}, known as the \defn{term algebra in} \ab{𝑆} \defn{over} \ab
X.  Terms are viewed as acting on other terms, so both the elements of the domain of \T{X}
and its basic operations are the terms themselves. That is, for each operation symbol \ab
f : \aof{∣~\ab{𝑆}~∣}, we denote by \ab f~\aof{̂}~\T{X} the operation on \Term{X} that maps
each tuple of terms, say, \ab t : \aof{∥~\ab{𝑆}~∥} \ab f \as{→} \Term{X}, to the formal
term \ab f \ab t.
%We let \T{X} denote the term algebra in \ab{𝑆} over \ab X; it has universe \Term{X} and
%operations \ab f \aof{̂} \T{X}, one for each symbol \ab f in \aof{∣~\ab{𝑆}~∣}.
We codify these notions in Agda as follows.

\begin{code}%
\>[0]\<%
\\
\>[0]\AgdaFunction{TermSetoid}\AgdaSpace{}%
\AgdaSymbol{:}\AgdaSpace{}%
\AgdaSymbol{(}\AgdaBound{X}\AgdaSpace{}%
\AgdaSymbol{:}\AgdaSpace{}%
\AgdaPrimitive{Type}\AgdaSpace{}%
\AgdaGeneralizable{χ}\AgdaSymbol{)}\AgdaSpace{}%
\AgdaSymbol{→}\AgdaSpace{}%
\AgdaRecord{Setoid}\AgdaSpace{}%
\AgdaSymbol{\AgdaUnderscore{}}\AgdaSpace{}%
\AgdaSymbol{\AgdaUnderscore{}}\<%
\\
\>[0]\AgdaFunction{TermSetoid}\AgdaSpace{}%
\AgdaBound{X}\AgdaSpace{}%
\AgdaSymbol{=}\AgdaSpace{}%
\AgdaKeyword{record}\AgdaSpace{}%
\AgdaSymbol{\{}\AgdaSpace{}%
\AgdaField{Carrier}\AgdaSpace{}%
\AgdaSymbol{=}\AgdaSpace{}%
\AgdaDatatype{Term}\AgdaSpace{}%
\AgdaBound{X}\AgdaSpace{}%
\AgdaSymbol{;}\AgdaSpace{}%
\AgdaOperator{\AgdaField{\AgdaUnderscore{}≈\AgdaUnderscore{}}}\AgdaSpace{}%
\AgdaSymbol{=}\AgdaSpace{}%
\AgdaOperator{\AgdaDatatype{\AgdaUnderscore{}≃\AgdaUnderscore{}}}\AgdaSpace{}%
\AgdaSymbol{;}\AgdaSpace{}%
\AgdaField{isEquivalence}\AgdaSpace{}%
\AgdaSymbol{=}\AgdaSpace{}%
\AgdaFunction{≃-isEquiv}\AgdaSpace{}%
\AgdaSymbol{\}}\<%
\\
\\[\AgdaEmptyExtraSkip]%
\>[0]\AgdaFunction{𝑻}\AgdaSpace{}%
\AgdaSymbol{:}\AgdaSpace{}%
\AgdaSymbol{(}\AgdaBound{X}\AgdaSpace{}%
\AgdaSymbol{:}\AgdaSpace{}%
\AgdaPrimitive{Type}\AgdaSpace{}%
\AgdaGeneralizable{χ}\AgdaSymbol{)}\AgdaSpace{}%
\AgdaSymbol{→}\AgdaSpace{}%
\AgdaRecord{Algebra}\AgdaSpace{}%
\AgdaSymbol{(}\AgdaFunction{ov}\AgdaSpace{}%
\AgdaGeneralizable{χ}\AgdaSymbol{)}\AgdaSpace{}%
\AgdaSymbol{(}\AgdaFunction{ov}\AgdaSpace{}%
\AgdaGeneralizable{χ}\AgdaSymbol{)}\<%
\\
\>[0]\AgdaField{Algebra.Domain}\AgdaSpace{}%
\AgdaSymbol{(}\AgdaFunction{𝑻}\AgdaSpace{}%
\AgdaBound{X}\AgdaSymbol{)}\AgdaSpace{}%
\AgdaSymbol{=}\AgdaSpace{}%
\AgdaFunction{TermSetoid}\AgdaSpace{}%
\AgdaBound{X}\<%
\\
\>[0]\AgdaField{Algebra.Interp}\AgdaSpace{}%
\AgdaSymbol{(}\AgdaFunction{𝑻}\AgdaSpace{}%
\AgdaBound{X}\AgdaSymbol{)}\AgdaSpace{}%
\AgdaOperator{\AgdaField{⟨\$⟩}}\AgdaSpace{}%
\AgdaSymbol{(}\AgdaBound{f}\AgdaSpace{}%
\AgdaOperator{\AgdaInductiveConstructor{,}}\AgdaSpace{}%
\AgdaBound{ts}\AgdaSymbol{)}\AgdaSpace{}%
\AgdaSymbol{=}\AgdaSpace{}%
\AgdaInductiveConstructor{node}\AgdaSpace{}%
\AgdaBound{f}\AgdaSpace{}%
\AgdaBound{ts}\<%
\\
\>[0]\AgdaField{cong}\AgdaSpace{}%
\AgdaSymbol{(}\AgdaField{Algebra.Interp}\AgdaSpace{}%
\AgdaSymbol{(}\AgdaFunction{𝑻}\AgdaSpace{}%
\AgdaBound{X}\AgdaSymbol{))}\AgdaSpace{}%
\AgdaSymbol{(}\AgdaInductiveConstructor{≡.refl}\AgdaSpace{}%
\AgdaOperator{\AgdaInductiveConstructor{,}}\AgdaSpace{}%
\AgdaBound{ss≃ts}\AgdaSymbol{)}\AgdaSpace{}%
\AgdaSymbol{=}\AgdaSpace{}%
\AgdaInductiveConstructor{gnl}\AgdaSpace{}%
\AgdaBound{ss≃ts}\<%
\end{code}

\paragraph*{Substitution, environments and interpretation of terms}
In this section, we formalize the notions of \emph{substitution}, \emph{environment}, and
\emph{interpretation of terms} in an algebra. The approach to formalizing these concepts,
and the Agda code presented in this subsection, is based on similar code developed by
Andreas Abel to formalize Birkhoff's completeness theorem~\cite{Abel:2021}.

\ifshort\else
Recall that the domain of an algebra \ab{𝑨} is a setoid, which we denote by
\af{𝔻[~\ab{𝑨}~]}, whose \afld{Carrier} is the carrier of the algebra, \af{𝕌[~\ab{𝑨}~]},
and whose equivalence relation represents equality of elements in \af{𝕌[~\ab{𝑨}~]}.
\fi

%Before defining the \af{Env} function (which will depend on a specific algebra) we first
The function \af{Sub} performs substitution from one context to
another.  Specifically, if \ab X and \ab Y are contexts, then \af{Sub} \ab X \ab Y
assigns a term in \ab X to each symbol in \ab Y.
The definition of \af{Sub} is a slight modification of the one given by Andreas Abel
(\textit{op.~cit.}), as is the recursive definition of \af{[~\ab{σ}~]} \ab t,
which denotes a substitution applied to a term.

\begin{code}%
\>[0]\<%
\\
\>[0]\AgdaFunction{Sub}\AgdaSpace{}%
\AgdaSymbol{:}\AgdaSpace{}%
\AgdaPrimitive{Type}\AgdaSpace{}%
\AgdaGeneralizable{χ}\AgdaSpace{}%
\AgdaSymbol{→}\AgdaSpace{}%
\AgdaPrimitive{Type}\AgdaSpace{}%
\AgdaGeneralizable{χ}\AgdaSpace{}%
\AgdaSymbol{→}\AgdaSpace{}%
\AgdaPrimitive{Type}\AgdaSpace{}%
\AgdaSymbol{\AgdaUnderscore{}}\<%
\\
\>[0]\AgdaFunction{Sub}\AgdaSpace{}%
\AgdaBound{X}\AgdaSpace{}%
\AgdaBound{Y}\AgdaSpace{}%
\AgdaSymbol{=}\AgdaSpace{}%
\AgdaSymbol{(}\AgdaBound{y}\AgdaSpace{}%
\AgdaSymbol{:}\AgdaSpace{}%
\AgdaBound{Y}\AgdaSymbol{)}\AgdaSpace{}%
\AgdaSymbol{→}\AgdaSpace{}%
\AgdaDatatype{Term}\AgdaSpace{}%
\AgdaBound{X}\<%
\\
\\[\AgdaEmptyExtraSkip]%
\>[0]\AgdaOperator{\AgdaFunction{[\AgdaUnderscore{}]\AgdaUnderscore{}}}\AgdaSpace{}%
\AgdaSymbol{:}\AgdaSpace{}%
\AgdaSymbol{\{}\AgdaBound{X}\AgdaSpace{}%
\AgdaBound{Y}\AgdaSpace{}%
\AgdaSymbol{:}\AgdaSpace{}%
\AgdaPrimitive{Type}\AgdaSpace{}%
\AgdaGeneralizable{χ}\AgdaSymbol{\}}\AgdaSpace{}%
\AgdaSymbol{→}\AgdaSpace{}%
\AgdaFunction{Sub}\AgdaSpace{}%
\AgdaBound{X}\AgdaSpace{}%
\AgdaBound{Y}\AgdaSpace{}%
\AgdaSymbol{→}\AgdaSpace{}%
\AgdaDatatype{Term}\AgdaSpace{}%
\AgdaBound{Y}\AgdaSpace{}%
\AgdaSymbol{→}\AgdaSpace{}%
\AgdaDatatype{Term}\AgdaSpace{}%
\AgdaBound{X}\<%
\\
\>[0]\AgdaOperator{\AgdaFunction{[}}\AgdaSpace{}%
\AgdaBound{σ}\AgdaSpace{}%
\AgdaOperator{\AgdaFunction{]}}\AgdaSpace{}%
\AgdaSymbol{(}\AgdaInductiveConstructor{ℊ}\AgdaSpace{}%
\AgdaBound{x}\AgdaSymbol{)}\AgdaSpace{}%
\AgdaSymbol{=}\AgdaSpace{}%
\AgdaBound{σ}\AgdaSpace{}%
\AgdaBound{x}\<%
\\
\>[0]\AgdaOperator{\AgdaFunction{[}}\AgdaSpace{}%
\AgdaBound{σ}\AgdaSpace{}%
\AgdaOperator{\AgdaFunction{]}}\AgdaSpace{}%
\AgdaSymbol{(}\AgdaInductiveConstructor{node}\AgdaSpace{}%
\AgdaBound{f}\AgdaSpace{}%
\AgdaBound{ts}\AgdaSymbol{)}\AgdaSpace{}%
\AgdaSymbol{=}\AgdaSpace{}%
\AgdaInductiveConstructor{node}\AgdaSpace{}%
\AgdaBound{f}\AgdaSpace{}%
\AgdaSymbol{(λ}\AgdaSpace{}%
\AgdaBound{i}\AgdaSpace{}%
\AgdaSymbol{→}\AgdaSpace{}%
\AgdaOperator{\AgdaFunction{[}}\AgdaSpace{}%
\AgdaBound{σ}\AgdaSpace{}%
\AgdaOperator{\AgdaFunction{]}}\AgdaSpace{}%
\AgdaSymbol{(}\AgdaBound{ts}\AgdaSpace{}%
\AgdaBound{i}\AgdaSymbol{))}\<%
\\
\>[0]\<%
\end{code}

Fix a signature \ab{𝑆}, a context \ab X, and an \ab{𝑆}-algebra \ab{𝑨}.
An \defn{environment} for these data consists of the function type \ab X \as{→}
\af{𝕌[~\ab{𝑨}~]} along with an equality on this type.
The function \af{Env} manifests this notion by taking an \ab{𝑆}-algebra \ab{𝑨} and a
context \ab{X} and returning a setoid whose \afld{Carrier} is the type
\ab X~\as{→}~\af{𝕌[~\ab{𝑨}~]} and whose equivalence relation
is pointwise equality of functions in \ab X \as{→} \af{𝕌[~\ab{𝑨}~]}.

\begin{code}%
\>[0]\<%
\\
\>[0]\AgdaKeyword{module}\AgdaSpace{}%
\AgdaModule{Environment}\AgdaSpace{}%
\AgdaSymbol{(}\AgdaBound{𝑨}\AgdaSpace{}%
\AgdaSymbol{:}\AgdaSpace{}%
\AgdaRecord{Algebra}\AgdaSpace{}%
\AgdaGeneralizable{α}\AgdaSpace{}%
\AgdaGeneralizable{ℓ}\AgdaSymbol{)}\AgdaSpace{}%
\AgdaKeyword{where}\<%
\\
\>[0][@{}l@{\AgdaIndent{0}}]%
\>[1]\AgdaKeyword{open}\AgdaSpace{}%
\AgdaModule{Setoid}\AgdaSpace{}%
\AgdaOperator{\AgdaFunction{𝔻[}}\AgdaSpace{}%
\AgdaBound{𝑨}\AgdaSpace{}%
\AgdaOperator{\AgdaFunction{]}}\AgdaSpace{}%
\AgdaKeyword{using}\AgdaSpace{}%
\AgdaSymbol{(}\AgdaSpace{}%
\AgdaOperator{\AgdaField{\AgdaUnderscore{}≈\AgdaUnderscore{}}}\AgdaSpace{}%
\AgdaSymbol{;}\AgdaSpace{}%
\AgdaFunction{refl}\AgdaSpace{}%
\AgdaSymbol{;}\AgdaSpace{}%
\AgdaFunction{sym}\AgdaSpace{}%
\AgdaSymbol{;}\AgdaSpace{}%
\AgdaFunction{trans}\AgdaSpace{}%
\AgdaSymbol{)}\<%
\\
\>[1]\AgdaFunction{Env}\AgdaSpace{}%
\AgdaSymbol{:}\AgdaSpace{}%
\AgdaPrimitive{Type}\AgdaSpace{}%
\AgdaGeneralizable{χ}\AgdaSpace{}%
\AgdaSymbol{→}\AgdaSpace{}%
\AgdaRecord{Setoid}\AgdaSpace{}%
\AgdaSymbol{\AgdaUnderscore{}}\AgdaSpace{}%
\AgdaSymbol{\AgdaUnderscore{}}\<%
\\
\>[1]\AgdaFunction{Env}\AgdaSpace{}%
\AgdaBound{X}\AgdaSpace{}%
\AgdaSymbol{=}\AgdaSpace{}%
\AgdaKeyword{record}%
\>[17]\AgdaSymbol{\{}\AgdaSpace{}%
\AgdaField{Carrier}\AgdaSpace{}%
\AgdaSymbol{=}\AgdaSpace{}%
\AgdaBound{X}\AgdaSpace{}%
\AgdaSymbol{→}\AgdaSpace{}%
\AgdaOperator{\AgdaFunction{𝕌[}}\AgdaSpace{}%
\AgdaBound{𝑨}\AgdaSpace{}%
\AgdaOperator{\AgdaFunction{]}}\<%
\\
\>[17]\AgdaSymbol{;}\AgdaSpace{}%
\AgdaOperator{\AgdaField{\AgdaUnderscore{}≈\AgdaUnderscore{}}}\AgdaSpace{}%
\AgdaSymbol{=}\AgdaSpace{}%
\AgdaSymbol{λ}\AgdaSpace{}%
\AgdaBound{ρ}\AgdaSpace{}%
\AgdaBound{τ}\AgdaSpace{}%
\AgdaSymbol{→}\AgdaSpace{}%
\AgdaSymbol{(}\AgdaBound{x}\AgdaSpace{}%
\AgdaSymbol{:}\AgdaSpace{}%
\AgdaBound{X}\AgdaSymbol{)}\AgdaSpace{}%
\AgdaSymbol{→}\AgdaSpace{}%
\AgdaBound{ρ}\AgdaSpace{}%
\AgdaBound{x}\AgdaSpace{}%
\AgdaOperator{\AgdaFunction{≈}}\AgdaSpace{}%
\AgdaBound{τ}\AgdaSpace{}%
\AgdaBound{x}\<%
\\
\>[17]\AgdaSymbol{;}\AgdaSpace{}%
\AgdaField{isEquivalence}\AgdaSpace{}%
\AgdaSymbol{=}\AgdaSpace{}%
\AgdaKeyword{record}%
\>[43]\AgdaSymbol{\{}\AgdaSpace{}%
\AgdaField{refl}%
\>[52]\AgdaSymbol{=}\AgdaSpace{}%
\AgdaSymbol{λ}\AgdaSpace{}%
\AgdaBound{\AgdaUnderscore{}}%
\>[63]\AgdaSymbol{→}\AgdaSpace{}%
\AgdaFunction{refl}\<%
\\
\>[43]\AgdaSymbol{;}\AgdaSpace{}%
\AgdaField{sym}%
\>[52]\AgdaSymbol{=}\AgdaSpace{}%
\AgdaSymbol{λ}\AgdaSpace{}%
\AgdaBound{h}\AgdaSpace{}%
\AgdaBound{x}%
\>[63]\AgdaSymbol{→}\AgdaSpace{}%
\AgdaFunction{sym}\AgdaSpace{}%
\AgdaSymbol{(}\AgdaBound{h}\AgdaSpace{}%
\AgdaBound{x}\AgdaSymbol{)}\<%
\\
\>[43]\AgdaSymbol{;}\AgdaSpace{}%
\AgdaField{trans}%
\>[52]\AgdaSymbol{=}\AgdaSpace{}%
\AgdaSymbol{λ}\AgdaSpace{}%
\AgdaBound{g}\AgdaSpace{}%
\AgdaBound{h}\AgdaSpace{}%
\AgdaBound{x}%
\>[63]\AgdaSymbol{→}\AgdaSpace{}%
\AgdaFunction{trans}\AgdaSpace{}%
\AgdaSymbol{(}\AgdaBound{g}\AgdaSpace{}%
\AgdaBound{x}\AgdaSymbol{)(}\AgdaBound{h}\AgdaSpace{}%
\AgdaBound{x}\AgdaSymbol{)}\AgdaSpace{}%
\AgdaSymbol{\}\}}\<%
\\
\>[0]\<%
\end{code}
\ifshort\else
Notice that this definition, as well as the next, are relative to a certain fixed algebra,
so we put them inside a submodule called \am{Environment}. This allows us to load the
submodule and associate its definitions with a number of different algebras simultaneously.
\fi

Next, the recursive function \af{⟦\au{}⟧} denotes \defn{interpretation} of
a term in a given algebra, \emph{evaluated} in a given environment.

\begin{code}%
\>[0]\<%
\\
\>[0][@{}l@{\AgdaIndent{1}}]%
\>[1]\AgdaOperator{\AgdaFunction{⟦\AgdaUnderscore{}⟧}}\AgdaSpace{}%
\AgdaSymbol{:}\AgdaSpace{}%
\AgdaSymbol{\{}\AgdaBound{X}\AgdaSpace{}%
\AgdaSymbol{:}\AgdaSpace{}%
\AgdaPrimitive{Type}\AgdaSpace{}%
\AgdaGeneralizable{χ}\AgdaSymbol{\}(}\AgdaBound{t}\AgdaSpace{}%
\AgdaSymbol{:}\AgdaSpace{}%
\AgdaDatatype{Term}\AgdaSpace{}%
\AgdaBound{X}\AgdaSymbol{)}\AgdaSpace{}%
\AgdaSymbol{→}\AgdaSpace{}%
\AgdaSymbol{(}\AgdaFunction{Env}\AgdaSpace{}%
\AgdaBound{X}\AgdaSymbol{)}\AgdaSpace{}%
\AgdaOperator{\AgdaRecord{⟶}}\AgdaSpace{}%
\AgdaOperator{\AgdaFunction{𝔻[}}\AgdaSpace{}%
\AgdaBound{𝑨}\AgdaSpace{}%
\AgdaOperator{\AgdaFunction{]}}\<%
\\
\>[1]\AgdaOperator{\AgdaFunction{⟦}}\AgdaSpace{}%
\AgdaInductiveConstructor{ℊ}\AgdaSpace{}%
\AgdaBound{x}\AgdaSpace{}%
\AgdaOperator{\AgdaFunction{⟧}}%
\>[18]\AgdaOperator{\AgdaField{⟨\$⟩}}\AgdaSpace{}%
\AgdaBound{ρ}%
\>[27]\AgdaSymbol{=}\AgdaSpace{}%
\AgdaBound{ρ}\AgdaSpace{}%
\AgdaBound{x}\<%
\\
\>[1]\AgdaOperator{\AgdaFunction{⟦}}\AgdaSpace{}%
\AgdaInductiveConstructor{node}\AgdaSpace{}%
\AgdaBound{f}\AgdaSpace{}%
\AgdaBound{args}\AgdaSpace{}%
\AgdaOperator{\AgdaFunction{⟧}}%
\>[18]\AgdaOperator{\AgdaField{⟨\$⟩}}\AgdaSpace{}%
\AgdaBound{ρ}%
\>[27]\AgdaSymbol{=}\AgdaSpace{}%
\AgdaSymbol{(}\AgdaField{Interp}\AgdaSpace{}%
\AgdaBound{𝑨}\AgdaSymbol{)}\AgdaSpace{}%
\AgdaOperator{\AgdaField{⟨\$⟩}}\AgdaSpace{}%
\AgdaSymbol{(}\AgdaBound{f}\AgdaSpace{}%
\AgdaOperator{\AgdaInductiveConstructor{,}}\AgdaSpace{}%
\AgdaSymbol{λ}\AgdaSpace{}%
\AgdaBound{i}\AgdaSpace{}%
\AgdaSymbol{→}\AgdaSpace{}%
\AgdaOperator{\AgdaFunction{⟦}}\AgdaSpace{}%
\AgdaBound{args}\AgdaSpace{}%
\AgdaBound{i}\AgdaSpace{}%
\AgdaOperator{\AgdaFunction{⟧}}\AgdaSpace{}%
\AgdaOperator{\AgdaField{⟨\$⟩}}\AgdaSpace{}%
\AgdaBound{ρ}\AgdaSymbol{)}\<%
\\
\>[1]\AgdaField{cong}\AgdaSpace{}%
\AgdaOperator{\AgdaFunction{⟦}}\AgdaSpace{}%
\AgdaInductiveConstructor{ℊ}\AgdaSpace{}%
\AgdaBound{x}\AgdaSpace{}%
\AgdaOperator{\AgdaFunction{⟧}}\AgdaSpace{}%
\AgdaBound{u≈v}%
\>[27]\AgdaSymbol{=}\AgdaSpace{}%
\AgdaBound{u≈v}\AgdaSpace{}%
\AgdaBound{x}\<%
\\
\>[1]\AgdaField{cong}\AgdaSpace{}%
\AgdaOperator{\AgdaFunction{⟦}}\AgdaSpace{}%
\AgdaInductiveConstructor{node}\AgdaSpace{}%
\AgdaBound{f}\AgdaSpace{}%
\AgdaBound{args}\AgdaSpace{}%
\AgdaOperator{\AgdaFunction{⟧}}\AgdaSpace{}%
\AgdaBound{x≈y}%
\>[27]\AgdaSymbol{=}\AgdaSpace{}%
\AgdaField{cong}\AgdaSpace{}%
\AgdaSymbol{(}\AgdaField{Interp}\AgdaSpace{}%
\AgdaBound{𝑨}\AgdaSymbol{)(}\AgdaInductiveConstructor{≡.refl}\AgdaSpace{}%
\AgdaOperator{\AgdaInductiveConstructor{,}}\AgdaSpace{}%
\AgdaSymbol{λ}\AgdaSpace{}%
\AgdaBound{i}\AgdaSpace{}%
\AgdaSymbol{→}\AgdaSpace{}%
\AgdaField{cong}\AgdaSpace{}%
\AgdaOperator{\AgdaFunction{⟦}}\AgdaSpace{}%
\AgdaBound{args}\AgdaSpace{}%
\AgdaBound{i}\AgdaSpace{}%
\AgdaOperator{\AgdaFunction{⟧}}\AgdaSpace{}%
\AgdaBound{x≈y}\AgdaSpace{}%
\AgdaSymbol{)}\<%
\\
\>[0]\<%
\end{code}

Two terms interpreted in \ab{𝑨} are proclaimed \defn{equal} if they are equal for all
environments.  This equivalence of terms%
\ifshort\else
, and proof that it is an equivalence relation,
\fi
~is formalized in Agda as follows.

\begin{code}%
\>[0]\<%
\\
\>[0][@{}l@{\AgdaIndent{1}}]%
\>[1]\AgdaFunction{Equal}\AgdaSpace{}%
\AgdaSymbol{:}\AgdaSpace{}%
\AgdaSymbol{\{}\AgdaBound{X}\AgdaSpace{}%
\AgdaSymbol{:}\AgdaSpace{}%
\AgdaPrimitive{Type}\AgdaSpace{}%
\AgdaGeneralizable{χ}\AgdaSymbol{\}(}\AgdaBound{s}\AgdaSpace{}%
\AgdaBound{t}\AgdaSpace{}%
\AgdaSymbol{:}\AgdaSpace{}%
\AgdaDatatype{Term}\AgdaSpace{}%
\AgdaBound{X}\AgdaSymbol{)}\AgdaSpace{}%
\AgdaSymbol{→}\AgdaSpace{}%
\AgdaPrimitive{Type}\AgdaSpace{}%
\AgdaSymbol{\AgdaUnderscore{}}\<%
\\
\>[1]\AgdaFunction{Equal}\AgdaSpace{}%
\AgdaSymbol{\{}\AgdaArgument{X}\AgdaSpace{}%
\AgdaSymbol{=}\AgdaSpace{}%
\AgdaBound{X}\AgdaSymbol{\}}\AgdaSpace{}%
\AgdaBound{s}\AgdaSpace{}%
\AgdaBound{t}\AgdaSpace{}%
\AgdaSymbol{=}\AgdaSpace{}%
\AgdaSymbol{∀}\AgdaSpace{}%
\AgdaSymbol{(}\AgdaBound{ρ}\AgdaSpace{}%
\AgdaSymbol{:}\AgdaSpace{}%
\AgdaField{Carrier}\AgdaSpace{}%
\AgdaSymbol{(}\AgdaFunction{Env}\AgdaSpace{}%
\AgdaBound{X}\AgdaSymbol{))}\AgdaSpace{}%
\AgdaSymbol{→}\AgdaSpace{}%
\AgdaOperator{\AgdaFunction{⟦}}\AgdaSpace{}%
\AgdaBound{s}\AgdaSpace{}%
\AgdaOperator{\AgdaFunction{⟧}}\AgdaSpace{}%
\AgdaOperator{\AgdaField{⟨\$⟩}}\AgdaSpace{}%
\AgdaBound{ρ}\AgdaSpace{}%
\AgdaOperator{\AgdaFunction{≈}}\AgdaSpace{}%
\AgdaOperator{\AgdaFunction{⟦}}\AgdaSpace{}%
\AgdaBound{t}\AgdaSpace{}%
\AgdaOperator{\AgdaFunction{⟧}}\AgdaSpace{}%
\AgdaOperator{\AgdaField{⟨\$⟩}}\AgdaSpace{}%
\AgdaBound{ρ}\<%
\\
\>[0]\<%
\end{code}
\ifshort
Proof that \af{Equal} is an equivalence relation, and that the implication \ab
s~\af{≃}~\ab t \as{→} \af{Equal} \ab s \ab t holds for all terms \ab s and \ab t, is
trivial (\seeshort for details).
We denote these facts by \af{EqualIsEquiv} and \af{≃→Equal} in the sequel.
\else
\begin{code}%
\>[0][@{}l@{\AgdaIndent{1}}]%
\>[1]\AgdaFunction{≃→Equal}\AgdaSpace{}%
\AgdaSymbol{:}\AgdaSpace{}%
\AgdaSymbol{\{}\AgdaBound{X}\AgdaSpace{}%
\AgdaSymbol{:}\AgdaSpace{}%
\AgdaPrimitive{Type}\AgdaSpace{}%
\AgdaGeneralizable{χ}\AgdaSymbol{\}(}\AgdaBound{s}\AgdaSpace{}%
\AgdaBound{t}\AgdaSpace{}%
\AgdaSymbol{:}\AgdaSpace{}%
\AgdaDatatype{Term}\AgdaSpace{}%
\AgdaBound{X}\AgdaSymbol{)}\AgdaSpace{}%
\AgdaSymbol{→}\AgdaSpace{}%
\AgdaBound{s}\AgdaSpace{}%
\AgdaOperator{\AgdaDatatype{≃}}\AgdaSpace{}%
\AgdaBound{t}\AgdaSpace{}%
\AgdaSymbol{→}\AgdaSpace{}%
\AgdaFunction{Equal}\AgdaSpace{}%
\AgdaBound{s}\AgdaSpace{}%
\AgdaBound{t}\<%
\\
\>[1]\AgdaFunction{≃→Equal}\AgdaSpace{}%
\AgdaDottedPattern{\AgdaSymbol{.(}}\AgdaDottedPattern{\AgdaInductiveConstructor{ℊ}}\AgdaSpace{}%
\AgdaDottedPattern{\AgdaSymbol{\AgdaUnderscore{})}}\AgdaSpace{}%
\AgdaDottedPattern{\AgdaSymbol{.(}}\AgdaDottedPattern{\AgdaInductiveConstructor{ℊ}}\AgdaSpace{}%
\AgdaDottedPattern{\AgdaSymbol{\AgdaUnderscore{})}}\AgdaSpace{}%
\AgdaSymbol{(}\AgdaInductiveConstructor{rfl}\AgdaSpace{}%
\AgdaInductiveConstructor{≡.refl}\AgdaSymbol{)}\AgdaSpace{}%
\AgdaSymbol{=}\AgdaSpace{}%
\AgdaSymbol{λ}\AgdaSpace{}%
\AgdaBound{\AgdaUnderscore{}}\AgdaSpace{}%
\AgdaSymbol{→}\AgdaSpace{}%
\AgdaFunction{refl}\<%
\\
\>[1]\AgdaFunction{≃→Equal}\AgdaSpace{}%
\AgdaSymbol{(}\AgdaInductiveConstructor{node}\AgdaSpace{}%
\AgdaSymbol{\AgdaUnderscore{}}\AgdaSpace{}%
\AgdaBound{s}\AgdaSymbol{)(}\AgdaInductiveConstructor{node}\AgdaSpace{}%
\AgdaSymbol{\AgdaUnderscore{}}\AgdaSpace{}%
\AgdaBound{t}\AgdaSymbol{)(}\AgdaInductiveConstructor{gnl}\AgdaSpace{}%
\AgdaBound{x}\AgdaSymbol{)}\AgdaSpace{}%
\AgdaSymbol{=}\<%
\\
\>[1][@{}l@{\AgdaIndent{0}}]%
\>[2]\AgdaSymbol{λ}\AgdaSpace{}%
\AgdaBound{ρ}\AgdaSpace{}%
\AgdaSymbol{→}\AgdaSpace{}%
\AgdaField{cong}\AgdaSpace{}%
\AgdaSymbol{(}\AgdaField{Interp}\AgdaSpace{}%
\AgdaBound{𝑨}\AgdaSymbol{)(}\AgdaInductiveConstructor{≡.refl}\AgdaSpace{}%
\AgdaOperator{\AgdaInductiveConstructor{,}}\AgdaSpace{}%
\AgdaSymbol{λ}\AgdaSpace{}%
\AgdaBound{i}\AgdaSpace{}%
\AgdaSymbol{→}\AgdaSpace{}%
\AgdaFunction{≃→Equal}\AgdaSymbol{(}\AgdaBound{s}\AgdaSpace{}%
\AgdaBound{i}\AgdaSymbol{)(}\AgdaBound{t}\AgdaSpace{}%
\AgdaBound{i}\AgdaSymbol{)(}\AgdaBound{x}\AgdaSpace{}%
\AgdaBound{i}\AgdaSymbol{)}\AgdaBound{ρ}\AgdaSpace{}%
\AgdaSymbol{)}\<%
\\
\\[\AgdaEmptyExtraSkip]%
\>[1]\AgdaFunction{EqualIsEquiv}\AgdaSpace{}%
\AgdaSymbol{:}\AgdaSpace{}%
\AgdaSymbol{\{}\AgdaBound{Γ}\AgdaSpace{}%
\AgdaSymbol{:}\AgdaSpace{}%
\AgdaPrimitive{Type}\AgdaSpace{}%
\AgdaGeneralizable{χ}\AgdaSymbol{\}}\AgdaSpace{}%
\AgdaSymbol{→}\AgdaSpace{}%
\AgdaRecord{IsEquivalence}\AgdaSpace{}%
\AgdaSymbol{(}\AgdaFunction{Equal}\AgdaSpace{}%
\AgdaSymbol{\{}\AgdaArgument{X}\AgdaSpace{}%
\AgdaSymbol{=}\AgdaSpace{}%
\AgdaBound{Γ}\AgdaSymbol{\})}\<%
\\
\>[1]\AgdaField{reflᵉ}%
\>[9]\AgdaFunction{EqualIsEquiv}\AgdaSpace{}%
\AgdaSymbol{=}\AgdaSpace{}%
\AgdaSymbol{λ}\AgdaSpace{}%
\AgdaBound{\AgdaUnderscore{}}%
\>[35]\AgdaSymbol{→}\AgdaSpace{}%
\AgdaFunction{refl}\<%
\\
\>[1]\AgdaField{symᵉ}%
\>[9]\AgdaFunction{EqualIsEquiv}\AgdaSpace{}%
\AgdaSymbol{=}\AgdaSpace{}%
\AgdaSymbol{λ}\AgdaSpace{}%
\AgdaBound{x=y}\AgdaSpace{}%
\AgdaBound{ρ}%
\>[35]\AgdaSymbol{→}\AgdaSpace{}%
\AgdaFunction{sym}\AgdaSpace{}%
\AgdaSymbol{(}\AgdaBound{x=y}\AgdaSpace{}%
\AgdaBound{ρ}\AgdaSymbol{)}\<%
\\
\>[1]\AgdaField{transᵉ}%
\>[9]\AgdaFunction{EqualIsEquiv}\AgdaSpace{}%
\AgdaSymbol{=}\AgdaSpace{}%
\AgdaSymbol{λ}\AgdaSpace{}%
\AgdaBound{ij}\AgdaSpace{}%
\AgdaBound{jk}\AgdaSpace{}%
\AgdaBound{ρ}%
\>[35]\AgdaSymbol{→}\AgdaSpace{}%
\AgdaFunction{trans}\AgdaSpace{}%
\AgdaSymbol{(}\AgdaBound{ij}\AgdaSpace{}%
\AgdaBound{ρ}\AgdaSymbol{)}\AgdaSpace{}%
\AgdaSymbol{(}\AgdaBound{jk}\AgdaSpace{}%
\AgdaBound{ρ}\AgdaSymbol{)}\<%
\\
\>[0]\<%
\end{code}
\fi

The next lemma says that applying a substitution \ab{σ} to a term \ab{t}
and evaluating the result in the environment \ab{ρ} has the same effect as evaluating
\ab{t} the a new environment, specifically, in the environment \as{λ} \ab x \as{→} \aof{⟦~\ab{σ}~\ab{x}~⟧}~\aofld{⟨\$⟩}
\ab{ρ} (see~\cite{Abel:2021} or~\cite[Lem.~3.3.11]{Mitchell:1996}).

\begin{code}%
\>[0]\<%
\\
\>[0][@{}l@{\AgdaIndent{1}}]%
\>[1]\AgdaFunction{substitution}\AgdaSpace{}%
\AgdaSymbol{:}%
\>[17]\AgdaSymbol{\{}\AgdaBound{X}\AgdaSpace{}%
\AgdaBound{Y}\AgdaSpace{}%
\AgdaSymbol{:}\AgdaSpace{}%
\AgdaPrimitive{Type}\AgdaSpace{}%
\AgdaGeneralizable{χ}\AgdaSymbol{\}}\AgdaSpace{}%
\AgdaSymbol{→}\AgdaSpace{}%
\AgdaSymbol{(}\AgdaBound{t}\AgdaSpace{}%
\AgdaSymbol{:}\AgdaSpace{}%
\AgdaDatatype{Term}\AgdaSpace{}%
\AgdaBound{Y}\AgdaSymbol{)}\AgdaSpace{}%
\AgdaSymbol{(}\AgdaBound{σ}\AgdaSpace{}%
\AgdaSymbol{:}\AgdaSpace{}%
\AgdaFunction{Sub}\AgdaSpace{}%
\AgdaBound{X}\AgdaSpace{}%
\AgdaBound{Y}\AgdaSymbol{)}\AgdaSpace{}%
\AgdaSymbol{(}\AgdaBound{ρ}\AgdaSpace{}%
\AgdaSymbol{:}\AgdaSpace{}%
\AgdaField{Carrier}\AgdaSymbol{(}\AgdaSpace{}%
\AgdaFunction{Env}\AgdaSpace{}%
\AgdaBound{X}\AgdaSpace{}%
\AgdaSymbol{)}\AgdaSpace{}%
\AgdaSymbol{)}\<%
\\
\>[1][@{}l@{\AgdaIndent{0}}]%
\>[2]\AgdaSymbol{→}%
\>[17]\AgdaOperator{\AgdaFunction{⟦}}\AgdaSpace{}%
\AgdaOperator{\AgdaFunction{[}}\AgdaSpace{}%
\AgdaBound{σ}\AgdaSpace{}%
\AgdaOperator{\AgdaFunction{]}}\AgdaSpace{}%
\AgdaBound{t}\AgdaSpace{}%
\AgdaOperator{\AgdaFunction{⟧}}\AgdaSpace{}%
\AgdaOperator{\AgdaField{⟨\$⟩}}\AgdaSpace{}%
\AgdaBound{ρ}\AgdaSpace{}%
\AgdaOperator{\AgdaFunction{≈}}\AgdaSpace{}%
\AgdaOperator{\AgdaFunction{⟦}}\AgdaSpace{}%
\AgdaBound{t}\AgdaSpace{}%
\AgdaOperator{\AgdaFunction{⟧}}\AgdaSpace{}%
\AgdaOperator{\AgdaField{⟨\$⟩}}\AgdaSpace{}%
\AgdaSymbol{(λ}\AgdaSpace{}%
\AgdaBound{x}\AgdaSpace{}%
\AgdaSymbol{→}\AgdaSpace{}%
\AgdaOperator{\AgdaFunction{⟦}}\AgdaSpace{}%
\AgdaBound{σ}\AgdaSpace{}%
\AgdaBound{x}\AgdaSpace{}%
\AgdaOperator{\AgdaFunction{⟧}}\AgdaSpace{}%
\AgdaOperator{\AgdaField{⟨\$⟩}}\AgdaSpace{}%
\AgdaBound{ρ}\AgdaSymbol{)}\<%
\\
\>[1]\AgdaFunction{substitution}\AgdaSpace{}%
\AgdaSymbol{(}\AgdaInductiveConstructor{ℊ}\AgdaSpace{}%
\AgdaBound{x}\AgdaSymbol{)}%
\>[27]\AgdaBound{σ}\AgdaSpace{}%
\AgdaBound{ρ}\AgdaSpace{}%
\AgdaSymbol{=}\AgdaSpace{}%
\AgdaFunction{refl}\<%
\\
\>[1]\AgdaFunction{substitution}\AgdaSpace{}%
\AgdaSymbol{(}\AgdaInductiveConstructor{node}\AgdaSpace{}%
\AgdaBound{f}\AgdaSpace{}%
\AgdaBound{ts}\AgdaSymbol{)}%
\>[27]\AgdaBound{σ}\AgdaSpace{}%
\AgdaBound{ρ}\AgdaSpace{}%
\AgdaSymbol{=}\AgdaSpace{}%
\AgdaField{cong}\AgdaSpace{}%
\AgdaSymbol{(}\AgdaField{Interp}\AgdaSpace{}%
\AgdaBound{𝑨}\AgdaSymbol{)(}\AgdaInductiveConstructor{≡.refl}\AgdaSpace{}%
\AgdaOperator{\AgdaInductiveConstructor{,}}\AgdaSpace{}%
\AgdaSymbol{λ}\AgdaSpace{}%
\AgdaBound{i}\AgdaSpace{}%
\AgdaSymbol{→}\AgdaSpace{}%
\AgdaFunction{substitution}\AgdaSpace{}%
\AgdaSymbol{(}\AgdaBound{ts}\AgdaSpace{}%
\AgdaBound{i}\AgdaSymbol{)}\AgdaSpace{}%
\AgdaBound{σ}\AgdaSpace{}%
\AgdaBound{ρ}\AgdaSymbol{)}\<%
\\
\>[0]\<%
\end{code}
This concludes the definition of the \am{Environment} module based on~\cite{Abel:2021}.

\ifshort\else
\paragraph*{Compatibility of terms}
\fi
We will need two more facts about term operations.  The first, called
\af{comm-hom-term}, asserts that every term commutes with every homomorphism.  The second,
\af{interp-prod}, shows how to express the interpretation of a term in a product algebra.
\ifshort
We omit the formalization of these facts, but \seeshort for details.
\else

\begin{code}%
\>[0]\<%
\\
\>[0]\AgdaKeyword{module}\AgdaSpace{}%
\AgdaModule{\AgdaUnderscore{}}\AgdaSpace{}%
\AgdaSymbol{\{}\AgdaBound{X}\AgdaSpace{}%
\AgdaSymbol{:}\AgdaSpace{}%
\AgdaPrimitive{Type}\AgdaSpace{}%
\AgdaGeneralizable{χ}\AgdaSymbol{\}\{}\AgdaBound{𝑨}\AgdaSpace{}%
\AgdaSymbol{:}\AgdaSpace{}%
\AgdaRecord{Algebra}\AgdaSpace{}%
\AgdaGeneralizable{α}\AgdaSpace{}%
\AgdaGeneralizable{ρᵃ}\AgdaSymbol{\}\{}\AgdaBound{𝑩}\AgdaSpace{}%
\AgdaSymbol{:}\AgdaSpace{}%
\AgdaRecord{Algebra}\AgdaSpace{}%
\AgdaGeneralizable{β}\AgdaSpace{}%
\AgdaGeneralizable{ρᵇ}\AgdaSymbol{\}(}\AgdaBound{hh}\AgdaSpace{}%
\AgdaSymbol{:}\AgdaSpace{}%
\AgdaFunction{hom}\AgdaSpace{}%
\AgdaBound{𝑨}\AgdaSpace{}%
\AgdaBound{𝑩}\AgdaSymbol{)}\AgdaSpace{}%
\AgdaKeyword{where}\<%
\\
\>[0][@{}l@{\AgdaIndent{0}}]%
\>[1]\AgdaKeyword{open}\AgdaSpace{}%
\AgdaModule{Environment}\AgdaSpace{}%
\AgdaBound{𝑨}\AgdaSpace{}%
\AgdaKeyword{using}\AgdaSpace{}%
\AgdaSymbol{(}\AgdaSpace{}%
\AgdaOperator{\AgdaFunction{⟦\AgdaUnderscore{}⟧}}\AgdaSpace{}%
\AgdaSymbol{)}\<%
\\
\>[1]\AgdaKeyword{open}\AgdaSpace{}%
\AgdaModule{Environment}\AgdaSpace{}%
\AgdaBound{𝑩}\AgdaSpace{}%
\AgdaKeyword{using}\AgdaSpace{}%
\AgdaSymbol{()}\AgdaSpace{}%
\AgdaKeyword{renaming}\AgdaSpace{}%
\AgdaSymbol{(}\AgdaSpace{}%
\AgdaOperator{\AgdaFunction{⟦\AgdaUnderscore{}⟧}}\AgdaSpace{}%
\AgdaSymbol{to}\AgdaSpace{}%
\AgdaOperator{\AgdaFunction{⟦\AgdaUnderscore{}⟧ᴮ}}\AgdaSpace{}%
\AgdaSymbol{)}\<%
\\
\>[1]\AgdaKeyword{open}\AgdaSpace{}%
\AgdaModule{Setoid}\AgdaSpace{}%
\AgdaOperator{\AgdaFunction{𝔻[}}\AgdaSpace{}%
\AgdaBound{𝑩}\AgdaSpace{}%
\AgdaOperator{\AgdaFunction{]}}\AgdaSpace{}%
\AgdaKeyword{using}\AgdaSpace{}%
\AgdaSymbol{(}\AgdaSpace{}%
\AgdaOperator{\AgdaField{\AgdaUnderscore{}≈\AgdaUnderscore{}}}\AgdaSpace{}%
\AgdaSymbol{;}\AgdaSpace{}%
\AgdaFunction{refl}\AgdaSpace{}%
\AgdaSymbol{)}\<%
\\
\>[1]\AgdaKeyword{private}\AgdaSpace{}%
\AgdaFunction{hfunc}\AgdaSpace{}%
\AgdaSymbol{=}\AgdaSpace{}%
\AgdaOperator{\AgdaFunction{∣}}\AgdaSpace{}%
\AgdaBound{hh}\AgdaSpace{}%
\AgdaOperator{\AgdaFunction{∣}}\AgdaSpace{}%
\AgdaSymbol{;}\AgdaSpace{}%
\AgdaFunction{h}\AgdaSpace{}%
\AgdaSymbol{=}\AgdaSpace{}%
\AgdaOperator{\AgdaField{\AgdaUnderscore{}⟨\$⟩\AgdaUnderscore{}}}\AgdaSpace{}%
\AgdaFunction{hfunc}\<%
\\
\>[1]\AgdaFunction{comm-hom-term}\AgdaSpace{}%
\AgdaSymbol{:}\AgdaSpace{}%
\AgdaSymbol{(}\AgdaBound{t}\AgdaSpace{}%
\AgdaSymbol{:}\AgdaSpace{}%
\AgdaDatatype{Term}\AgdaSpace{}%
\AgdaBound{X}\AgdaSymbol{)}\AgdaSpace{}%
\AgdaSymbol{(}\AgdaBound{a}\AgdaSpace{}%
\AgdaSymbol{:}\AgdaSpace{}%
\AgdaBound{X}\AgdaSpace{}%
\AgdaSymbol{→}\AgdaSpace{}%
\AgdaOperator{\AgdaFunction{𝕌[}}\AgdaSpace{}%
\AgdaBound{𝑨}\AgdaSpace{}%
\AgdaOperator{\AgdaFunction{]}}\AgdaSymbol{)}\AgdaSpace{}%
\AgdaSymbol{→}\AgdaSpace{}%
\AgdaFunction{h}\AgdaSpace{}%
\AgdaSymbol{(}\AgdaOperator{\AgdaFunction{⟦}}\AgdaSpace{}%
\AgdaBound{t}\AgdaSpace{}%
\AgdaOperator{\AgdaFunction{⟧}}\AgdaSpace{}%
\AgdaOperator{\AgdaField{⟨\$⟩}}\AgdaSpace{}%
\AgdaBound{a}\AgdaSymbol{)}\AgdaSpace{}%
\AgdaOperator{\AgdaFunction{≈}}\AgdaSpace{}%
\AgdaOperator{\AgdaFunction{⟦}}\AgdaSpace{}%
\AgdaBound{t}\AgdaSpace{}%
\AgdaOperator{\AgdaFunction{⟧ᴮ}}\AgdaSpace{}%
\AgdaOperator{\AgdaField{⟨\$⟩}}\AgdaSpace{}%
\AgdaSymbol{(}\AgdaFunction{h}\AgdaSpace{}%
\AgdaOperator{\AgdaFunction{∘}}\AgdaSpace{}%
\AgdaBound{a}\AgdaSymbol{)}\<%
\\
\>[1]\AgdaFunction{comm-hom-term}\AgdaSpace{}%
\AgdaSymbol{(}\AgdaInductiveConstructor{ℊ}\AgdaSpace{}%
\AgdaBound{x}\AgdaSymbol{)}\AgdaSpace{}%
\AgdaBound{a}%
\>[29]\AgdaSymbol{=}\AgdaSpace{}%
\AgdaFunction{refl}\<%
\\
\>[1]\AgdaFunction{comm-hom-term}\AgdaSpace{}%
\AgdaSymbol{(}\AgdaInductiveConstructor{node}\AgdaSpace{}%
\AgdaBound{f}\AgdaSpace{}%
\AgdaBound{t}\AgdaSymbol{)}\AgdaSpace{}%
\AgdaBound{a}%
\>[29]\AgdaSymbol{=}\<%
\\
\>[1][@{}l@{\AgdaIndent{0}}]%
\>[2]\AgdaOperator{\AgdaFunction{begin}}\<%
\\
\>[2][@{}l@{\AgdaIndent{0}}]%
\>[3]\AgdaFunction{h}\AgdaSymbol{(}\AgdaOperator{\AgdaFunction{⟦}}\AgdaSpace{}%
\AgdaInductiveConstructor{node}\AgdaSpace{}%
\AgdaBound{f}\AgdaSpace{}%
\AgdaBound{t}\AgdaSpace{}%
\AgdaOperator{\AgdaFunction{⟧}}\AgdaSpace{}%
\AgdaOperator{\AgdaField{⟨\$⟩}}\AgdaSpace{}%
\AgdaBound{a}\AgdaSymbol{)}%
\>[36]\AgdaFunction{≈⟨}\AgdaSpace{}%
\AgdaField{compatible}\AgdaSpace{}%
\AgdaOperator{\AgdaFunction{∥}}\AgdaSpace{}%
\AgdaBound{hh}\AgdaSpace{}%
\AgdaOperator{\AgdaFunction{∥}}\AgdaSpace{}%
\AgdaFunction{⟩}\<%
\\
\>[3]\AgdaSymbol{(}\AgdaBound{f}\AgdaSpace{}%
\AgdaOperator{\AgdaFunction{̂}}\AgdaSpace{}%
\AgdaBound{𝑩}\AgdaSymbol{)(λ}\AgdaSpace{}%
\AgdaBound{i}\AgdaSpace{}%
\AgdaSymbol{→}\AgdaSpace{}%
\AgdaFunction{h}\AgdaSymbol{(}\AgdaOperator{\AgdaFunction{⟦}}\AgdaSpace{}%
\AgdaBound{t}\AgdaSpace{}%
\AgdaBound{i}\AgdaSpace{}%
\AgdaOperator{\AgdaFunction{⟧}}\AgdaSpace{}%
\AgdaOperator{\AgdaField{⟨\$⟩}}\AgdaSpace{}%
\AgdaBound{a}\AgdaSymbol{))}%
\>[36]\AgdaFunction{≈⟨}\AgdaSpace{}%
\AgdaField{cong}\AgdaSymbol{(}\AgdaField{Interp}\AgdaSpace{}%
\AgdaBound{𝑩}\AgdaSymbol{)(}\AgdaInductiveConstructor{≡.refl}\AgdaSpace{}%
\AgdaOperator{\AgdaInductiveConstructor{,}}\AgdaSpace{}%
\AgdaSymbol{λ}\AgdaSpace{}%
\AgdaBound{i}\AgdaSpace{}%
\AgdaSymbol{→}\AgdaSpace{}%
\AgdaFunction{comm-hom-term}\AgdaSpace{}%
\AgdaSymbol{(}\AgdaBound{t}\AgdaSpace{}%
\AgdaBound{i}\AgdaSymbol{)}\AgdaSpace{}%
\AgdaBound{a}\AgdaSymbol{)}\AgdaFunction{⟩}\<%
\\
\>[3]\AgdaOperator{\AgdaFunction{⟦}}\AgdaSpace{}%
\AgdaInductiveConstructor{node}\AgdaSpace{}%
\AgdaBound{f}\AgdaSpace{}%
\AgdaBound{t}\AgdaSpace{}%
\AgdaOperator{\AgdaFunction{⟧ᴮ}}\AgdaSpace{}%
\AgdaOperator{\AgdaField{⟨\$⟩}}\AgdaSpace{}%
\AgdaSymbol{(}\AgdaFunction{h}\AgdaSpace{}%
\AgdaOperator{\AgdaFunction{∘}}\AgdaSpace{}%
\AgdaBound{a}\AgdaSymbol{)}%
\>[36]\AgdaOperator{\AgdaFunction{∎}}\AgdaSpace{}%
\AgdaKeyword{where}%
\>[45]\AgdaKeyword{open}\AgdaSpace{}%
\AgdaModule{SetoidReasoning}\AgdaSpace{}%
\AgdaOperator{\AgdaFunction{𝔻[}}\AgdaSpace{}%
\AgdaBound{𝑩}\AgdaSpace{}%
\AgdaOperator{\AgdaFunction{]}}\<%
\\
\\[\AgdaEmptyExtraSkip]%
\>[0]\AgdaKeyword{module}\AgdaSpace{}%
\AgdaModule{\AgdaUnderscore{}}\AgdaSpace{}%
\AgdaSymbol{\{}\AgdaBound{X}\AgdaSpace{}%
\AgdaSymbol{:}\AgdaSpace{}%
\AgdaPrimitive{Type}\AgdaSpace{}%
\AgdaGeneralizable{χ}\AgdaSymbol{\}\{}\AgdaBound{ι}\AgdaSpace{}%
\AgdaSymbol{:}\AgdaSpace{}%
\AgdaPostulate{Level}\AgdaSymbol{\}}\AgdaSpace{}%
\AgdaSymbol{\{}\AgdaBound{I}\AgdaSpace{}%
\AgdaSymbol{:}\AgdaSpace{}%
\AgdaPrimitive{Type}\AgdaSpace{}%
\AgdaBound{ι}\AgdaSymbol{\}}\AgdaSpace{}%
\AgdaSymbol{(}\AgdaBound{𝒜}\AgdaSpace{}%
\AgdaSymbol{:}\AgdaSpace{}%
\AgdaBound{I}\AgdaSpace{}%
\AgdaSymbol{→}\AgdaSpace{}%
\AgdaRecord{Algebra}\AgdaSpace{}%
\AgdaGeneralizable{α}\AgdaSpace{}%
\AgdaGeneralizable{ρᵃ}\AgdaSymbol{)}\AgdaSpace{}%
\AgdaKeyword{where}\<%
\\
\>[0][@{}l@{\AgdaIndent{0}}]%
\>[1]\AgdaKeyword{open}\AgdaSpace{}%
\AgdaModule{Setoid}\AgdaSpace{}%
\AgdaOperator{\AgdaFunction{𝔻[}}\AgdaSpace{}%
\AgdaFunction{⨅}\AgdaSpace{}%
\AgdaBound{𝒜}\AgdaSpace{}%
\AgdaOperator{\AgdaFunction{]}}%
\>[23]\AgdaKeyword{using}\AgdaSpace{}%
\AgdaSymbol{(}\AgdaSpace{}%
\AgdaOperator{\AgdaField{\AgdaUnderscore{}≈\AgdaUnderscore{}}}\AgdaSpace{}%
\AgdaSymbol{)}\<%
\\
\>[1]\AgdaKeyword{open}\AgdaSpace{}%
\AgdaModule{Environment}%
\>[23]\AgdaKeyword{using}\AgdaSpace{}%
\AgdaSymbol{(}\AgdaSpace{}%
\AgdaOperator{\AgdaFunction{⟦\AgdaUnderscore{}⟧}}\AgdaSpace{}%
\AgdaSymbol{;}\AgdaSpace{}%
\AgdaFunction{≃→Equal}\AgdaSpace{}%
\AgdaSymbol{)}\<%
\\
\>[1]\AgdaFunction{interp-prod}\AgdaSpace{}%
\AgdaSymbol{:}\AgdaSpace{}%
\AgdaSymbol{(}\AgdaBound{p}\AgdaSpace{}%
\AgdaSymbol{:}\AgdaSpace{}%
\AgdaDatatype{Term}\AgdaSpace{}%
\AgdaBound{X}\AgdaSymbol{)}\AgdaSpace{}%
\AgdaSymbol{→}\AgdaSpace{}%
\AgdaSymbol{∀}\AgdaSpace{}%
\AgdaBound{ρ}\AgdaSpace{}%
\AgdaSymbol{→}%
\>[37]\AgdaSymbol{(}\AgdaOperator{\AgdaFunction{⟦}}\AgdaSpace{}%
\AgdaFunction{⨅}\AgdaSpace{}%
\AgdaBound{𝒜}\AgdaSpace{}%
\AgdaOperator{\AgdaFunction{⟧}}\AgdaSpace{}%
\AgdaBound{p}\AgdaSymbol{)}\AgdaSpace{}%
\AgdaOperator{\AgdaField{⟨\$⟩}}\AgdaSpace{}%
\AgdaBound{ρ}%
\>[57]\AgdaOperator{\AgdaFunction{≈}}%
\>[61]\AgdaSymbol{λ}\AgdaSpace{}%
\AgdaBound{i}\AgdaSpace{}%
\AgdaSymbol{→}\AgdaSpace{}%
\AgdaSymbol{(}\AgdaOperator{\AgdaFunction{⟦}}\AgdaSpace{}%
\AgdaBound{𝒜}\AgdaSpace{}%
\AgdaBound{i}\AgdaSpace{}%
\AgdaOperator{\AgdaFunction{⟧}}\AgdaSpace{}%
\AgdaBound{p}\AgdaSymbol{)}\AgdaSpace{}%
\AgdaOperator{\AgdaField{⟨\$⟩}}\AgdaSpace{}%
\AgdaSymbol{λ}\AgdaSpace{}%
\AgdaBound{x}\AgdaSpace{}%
\AgdaSymbol{→}\AgdaSpace{}%
\AgdaSymbol{(}\AgdaBound{ρ}\AgdaSpace{}%
\AgdaBound{x}\AgdaSymbol{)}\AgdaSpace{}%
\AgdaBound{i}\<%
\\
\>[1]\AgdaFunction{interp-prod}\AgdaSpace{}%
\AgdaSymbol{(}\AgdaInductiveConstructor{ℊ}\AgdaSpace{}%
\AgdaBound{x}\AgdaSymbol{)}%
\>[25]\AgdaSymbol{=}\AgdaSpace{}%
\AgdaSymbol{λ}\AgdaSpace{}%
\AgdaBound{ρ}\AgdaSpace{}%
\AgdaBound{i}%
\>[34]\AgdaSymbol{→}\AgdaSpace{}%
\AgdaFunction{≃→Equal}\AgdaSpace{}%
\AgdaSymbol{(}\AgdaBound{𝒜}\AgdaSpace{}%
\AgdaBound{i}\AgdaSymbol{)}\AgdaSpace{}%
\AgdaSymbol{(}\AgdaInductiveConstructor{ℊ}\AgdaSpace{}%
\AgdaBound{x}\AgdaSymbol{)}\AgdaSpace{}%
\AgdaSymbol{(}\AgdaInductiveConstructor{ℊ}\AgdaSpace{}%
\AgdaBound{x}\AgdaSymbol{)}\AgdaSpace{}%
\AgdaFunction{≃-isRefl}\AgdaSpace{}%
\AgdaSymbol{λ}\AgdaSpace{}%
\AgdaBound{\AgdaUnderscore{}}\AgdaSpace{}%
\AgdaSymbol{→}\AgdaSpace{}%
\AgdaSymbol{(}\AgdaBound{ρ}\AgdaSpace{}%
\AgdaBound{x}\AgdaSymbol{)}\AgdaSpace{}%
\AgdaBound{i}\<%
\\
\>[1]\AgdaFunction{interp-prod}\AgdaSpace{}%
\AgdaSymbol{(}\AgdaInductiveConstructor{node}\AgdaSpace{}%
\AgdaBound{f}\AgdaSpace{}%
\AgdaBound{t}\AgdaSymbol{)}%
\>[25]\AgdaSymbol{=}\AgdaSpace{}%
\AgdaSymbol{λ}\AgdaSpace{}%
\AgdaBound{ρ}%
\>[34]\AgdaSymbol{→}\AgdaSpace{}%
\AgdaField{cong}\AgdaSpace{}%
\AgdaSymbol{(}\AgdaField{Interp}\AgdaSpace{}%
\AgdaSymbol{(}\AgdaFunction{⨅}\AgdaSpace{}%
\AgdaBound{𝒜}\AgdaSymbol{))}\AgdaSpace{}%
\AgdaSymbol{(}\AgdaSpace{}%
\AgdaInductiveConstructor{≡.refl}\AgdaSpace{}%
\AgdaOperator{\AgdaInductiveConstructor{,}}\AgdaSpace{}%
\AgdaSymbol{λ}\AgdaSpace{}%
\AgdaBound{j}\AgdaSpace{}%
\AgdaBound{k}\AgdaSpace{}%
\AgdaSymbol{→}\AgdaSpace{}%
\AgdaFunction{interp-prod}\AgdaSpace{}%
\AgdaSymbol{(}\AgdaBound{t}\AgdaSpace{}%
\AgdaBound{j}\AgdaSymbol{)}\AgdaSpace{}%
\AgdaBound{ρ}\AgdaSpace{}%
\AgdaBound{k}\AgdaSpace{}%
\AgdaSymbol{)}\<%
\end{code}
\fi

\section{Equational Logic}
\label{equational-logic}

\paragraph*{Term identities, equational theories, and the ⊧ relation}
Given a signature \ab{𝑆} and a context \ab X, an \ab{𝑆}-\defn{term equation} or \ab{𝑆}-\defn{term identity}
is an ordered pair (\ab p , \ab q) of 𝑆-terms. For instance, if the context is \ab X :
\ap{Type} \ab{χ}, then a term equation is a pair inhabiting the Cartesian product type
\ad{Term}~\ab{X} \aof{×} \ad{Term}~\ab{X}. Such pairs of terms are also denoted by \ab p \af{≈} \ab
q and are often simply called equations or identities, especially when the signature \ab{𝑆} is obvious.

We define an \defn{equational theory} (or \defn{algebraic theory}) to be a pair \ab{T} =
(\ab{𝑆} , \ab{ℰᵀ}) consisting of a signature \ab{𝑆} and a collection \ab{ℰᵀ} of
\ab{𝑆}-term equations. Some authors reserve the term \defn{theory} for
a \emph{deductively closed} set of equations, that is, a set of equations that is closed
under \emph{entailment} (defined below).

We say that the algebra \ab{𝑨} \emph{satisfies} the equation \ab p \af{≈} \ab q if,
for all \ab{ρ} : \ab X \as{→} \aof{𝔻[~\ab{𝑨}~]},
%(assigning values in the domain of \ab{𝑨} to variable symbols in \ab X)
we have \aof{⟦~\ab{p}~⟧} \aofld{⟨\$⟩} \ab{ρ} \af{≈} \aof{⟦~\ab{q}~⟧} \aofld{⟨\$⟩} \ab{ρ}.
In other words, when they are interpreted in the algebra \ab{𝑨},
the terms \ab{p} and \ab{q} are equal no matter what values in \ab{𝑨} are assigned to variable symbols in \ab{X}.
In this situation, we write
\ab{𝑨}~\aof{⊧}~\ab{p}~\aof{≈}~\ab{q} and say that \ab{𝑨} \defn{models} \ab{p}~\af{≈}~\ab{q},
or that \ab{𝑨} is a \defn{model} of \ab{p}~\af{≈}~\ab{q}.
If \ab{𝒦} is a class of algebras, all of the same signature, we write \ab{𝒦}~\aof{⊫}~\ab{p}~\aof{≈}~\ab{q}
and say that \ab{𝒦} \defn{models} the identity \ab{p}~\af{≈}~\ab{q} provided for every \ab{𝑨} \aof{∈} \ab{𝒦},
we have \ab{𝑨}~\aof{⊧}~\ab{p}~\aof{≈}~\ab{q}.

\ifshort\else
\begin{code}%
\>[0]\AgdaKeyword{module}\AgdaSpace{}%
\AgdaModule{\AgdaUnderscore{}}\AgdaSpace{}%
\AgdaSymbol{\{}\AgdaBound{X}\AgdaSpace{}%
\AgdaSymbol{:}\AgdaSpace{}%
\AgdaPrimitive{Type}\AgdaSpace{}%
\AgdaGeneralizable{χ}\AgdaSymbol{\}}\AgdaSpace{}%
\AgdaKeyword{where}\<%
\end{code}
\fi
\begin{code}%
\>[0]\<%
\\
\>[0][@{}l@{\AgdaIndent{1}}]%
\>[1]\AgdaOperator{\AgdaFunction{\AgdaUnderscore{}⊧\AgdaUnderscore{}≈\AgdaUnderscore{}}}\AgdaSpace{}%
\AgdaSymbol{:}\AgdaSpace{}%
\AgdaRecord{Algebra}\AgdaSpace{}%
\AgdaGeneralizable{α}\AgdaSpace{}%
\AgdaGeneralizable{ρᵃ}\AgdaSpace{}%
\AgdaSymbol{→}\AgdaSpace{}%
\AgdaDatatype{Term}\AgdaSpace{}%
\AgdaBound{X}\AgdaSpace{}%
\AgdaSymbol{→}\AgdaSpace{}%
\AgdaDatatype{Term}\AgdaSpace{}%
\AgdaBound{X}\AgdaSpace{}%
\AgdaSymbol{→}\AgdaSpace{}%
\AgdaPrimitive{Type}\AgdaSpace{}%
\AgdaSymbol{\AgdaUnderscore{}}\<%
\\
\>[1]\AgdaBound{𝑨}\AgdaSpace{}%
\AgdaOperator{\AgdaFunction{⊧}}\AgdaSpace{}%
\AgdaBound{p}\AgdaSpace{}%
\AgdaOperator{\AgdaFunction{≈}}\AgdaSpace{}%
\AgdaBound{q}\AgdaSpace{}%
\AgdaSymbol{=}\AgdaSpace{}%
\AgdaFunction{Equal}\AgdaSpace{}%
\AgdaBound{p}\AgdaSpace{}%
\AgdaBound{q}\AgdaSpace{}%
\AgdaKeyword{where}\AgdaSpace{}%
\AgdaKeyword{open}\AgdaSpace{}%
\AgdaModule{Environment}\AgdaSpace{}%
\AgdaBound{𝑨}\<%
\\
\\[\AgdaEmptyExtraSkip]%
\>[1]\AgdaOperator{\AgdaFunction{\AgdaUnderscore{}⊫\AgdaUnderscore{}≈\AgdaUnderscore{}}}\AgdaSpace{}%
\AgdaSymbol{:}\AgdaSpace{}%
\AgdaFunction{Pred}\AgdaSpace{}%
\AgdaSymbol{(}\AgdaRecord{Algebra}\AgdaSpace{}%
\AgdaGeneralizable{α}\AgdaSpace{}%
\AgdaGeneralizable{ρᵃ}\AgdaSymbol{)}\AgdaSpace{}%
\AgdaGeneralizable{ℓ}\AgdaSpace{}%
\AgdaSymbol{→}\AgdaSpace{}%
\AgdaDatatype{Term}\AgdaSpace{}%
\AgdaBound{X}\AgdaSpace{}%
\AgdaSymbol{→}\AgdaSpace{}%
\AgdaDatatype{Term}\AgdaSpace{}%
\AgdaBound{X}\AgdaSpace{}%
\AgdaSymbol{→}\AgdaSpace{}%
\AgdaPrimitive{Type}\AgdaSpace{}%
\AgdaSymbol{\AgdaUnderscore{}}\<%
\\
\>[1]\AgdaBound{𝒦}\AgdaSpace{}%
\AgdaOperator{\AgdaFunction{⊫}}\AgdaSpace{}%
\AgdaBound{p}\AgdaSpace{}%
\AgdaOperator{\AgdaFunction{≈}}\AgdaSpace{}%
\AgdaBound{q}\AgdaSpace{}%
\AgdaSymbol{=}\AgdaSpace{}%
\AgdaSymbol{∀}\AgdaSpace{}%
\AgdaBound{𝑨}\AgdaSpace{}%
\AgdaSymbol{→}\AgdaSpace{}%
\AgdaBound{𝒦}\AgdaSpace{}%
\AgdaBound{𝑨}\AgdaSpace{}%
\AgdaSymbol{→}\AgdaSpace{}%
\AgdaBound{𝑨}\AgdaSpace{}%
\AgdaOperator{\AgdaFunction{⊧}}\AgdaSpace{}%
\AgdaBound{p}\AgdaSpace{}%
\AgdaOperator{\AgdaFunction{≈}}\AgdaSpace{}%
\AgdaBound{q}\<%
\\
\>[0]\<%
\end{code}
We represent a set of identities as a predicate over pairs of
terms, say, \ab{ℰ} : \af{Pred}(\ad{Term} \ab{X} \af{×} \ad{Term} \ab{X})~\au{}  and we denote by
\ab{𝑨}~\aof{⊨}~\ab{ℰ} the assertion that the algebra \ab{𝑨} models \ab{p}~\af{≈}~\ab{q}
for all (\ab{p} , \ab{q}) \af{∈} \ab{ℰ}.\footnote{Notice that \af{⊨} is
a stretched version of the models symbol, \af{⊧};
\ifshort\else
this makes it possible for Agda to distinguish and parse expressions involving the types
\af{\au{}⊨\au{}} and \af{\au{}⊧\au{}≈\au{}}.
\fi
In Emacs \texttt{agda2-mode}, the symbol \af{⊨} is produced by typing
\textbackslash\textbar{}=, while \af{⊧} is
produced with \textbackslash{}models.}

\begin{code}%
\>[0]\<%
\\
\>[0][@{}l@{\AgdaIndent{1}}]%
\>[1]\AgdaOperator{\AgdaFunction{\AgdaUnderscore{}⊨\AgdaUnderscore{}}}\AgdaSpace{}%
\AgdaSymbol{:}\AgdaSpace{}%
\AgdaSymbol{(}\AgdaBound{𝑨}\AgdaSpace{}%
\AgdaSymbol{:}\AgdaSpace{}%
\AgdaRecord{Algebra}\AgdaSpace{}%
\AgdaGeneralizable{α}\AgdaSpace{}%
\AgdaGeneralizable{ρᵃ}\AgdaSymbol{)}\AgdaSpace{}%
\AgdaSymbol{→}\AgdaSpace{}%
\AgdaFunction{Pred}\AgdaSymbol{(}\AgdaDatatype{Term}\AgdaSpace{}%
\AgdaBound{X}\AgdaSpace{}%
\AgdaOperator{\AgdaFunction{×}}\AgdaSpace{}%
\AgdaDatatype{Term}\AgdaSpace{}%
\AgdaBound{X}\AgdaSymbol{)(}\AgdaFunction{ov}\AgdaSpace{}%
\AgdaBound{χ}\AgdaSymbol{)}\AgdaSpace{}%
\AgdaSymbol{→}\AgdaSpace{}%
\AgdaPrimitive{Type}\AgdaSpace{}%
\AgdaSymbol{\AgdaUnderscore{}}\<%
\\
\>[1]\AgdaBound{𝑨}\AgdaSpace{}%
\AgdaOperator{\AgdaFunction{⊨}}\AgdaSpace{}%
\AgdaBound{ℰ}\AgdaSpace{}%
\AgdaSymbol{=}\AgdaSpace{}%
\AgdaSymbol{∀}\AgdaSpace{}%
\AgdaSymbol{\{}\AgdaBound{p}\AgdaSpace{}%
\AgdaBound{q}\AgdaSymbol{\}}\AgdaSpace{}%
\AgdaSymbol{→}\AgdaSpace{}%
\AgdaSymbol{(}\AgdaBound{p}\AgdaSpace{}%
\AgdaOperator{\AgdaInductiveConstructor{,}}\AgdaSpace{}%
\AgdaBound{q}\AgdaSymbol{)}\AgdaSpace{}%
\AgdaOperator{\AgdaFunction{∈}}\AgdaSpace{}%
\AgdaBound{ℰ}\AgdaSpace{}%
\AgdaSymbol{→}\AgdaSpace{}%
\AgdaFunction{Equal}\AgdaSpace{}%
\AgdaBound{p}\AgdaSpace{}%
\AgdaBound{q}\AgdaSpace{}%
\AgdaKeyword{where}\AgdaSpace{}%
\AgdaKeyword{open}\AgdaSpace{}%
\AgdaModule{Environment}\AgdaSpace{}%
\AgdaBound{𝑨}\<%
\\
\>[0]\<%
\end{code}

If \ab{𝒦} is a class of structures and \ab{ℰ} a set of term identities, then the set of
term equations modeled by \ab{𝒦} is denoted by \af{Th}~\ab{𝒦} and is called the
\defn{equational theory} of \ab{𝒦}, while the class of structures modeling \ab{ℰ} is
denoted by \af{Mod}~\ab{ℰ} and is called the \defn{equational class axiomatized} by
\ab{ℰ}. We formalize these concepts in Agda with the following types.

\begin{code}%
\>[0]\<%
\\
\>[0]\AgdaFunction{Th}\AgdaSpace{}%
\AgdaSymbol{:}\AgdaSpace{}%
\AgdaSymbol{\{}\AgdaBound{X}\AgdaSpace{}%
\AgdaSymbol{:}\AgdaSpace{}%
\AgdaPrimitive{Type}\AgdaSpace{}%
\AgdaGeneralizable{χ}\AgdaSymbol{\}}\AgdaSpace{}%
\AgdaSymbol{→}\AgdaSpace{}%
\AgdaFunction{Pred}\AgdaSpace{}%
\AgdaSymbol{(}\AgdaRecord{Algebra}\AgdaSpace{}%
\AgdaGeneralizable{α}\AgdaSpace{}%
\AgdaGeneralizable{ρᵃ}\AgdaSymbol{)}\AgdaSpace{}%
\AgdaGeneralizable{ℓ}\AgdaSpace{}%
\AgdaSymbol{→}\AgdaSpace{}%
\AgdaFunction{Pred}\AgdaSymbol{(}\AgdaDatatype{Term}\AgdaSpace{}%
\AgdaBound{X}\AgdaSpace{}%
\AgdaOperator{\AgdaFunction{×}}\AgdaSpace{}%
\AgdaDatatype{Term}\AgdaSpace{}%
\AgdaBound{X}\AgdaSymbol{)}\AgdaSpace{}%
\AgdaSymbol{\AgdaUnderscore{}}\<%
\\
\>[0]\AgdaFunction{Th}\AgdaSpace{}%
\AgdaBound{𝒦}\AgdaSpace{}%
\AgdaSymbol{=}\AgdaSpace{}%
\AgdaSymbol{λ}\AgdaSpace{}%
\AgdaSymbol{(}\AgdaBound{p}\AgdaSpace{}%
\AgdaOperator{\AgdaInductiveConstructor{,}}\AgdaSpace{}%
\AgdaBound{q}\AgdaSymbol{)}\AgdaSpace{}%
\AgdaSymbol{→}\AgdaSpace{}%
\AgdaBound{𝒦}\AgdaSpace{}%
\AgdaOperator{\AgdaFunction{⊫}}\AgdaSpace{}%
\AgdaBound{p}\AgdaSpace{}%
\AgdaOperator{\AgdaFunction{≈}}\AgdaSpace{}%
\AgdaBound{q}\<%
\\
\\[\AgdaEmptyExtraSkip]%
\>[0]\AgdaFunction{Mod}\AgdaSpace{}%
\AgdaSymbol{:}\AgdaSpace{}%
\AgdaSymbol{\{}\AgdaBound{X}\AgdaSpace{}%
\AgdaSymbol{:}\AgdaSpace{}%
\AgdaPrimitive{Type}\AgdaSpace{}%
\AgdaGeneralizable{χ}\AgdaSymbol{\}}\AgdaSpace{}%
\AgdaSymbol{→}\AgdaSpace{}%
\AgdaFunction{Pred}\AgdaSymbol{(}\AgdaDatatype{Term}\AgdaSpace{}%
\AgdaBound{X}\AgdaSpace{}%
\AgdaOperator{\AgdaFunction{×}}\AgdaSpace{}%
\AgdaDatatype{Term}\AgdaSpace{}%
\AgdaBound{X}\AgdaSymbol{)}\AgdaSpace{}%
\AgdaGeneralizable{ℓ}\AgdaSpace{}%
\AgdaSymbol{→}\AgdaSpace{}%
\AgdaFunction{Pred}\AgdaSpace{}%
\AgdaSymbol{(}\AgdaRecord{Algebra}\AgdaSpace{}%
\AgdaGeneralizable{α}\AgdaSpace{}%
\AgdaGeneralizable{ρᵃ}\AgdaSymbol{)}\AgdaSpace{}%
\AgdaSymbol{\AgdaUnderscore{}}\<%
\\
\>[0]\AgdaFunction{Mod}\AgdaSpace{}%
\AgdaBound{ℰ}\AgdaSpace{}%
\AgdaBound{𝑨}\AgdaSpace{}%
\AgdaSymbol{=}\AgdaSpace{}%
\AgdaSymbol{∀}\AgdaSpace{}%
\AgdaSymbol{\{}\AgdaBound{p}\AgdaSpace{}%
\AgdaBound{q}\AgdaSymbol{\}}\AgdaSpace{}%
\AgdaSymbol{→}\AgdaSpace{}%
\AgdaSymbol{(}\AgdaBound{p}\AgdaSpace{}%
\AgdaOperator{\AgdaInductiveConstructor{,}}\AgdaSpace{}%
\AgdaBound{q}\AgdaSymbol{)}\AgdaSpace{}%
\AgdaOperator{\AgdaFunction{∈}}\AgdaSpace{}%
\AgdaBound{ℰ}\AgdaSpace{}%
\AgdaSymbol{→}\AgdaSpace{}%
\AgdaFunction{Equal}\AgdaSpace{}%
\AgdaBound{p}\AgdaSpace{}%
\AgdaBound{q}\AgdaSpace{}%
\AgdaKeyword{where}\AgdaSpace{}%
\AgdaKeyword{open}\AgdaSpace{}%
\AgdaModule{Environment}\AgdaSpace{}%
\AgdaBound{𝑨}\<%
\end{code}

\paragraph*{Entailment}

If \ab{ℰ} is a set of \ab{𝑆}-term equations and \ab{p} and \ab{q} are \ab{𝑆}-terms,
we say that \ab{ℰ} \defn{entails} the equation \ab{p}~\aof{≈}~\ab{q}, and we write
\ab{ℰ}~\ad{⊢}~\ab{p}~\ad{≈}~\ab{q}, just in case every model of \ab{ℰ} also models
\ab{p}~\aof{≈}~\ab{q}.
We represent entailment in type theory using an inductive type that is similar to
the one defined by Abel in~\cite{Abel:2021}.  We call this the \defn{entailment type}
and define it as follows.

\begin{code}%
\>[0]\<%
\\
\>[0]\AgdaKeyword{data}\AgdaSpace{}%
\AgdaOperator{\AgdaDatatype{\AgdaUnderscore{}⊢\AgdaUnderscore{}▹\AgdaUnderscore{}≈\AgdaUnderscore{}}}%
\>[14]\AgdaSymbol{(}\AgdaBound{ℰ}\AgdaSpace{}%
\AgdaSymbol{:}\AgdaSpace{}%
\AgdaSymbol{\{}\AgdaBound{Y}\AgdaSpace{}%
\AgdaSymbol{:}\AgdaSpace{}%
\AgdaPrimitive{Type}\AgdaSpace{}%
\AgdaGeneralizable{χ}\AgdaSymbol{\}}\AgdaSpace{}%
\AgdaSymbol{→}\AgdaSpace{}%
\AgdaFunction{Pred}\AgdaSymbol{(}\AgdaDatatype{Term}\AgdaSpace{}%
\AgdaBound{Y}\AgdaSpace{}%
\AgdaOperator{\AgdaFunction{×}}\AgdaSpace{}%
\AgdaDatatype{Term}\AgdaSpace{}%
\AgdaBound{Y}\AgdaSymbol{)}\AgdaSpace{}%
\AgdaSymbol{(}\AgdaFunction{ov}\AgdaSpace{}%
\AgdaGeneralizable{χ}\AgdaSymbol{))}\AgdaSpace{}%
\AgdaSymbol{:}\<%
\\
\>[14]\AgdaSymbol{(}\AgdaBound{X}\AgdaSpace{}%
\AgdaSymbol{:}\AgdaSpace{}%
\AgdaPrimitive{Type}\AgdaSpace{}%
\AgdaBound{χ}\AgdaSymbol{)(}\AgdaBound{p}\AgdaSpace{}%
\AgdaBound{q}\AgdaSpace{}%
\AgdaSymbol{:}\AgdaSpace{}%
\AgdaDatatype{Term}\AgdaSpace{}%
\AgdaBound{X}\AgdaSymbol{)}\AgdaSpace{}%
\AgdaSymbol{→}\AgdaSpace{}%
\AgdaPrimitive{Type}\AgdaSpace{}%
\AgdaSymbol{(}\AgdaFunction{ov}\AgdaSpace{}%
\AgdaBound{χ}\AgdaSymbol{)}\AgdaSpace{}%
\AgdaKeyword{where}\<%
\\
\\[\AgdaEmptyExtraSkip]%
\>[0][@{}l@{\AgdaIndent{0}}]%
\>[1]\AgdaInductiveConstructor{hyp}%
\>[13]\AgdaSymbol{:}%
\>[16]\AgdaSymbol{∀\{}\AgdaBound{Y}\AgdaSymbol{\}\{}\AgdaBound{p}\AgdaSpace{}%
\AgdaBound{q}\AgdaSpace{}%
\AgdaSymbol{:}\AgdaSpace{}%
\AgdaDatatype{Term}\AgdaSpace{}%
\AgdaBound{Y}\AgdaSymbol{\}}\AgdaSpace{}%
\AgdaSymbol{→}\AgdaSpace{}%
\AgdaSymbol{(}\AgdaBound{p}\AgdaSpace{}%
\AgdaOperator{\AgdaInductiveConstructor{,}}\AgdaSpace{}%
\AgdaBound{q}\AgdaSymbol{)}\AgdaSpace{}%
\AgdaOperator{\AgdaFunction{∈}}\AgdaSpace{}%
\AgdaBound{ℰ}\AgdaSpace{}%
\AgdaSymbol{→}\AgdaSpace{}%
\AgdaBound{ℰ}\AgdaSpace{}%
\AgdaOperator{\AgdaDatatype{⊢}}\AgdaSpace{}%
\AgdaSymbol{\AgdaUnderscore{}}\AgdaSpace{}%
\AgdaOperator{\AgdaDatatype{▹}}\AgdaSpace{}%
\AgdaBound{p}\AgdaSpace{}%
\AgdaOperator{\AgdaDatatype{≈}}\AgdaSpace{}%
\AgdaBound{q}\<%
\\
\>[1]\AgdaInductiveConstructor{app}%
\>[13]\AgdaSymbol{:}%
\>[16]\AgdaSymbol{∀\{}\AgdaBound{Y}\AgdaSymbol{\}\{}\AgdaBound{ps}%
\>[4291I]\AgdaBound{qs}\AgdaSpace{}%
\AgdaSymbol{:}\AgdaSpace{}%
\AgdaOperator{\AgdaFunction{∥}}\AgdaSpace{}%
\AgdaBound{𝑆}\AgdaSpace{}%
\AgdaOperator{\AgdaFunction{∥}}\AgdaSpace{}%
\AgdaGeneralizable{𝑓}\AgdaSpace{}%
\AgdaSymbol{→}\AgdaSpace{}%
\AgdaDatatype{Term}\AgdaSpace{}%
\AgdaBound{Y}\AgdaSymbol{\}}\<%
\\
\>[4291I][@{}l@{\AgdaIndent{0}}]%
\>[26]\AgdaSymbol{→}\AgdaSpace{}%
\AgdaSymbol{(∀}\AgdaSpace{}%
\AgdaBound{i}\AgdaSpace{}%
\AgdaSymbol{→}\AgdaSpace{}%
\AgdaBound{ℰ}\AgdaSpace{}%
\AgdaOperator{\AgdaDatatype{⊢}}\AgdaSpace{}%
\AgdaBound{Y}\AgdaSpace{}%
\AgdaOperator{\AgdaDatatype{▹}}\AgdaSpace{}%
\AgdaBound{ps}\AgdaSpace{}%
\AgdaBound{i}\AgdaSpace{}%
\AgdaOperator{\AgdaDatatype{≈}}\AgdaSpace{}%
\AgdaBound{qs}\AgdaSpace{}%
\AgdaBound{i}\AgdaSymbol{)}\AgdaSpace{}%
\AgdaSymbol{→}\AgdaSpace{}%
\AgdaBound{ℰ}\AgdaSpace{}%
\AgdaOperator{\AgdaDatatype{⊢}}\AgdaSpace{}%
\AgdaBound{Y}\AgdaSpace{}%
\AgdaOperator{\AgdaDatatype{▹}}\AgdaSpace{}%
\AgdaSymbol{(}\AgdaInductiveConstructor{node}\AgdaSpace{}%
\AgdaGeneralizable{𝑓}\AgdaSpace{}%
\AgdaBound{ps}\AgdaSymbol{)}\AgdaSpace{}%
\AgdaOperator{\AgdaDatatype{≈}}\AgdaSpace{}%
\AgdaSymbol{(}\AgdaInductiveConstructor{node}\AgdaSpace{}%
\AgdaGeneralizable{𝑓}\AgdaSpace{}%
\AgdaBound{qs}\AgdaSymbol{)}\<%
\\
\>[1]\AgdaInductiveConstructor{sub}%
\>[13]\AgdaSymbol{:}%
\>[16]\AgdaSymbol{∀\{}\AgdaBound{p}\AgdaSpace{}%
\AgdaBound{q}\AgdaSymbol{\}}%
\>[26]\AgdaSymbol{→}\AgdaSpace{}%
\AgdaBound{ℰ}\AgdaSpace{}%
\AgdaOperator{\AgdaDatatype{⊢}}\AgdaSpace{}%
\AgdaGeneralizable{Γ}\AgdaSpace{}%
\AgdaOperator{\AgdaDatatype{▹}}\AgdaSpace{}%
\AgdaBound{p}\AgdaSpace{}%
\AgdaOperator{\AgdaDatatype{≈}}\AgdaSpace{}%
\AgdaBound{q}\AgdaSpace{}%
\AgdaSymbol{→}\AgdaSpace{}%
\AgdaSymbol{(}\AgdaBound{σ}\AgdaSpace{}%
\AgdaSymbol{:}\AgdaSpace{}%
\AgdaFunction{Sub}\AgdaSpace{}%
\AgdaGeneralizable{Δ}\AgdaSpace{}%
\AgdaGeneralizable{Γ}\AgdaSymbol{)}\AgdaSpace{}%
\AgdaSymbol{→}\AgdaSpace{}%
\AgdaBound{ℰ}\AgdaSpace{}%
\AgdaOperator{\AgdaDatatype{⊢}}\AgdaSpace{}%
\AgdaGeneralizable{Δ}\AgdaSpace{}%
\AgdaOperator{\AgdaDatatype{▹}}\AgdaSpace{}%
\AgdaSymbol{(}\AgdaOperator{\AgdaFunction{[}}\AgdaSpace{}%
\AgdaBound{σ}\AgdaSpace{}%
\AgdaOperator{\AgdaFunction{]}}\AgdaSpace{}%
\AgdaBound{p}\AgdaSymbol{)}\AgdaSpace{}%
\AgdaOperator{\AgdaDatatype{≈}}\AgdaSpace{}%
\AgdaSymbol{(}\AgdaOperator{\AgdaFunction{[}}\AgdaSpace{}%
\AgdaBound{σ}\AgdaSpace{}%
\AgdaOperator{\AgdaFunction{]}}\AgdaSpace{}%
\AgdaBound{q}\AgdaSymbol{)}\<%
\\
\>[1]\AgdaInductiveConstructor{reflexive}%
\>[13]\AgdaSymbol{:}%
\>[16]\AgdaSymbol{∀\{}\AgdaBound{p}\AgdaSymbol{\}}%
\>[26]\AgdaSymbol{→}\AgdaSpace{}%
\AgdaBound{ℰ}\AgdaSpace{}%
\AgdaOperator{\AgdaDatatype{⊢}}\AgdaSpace{}%
\AgdaGeneralizable{Γ}\AgdaSpace{}%
\AgdaOperator{\AgdaDatatype{▹}}\AgdaSpace{}%
\AgdaBound{p}\AgdaSpace{}%
\AgdaOperator{\AgdaDatatype{≈}}\AgdaSpace{}%
\AgdaBound{p}\<%
\\
\>[1]\AgdaInductiveConstructor{symmetric}%
\>[13]\AgdaSymbol{:}%
\>[16]\AgdaSymbol{∀\{}\AgdaBound{p}\AgdaSpace{}%
\AgdaBound{q}\AgdaSymbol{\}}%
\>[26]\AgdaSymbol{→}\AgdaSpace{}%
\AgdaBound{ℰ}\AgdaSpace{}%
\AgdaOperator{\AgdaDatatype{⊢}}\AgdaSpace{}%
\AgdaGeneralizable{Γ}\AgdaSpace{}%
\AgdaOperator{\AgdaDatatype{▹}}\AgdaSpace{}%
\AgdaBound{p}\AgdaSpace{}%
\AgdaOperator{\AgdaDatatype{≈}}\AgdaSpace{}%
\AgdaBound{q}\AgdaSpace{}%
\AgdaSymbol{→}\AgdaSpace{}%
\AgdaBound{ℰ}\AgdaSpace{}%
\AgdaOperator{\AgdaDatatype{⊢}}\AgdaSpace{}%
\AgdaGeneralizable{Γ}\AgdaSpace{}%
\AgdaOperator{\AgdaDatatype{▹}}\AgdaSpace{}%
\AgdaBound{q}\AgdaSpace{}%
\AgdaOperator{\AgdaDatatype{≈}}\AgdaSpace{}%
\AgdaBound{p}\<%
\\
\>[1]\AgdaInductiveConstructor{transitive}%
\>[13]\AgdaSymbol{:}%
\>[16]\AgdaSymbol{∀\{}\AgdaBound{p}\AgdaSpace{}%
\AgdaBound{q}\AgdaSpace{}%
\AgdaBound{r}\AgdaSymbol{\}}%
\>[26]\AgdaSymbol{→}\AgdaSpace{}%
\AgdaBound{ℰ}\AgdaSpace{}%
\AgdaOperator{\AgdaDatatype{⊢}}\AgdaSpace{}%
\AgdaGeneralizable{Γ}\AgdaSpace{}%
\AgdaOperator{\AgdaDatatype{▹}}\AgdaSpace{}%
\AgdaBound{p}\AgdaSpace{}%
\AgdaOperator{\AgdaDatatype{≈}}\AgdaSpace{}%
\AgdaBound{q}\AgdaSpace{}%
\AgdaSymbol{→}\AgdaSpace{}%
\AgdaBound{ℰ}\AgdaSpace{}%
\AgdaOperator{\AgdaDatatype{⊢}}\AgdaSpace{}%
\AgdaGeneralizable{Γ}\AgdaSpace{}%
\AgdaOperator{\AgdaDatatype{▹}}\AgdaSpace{}%
\AgdaBound{q}\AgdaSpace{}%
\AgdaOperator{\AgdaDatatype{≈}}\AgdaSpace{}%
\AgdaBound{r}\AgdaSpace{}%
\AgdaSymbol{→}\AgdaSpace{}%
\AgdaBound{ℰ}\AgdaSpace{}%
\AgdaOperator{\AgdaDatatype{⊢}}\AgdaSpace{}%
\AgdaGeneralizable{Γ}\AgdaSpace{}%
\AgdaOperator{\AgdaDatatype{▹}}\AgdaSpace{}%
\AgdaBound{p}\AgdaSpace{}%
\AgdaOperator{\AgdaDatatype{≈}}\AgdaSpace{}%
\AgdaBound{r}\<%
\\
\>[0]\<%
\end{code}

The fact that this type represents the informal semantic notion of entailment
given at the start of this subsection is called \defn{soundness} and
\defn{completeness}.
More precisely, \defn{the entailment type is sound} means the following:
if \ab{ℰ}~\ad{⊢}~\ab{X}~\ad{▹}~\ab p~\ad{≈}~\ab q, then \ab p \aof{≈} \ab q holds in
every model of \ab{ℰ}.
\defn{The entailment type is complete} means the following:
if \ab p \aof{≈} \ab q holds in every model of \ab{ℰ},
then \ab{ℰ}~\ad{⊢}~\ab{X}~\ad{▹}~\ab p~\aof{≈}~\ab q.
Soundness and completeness of an entailment type similar to the one defined above was
proved by Abel in~\cite{Abel:2021}.  We will invoke soundness of the entailment type only once below%
\ifshort
~(by the name \af{sound}), so we omit its proof, but see~\cite{Abel:2021}
or~\cite{DeMeo:2021c} for the complete formalization.
\else
; nonetheless, here is its formalization (due to Abel, \textit{op. cit.}):

\begin{code}%
\>[0]\<%
\\
\>[0]\AgdaKeyword{module}\AgdaSpace{}%
\AgdaModule{Soundness}%
\>[18]\AgdaSymbol{(}\AgdaBound{ℰ}\AgdaSpace{}%
\AgdaSymbol{:}\AgdaSpace{}%
\AgdaSymbol{\{}\AgdaBound{Y}\AgdaSpace{}%
\AgdaSymbol{:}\AgdaSpace{}%
\AgdaPrimitive{Type}\AgdaSpace{}%
\AgdaGeneralizable{χ}\AgdaSymbol{\}}\AgdaSpace{}%
\AgdaSymbol{→}\AgdaSpace{}%
\AgdaFunction{Pred}\AgdaSymbol{(}\AgdaDatatype{Term}\AgdaSpace{}%
\AgdaBound{Y}\AgdaSpace{}%
\AgdaOperator{\AgdaFunction{×}}\AgdaSpace{}%
\AgdaDatatype{Term}\AgdaSpace{}%
\AgdaBound{Y}\AgdaSymbol{)}\AgdaSpace{}%
\AgdaSymbol{(}\AgdaFunction{ov}\AgdaSpace{}%
\AgdaGeneralizable{χ}\AgdaSymbol{))}\<%
\\
\>[18]\AgdaSymbol{(}\AgdaBound{𝑨}\AgdaSpace{}%
\AgdaSymbol{:}\AgdaSpace{}%
\AgdaRecord{Algebra}\AgdaSpace{}%
\AgdaGeneralizable{α}\AgdaSpace{}%
\AgdaGeneralizable{ρᵃ}\AgdaSymbol{)}%
\>[52]\AgdaComment{--\ We\ assume\ an\ algebra\ 𝑨}\<%
\\
\>[18]\AgdaSymbol{(}\AgdaBound{V}\AgdaSpace{}%
\AgdaSymbol{:}\AgdaSpace{}%
\AgdaSymbol{∀\{}\AgdaBound{Y}\AgdaSymbol{\}}\AgdaSpace{}%
\AgdaSymbol{→}\AgdaSpace{}%
\AgdaOperator{\AgdaFunction{\AgdaUnderscore{}⊨\AgdaUnderscore{}}}\AgdaSymbol{\{}\AgdaArgument{χ}\AgdaSpace{}%
\AgdaSymbol{=}\AgdaSpace{}%
\AgdaGeneralizable{χ}\AgdaSymbol{\}}\AgdaSpace{}%
\AgdaBound{𝑨}\AgdaSpace{}%
\AgdaSymbol{(}\AgdaBound{ℰ}\AgdaSymbol{\{}\AgdaBound{Y}\AgdaSymbol{\}))}%
\>[52]\AgdaComment{--\ that\ models\ all\ equations\ in\ ℰ.}\<%
\\
\>[18]\AgdaKeyword{where}\<%
\\
\>[0][@{}l@{\AgdaIndent{0}}]%
\>[1]\AgdaKeyword{open}\AgdaSpace{}%
\AgdaModule{SetoidReasoning}\AgdaSpace{}%
\AgdaOperator{\AgdaFunction{𝔻[}}\AgdaSpace{}%
\AgdaBound{𝑨}\AgdaSpace{}%
\AgdaOperator{\AgdaFunction{]}}\<%
\\
\>[1]\AgdaKeyword{open}\AgdaSpace{}%
\AgdaModule{Environment}\AgdaSpace{}%
\AgdaBound{𝑨}\<%
\\
\>[1]\AgdaFunction{sound}\AgdaSpace{}%
\AgdaSymbol{:}\AgdaSpace{}%
\AgdaSymbol{∀}\AgdaSpace{}%
\AgdaSymbol{\{}\AgdaBound{p}\AgdaSpace{}%
\AgdaBound{q}\AgdaSymbol{\}}\AgdaSpace{}%
\AgdaSymbol{→}\AgdaSpace{}%
\AgdaBound{ℰ}\AgdaSpace{}%
\AgdaOperator{\AgdaDatatype{⊢}}\AgdaSpace{}%
\AgdaGeneralizable{Γ}\AgdaSpace{}%
\AgdaOperator{\AgdaDatatype{▹}}\AgdaSpace{}%
\AgdaBound{p}\AgdaSpace{}%
\AgdaOperator{\AgdaDatatype{≈}}\AgdaSpace{}%
\AgdaBound{q}\AgdaSpace{}%
\AgdaSymbol{→}\AgdaSpace{}%
\AgdaBound{𝑨}\AgdaSpace{}%
\AgdaOperator{\AgdaFunction{⊧}}\AgdaSpace{}%
\AgdaBound{p}\AgdaSpace{}%
\AgdaOperator{\AgdaFunction{≈}}\AgdaSpace{}%
\AgdaBound{q}\<%
\\
\>[1]\AgdaFunction{sound}\AgdaSpace{}%
\AgdaSymbol{(}\AgdaInductiveConstructor{hyp}\AgdaSpace{}%
\AgdaBound{i}\AgdaSymbol{)}\AgdaSpace{}%
\AgdaSymbol{=}\AgdaSpace{}%
\AgdaBound{V}\AgdaSpace{}%
\AgdaBound{i}\<%
\\
\>[1]\AgdaFunction{sound}\AgdaSpace{}%
\AgdaSymbol{(}\AgdaInductiveConstructor{app}\AgdaSpace{}%
\AgdaBound{es}\AgdaSymbol{)}\AgdaSpace{}%
\AgdaBound{ρ}\AgdaSpace{}%
\AgdaSymbol{=}\AgdaSpace{}%
\AgdaField{cong}\AgdaSpace{}%
\AgdaSymbol{(}\AgdaField{Interp}\AgdaSpace{}%
\AgdaBound{𝑨}\AgdaSymbol{)}\AgdaSpace{}%
\AgdaSymbol{(}\AgdaInductiveConstructor{≡.refl}\AgdaSpace{}%
\AgdaOperator{\AgdaInductiveConstructor{,}}\AgdaSpace{}%
\AgdaSymbol{λ}\AgdaSpace{}%
\AgdaBound{i}\AgdaSpace{}%
\AgdaSymbol{→}\AgdaSpace{}%
\AgdaFunction{sound}\AgdaSpace{}%
\AgdaSymbol{(}\AgdaBound{es}\AgdaSpace{}%
\AgdaBound{i}\AgdaSymbol{)}\AgdaSpace{}%
\AgdaBound{ρ}\AgdaSymbol{)}\<%
\\
\>[1]\AgdaFunction{sound}\AgdaSpace{}%
\AgdaSymbol{(}\AgdaInductiveConstructor{sub}\AgdaSpace{}%
\AgdaSymbol{\{}\AgdaArgument{p}\AgdaSpace{}%
\AgdaSymbol{=}\AgdaSpace{}%
\AgdaBound{p}\AgdaSymbol{\}\{}\AgdaBound{q}\AgdaSymbol{\}}\AgdaSpace{}%
\AgdaBound{Epq}\AgdaSpace{}%
\AgdaBound{σ}\AgdaSymbol{)}\AgdaSpace{}%
\AgdaBound{ρ}\AgdaSpace{}%
\AgdaSymbol{=}\<%
\\
\>[1][@{}l@{\AgdaIndent{0}}]%
\>[2]\AgdaOperator{\AgdaFunction{begin}}\<%
\\
\>[2][@{}l@{\AgdaIndent{0}}]%
\>[3]\AgdaOperator{\AgdaFunction{⟦}}\AgdaSpace{}%
\AgdaOperator{\AgdaFunction{[}}\AgdaSpace{}%
\AgdaBound{σ}\AgdaSpace{}%
\AgdaOperator{\AgdaFunction{]}}\AgdaSpace{}%
\AgdaBound{p}%
\>[14]\AgdaOperator{\AgdaFunction{⟧}}\AgdaSpace{}%
\AgdaOperator{\AgdaField{⟨\$⟩}}%
\>[40]\AgdaBound{ρ}%
\>[44]\AgdaFunction{≈⟨}%
\>[49]\AgdaFunction{substitution}\AgdaSpace{}%
\AgdaBound{p}\AgdaSpace{}%
\AgdaBound{σ}\AgdaSpace{}%
\AgdaBound{ρ}%
\>[82]\AgdaFunction{⟩}\<%
\\
\>[3]\AgdaOperator{\AgdaFunction{⟦}}\AgdaSpace{}%
\AgdaBound{p}%
\>[14]\AgdaOperator{\AgdaFunction{⟧}}\AgdaSpace{}%
\AgdaOperator{\AgdaField{⟨\$⟩}}\AgdaSpace{}%
\AgdaSymbol{(λ}\AgdaSpace{}%
\AgdaBound{x}\AgdaSpace{}%
\AgdaSymbol{→}\AgdaSpace{}%
\AgdaOperator{\AgdaFunction{⟦}}\AgdaSpace{}%
\AgdaBound{σ}\AgdaSpace{}%
\AgdaBound{x}\AgdaSpace{}%
\AgdaOperator{\AgdaFunction{⟧}}\AgdaSpace{}%
\AgdaOperator{\AgdaField{⟨\$⟩}}%
\>[40]\AgdaBound{ρ}\AgdaSymbol{)}%
\>[44]\AgdaFunction{≈⟨}%
\>[49]\AgdaFunction{sound}\AgdaSpace{}%
\AgdaBound{Epq}\AgdaSpace{}%
\AgdaSymbol{(λ}\AgdaSpace{}%
\AgdaBound{x}\AgdaSpace{}%
\AgdaSymbol{→}\AgdaSpace{}%
\AgdaOperator{\AgdaFunction{⟦}}\AgdaSpace{}%
\AgdaBound{σ}\AgdaSpace{}%
\AgdaBound{x}\AgdaSpace{}%
\AgdaOperator{\AgdaFunction{⟧}}\AgdaSpace{}%
\AgdaOperator{\AgdaField{⟨\$⟩}}\AgdaSpace{}%
\AgdaBound{ρ}\AgdaSymbol{)}%
\>[82]\AgdaFunction{⟩}\<%
\\
\>[3]\AgdaOperator{\AgdaFunction{⟦}}\AgdaSpace{}%
\AgdaBound{q}%
\>[14]\AgdaOperator{\AgdaFunction{⟧}}\AgdaSpace{}%
\AgdaOperator{\AgdaField{⟨\$⟩}}\AgdaSpace{}%
\AgdaSymbol{(λ}\AgdaSpace{}%
\AgdaBound{x}\AgdaSpace{}%
\AgdaSymbol{→}\AgdaSpace{}%
\AgdaOperator{\AgdaFunction{⟦}}\AgdaSpace{}%
\AgdaBound{σ}\AgdaSpace{}%
\AgdaBound{x}\AgdaSpace{}%
\AgdaOperator{\AgdaFunction{⟧}}\AgdaSpace{}%
\AgdaOperator{\AgdaField{⟨\$⟩}}%
\>[40]\AgdaBound{ρ}\AgdaSymbol{)}%
\>[44]\AgdaFunction{≈˘⟨}%
\>[49]\AgdaFunction{substitution}\AgdaSpace{}%
\AgdaBound{q}\AgdaSpace{}%
\AgdaBound{σ}\AgdaSpace{}%
\AgdaBound{ρ}%
\>[82]\AgdaFunction{⟩}\<%
\\
\>[3]\AgdaOperator{\AgdaFunction{⟦}}\AgdaSpace{}%
\AgdaOperator{\AgdaFunction{[}}\AgdaSpace{}%
\AgdaBound{σ}\AgdaSpace{}%
\AgdaOperator{\AgdaFunction{]}}\AgdaSpace{}%
\AgdaBound{q}%
\>[14]\AgdaOperator{\AgdaFunction{⟧}}\AgdaSpace{}%
\AgdaOperator{\AgdaField{⟨\$⟩}}%
\>[40]\AgdaBound{ρ}%
\>[44]\AgdaOperator{\AgdaFunction{∎}}\<%
\\
\>[1]\AgdaFunction{sound}\AgdaSpace{}%
\AgdaSymbol{(}\AgdaInductiveConstructor{reflexive}%
\>[20]\AgdaSymbol{\{}\AgdaArgument{p}\AgdaSpace{}%
\AgdaSymbol{=}\AgdaSpace{}%
\AgdaBound{p}\AgdaSymbol{\}}%
\>[44]\AgdaSymbol{)}\AgdaSpace{}%
\AgdaSymbol{=}\AgdaSpace{}%
\AgdaField{reflᵉ}%
\>[56]\AgdaFunction{EqualIsEquiv}\AgdaSpace{}%
\AgdaSymbol{\{}\AgdaArgument{x}\AgdaSpace{}%
\AgdaSymbol{=}\AgdaSpace{}%
\AgdaBound{p}\AgdaSymbol{\}}\<%
\\
\>[1]\AgdaFunction{sound}\AgdaSpace{}%
\AgdaSymbol{(}\AgdaInductiveConstructor{symmetric}%
\>[20]\AgdaSymbol{\{}\AgdaArgument{p}\AgdaSpace{}%
\AgdaSymbol{=}\AgdaSpace{}%
\AgdaBound{p}\AgdaSymbol{\}\{}\AgdaBound{q}\AgdaSymbol{\}}%
\>[35]\AgdaBound{Epq}%
\>[44]\AgdaSymbol{)}\AgdaSpace{}%
\AgdaSymbol{=}\AgdaSpace{}%
\AgdaField{symᵉ}%
\>[56]\AgdaFunction{EqualIsEquiv}\AgdaSpace{}%
\AgdaSymbol{\{}\AgdaArgument{x}\AgdaSpace{}%
\AgdaSymbol{=}\AgdaSpace{}%
\AgdaBound{p}\AgdaSymbol{\}\{}\AgdaBound{q}\AgdaSymbol{\}}%
\>[84]\AgdaSymbol{(}\AgdaFunction{sound}\AgdaSpace{}%
\AgdaBound{Epq}\AgdaSymbol{)}\<%
\\
\>[1]\AgdaFunction{sound}\AgdaSpace{}%
\AgdaSymbol{(}\AgdaInductiveConstructor{transitive}%
\>[20]\AgdaSymbol{\{}\AgdaArgument{p}\AgdaSpace{}%
\AgdaSymbol{=}\AgdaSpace{}%
\AgdaBound{p}\AgdaSymbol{\}\{}\AgdaBound{q}\AgdaSymbol{\}\{}\AgdaBound{r}\AgdaSymbol{\}}%
\>[35]\AgdaBound{Epq}\AgdaSpace{}%
\AgdaBound{Eqr}%
\>[44]\AgdaSymbol{)}\AgdaSpace{}%
\AgdaSymbol{=}\AgdaSpace{}%
\AgdaField{transᵉ}%
\>[56]\AgdaFunction{EqualIsEquiv}\AgdaSpace{}%
\AgdaSymbol{\{}\AgdaArgument{i}\AgdaSpace{}%
\AgdaSymbol{=}\AgdaSpace{}%
\AgdaBound{p}\AgdaSymbol{\}\{}\AgdaBound{q}\AgdaSymbol{\}\{}\AgdaBound{r}\AgdaSymbol{\}}%
\>[84]\AgdaSymbol{(}\AgdaFunction{sound}\AgdaSpace{}%
\AgdaBound{Epq}\AgdaSymbol{)(}\AgdaFunction{sound}\AgdaSpace{}%
\AgdaBound{Eqr}\AgdaSymbol{)}\<%
\end{code}
\fi

\paragraph*{The Closure Operators H, S, P and V}
Fix a signature \ab{𝑆}, let \ab{𝒦} be a class of \ab{𝑆}-algebras, and define
\begin{itemize}
\item \af H \ab{𝒦} = algebras isomorphic to homomorphic images of members of \ab{𝒦};
\item \af S \ab{𝒦} = algebras isomorphic to subalgebras of a members of \ab{𝒦};
\item \af P \ab{𝒦} = algebras isomorphic to products of members of \ab{𝒦}.
\end{itemize}
\ifshort\else
A straight-forward verification confirms that
\fi
\af H, \af S, and \af P are \emph{closure operators} (expansive, monotone, and
idempotent).  A class \ab{𝒦} of \ab{𝑆}-algebras is said to be \emph{closed under
the taking of homomorphic images} provided \af H \ab{𝒦} \aof{⊆} \ab{𝒦}. Similarly, \ab{𝒦} is
\emph{closed under the taking of subalgebras} (resp., \emph{arbitrary products}) provided
\af S \ab{𝒦} \aof{⊆} \ab{𝒦} (resp., \af P \ab{𝒦} \aof{⊆} \ab{𝒦}). The operators \af H, \af
S, and \af P can be composed with one another repeatedly, forming yet more closure
operators.

% An algebra is a homomorphic image (resp., subalgebra; resp., product) of every algebra to which it is isomorphic. Thus, the class \af H \ab{𝒦} (resp., \af S \ab{𝒦}; resp., \af P \ab{𝒦}) is closed under isomorphism.

A \emph{variety} is a class of \ab{𝑆}-algebras that is closed under the taking of
homomorphic images, subalgebras, and arbitrary products.  To represent varieties
we define types for the closure operators \af H, \af S, and \af P that are composable; we
then define a type \af V which represents closure under all three of these operators.
Thus, if \ab{𝒦} is a class of \ab{𝑆}-algebras, then
\af V \ab{𝒦} := \af H (\af S (\af P \ab{𝒦})), and \ab{𝒦} is a variety iff \af V \ab{𝒦} \aof{⊆} \ab{𝒦}.
\ifshort\else

We now define the type \af H to represent classes of algebras that include all homomorphic images of algebras in the class---i.e., classes that are closed under the taking of homomorphic images---the type \af S to represent classes of algebras that closed under the taking of subalgebras, and the type \af P to represent classes of algebras closed under the taking of arbitrary products.

\begin{code}%
\>[0]\<%
\\
\>[0]\AgdaKeyword{module}\AgdaSpace{}%
\AgdaModule{\AgdaUnderscore{}}\AgdaSpace{}%
\AgdaSymbol{\{}\AgdaBound{α}\AgdaSpace{}%
\AgdaBound{ρᵃ}\AgdaSpace{}%
\AgdaBound{β}\AgdaSpace{}%
\AgdaBound{ρᵇ}\AgdaSpace{}%
\AgdaSymbol{:}\AgdaSpace{}%
\AgdaPostulate{Level}\AgdaSymbol{\}}\AgdaSpace{}%
\AgdaKeyword{where}\<%
\end{code}
\fi
\begin{code}%
\>[0]\<%
\\
\>[0][@{}l@{\AgdaIndent{1}}]%
\>[1]\AgdaKeyword{private}\AgdaSpace{}%
\AgdaFunction{a}\AgdaSpace{}%
\AgdaSymbol{=}\AgdaSpace{}%
\AgdaBound{α}\AgdaSpace{}%
\AgdaOperator{\AgdaPrimitive{⊔}}\AgdaSpace{}%
\AgdaBound{ρᵃ}\<%
\\
\>[1]\AgdaFunction{H}\AgdaSpace{}%
\AgdaSymbol{:}\AgdaSpace{}%
\AgdaSymbol{∀}\AgdaSpace{}%
\AgdaBound{ℓ}\AgdaSpace{}%
\AgdaSymbol{→}\AgdaSpace{}%
\AgdaFunction{Pred}\AgdaSymbol{(}\AgdaRecord{Algebra}\AgdaSpace{}%
\AgdaBound{α}\AgdaSpace{}%
\AgdaBound{ρᵃ}\AgdaSymbol{)}\AgdaSpace{}%
\AgdaSymbol{(}\AgdaFunction{a}\AgdaSpace{}%
\AgdaOperator{\AgdaPrimitive{⊔}}\AgdaSpace{}%
\AgdaFunction{ov}\AgdaSpace{}%
\AgdaBound{ℓ}\AgdaSymbol{)}\AgdaSpace{}%
\AgdaSymbol{→}\AgdaSpace{}%
\AgdaFunction{Pred}\AgdaSymbol{(}\AgdaRecord{Algebra}\AgdaSpace{}%
\AgdaBound{β}\AgdaSpace{}%
\AgdaBound{ρᵇ}\AgdaSymbol{)}\AgdaSpace{}%
\AgdaSymbol{\AgdaUnderscore{}}\<%
\\
\>[1]\AgdaFunction{H}\AgdaSpace{}%
\AgdaSymbol{\AgdaUnderscore{}}\AgdaSpace{}%
\AgdaBound{𝒦}\AgdaSpace{}%
\AgdaBound{𝑩}\AgdaSpace{}%
\AgdaSymbol{=}\AgdaSpace{}%
\AgdaFunction{Σ[}\AgdaSpace{}%
\AgdaBound{𝑨}\AgdaSpace{}%
\AgdaFunction{∈}\AgdaSpace{}%
\AgdaRecord{Algebra}\AgdaSpace{}%
\AgdaBound{α}\AgdaSpace{}%
\AgdaBound{ρᵃ}\AgdaSpace{}%
\AgdaFunction{]}\AgdaSpace{}%
\AgdaBound{𝑨}\AgdaSpace{}%
\AgdaOperator{\AgdaFunction{∈}}\AgdaSpace{}%
\AgdaBound{𝒦}\AgdaSpace{}%
\AgdaOperator{\AgdaFunction{×}}\AgdaSpace{}%
\AgdaBound{𝑩}\AgdaSpace{}%
\AgdaOperator{\AgdaFunction{IsHomImageOf}}\AgdaSpace{}%
\AgdaBound{𝑨}\<%
\\
\\[\AgdaEmptyExtraSkip]%
\>[1]\AgdaFunction{S}\AgdaSpace{}%
\AgdaSymbol{:}\AgdaSpace{}%
\AgdaSymbol{∀}\AgdaSpace{}%
\AgdaBound{ℓ}\AgdaSpace{}%
\AgdaSymbol{→}\AgdaSpace{}%
\AgdaFunction{Pred}\AgdaSymbol{(}\AgdaRecord{Algebra}\AgdaSpace{}%
\AgdaBound{α}\AgdaSpace{}%
\AgdaBound{ρᵃ}\AgdaSymbol{)}\AgdaSpace{}%
\AgdaSymbol{(}\AgdaFunction{a}\AgdaSpace{}%
\AgdaOperator{\AgdaPrimitive{⊔}}\AgdaSpace{}%
\AgdaFunction{ov}\AgdaSpace{}%
\AgdaBound{ℓ}\AgdaSymbol{)}\AgdaSpace{}%
\AgdaSymbol{→}\AgdaSpace{}%
\AgdaFunction{Pred}\AgdaSymbol{(}\AgdaRecord{Algebra}\AgdaSpace{}%
\AgdaBound{β}\AgdaSpace{}%
\AgdaBound{ρᵇ}\AgdaSymbol{)}\AgdaSpace{}%
\AgdaSymbol{\AgdaUnderscore{}}\<%
\\
\>[1]\AgdaFunction{S}\AgdaSpace{}%
\AgdaSymbol{\AgdaUnderscore{}}\AgdaSpace{}%
\AgdaBound{𝒦}\AgdaSpace{}%
\AgdaBound{𝑩}\AgdaSpace{}%
\AgdaSymbol{=}\AgdaSpace{}%
\AgdaFunction{Σ[}\AgdaSpace{}%
\AgdaBound{𝑨}\AgdaSpace{}%
\AgdaFunction{∈}\AgdaSpace{}%
\AgdaRecord{Algebra}\AgdaSpace{}%
\AgdaBound{α}\AgdaSpace{}%
\AgdaBound{ρᵃ}\AgdaSpace{}%
\AgdaFunction{]}\AgdaSpace{}%
\AgdaBound{𝑨}\AgdaSpace{}%
\AgdaOperator{\AgdaFunction{∈}}\AgdaSpace{}%
\AgdaBound{𝒦}\AgdaSpace{}%
\AgdaOperator{\AgdaFunction{×}}\AgdaSpace{}%
\AgdaBound{𝑩}\AgdaSpace{}%
\AgdaOperator{\AgdaFunction{≤}}\AgdaSpace{}%
\AgdaBound{𝑨}\<%
\\
\\[\AgdaEmptyExtraSkip]%
\>[1]\AgdaFunction{P}\AgdaSpace{}%
\AgdaSymbol{:}\AgdaSpace{}%
\AgdaSymbol{∀}\AgdaSpace{}%
\AgdaBound{ℓ}\AgdaSpace{}%
\AgdaBound{ι}\AgdaSpace{}%
\AgdaSymbol{→}\AgdaSpace{}%
\AgdaFunction{Pred}\AgdaSymbol{(}\AgdaRecord{Algebra}\AgdaSpace{}%
\AgdaBound{α}\AgdaSpace{}%
\AgdaBound{ρᵃ}\AgdaSymbol{)}\AgdaSpace{}%
\AgdaSymbol{(}\AgdaFunction{a}\AgdaSpace{}%
\AgdaOperator{\AgdaPrimitive{⊔}}\AgdaSpace{}%
\AgdaFunction{ov}\AgdaSpace{}%
\AgdaBound{ℓ}\AgdaSymbol{)}\AgdaSpace{}%
\AgdaSymbol{→}\AgdaSpace{}%
\AgdaFunction{Pred}\AgdaSymbol{(}\AgdaRecord{Algebra}\AgdaSpace{}%
\AgdaBound{β}\AgdaSpace{}%
\AgdaBound{ρᵇ}\AgdaSymbol{)}\AgdaSpace{}%
\AgdaSymbol{\AgdaUnderscore{}}\<%
\\
\>[1]\AgdaFunction{P}\AgdaSpace{}%
\AgdaSymbol{\AgdaUnderscore{}}\AgdaSpace{}%
\AgdaBound{ι}\AgdaSpace{}%
\AgdaBound{𝒦}\AgdaSpace{}%
\AgdaBound{𝑩}\AgdaSpace{}%
\AgdaSymbol{=}\AgdaSpace{}%
\AgdaFunction{Σ[}\AgdaSpace{}%
\AgdaBound{I}\AgdaSpace{}%
\AgdaFunction{∈}\AgdaSpace{}%
\AgdaPrimitive{Type}\AgdaSpace{}%
\AgdaBound{ι}\AgdaSpace{}%
\AgdaFunction{]}\AgdaSpace{}%
\AgdaSymbol{(}\AgdaFunction{Σ[}\AgdaSpace{}%
\AgdaBound{𝒜}\AgdaSpace{}%
\AgdaFunction{∈}\AgdaSpace{}%
\AgdaSymbol{(}\AgdaBound{I}\AgdaSpace{}%
\AgdaSymbol{→}\AgdaSpace{}%
\AgdaRecord{Algebra}\AgdaSpace{}%
\AgdaBound{α}\AgdaSpace{}%
\AgdaBound{ρᵃ}\AgdaSymbol{)}\AgdaSpace{}%
\AgdaFunction{]}\AgdaSpace{}%
\AgdaSymbol{(∀}\AgdaSpace{}%
\AgdaBound{i}\AgdaSpace{}%
\AgdaSymbol{→}\AgdaSpace{}%
\AgdaBound{𝒜}\AgdaSpace{}%
\AgdaBound{i}\AgdaSpace{}%
\AgdaOperator{\AgdaFunction{∈}}\AgdaSpace{}%
\AgdaBound{𝒦}\AgdaSymbol{)}\AgdaSpace{}%
\AgdaOperator{\AgdaFunction{×}}\AgdaSpace{}%
\AgdaSymbol{(}\AgdaBound{𝑩}\AgdaSpace{}%
\AgdaOperator{\AgdaRecord{≅}}\AgdaSpace{}%
\AgdaFunction{⨅}\AgdaSpace{}%
\AgdaBound{𝒜}\AgdaSymbol{))}\<%
\\
\>[0]\<%
\end{code}
Finally, we define the \defn{varietal closure} of a class \ab{𝒦} to be the class \af{V}
\ab{𝒦} := \af{H} (\af{S} (\af{P} \ab{𝒦})).
\ifshort\else
\begin{code}%
\>[0]\<%
\\
\>[0]\AgdaKeyword{module}\AgdaSpace{}%
\AgdaModule{\AgdaUnderscore{}}%
\>[10]\AgdaSymbol{\{}\AgdaBound{α}\AgdaSpace{}%
\AgdaBound{ρᵃ}\AgdaSpace{}%
\AgdaBound{β}\AgdaSpace{}%
\AgdaBound{ρᵇ}\AgdaSpace{}%
\AgdaBound{γ}\AgdaSpace{}%
\AgdaBound{ρᶜ}\AgdaSpace{}%
\AgdaBound{δ}\AgdaSpace{}%
\AgdaBound{ρᵈ}\AgdaSpace{}%
\AgdaSymbol{:}\AgdaSpace{}%
\AgdaPostulate{Level}\AgdaSymbol{\}}\AgdaSpace{}%
\AgdaKeyword{where}\<%
\end{code}
\fi
\begin{code}%
\>[0]\<%
\\
\>[0][@{}l@{\AgdaIndent{1}}]%
\>[1]\AgdaKeyword{private}\AgdaSpace{}%
\AgdaFunction{a}\AgdaSpace{}%
\AgdaSymbol{=}\AgdaSpace{}%
\AgdaBound{α}\AgdaSpace{}%
\AgdaOperator{\AgdaPrimitive{⊔}}\AgdaSpace{}%
\AgdaBound{ρᵃ}\AgdaSpace{}%
\AgdaSymbol{;}\AgdaSpace{}%
\AgdaFunction{b}\AgdaSpace{}%
\AgdaSymbol{=}\AgdaSpace{}%
\AgdaBound{β}\AgdaSpace{}%
\AgdaOperator{\AgdaPrimitive{⊔}}\AgdaSpace{}%
\AgdaBound{ρᵇ}\<%
\\
\>[1]\AgdaFunction{V}\AgdaSpace{}%
\AgdaSymbol{:}\AgdaSpace{}%
\AgdaSymbol{∀}\AgdaSpace{}%
\AgdaBound{ℓ}\AgdaSpace{}%
\AgdaBound{ι}\AgdaSpace{}%
\AgdaSymbol{→}\AgdaSpace{}%
\AgdaFunction{Pred}\AgdaSymbol{(}\AgdaRecord{Algebra}\AgdaSpace{}%
\AgdaBound{α}\AgdaSpace{}%
\AgdaBound{ρᵃ}\AgdaSymbol{)}\AgdaSpace{}%
\AgdaSymbol{(}\AgdaFunction{a}\AgdaSpace{}%
\AgdaOperator{\AgdaPrimitive{⊔}}\AgdaSpace{}%
\AgdaFunction{ov}\AgdaSpace{}%
\AgdaBound{ℓ}\AgdaSymbol{)}\AgdaSpace{}%
\AgdaSymbol{→}%
\>[46]\AgdaFunction{Pred}\AgdaSymbol{(}\AgdaRecord{Algebra}\AgdaSpace{}%
\AgdaBound{δ}\AgdaSpace{}%
\AgdaBound{ρᵈ}\AgdaSymbol{)}\AgdaSpace{}%
\AgdaSymbol{\AgdaUnderscore{}}\<%
\\
\>[1]\AgdaFunction{V}\AgdaSpace{}%
\AgdaBound{ℓ}\AgdaSpace{}%
\AgdaBound{ι}\AgdaSpace{}%
\AgdaBound{𝒦}\AgdaSpace{}%
\AgdaSymbol{=}\AgdaSpace{}%
\AgdaFunction{H}\AgdaSymbol{\{}\AgdaBound{γ}\AgdaSymbol{\}\{}\AgdaBound{ρᶜ}\AgdaSymbol{\}\{}\AgdaBound{δ}\AgdaSymbol{\}\{}\AgdaBound{ρᵈ}\AgdaSymbol{\}}\AgdaSpace{}%
\AgdaSymbol{(}\AgdaFunction{a}\AgdaSpace{}%
\AgdaOperator{\AgdaPrimitive{⊔}}\AgdaSpace{}%
\AgdaFunction{b}\AgdaSpace{}%
\AgdaOperator{\AgdaPrimitive{⊔}}\AgdaSpace{}%
\AgdaBound{ℓ}\AgdaSpace{}%
\AgdaOperator{\AgdaPrimitive{⊔}}\AgdaSpace{}%
\AgdaBound{ι}\AgdaSymbol{)}\AgdaSpace{}%
\AgdaSymbol{(}\AgdaFunction{S}\AgdaSymbol{\{}\AgdaBound{β}\AgdaSymbol{\}\{}\AgdaBound{ρᵇ}\AgdaSymbol{\}}\AgdaSpace{}%
\AgdaSymbol{(}\AgdaFunction{a}\AgdaSpace{}%
\AgdaOperator{\AgdaPrimitive{⊔}}\AgdaSpace{}%
\AgdaBound{ℓ}\AgdaSpace{}%
\AgdaOperator{\AgdaPrimitive{⊔}}\AgdaSpace{}%
\AgdaBound{ι}\AgdaSymbol{)}\AgdaSpace{}%
\AgdaSymbol{(}\AgdaFunction{P}\AgdaSpace{}%
\AgdaBound{ℓ}\AgdaSpace{}%
\AgdaBound{ι}\AgdaSpace{}%
\AgdaBound{𝒦}\AgdaSymbol{))}\<%
\\
\>[0]\<%
\end{code}

An important property of the binary relation \aof{⊧} is \emph{algebraic invariance} (i.e.,
invariance under isomorphism).
\ifshort
Here is the formal statement of this property, without proof.
\else
We formalize this property as follows.

\begin{code}%
\>[0]\<%
\\
\>[0]\AgdaKeyword{module}\AgdaSpace{}%
\AgdaModule{\AgdaUnderscore{}}\AgdaSpace{}%
\AgdaSymbol{\{}\AgdaBound{X}\AgdaSpace{}%
\AgdaSymbol{:}\AgdaSpace{}%
\AgdaPrimitive{Type}\AgdaSpace{}%
\AgdaGeneralizable{χ}\AgdaSymbol{\}\{}\AgdaBound{𝑨}\AgdaSpace{}%
\AgdaSymbol{:}\AgdaSpace{}%
\AgdaRecord{Algebra}\AgdaSpace{}%
\AgdaGeneralizable{α}\AgdaSpace{}%
\AgdaGeneralizable{ρᵃ}\AgdaSymbol{\}(}\AgdaBound{𝑩}\AgdaSpace{}%
\AgdaSymbol{:}\AgdaSpace{}%
\AgdaRecord{Algebra}\AgdaSpace{}%
\AgdaGeneralizable{β}\AgdaSpace{}%
\AgdaGeneralizable{ρᵇ}\AgdaSymbol{)(}\AgdaBound{p}\AgdaSpace{}%
\AgdaBound{q}\AgdaSpace{}%
\AgdaSymbol{:}\AgdaSpace{}%
\AgdaDatatype{Term}\AgdaSpace{}%
\AgdaBound{X}\AgdaSymbol{)}\AgdaSpace{}%
\AgdaKeyword{where}\<%
\end{code}
\fi
\begin{code}%
\>[0]\<%
\\
\>[0][@{}l@{\AgdaIndent{1}}]%
\>[1]\AgdaFunction{⊧-I-invar}\AgdaSpace{}%
\AgdaSymbol{:}\AgdaSpace{}%
\AgdaBound{𝑨}\AgdaSpace{}%
\AgdaOperator{\AgdaFunction{⊧}}\AgdaSpace{}%
\AgdaBound{p}\AgdaSpace{}%
\AgdaOperator{\AgdaFunction{≈}}\AgdaSpace{}%
\AgdaBound{q}%
\>[24]\AgdaSymbol{→}%
\>[27]\AgdaBound{𝑨}\AgdaSpace{}%
\AgdaOperator{\AgdaRecord{≅}}\AgdaSpace{}%
\AgdaBound{𝑩}%
\>[34]\AgdaSymbol{→}%
\>[37]\AgdaBound{𝑩}\AgdaSpace{}%
\AgdaOperator{\AgdaFunction{⊧}}\AgdaSpace{}%
\AgdaBound{p}\AgdaSpace{}%
\AgdaOperator{\AgdaFunction{≈}}\AgdaSpace{}%
\AgdaBound{q}\<%
\\
\>[0]\<%
\end{code}
\ifshort\else
\begin{code}%
\>[0][@{}l@{\AgdaIndent{1}}]%
\>[1]\AgdaFunction{⊧-I-invar}\AgdaSpace{}%
\AgdaBound{Apq}\AgdaSpace{}%
\AgdaSymbol{(}\AgdaInductiveConstructor{mkiso}\AgdaSpace{}%
\AgdaBound{fh}\AgdaSpace{}%
\AgdaBound{gh}\AgdaSpace{}%
\AgdaBound{f∼g}\AgdaSpace{}%
\AgdaBound{g∼f}\AgdaSymbol{)}\AgdaSpace{}%
\AgdaBound{ρ}\AgdaSpace{}%
\AgdaSymbol{=}\<%
\\
\>[1][@{}l@{\AgdaIndent{0}}]%
\>[2]\AgdaOperator{\AgdaFunction{begin}}\<%
\\
\>[2][@{}l@{\AgdaIndent{0}}]%
\>[6]\AgdaOperator{\AgdaFunction{⟦}}\AgdaSpace{}%
\AgdaBound{p}\AgdaSpace{}%
\AgdaOperator{\AgdaFunction{⟧}}%
\>[14]\AgdaOperator{\AgdaField{⟨\$⟩}}%
\>[32]\AgdaBound{ρ}%
\>[37]\AgdaFunction{≈˘⟨}%
\>[42]\AgdaField{cong}\AgdaSpace{}%
\AgdaOperator{\AgdaFunction{⟦}}\AgdaSpace{}%
\AgdaBound{p}\AgdaSpace{}%
\AgdaOperator{\AgdaFunction{⟧}}\AgdaSpace{}%
\AgdaSymbol{(}\AgdaBound{f∼g}\AgdaSpace{}%
\AgdaOperator{\AgdaFunction{∘}}\AgdaSpace{}%
\AgdaBound{ρ}\AgdaSymbol{)}%
\>[70]\AgdaFunction{⟩}\<%
\\
\>[6]\AgdaOperator{\AgdaFunction{⟦}}\AgdaSpace{}%
\AgdaBound{p}\AgdaSpace{}%
\AgdaOperator{\AgdaFunction{⟧}}%
\>[14]\AgdaOperator{\AgdaField{⟨\$⟩}}\AgdaSpace{}%
\AgdaSymbol{(}\AgdaFunction{f}%
\>[22]\AgdaOperator{\AgdaFunction{∘}}%
\>[25]\AgdaSymbol{(}\AgdaFunction{g}%
\>[29]\AgdaOperator{\AgdaFunction{∘}}%
\>[32]\AgdaBound{ρ}\AgdaSymbol{))}%
\>[37]\AgdaFunction{≈˘⟨}%
\>[42]\AgdaFunction{comm-hom-term}\AgdaSpace{}%
\AgdaBound{fh}\AgdaSpace{}%
\AgdaBound{p}\AgdaSpace{}%
\AgdaSymbol{(}\AgdaFunction{g}\AgdaSpace{}%
\AgdaOperator{\AgdaFunction{∘}}\AgdaSpace{}%
\AgdaBound{ρ}\AgdaSymbol{)}%
\>[70]\AgdaFunction{⟩}\<%
\\
\>[2][@{}l@{\AgdaIndent{0}}]%
\>[4]\AgdaFunction{f}\AgdaSymbol{(}\AgdaOperator{\AgdaFunction{⟦}}\AgdaSpace{}%
\AgdaBound{p}\AgdaSpace{}%
\AgdaOperator{\AgdaFunction{⟧ᴬ}}%
\>[14]\AgdaOperator{\AgdaField{⟨\$⟩}}%
\>[25]\AgdaSymbol{(}\AgdaFunction{g}%
\>[29]\AgdaOperator{\AgdaFunction{∘}}%
\>[32]\AgdaBound{ρ}\AgdaSymbol{))}%
\>[37]\AgdaFunction{≈⟨}%
\>[42]\AgdaField{cong}\AgdaSpace{}%
\AgdaOperator{\AgdaFunction{∣}}\AgdaSpace{}%
\AgdaBound{fh}\AgdaSpace{}%
\AgdaOperator{\AgdaFunction{∣}}\AgdaSpace{}%
\AgdaSymbol{(}\AgdaBound{Apq}\AgdaSpace{}%
\AgdaSymbol{(}\AgdaFunction{g}\AgdaSpace{}%
\AgdaOperator{\AgdaFunction{∘}}\AgdaSpace{}%
\AgdaBound{ρ}\AgdaSymbol{))}%
\>[70]\AgdaFunction{⟩}\<%
\\
\>[4]\AgdaFunction{f}\AgdaSymbol{(}\AgdaOperator{\AgdaFunction{⟦}}\AgdaSpace{}%
\AgdaBound{q}\AgdaSpace{}%
\AgdaOperator{\AgdaFunction{⟧ᴬ}}%
\>[14]\AgdaOperator{\AgdaField{⟨\$⟩}}%
\>[25]\AgdaSymbol{(}\AgdaFunction{g}%
\>[29]\AgdaOperator{\AgdaFunction{∘}}%
\>[32]\AgdaBound{ρ}\AgdaSymbol{))}%
\>[37]\AgdaFunction{≈⟨}%
\>[42]\AgdaFunction{comm-hom-term}\AgdaSpace{}%
\AgdaBound{fh}\AgdaSpace{}%
\AgdaBound{q}\AgdaSpace{}%
\AgdaSymbol{(}\AgdaFunction{g}\AgdaSpace{}%
\AgdaOperator{\AgdaFunction{∘}}\AgdaSpace{}%
\AgdaBound{ρ}\AgdaSymbol{)}%
\>[70]\AgdaFunction{⟩}\<%
\\
\>[4][@{}l@{\AgdaIndent{0}}]%
\>[6]\AgdaOperator{\AgdaFunction{⟦}}\AgdaSpace{}%
\AgdaBound{q}\AgdaSpace{}%
\AgdaOperator{\AgdaFunction{⟧}}%
\>[14]\AgdaOperator{\AgdaField{⟨\$⟩}}\AgdaSpace{}%
\AgdaSymbol{(}\AgdaFunction{f}%
\>[22]\AgdaOperator{\AgdaFunction{∘}}%
\>[25]\AgdaSymbol{(}\AgdaFunction{g}%
\>[29]\AgdaOperator{\AgdaFunction{∘}}%
\>[32]\AgdaBound{ρ}\AgdaSymbol{))}%
\>[37]\AgdaFunction{≈⟨}%
\>[42]\AgdaField{cong}\AgdaSpace{}%
\AgdaOperator{\AgdaFunction{⟦}}\AgdaSpace{}%
\AgdaBound{q}\AgdaSpace{}%
\AgdaOperator{\AgdaFunction{⟧}}\AgdaSpace{}%
\AgdaSymbol{(}\AgdaBound{f∼g}\AgdaSpace{}%
\AgdaOperator{\AgdaFunction{∘}}\AgdaSpace{}%
\AgdaBound{ρ}\AgdaSymbol{)}%
\>[70]\AgdaFunction{⟩}\<%
\\
\>[6]\AgdaOperator{\AgdaFunction{⟦}}\AgdaSpace{}%
\AgdaBound{q}\AgdaSpace{}%
\AgdaOperator{\AgdaFunction{⟧}}%
\>[14]\AgdaOperator{\AgdaField{⟨\$⟩}}%
\>[32]\AgdaBound{ρ}%
\>[37]\AgdaOperator{\AgdaFunction{∎}}\<%
\\
\>[2]\AgdaKeyword{where}\<%
\\
\>[2]\AgdaKeyword{private}\AgdaSpace{}%
\AgdaFunction{f}\AgdaSpace{}%
\AgdaSymbol{=}\AgdaSpace{}%
\AgdaOperator{\AgdaField{\AgdaUnderscore{}⟨\$⟩\AgdaUnderscore{}}}\AgdaSpace{}%
\AgdaOperator{\AgdaFunction{∣}}\AgdaSpace{}%
\AgdaBound{fh}\AgdaSpace{}%
\AgdaOperator{\AgdaFunction{∣}}\AgdaSpace{}%
\AgdaSymbol{;}\AgdaSpace{}%
\AgdaFunction{g}\AgdaSpace{}%
\AgdaSymbol{=}\AgdaSpace{}%
\AgdaOperator{\AgdaField{\AgdaUnderscore{}⟨\$⟩\AgdaUnderscore{}}}\AgdaSpace{}%
\AgdaOperator{\AgdaFunction{∣}}\AgdaSpace{}%
\AgdaBound{gh}\AgdaSpace{}%
\AgdaOperator{\AgdaFunction{∣}}\<%
\\
\>[2]\AgdaKeyword{open}\AgdaSpace{}%
\AgdaModule{Environment}\AgdaSpace{}%
\AgdaBound{𝑨}%
\>[25]\AgdaKeyword{using}\AgdaSpace{}%
\AgdaSymbol{()}\AgdaSpace{}%
\AgdaKeyword{renaming}\AgdaSpace{}%
\AgdaSymbol{(}\AgdaSpace{}%
\AgdaOperator{\AgdaFunction{⟦\AgdaUnderscore{}⟧}}\AgdaSpace{}%
\AgdaSymbol{to}\AgdaSpace{}%
\AgdaOperator{\AgdaFunction{⟦\AgdaUnderscore{}⟧ᴬ}}\AgdaSpace{}%
\AgdaSymbol{)}\<%
\\
\>[2]\AgdaKeyword{open}\AgdaSpace{}%
\AgdaModule{Environment}\AgdaSpace{}%
\AgdaBound{𝑩}%
\>[25]\AgdaKeyword{using}\AgdaSpace{}%
\AgdaSymbol{(}\AgdaSpace{}%
\AgdaOperator{\AgdaFunction{⟦\AgdaUnderscore{}⟧}}\AgdaSpace{}%
\AgdaSymbol{)}\<%
\\
\>[2]\AgdaKeyword{open}\AgdaSpace{}%
\AgdaModule{SetoidReasoning}\AgdaSpace{}%
\AgdaOperator{\AgdaFunction{𝔻[}}\AgdaSpace{}%
\AgdaBound{𝑩}\AgdaSpace{}%
\AgdaOperator{\AgdaFunction{]}}\<%
\\
\>[0]\<%
\end{code}
\fi
Identities modeled by an algebra \ab{𝑨} are also modeled by every homomorphic image of
\ab{𝑨} and by every subalgebra of \ab{𝑨}.
\ifshort
We refer to these facts as \af{⊧-H-invar} and \af{⊧-S-invar}, but omit their formal
statements and proofs, which are analogous to those of \af{⊧-I-invar}.
\else
These facts are formalized in Agda as follows.

\begin{code}%
\>[0]\<%
\\
\>[0]\AgdaKeyword{module}\AgdaSpace{}%
\AgdaModule{\AgdaUnderscore{}}\AgdaSpace{}%
\AgdaSymbol{\{}\AgdaBound{X}\AgdaSpace{}%
\AgdaSymbol{:}\AgdaSpace{}%
\AgdaPrimitive{Type}\AgdaSpace{}%
\AgdaGeneralizable{χ}\AgdaSymbol{\}\{}\AgdaBound{𝑨}\AgdaSpace{}%
\AgdaSymbol{:}\AgdaSpace{}%
\AgdaRecord{Algebra}\AgdaSpace{}%
\AgdaGeneralizable{α}\AgdaSpace{}%
\AgdaGeneralizable{ρᵃ}\AgdaSymbol{\}\{}\AgdaBound{𝑩}\AgdaSpace{}%
\AgdaSymbol{:}\AgdaSpace{}%
\AgdaRecord{Algebra}\AgdaSpace{}%
\AgdaGeneralizable{β}\AgdaSpace{}%
\AgdaGeneralizable{ρᵇ}\AgdaSymbol{\}\{}\AgdaBound{p}\AgdaSpace{}%
\AgdaBound{q}\AgdaSpace{}%
\AgdaSymbol{:}\AgdaSpace{}%
\AgdaDatatype{Term}\AgdaSpace{}%
\AgdaBound{X}\AgdaSymbol{\}}\AgdaSpace{}%
\AgdaKeyword{where}\<%
\\
\\[\AgdaEmptyExtraSkip]%
\>[0][@{}l@{\AgdaIndent{0}}]%
\>[1]\AgdaFunction{⊧-H-invar}\AgdaSpace{}%
\AgdaSymbol{:}\AgdaSpace{}%
\AgdaBound{𝑨}\AgdaSpace{}%
\AgdaOperator{\AgdaFunction{⊧}}\AgdaSpace{}%
\AgdaBound{p}\AgdaSpace{}%
\AgdaOperator{\AgdaFunction{≈}}\AgdaSpace{}%
\AgdaBound{q}\AgdaSpace{}%
\AgdaSymbol{→}\AgdaSpace{}%
\AgdaBound{𝑩}\AgdaSpace{}%
\AgdaOperator{\AgdaFunction{IsHomImageOf}}\AgdaSpace{}%
\AgdaBound{𝑨}\AgdaSpace{}%
\AgdaSymbol{→}\AgdaSpace{}%
\AgdaBound{𝑩}\AgdaSpace{}%
\AgdaOperator{\AgdaFunction{⊧}}\AgdaSpace{}%
\AgdaBound{p}\AgdaSpace{}%
\AgdaOperator{\AgdaFunction{≈}}\AgdaSpace{}%
\AgdaBound{q}\<%
\\
\>[1]\AgdaFunction{⊧-H-invar}\AgdaSpace{}%
\AgdaBound{Apq}\AgdaSpace{}%
\AgdaSymbol{(}\AgdaBound{φh}\AgdaSpace{}%
\AgdaOperator{\AgdaInductiveConstructor{,}}\AgdaSpace{}%
\AgdaBound{φE}\AgdaSymbol{)}\AgdaSpace{}%
\AgdaBound{ρ}\AgdaSpace{}%
\AgdaSymbol{=}\<%
\\
\>[1][@{}l@{\AgdaIndent{0}}]%
\>[2]\AgdaOperator{\AgdaFunction{begin}}\<%
\\
\>[2][@{}l@{\AgdaIndent{0}}]%
\>[7]\AgdaOperator{\AgdaFunction{⟦}}\AgdaSpace{}%
\AgdaBound{p}\AgdaSpace{}%
\AgdaOperator{\AgdaFunction{⟧}}%
\>[15]\AgdaOperator{\AgdaField{⟨\$⟩}}%
\>[33]\AgdaBound{ρ}%
\>[38]\AgdaFunction{≈˘⟨}%
\>[43]\AgdaField{cong}\AgdaSpace{}%
\AgdaOperator{\AgdaFunction{⟦}}\AgdaSpace{}%
\AgdaBound{p}\AgdaSpace{}%
\AgdaOperator{\AgdaFunction{⟧}}\AgdaSymbol{(λ}\AgdaSpace{}%
\AgdaBound{\AgdaUnderscore{}}\AgdaSpace{}%
\AgdaSymbol{→}\AgdaSpace{}%
\AgdaFunction{InvIsInverseʳ}\AgdaSpace{}%
\AgdaBound{φE}\AgdaSymbol{)}%
\>[79]\AgdaFunction{⟩}\<%
\\
\>[7]\AgdaOperator{\AgdaFunction{⟦}}\AgdaSpace{}%
\AgdaBound{p}\AgdaSpace{}%
\AgdaOperator{\AgdaFunction{⟧}}%
\>[15]\AgdaOperator{\AgdaField{⟨\$⟩}}\AgdaSpace{}%
\AgdaSymbol{(}\AgdaFunction{φ}\AgdaSpace{}%
\AgdaOperator{\AgdaFunction{∘}}%
\>[25]\AgdaFunction{φ⁻¹}%
\>[30]\AgdaOperator{\AgdaFunction{∘}}%
\>[33]\AgdaBound{ρ}\AgdaSymbol{)}%
\>[38]\AgdaFunction{≈˘⟨}%
\>[43]\AgdaFunction{comm-hom-term}\AgdaSpace{}%
\AgdaBound{φh}\AgdaSpace{}%
\AgdaBound{p}\AgdaSpace{}%
\AgdaSymbol{(}\AgdaFunction{φ⁻¹}\AgdaSpace{}%
\AgdaOperator{\AgdaFunction{∘}}\AgdaSpace{}%
\AgdaBound{ρ}\AgdaSymbol{)}%
\>[79]\AgdaFunction{⟩}\<%
\\
\>[2][@{}l@{\AgdaIndent{0}}]%
\>[3]\AgdaFunction{φ}\AgdaSymbol{(}%
\>[7]\AgdaOperator{\AgdaFunction{⟦}}\AgdaSpace{}%
\AgdaBound{p}\AgdaSpace{}%
\AgdaOperator{\AgdaFunction{⟧ᴬ}}%
\>[15]\AgdaOperator{\AgdaField{⟨\$⟩}}\AgdaSpace{}%
\AgdaSymbol{(}%
\>[25]\AgdaFunction{φ⁻¹}%
\>[30]\AgdaOperator{\AgdaFunction{∘}}%
\>[33]\AgdaBound{ρ}\AgdaSymbol{))}%
\>[38]\AgdaFunction{≈⟨}%
\>[43]\AgdaField{cong}\AgdaSpace{}%
\AgdaOperator{\AgdaFunction{∣}}\AgdaSpace{}%
\AgdaBound{φh}\AgdaSpace{}%
\AgdaOperator{\AgdaFunction{∣}}\AgdaSpace{}%
\AgdaSymbol{(}\AgdaBound{Apq}\AgdaSpace{}%
\AgdaSymbol{(}\AgdaFunction{φ⁻¹}\AgdaSpace{}%
\AgdaOperator{\AgdaFunction{∘}}\AgdaSpace{}%
\AgdaBound{ρ}\AgdaSymbol{))}%
\>[79]\AgdaFunction{⟩}\<%
\\
\>[3]\AgdaFunction{φ}\AgdaSymbol{(}%
\>[7]\AgdaOperator{\AgdaFunction{⟦}}\AgdaSpace{}%
\AgdaBound{q}\AgdaSpace{}%
\AgdaOperator{\AgdaFunction{⟧ᴬ}}%
\>[15]\AgdaOperator{\AgdaField{⟨\$⟩}}\AgdaSpace{}%
\AgdaSymbol{(}%
\>[25]\AgdaFunction{φ⁻¹}%
\>[30]\AgdaOperator{\AgdaFunction{∘}}%
\>[33]\AgdaBound{ρ}\AgdaSymbol{))}%
\>[38]\AgdaFunction{≈⟨}%
\>[43]\AgdaFunction{comm-hom-term}\AgdaSpace{}%
\AgdaBound{φh}\AgdaSpace{}%
\AgdaBound{q}\AgdaSpace{}%
\AgdaSymbol{(}\AgdaFunction{φ⁻¹}\AgdaSpace{}%
\AgdaOperator{\AgdaFunction{∘}}\AgdaSpace{}%
\AgdaBound{ρ}\AgdaSymbol{)}%
\>[79]\AgdaFunction{⟩}\<%
\\
\>[7]\AgdaOperator{\AgdaFunction{⟦}}\AgdaSpace{}%
\AgdaBound{q}\AgdaSpace{}%
\AgdaOperator{\AgdaFunction{⟧}}%
\>[15]\AgdaOperator{\AgdaField{⟨\$⟩}}\AgdaSpace{}%
\AgdaSymbol{(}\AgdaFunction{φ}\AgdaSpace{}%
\AgdaOperator{\AgdaFunction{∘}}%
\>[25]\AgdaFunction{φ⁻¹}%
\>[30]\AgdaOperator{\AgdaFunction{∘}}%
\>[33]\AgdaBound{ρ}\AgdaSymbol{)}%
\>[38]\AgdaFunction{≈⟨}%
\>[43]\AgdaField{cong}\AgdaSpace{}%
\AgdaOperator{\AgdaFunction{⟦}}\AgdaSpace{}%
\AgdaBound{q}\AgdaSpace{}%
\AgdaOperator{\AgdaFunction{⟧}}\AgdaSymbol{(λ}\AgdaSpace{}%
\AgdaBound{\AgdaUnderscore{}}\AgdaSpace{}%
\AgdaSymbol{→}\AgdaSpace{}%
\AgdaFunction{InvIsInverseʳ}\AgdaSpace{}%
\AgdaBound{φE}\AgdaSymbol{)}%
\>[79]\AgdaFunction{⟩}\<%
\\
\>[7]\AgdaOperator{\AgdaFunction{⟦}}\AgdaSpace{}%
\AgdaBound{q}\AgdaSpace{}%
\AgdaOperator{\AgdaFunction{⟧}}%
\>[15]\AgdaOperator{\AgdaField{⟨\$⟩}}%
\>[33]\AgdaBound{ρ}%
\>[38]\AgdaOperator{\AgdaFunction{∎}}\<%
\\
\>[2]\AgdaKeyword{where}\<%
\\
\>[2]\AgdaFunction{φ⁻¹}\AgdaSpace{}%
\AgdaSymbol{:}\AgdaSpace{}%
\AgdaOperator{\AgdaFunction{𝕌[}}\AgdaSpace{}%
\AgdaBound{𝑩}\AgdaSpace{}%
\AgdaOperator{\AgdaFunction{]}}\AgdaSpace{}%
\AgdaSymbol{→}\AgdaSpace{}%
\AgdaOperator{\AgdaFunction{𝕌[}}\AgdaSpace{}%
\AgdaBound{𝑨}\AgdaSpace{}%
\AgdaOperator{\AgdaFunction{]}}\<%
\\
\>[2]\AgdaFunction{φ⁻¹}\AgdaSpace{}%
\AgdaSymbol{=}\AgdaSpace{}%
\AgdaFunction{SurjInv}\AgdaSpace{}%
\AgdaOperator{\AgdaFunction{∣}}\AgdaSpace{}%
\AgdaBound{φh}\AgdaSpace{}%
\AgdaOperator{\AgdaFunction{∣}}\AgdaSpace{}%
\AgdaBound{φE}\<%
\\
\>[2]\AgdaKeyword{private}\AgdaSpace{}%
\AgdaFunction{φ}\AgdaSpace{}%
\AgdaSymbol{=}\AgdaSpace{}%
\AgdaSymbol{(}\AgdaOperator{\AgdaField{\AgdaUnderscore{}⟨\$⟩\AgdaUnderscore{}}}\AgdaSpace{}%
\AgdaOperator{\AgdaFunction{∣}}\AgdaSpace{}%
\AgdaBound{φh}\AgdaSpace{}%
\AgdaOperator{\AgdaFunction{∣}}\AgdaSymbol{)}\<%
\\
\>[2]\AgdaKeyword{open}\AgdaSpace{}%
\AgdaModule{Environment}\AgdaSpace{}%
\AgdaBound{𝑨}%
\>[22]\AgdaKeyword{using}\AgdaSpace{}%
\AgdaSymbol{()}\AgdaSpace{}%
\AgdaKeyword{renaming}\AgdaSpace{}%
\AgdaSymbol{(}\AgdaSpace{}%
\AgdaOperator{\AgdaFunction{⟦\AgdaUnderscore{}⟧}}\AgdaSpace{}%
\AgdaSymbol{to}\AgdaSpace{}%
\AgdaOperator{\AgdaFunction{⟦\AgdaUnderscore{}⟧ᴬ}}\AgdaSymbol{)}\<%
\\
\>[2]\AgdaKeyword{open}\AgdaSpace{}%
\AgdaModule{Environment}\AgdaSpace{}%
\AgdaBound{𝑩}%
\>[22]\AgdaKeyword{using}\AgdaSpace{}%
\AgdaSymbol{(}\AgdaSpace{}%
\AgdaOperator{\AgdaFunction{⟦\AgdaUnderscore{}⟧}}\AgdaSpace{}%
\AgdaSymbol{)}\<%
\\
\>[2]\AgdaKeyword{open}\AgdaSpace{}%
\AgdaModule{SetoidReasoning}\AgdaSpace{}%
\AgdaOperator{\AgdaFunction{𝔻[}}\AgdaSpace{}%
\AgdaBound{𝑩}\AgdaSpace{}%
\AgdaOperator{\AgdaFunction{]}}\<%
\\
\\[\AgdaEmptyExtraSkip]%
\>[1]\AgdaFunction{⊧-S-invar}\AgdaSpace{}%
\AgdaSymbol{:}\AgdaSpace{}%
\AgdaBound{𝑨}\AgdaSpace{}%
\AgdaOperator{\AgdaFunction{⊧}}\AgdaSpace{}%
\AgdaBound{p}\AgdaSpace{}%
\AgdaOperator{\AgdaFunction{≈}}\AgdaSpace{}%
\AgdaBound{q}\AgdaSpace{}%
\AgdaSymbol{→}\AgdaSpace{}%
\AgdaBound{𝑩}\AgdaSpace{}%
\AgdaOperator{\AgdaFunction{≤}}\AgdaSpace{}%
\AgdaBound{𝑨}\AgdaSpace{}%
\AgdaSymbol{→}\AgdaSpace{}%
\AgdaBound{𝑩}\AgdaSpace{}%
\AgdaOperator{\AgdaFunction{⊧}}\AgdaSpace{}%
\AgdaBound{p}\AgdaSpace{}%
\AgdaOperator{\AgdaFunction{≈}}\AgdaSpace{}%
\AgdaBound{q}\<%
\\
\>[1]\AgdaFunction{⊧-S-invar}\AgdaSpace{}%
\AgdaBound{Apq}\AgdaSpace{}%
\AgdaBound{B≤A}\AgdaSpace{}%
\AgdaBound{b}\AgdaSpace{}%
\AgdaSymbol{=}\AgdaSpace{}%
\AgdaOperator{\AgdaFunction{∥}}\AgdaSpace{}%
\AgdaBound{B≤A}\AgdaSpace{}%
\AgdaOperator{\AgdaFunction{∥}}\<%
\\
\>[1][@{}l@{\AgdaIndent{0}}]%
\>[2]\AgdaSymbol{(}%
\>[5004I]\AgdaOperator{\AgdaFunction{begin}}\<%
\\
\>[.][@{}l@{}]\<[5004I]%
\>[4]\AgdaFunction{h}\AgdaSpace{}%
\AgdaSymbol{(}%
\>[9]\AgdaOperator{\AgdaFunction{⟦}}\AgdaSpace{}%
\AgdaBound{p}\AgdaSpace{}%
\AgdaOperator{\AgdaFunction{⟧}}%
\>[17]\AgdaOperator{\AgdaField{⟨\$⟩}}%
\>[27]\AgdaBound{b}\AgdaSymbol{)}%
\>[31]\AgdaFunction{≈⟨}%
\>[36]\AgdaFunction{comm-hom-term}\AgdaSpace{}%
\AgdaFunction{hh}\AgdaSpace{}%
\AgdaBound{p}\AgdaSpace{}%
\AgdaBound{b}%
\>[58]\AgdaFunction{⟩}\<%
\\
\>[9]\AgdaOperator{\AgdaFunction{⟦}}\AgdaSpace{}%
\AgdaBound{p}\AgdaSpace{}%
\AgdaOperator{\AgdaFunction{⟧ᴬ}}%
\>[17]\AgdaOperator{\AgdaField{⟨\$⟩}}\AgdaSpace{}%
\AgdaSymbol{(}\AgdaFunction{h}\AgdaSpace{}%
\AgdaOperator{\AgdaFunction{∘}}%
\>[27]\AgdaBound{b}\AgdaSymbol{)}%
\>[31]\AgdaFunction{≈⟨}%
\>[36]\AgdaBound{Apq}\AgdaSpace{}%
\AgdaSymbol{(}\AgdaFunction{h}\AgdaSpace{}%
\AgdaOperator{\AgdaFunction{∘}}\AgdaSpace{}%
\AgdaBound{b}\AgdaSymbol{)}%
\>[58]\AgdaFunction{⟩}\<%
\\
\>[9]\AgdaOperator{\AgdaFunction{⟦}}\AgdaSpace{}%
\AgdaBound{q}\AgdaSpace{}%
\AgdaOperator{\AgdaFunction{⟧ᴬ}}%
\>[17]\AgdaOperator{\AgdaField{⟨\$⟩}}\AgdaSpace{}%
\AgdaSymbol{(}\AgdaFunction{h}\AgdaSpace{}%
\AgdaOperator{\AgdaFunction{∘}}%
\>[27]\AgdaBound{b}\AgdaSymbol{)}%
\>[31]\AgdaFunction{≈˘⟨}%
\>[36]\AgdaFunction{comm-hom-term}\AgdaSpace{}%
\AgdaFunction{hh}\AgdaSpace{}%
\AgdaBound{q}\AgdaSpace{}%
\AgdaBound{b}%
\>[58]\AgdaFunction{⟩}\<%
\\
\>[4]\AgdaFunction{h}\AgdaSpace{}%
\AgdaSymbol{(}%
\>[9]\AgdaOperator{\AgdaFunction{⟦}}\AgdaSpace{}%
\AgdaBound{q}\AgdaSpace{}%
\AgdaOperator{\AgdaFunction{⟧}}%
\>[17]\AgdaOperator{\AgdaField{⟨\$⟩}}%
\>[27]\AgdaBound{b}\AgdaSymbol{)}%
\>[31]\AgdaOperator{\AgdaFunction{∎}}\AgdaSpace{}%
\AgdaSymbol{)}\<%
\\
\>[2]\AgdaKeyword{where}\<%
\\
\>[2]\AgdaKeyword{open}\AgdaSpace{}%
\AgdaModule{SetoidReasoning}\AgdaSpace{}%
\AgdaOperator{\AgdaFunction{𝔻[}}\AgdaSpace{}%
\AgdaBound{𝑨}\AgdaSpace{}%
\AgdaOperator{\AgdaFunction{]}}\<%
\\
\>[2]\AgdaKeyword{open}\AgdaSpace{}%
\AgdaModule{Setoid}\AgdaSpace{}%
\AgdaOperator{\AgdaFunction{𝔻[}}\AgdaSpace{}%
\AgdaBound{𝑨}\AgdaSpace{}%
\AgdaOperator{\AgdaFunction{]}}%
\>[22]\AgdaKeyword{using}\AgdaSpace{}%
\AgdaSymbol{(}\AgdaSpace{}%
\AgdaOperator{\AgdaField{\AgdaUnderscore{}≈\AgdaUnderscore{}}}\AgdaSpace{}%
\AgdaSymbol{)}\<%
\\
\>[2]\AgdaKeyword{open}\AgdaSpace{}%
\AgdaModule{Environment}\AgdaSpace{}%
\AgdaBound{𝑨}%
\>[22]\AgdaKeyword{using}\AgdaSpace{}%
\AgdaSymbol{()}\AgdaSpace{}%
\AgdaKeyword{renaming}\AgdaSpace{}%
\AgdaSymbol{(}\AgdaSpace{}%
\AgdaOperator{\AgdaFunction{⟦\AgdaUnderscore{}⟧}}\AgdaSpace{}%
\AgdaSymbol{to}\AgdaSpace{}%
\AgdaOperator{\AgdaFunction{⟦\AgdaUnderscore{}⟧ᴬ}}\AgdaSpace{}%
\AgdaSymbol{)}\<%
\\
\>[2]\AgdaKeyword{open}\AgdaSpace{}%
\AgdaModule{Environment}\AgdaSpace{}%
\AgdaBound{𝑩}%
\>[22]\AgdaKeyword{using}\AgdaSpace{}%
\AgdaSymbol{(}\AgdaSpace{}%
\AgdaOperator{\AgdaFunction{⟦\AgdaUnderscore{}⟧}}\AgdaSpace{}%
\AgdaSymbol{)}\<%
\\
\>[2]\AgdaKeyword{private}\AgdaSpace{}%
\AgdaFunction{hh}\AgdaSpace{}%
\AgdaSymbol{=}\AgdaSpace{}%
\AgdaOperator{\AgdaFunction{∣}}\AgdaSpace{}%
\AgdaBound{B≤A}\AgdaSpace{}%
\AgdaOperator{\AgdaFunction{∣}}\AgdaSpace{}%
\AgdaSymbol{;}\AgdaSpace{}%
\AgdaFunction{h}\AgdaSpace{}%
\AgdaSymbol{=}\AgdaSpace{}%
\AgdaOperator{\AgdaField{\AgdaUnderscore{}⟨\$⟩\AgdaUnderscore{}}}\AgdaSpace{}%
\AgdaOperator{\AgdaFunction{∣}}\AgdaSpace{}%
\AgdaFunction{hh}\AgdaSpace{}%
\AgdaOperator{\AgdaFunction{∣}}\<%
\\
\>[0]\<%
\end{code}
\fi
An identity satisfied by all algebras in an indexed collection is
also satisfied by the product of algebras in the collection.
\ifshort
We refer to this fact as \af{⊧-P-invar}.
\else

\begin{code}%
\>[0]\<%
\\
\>[0]\AgdaKeyword{module}\AgdaSpace{}%
\AgdaModule{\AgdaUnderscore{}}\AgdaSpace{}%
\AgdaSymbol{\{}\AgdaBound{X}\AgdaSpace{}%
\AgdaSymbol{:}\AgdaSpace{}%
\AgdaPrimitive{Type}\AgdaSpace{}%
\AgdaGeneralizable{χ}\AgdaSymbol{\}\{}\AgdaBound{I}\AgdaSpace{}%
\AgdaSymbol{:}\AgdaSpace{}%
\AgdaPrimitive{Type}\AgdaSpace{}%
\AgdaGeneralizable{ℓ}\AgdaSymbol{\}(}\AgdaBound{𝒜}\AgdaSpace{}%
\AgdaSymbol{:}\AgdaSpace{}%
\AgdaBound{I}\AgdaSpace{}%
\AgdaSymbol{→}\AgdaSpace{}%
\AgdaRecord{Algebra}\AgdaSpace{}%
\AgdaGeneralizable{α}\AgdaSpace{}%
\AgdaGeneralizable{ρᵃ}\AgdaSymbol{)\{}\AgdaBound{p}\AgdaSpace{}%
\AgdaBound{q}\AgdaSpace{}%
\AgdaSymbol{:}\AgdaSpace{}%
\AgdaDatatype{Term}\AgdaSpace{}%
\AgdaBound{X}\AgdaSymbol{\}}\AgdaSpace{}%
\AgdaKeyword{where}\<%
\\
\>[0][@{}l@{\AgdaIndent{0}}]%
\>[1]\AgdaFunction{⊧-P-invar}\AgdaSpace{}%
\AgdaSymbol{:}\AgdaSpace{}%
\AgdaSymbol{(∀}\AgdaSpace{}%
\AgdaBound{i}\AgdaSpace{}%
\AgdaSymbol{→}\AgdaSpace{}%
\AgdaBound{𝒜}\AgdaSpace{}%
\AgdaBound{i}\AgdaSpace{}%
\AgdaOperator{\AgdaFunction{⊧}}\AgdaSpace{}%
\AgdaBound{p}\AgdaSpace{}%
\AgdaOperator{\AgdaFunction{≈}}\AgdaSpace{}%
\AgdaBound{q}\AgdaSymbol{)}\AgdaSpace{}%
\AgdaSymbol{→}\AgdaSpace{}%
\AgdaFunction{⨅}\AgdaSpace{}%
\AgdaBound{𝒜}\AgdaSpace{}%
\AgdaOperator{\AgdaFunction{⊧}}\AgdaSpace{}%
\AgdaBound{p}\AgdaSpace{}%
\AgdaOperator{\AgdaFunction{≈}}\AgdaSpace{}%
\AgdaBound{q}\<%
\\
\>[1]\AgdaFunction{⊧-P-invar}\AgdaSpace{}%
\AgdaBound{𝒜pq}\AgdaSpace{}%
\AgdaBound{a}\AgdaSpace{}%
\AgdaSymbol{=}\<%
\\
\>[1][@{}l@{\AgdaIndent{0}}]%
\>[2]\AgdaOperator{\AgdaFunction{begin}}\<%
\\
\>[2][@{}l@{\AgdaIndent{0}}]%
\>[3]\AgdaOperator{\AgdaFunction{⟦}}\AgdaSpace{}%
\AgdaBound{p}\AgdaSpace{}%
\AgdaOperator{\AgdaFunction{⟧₁}}%
\>[24]\AgdaOperator{\AgdaField{⟨\$⟩}}%
\>[29]\AgdaBound{a}%
\>[46]\AgdaFunction{≈⟨}%
\>[51]\AgdaFunction{interp-prod}\AgdaSpace{}%
\AgdaBound{𝒜}\AgdaSpace{}%
\AgdaBound{p}\AgdaSpace{}%
\AgdaBound{a}%
\>[70]\AgdaFunction{⟩}\<%
\\
\>[3]\AgdaSymbol{(}\AgdaSpace{}%
\AgdaSymbol{λ}\AgdaSpace{}%
\AgdaBound{i}\AgdaSpace{}%
\AgdaSymbol{→}\AgdaSpace{}%
\AgdaSymbol{(}\AgdaOperator{\AgdaFunction{⟦}}\AgdaSpace{}%
\AgdaBound{𝒜}\AgdaSpace{}%
\AgdaBound{i}\AgdaSpace{}%
\AgdaOperator{\AgdaFunction{⟧}}\AgdaSpace{}%
\AgdaBound{p}\AgdaSymbol{)}%
\>[24]\AgdaOperator{\AgdaField{⟨\$⟩}}%
\>[29]\AgdaSymbol{λ}\AgdaSpace{}%
\AgdaBound{x}\AgdaSpace{}%
\AgdaSymbol{→}\AgdaSpace{}%
\AgdaSymbol{(}\AgdaBound{a}\AgdaSpace{}%
\AgdaBound{x}\AgdaSymbol{)}\AgdaSpace{}%
\AgdaBound{i}\AgdaSpace{}%
\AgdaSymbol{)}%
\>[46]\AgdaFunction{≈⟨}\AgdaSpace{}%
\AgdaSymbol{(λ}\AgdaSpace{}%
\AgdaBound{i}\AgdaSpace{}%
\AgdaSymbol{→}\AgdaSpace{}%
\AgdaBound{𝒜pq}\AgdaSpace{}%
\AgdaBound{i}\AgdaSpace{}%
\AgdaSymbol{(λ}\AgdaSpace{}%
\AgdaBound{x}\AgdaSpace{}%
\AgdaSymbol{→}\AgdaSpace{}%
\AgdaSymbol{(}\AgdaBound{a}\AgdaSpace{}%
\AgdaBound{x}\AgdaSymbol{)}\AgdaSpace{}%
\AgdaBound{i}\AgdaSymbol{))}\AgdaSpace{}%
\AgdaFunction{⟩}\<%
\\
\>[3]\AgdaSymbol{(}\AgdaSpace{}%
\AgdaSymbol{λ}\AgdaSpace{}%
\AgdaBound{i}\AgdaSpace{}%
\AgdaSymbol{→}\AgdaSpace{}%
\AgdaSymbol{(}\AgdaOperator{\AgdaFunction{⟦}}\AgdaSpace{}%
\AgdaBound{𝒜}\AgdaSpace{}%
\AgdaBound{i}\AgdaSpace{}%
\AgdaOperator{\AgdaFunction{⟧}}\AgdaSpace{}%
\AgdaBound{q}\AgdaSymbol{)}%
\>[24]\AgdaOperator{\AgdaField{⟨\$⟩}}%
\>[29]\AgdaSymbol{λ}\AgdaSpace{}%
\AgdaBound{x}\AgdaSpace{}%
\AgdaSymbol{→}\AgdaSpace{}%
\AgdaSymbol{(}\AgdaBound{a}\AgdaSpace{}%
\AgdaBound{x}\AgdaSymbol{)}\AgdaSpace{}%
\AgdaBound{i}\AgdaSpace{}%
\AgdaSymbol{)}%
\>[46]\AgdaFunction{≈˘⟨}%
\>[51]\AgdaFunction{interp-prod}\AgdaSpace{}%
\AgdaBound{𝒜}\AgdaSpace{}%
\AgdaBound{q}\AgdaSpace{}%
\AgdaBound{a}%
\>[70]\AgdaFunction{⟩}\<%
\\
\>[3]\AgdaOperator{\AgdaFunction{⟦}}\AgdaSpace{}%
\AgdaBound{q}\AgdaSpace{}%
\AgdaOperator{\AgdaFunction{⟧₁}}%
\>[24]\AgdaOperator{\AgdaField{⟨\$⟩}}%
\>[29]\AgdaBound{a}%
\>[46]\AgdaOperator{\AgdaFunction{∎}}\<%
\\
\>[2]\AgdaKeyword{where}\<%
\\
\>[2]\AgdaKeyword{open}\AgdaSpace{}%
\AgdaModule{Environment}\AgdaSpace{}%
\AgdaSymbol{(}\AgdaFunction{⨅}\AgdaSpace{}%
\AgdaBound{𝒜}\AgdaSymbol{)}%
\>[26]\AgdaKeyword{using}\AgdaSpace{}%
\AgdaSymbol{()}\AgdaSpace{}%
\AgdaKeyword{renaming}\AgdaSpace{}%
\AgdaSymbol{(}\AgdaSpace{}%
\AgdaOperator{\AgdaFunction{⟦\AgdaUnderscore{}⟧}}\AgdaSpace{}%
\AgdaSymbol{to}\AgdaSpace{}%
\AgdaOperator{\AgdaFunction{⟦\AgdaUnderscore{}⟧₁}}\AgdaSpace{}%
\AgdaSymbol{)}\<%
\\
\>[2]\AgdaKeyword{open}\AgdaSpace{}%
\AgdaModule{Environment}%
\>[26]\AgdaKeyword{using}\AgdaSpace{}%
\AgdaSymbol{(}\AgdaSpace{}%
\AgdaOperator{\AgdaFunction{⟦\AgdaUnderscore{}⟧}}\AgdaSpace{}%
\AgdaSymbol{)}\<%
\\
\>[2]\AgdaKeyword{open}\AgdaSpace{}%
\AgdaModule{Setoid}\AgdaSpace{}%
\AgdaOperator{\AgdaFunction{𝔻[}}\AgdaSpace{}%
\AgdaFunction{⨅}\AgdaSpace{}%
\AgdaBound{𝒜}\AgdaSpace{}%
\AgdaOperator{\AgdaFunction{]}}%
\>[26]\AgdaKeyword{using}\AgdaSpace{}%
\AgdaSymbol{(}\AgdaSpace{}%
\AgdaOperator{\AgdaField{\AgdaUnderscore{}≈\AgdaUnderscore{}}}\AgdaSpace{}%
\AgdaSymbol{)}\<%
\\
\>[2]\AgdaKeyword{open}\AgdaSpace{}%
\AgdaModule{SetoidReasoning}\AgdaSpace{}%
\AgdaOperator{\AgdaFunction{𝔻[}}\AgdaSpace{}%
\AgdaFunction{⨅}\AgdaSpace{}%
\AgdaBound{𝒜}\AgdaSpace{}%
\AgdaOperator{\AgdaFunction{]}}\<%
\\
\>[0]\<%
\end{code}
\fi

The classes \af H \ab{𝒦}, \af S \ab{𝒦}, \af P \ab{𝒦}, and \af V \ab{𝒦} all satisfy the
same term identities.  We will only use a subset of the inclusions needed to prove this
assertion, and we present here only the facts we need.\footnote{For more details, see the
\ualmodule{Varieties.Func.Preservation} module of the \agdaalgebras library.}
First, the closure operator \af H preserves the identities modeled by the
given class; this follows almost immediately from the invariance lemma
\af{⊧-H-invar} proved above.

\begin{AgdaAlign}
\begin{code}%
\>[0]\<%
\\
\>[0]\AgdaKeyword{module}\AgdaSpace{}%
\AgdaModule{\AgdaUnderscore{}}%
\>[10]\AgdaSymbol{\{}\AgdaBound{X}\AgdaSpace{}%
\AgdaSymbol{:}\AgdaSpace{}%
\AgdaPrimitive{Type}\AgdaSpace{}%
\AgdaGeneralizable{χ}\AgdaSymbol{\}\{}\AgdaBound{𝒦}\AgdaSpace{}%
\AgdaSymbol{:}\AgdaSpace{}%
\AgdaFunction{Pred}\AgdaSymbol{(}\AgdaRecord{Algebra}\AgdaSpace{}%
\AgdaGeneralizable{α}\AgdaSpace{}%
\AgdaGeneralizable{ρᵃ}\AgdaSymbol{)}\AgdaSpace{}%
\AgdaSymbol{(}\AgdaGeneralizable{α}\AgdaSpace{}%
\AgdaOperator{\AgdaPrimitive{⊔}}\AgdaSpace{}%
\AgdaGeneralizable{ρᵃ}\AgdaSpace{}%
\AgdaOperator{\AgdaPrimitive{⊔}}\AgdaSpace{}%
\AgdaFunction{ov}\AgdaSpace{}%
\AgdaGeneralizable{ℓ}\AgdaSymbol{)\}\{}\AgdaBound{p}\AgdaSpace{}%
\AgdaBound{q}\AgdaSpace{}%
\AgdaSymbol{:}\AgdaSpace{}%
\AgdaDatatype{Term}\AgdaSpace{}%
\AgdaBound{X}\AgdaSymbol{\}}\AgdaSpace{}%
\AgdaKeyword{where}\<%
\\
\>[0][@{}l@{\AgdaIndent{0}}]%
\>[1]\AgdaFunction{H-id1}\AgdaSpace{}%
\AgdaSymbol{:}\AgdaSpace{}%
\AgdaBound{𝒦}\AgdaSpace{}%
\AgdaOperator{\AgdaFunction{⊫}}\AgdaSpace{}%
\AgdaBound{p}\AgdaSpace{}%
\AgdaOperator{\AgdaFunction{≈}}\AgdaSpace{}%
\AgdaBound{q}\AgdaSpace{}%
\AgdaSymbol{→}\AgdaSpace{}%
\AgdaFunction{H}\AgdaSymbol{\{}\AgdaArgument{β}\AgdaSpace{}%
\AgdaSymbol{=}\AgdaSpace{}%
\AgdaBound{α}\AgdaSymbol{\}\{}\AgdaBound{ρᵃ}\AgdaSymbol{\}}\AgdaBound{ℓ}\AgdaSpace{}%
\AgdaBound{𝒦}\AgdaSpace{}%
\AgdaOperator{\AgdaFunction{⊫}}\AgdaSpace{}%
\AgdaBound{p}\AgdaSpace{}%
\AgdaOperator{\AgdaFunction{≈}}\AgdaSpace{}%
\AgdaBound{q}\<%
\\
\>[1]\AgdaFunction{H-id1}\AgdaSpace{}%
\AgdaBound{σ}\AgdaSpace{}%
\AgdaBound{𝑩}\AgdaSpace{}%
\AgdaSymbol{(}\AgdaBound{𝑨}\AgdaSpace{}%
\AgdaOperator{\AgdaInductiveConstructor{,}}\AgdaSpace{}%
\AgdaBound{kA}\AgdaSpace{}%
\AgdaOperator{\AgdaInductiveConstructor{,}}\AgdaSpace{}%
\AgdaBound{BimgA}\AgdaSymbol{)}\AgdaSpace{}%
\AgdaSymbol{=}\AgdaSpace{}%
\AgdaFunction{⊧-H-invar}\AgdaSymbol{\{}\AgdaArgument{p}\AgdaSpace{}%
\AgdaSymbol{=}\AgdaSpace{}%
\AgdaBound{p}\AgdaSymbol{\}\{}\AgdaBound{q}\AgdaSymbol{\}}\AgdaSpace{}%
\AgdaSymbol{(}\AgdaBound{σ}\AgdaSpace{}%
\AgdaBound{𝑨}\AgdaSpace{}%
\AgdaBound{kA}\AgdaSymbol{)}\AgdaSpace{}%
\AgdaBound{BimgA}\<%
\\
\>[0]\<%
\end{code}
The analogous preservation result for \af S is a simple consequence of
the invariance lemma \af{⊧-S-invar}; the obvious converse, which we call
\af{S-id2}, has an equally straightforward proof.

\begin{code}%
\>[0]\<%
\\
\>[0][@{}l@{\AgdaIndent{1}}]%
\>[1]\AgdaFunction{S-id1}\AgdaSpace{}%
\AgdaSymbol{:}\AgdaSpace{}%
\AgdaBound{𝒦}\AgdaSpace{}%
\AgdaOperator{\AgdaFunction{⊫}}\AgdaSpace{}%
\AgdaBound{p}\AgdaSpace{}%
\AgdaOperator{\AgdaFunction{≈}}\AgdaSpace{}%
\AgdaBound{q}\AgdaSpace{}%
\AgdaSymbol{→}\AgdaSpace{}%
\AgdaFunction{S}\AgdaSymbol{\{}\AgdaArgument{β}\AgdaSpace{}%
\AgdaSymbol{=}\AgdaSpace{}%
\AgdaBound{α}\AgdaSymbol{\}\{}\AgdaBound{ρᵃ}\AgdaSymbol{\}}\AgdaBound{ℓ}\AgdaSpace{}%
\AgdaBound{𝒦}\AgdaSpace{}%
\AgdaOperator{\AgdaFunction{⊫}}\AgdaSpace{}%
\AgdaBound{p}\AgdaSpace{}%
\AgdaOperator{\AgdaFunction{≈}}\AgdaSpace{}%
\AgdaBound{q}\<%
\\
\>[1]\AgdaFunction{S-id1}\AgdaSpace{}%
\AgdaBound{σ}\AgdaSpace{}%
\AgdaBound{𝑩}\AgdaSpace{}%
\AgdaSymbol{(}\AgdaBound{𝑨}\AgdaSpace{}%
\AgdaOperator{\AgdaInductiveConstructor{,}}\AgdaSpace{}%
\AgdaBound{kA}\AgdaSpace{}%
\AgdaOperator{\AgdaInductiveConstructor{,}}\AgdaSpace{}%
\AgdaBound{B≤A}\AgdaSymbol{)}\AgdaSpace{}%
\AgdaSymbol{=}\AgdaSpace{}%
\AgdaFunction{⊧-S-invar}\AgdaSymbol{\{}\AgdaArgument{p}\AgdaSpace{}%
\AgdaSymbol{=}\AgdaSpace{}%
\AgdaBound{p}\AgdaSymbol{\}\{}\AgdaBound{q}\AgdaSymbol{\}}\AgdaSpace{}%
\AgdaSymbol{(}\AgdaBound{σ}\AgdaSpace{}%
\AgdaBound{𝑨}\AgdaSpace{}%
\AgdaBound{kA}\AgdaSymbol{)}\AgdaSpace{}%
\AgdaBound{B≤A}\<%
\\
\>[1]\AgdaFunction{S-id2}\AgdaSpace{}%
\AgdaSymbol{:}\AgdaSpace{}%
\AgdaFunction{S}\AgdaSpace{}%
\AgdaBound{ℓ}\AgdaSpace{}%
\AgdaBound{𝒦}\AgdaSpace{}%
\AgdaOperator{\AgdaFunction{⊫}}\AgdaSpace{}%
\AgdaBound{p}\AgdaSpace{}%
\AgdaOperator{\AgdaFunction{≈}}\AgdaSpace{}%
\AgdaBound{q}\AgdaSpace{}%
\AgdaSymbol{→}\AgdaSpace{}%
\AgdaBound{𝒦}\AgdaSpace{}%
\AgdaOperator{\AgdaFunction{⊫}}\AgdaSpace{}%
\AgdaBound{p}\AgdaSpace{}%
\AgdaOperator{\AgdaFunction{≈}}\AgdaSpace{}%
\AgdaBound{q}\<%
\\
\>[1]\AgdaFunction{S-id2}\AgdaSpace{}%
\AgdaBound{Spq}\AgdaSpace{}%
\AgdaBound{𝑨}\AgdaSpace{}%
\AgdaBound{kA}\AgdaSpace{}%
\AgdaSymbol{=}\AgdaSpace{}%
\AgdaBound{Spq}\AgdaSpace{}%
\AgdaBound{𝑨}\AgdaSpace{}%
\AgdaSymbol{(}\AgdaBound{𝑨}\AgdaSpace{}%
\AgdaOperator{\AgdaInductiveConstructor{,}}\AgdaSpace{}%
\AgdaSymbol{(}\AgdaBound{kA}\AgdaSpace{}%
\AgdaOperator{\AgdaInductiveConstructor{,}}\AgdaSpace{}%
\AgdaFunction{≤-reflexive}\AgdaSymbol{))}\<%
\\
\>[0]\<%
\end{code}
Finally, we have analogous pairs of implications for \af P and \af V,
\ifshort
called P-id1 and V-id1 in the \agdaalgebras library.
\else
In each case, we will only need the first implication, so we omit the others from this presentation.

\begin{code}%
\>[0]\<%
\\
\>[0][@{}l@{\AgdaIndent{1}}]%
\>[1]\AgdaFunction{P-id1}\AgdaSpace{}%
\AgdaSymbol{:}\AgdaSpace{}%
\AgdaSymbol{∀\{}\AgdaBound{ι}\AgdaSymbol{\}}\AgdaSpace{}%
\AgdaSymbol{→}\AgdaSpace{}%
\AgdaBound{𝒦}\AgdaSpace{}%
\AgdaOperator{\AgdaFunction{⊫}}\AgdaSpace{}%
\AgdaBound{p}\AgdaSpace{}%
\AgdaOperator{\AgdaFunction{≈}}\AgdaSpace{}%
\AgdaBound{q}\AgdaSpace{}%
\AgdaSymbol{→}\AgdaSpace{}%
\AgdaFunction{P}\AgdaSymbol{\{}\AgdaArgument{β}\AgdaSpace{}%
\AgdaSymbol{=}\AgdaSpace{}%
\AgdaBound{α}\AgdaSymbol{\}\{}\AgdaBound{ρᵃ}\AgdaSymbol{\}}\AgdaBound{ℓ}\AgdaSpace{}%
\AgdaBound{ι}\AgdaSpace{}%
\AgdaBound{𝒦}\AgdaSpace{}%
\AgdaOperator{\AgdaFunction{⊫}}\AgdaSpace{}%
\AgdaBound{p}\AgdaSpace{}%
\AgdaOperator{\AgdaFunction{≈}}\AgdaSpace{}%
\AgdaBound{q}\<%
\\
\>[1]\AgdaFunction{P-id1}\AgdaSpace{}%
\AgdaBound{σ}\AgdaSpace{}%
\AgdaBound{𝑨}\AgdaSpace{}%
\AgdaSymbol{(}\AgdaBound{I}\AgdaSpace{}%
\AgdaOperator{\AgdaInductiveConstructor{,}}\AgdaSpace{}%
\AgdaBound{𝒜}\AgdaSpace{}%
\AgdaOperator{\AgdaInductiveConstructor{,}}\AgdaSpace{}%
\AgdaBound{kA}\AgdaSpace{}%
\AgdaOperator{\AgdaInductiveConstructor{,}}\AgdaSpace{}%
\AgdaBound{A≅⨅A}\AgdaSymbol{)}\AgdaSpace{}%
\AgdaSymbol{=}\AgdaSpace{}%
\AgdaFunction{⊧-I-invar}\AgdaSpace{}%
\AgdaBound{𝑨}\AgdaSpace{}%
\AgdaBound{p}\AgdaSpace{}%
\AgdaBound{q}\AgdaSpace{}%
\AgdaFunction{IH}\AgdaSpace{}%
\AgdaSymbol{(}\AgdaFunction{≅-sym}\AgdaSpace{}%
\AgdaBound{A≅⨅A}\AgdaSymbol{)}\<%
\\
\>[1][@{}l@{\AgdaIndent{0}}]%
\>[2]\AgdaKeyword{where}\<%
\\
\>[2]\AgdaFunction{IH}\AgdaSpace{}%
\AgdaSymbol{:}\AgdaSpace{}%
\AgdaFunction{⨅}\AgdaSpace{}%
\AgdaBound{𝒜}\AgdaSpace{}%
\AgdaOperator{\AgdaFunction{⊧}}\AgdaSpace{}%
\AgdaBound{p}\AgdaSpace{}%
\AgdaOperator{\AgdaFunction{≈}}\AgdaSpace{}%
\AgdaBound{q}\<%
\\
\>[2]\AgdaFunction{IH}\AgdaSpace{}%
\AgdaSymbol{=}\AgdaSpace{}%
\AgdaFunction{⊧-P-invar}\AgdaSpace{}%
\AgdaBound{𝒜}\AgdaSpace{}%
\AgdaSymbol{\{}\AgdaBound{p}\AgdaSymbol{\}\{}\AgdaBound{q}\AgdaSymbol{\}}\AgdaSpace{}%
\AgdaSymbol{(λ}\AgdaSpace{}%
\AgdaBound{i}\AgdaSpace{}%
\AgdaSymbol{→}\AgdaSpace{}%
\AgdaBound{σ}\AgdaSpace{}%
\AgdaSymbol{(}\AgdaBound{𝒜}\AgdaSpace{}%
\AgdaBound{i}\AgdaSymbol{)}\AgdaSpace{}%
\AgdaSymbol{(}\AgdaBound{kA}\AgdaSpace{}%
\AgdaBound{i}\AgdaSymbol{))}\<%
\\
\\[\AgdaEmptyExtraSkip]%
\>[0]\AgdaKeyword{module}\AgdaSpace{}%
\AgdaModule{\AgdaUnderscore{}}\AgdaSpace{}%
\AgdaSymbol{\{}\AgdaBound{X}\AgdaSpace{}%
\AgdaSymbol{:}\AgdaSpace{}%
\AgdaPrimitive{Type}\AgdaSpace{}%
\AgdaGeneralizable{χ}\AgdaSymbol{\}\{}\AgdaBound{ι}\AgdaSpace{}%
\AgdaSymbol{:}\AgdaSpace{}%
\AgdaPostulate{Level}\AgdaSymbol{\}\{}\AgdaBound{𝒦}\AgdaSpace{}%
\AgdaSymbol{:}\AgdaSpace{}%
\AgdaFunction{Pred}\AgdaSymbol{(}\AgdaRecord{Algebra}\AgdaSpace{}%
\AgdaGeneralizable{α}\AgdaSpace{}%
\AgdaGeneralizable{ρᵃ}\AgdaSymbol{)(}\AgdaGeneralizable{α}\AgdaSpace{}%
\AgdaOperator{\AgdaPrimitive{⊔}}\AgdaSpace{}%
\AgdaGeneralizable{ρᵃ}\AgdaSpace{}%
\AgdaOperator{\AgdaPrimitive{⊔}}\AgdaSpace{}%
\AgdaFunction{ov}\AgdaSpace{}%
\AgdaGeneralizable{ℓ}\AgdaSymbol{)\}\{}\AgdaBound{p}\AgdaSpace{}%
\AgdaBound{q}\AgdaSpace{}%
\AgdaSymbol{:}\AgdaSpace{}%
\AgdaDatatype{Term}\AgdaSpace{}%
\AgdaBound{X}\AgdaSymbol{\}}\AgdaSpace{}%
\AgdaKeyword{where}\<%
\\
\>[0][@{}l@{\AgdaIndent{0}}]%
\>[1]\AgdaKeyword{private}\AgdaSpace{}%
\AgdaFunction{aℓι}\AgdaSpace{}%
\AgdaSymbol{=}\AgdaSpace{}%
\AgdaBound{α}\AgdaSpace{}%
\AgdaOperator{\AgdaPrimitive{⊔}}\AgdaSpace{}%
\AgdaBound{ρᵃ}\AgdaSpace{}%
\AgdaOperator{\AgdaPrimitive{⊔}}\AgdaSpace{}%
\AgdaBound{ℓ}\AgdaSpace{}%
\AgdaOperator{\AgdaPrimitive{⊔}}\AgdaSpace{}%
\AgdaBound{ι}\<%
\\
\\[\AgdaEmptyExtraSkip]%
\>[1]\AgdaFunction{V-id1}\AgdaSpace{}%
\AgdaSymbol{:}\AgdaSpace{}%
\AgdaBound{𝒦}\AgdaSpace{}%
\AgdaOperator{\AgdaFunction{⊫}}\AgdaSpace{}%
\AgdaBound{p}\AgdaSpace{}%
\AgdaOperator{\AgdaFunction{≈}}\AgdaSpace{}%
\AgdaBound{q}\AgdaSpace{}%
\AgdaSymbol{→}\AgdaSpace{}%
\AgdaFunction{V}\AgdaSpace{}%
\AgdaBound{ℓ}\AgdaSpace{}%
\AgdaBound{ι}\AgdaSpace{}%
\AgdaBound{𝒦}\AgdaSpace{}%
\AgdaOperator{\AgdaFunction{⊫}}\AgdaSpace{}%
\AgdaBound{p}\AgdaSpace{}%
\AgdaOperator{\AgdaFunction{≈}}\AgdaSpace{}%
\AgdaBound{q}\<%
\\
\>[1]\AgdaFunction{V-id1}\AgdaSpace{}%
\AgdaBound{σ}\AgdaSpace{}%
\AgdaBound{𝑩}\AgdaSpace{}%
\AgdaSymbol{(}\AgdaBound{𝑨}\AgdaSpace{}%
\AgdaOperator{\AgdaInductiveConstructor{,}}\AgdaSpace{}%
\AgdaSymbol{(}\AgdaBound{⨅A}\AgdaSpace{}%
\AgdaOperator{\AgdaInductiveConstructor{,}}\AgdaSpace{}%
\AgdaBound{p⨅A}\AgdaSpace{}%
\AgdaOperator{\AgdaInductiveConstructor{,}}\AgdaSpace{}%
\AgdaBound{A≤⨅A}\AgdaSymbol{)}\AgdaSpace{}%
\AgdaOperator{\AgdaInductiveConstructor{,}}\AgdaSpace{}%
\AgdaBound{BimgA}\AgdaSymbol{)}\AgdaSpace{}%
\AgdaSymbol{=}\<%
\\
\>[1][@{}l@{\AgdaIndent{0}}]%
\>[2]\AgdaFunction{H-id1}\AgdaSymbol{\{}\AgdaArgument{ℓ}\AgdaSpace{}%
\AgdaSymbol{=}\AgdaSpace{}%
\AgdaFunction{aℓι}\AgdaSymbol{\}\{}\AgdaArgument{𝒦}\AgdaSpace{}%
\AgdaSymbol{=}\AgdaSpace{}%
\AgdaFunction{S}\AgdaSpace{}%
\AgdaFunction{aℓι}\AgdaSpace{}%
\AgdaSymbol{(}\AgdaFunction{P}\AgdaSpace{}%
\AgdaSymbol{\{}\AgdaArgument{β}\AgdaSpace{}%
\AgdaSymbol{=}\AgdaSpace{}%
\AgdaBound{α}\AgdaSymbol{\}\{}\AgdaBound{ρᵃ}\AgdaSymbol{\}}\AgdaBound{ℓ}\AgdaSpace{}%
\AgdaBound{ι}\AgdaSpace{}%
\AgdaBound{𝒦}\AgdaSymbol{)\}\{}\AgdaArgument{p}\AgdaSpace{}%
\AgdaSymbol{=}\AgdaSpace{}%
\AgdaBound{p}\AgdaSymbol{\}\{}\AgdaBound{q}\AgdaSymbol{\}}\AgdaSpace{}%
\AgdaFunction{spK⊧pq}\AgdaSpace{}%
\AgdaBound{𝑩}\AgdaSpace{}%
\AgdaSymbol{(}\AgdaBound{𝑨}\AgdaSpace{}%
\AgdaOperator{\AgdaInductiveConstructor{,}}\AgdaSpace{}%
\AgdaSymbol{(}\AgdaFunction{spA}\AgdaSpace{}%
\AgdaOperator{\AgdaInductiveConstructor{,}}\AgdaSpace{}%
\AgdaBound{BimgA}\AgdaSymbol{))}\<%
\\
\>[2][@{}l@{\AgdaIndent{0}}]%
\>[3]\AgdaKeyword{where}\<%
\\
\>[3]\AgdaFunction{spA}\AgdaSpace{}%
\AgdaSymbol{:}\AgdaSpace{}%
\AgdaBound{𝑨}\AgdaSpace{}%
\AgdaOperator{\AgdaFunction{∈}}\AgdaSpace{}%
\AgdaFunction{S}\AgdaSpace{}%
\AgdaFunction{aℓι}\AgdaSpace{}%
\AgdaSymbol{(}\AgdaFunction{P}\AgdaSpace{}%
\AgdaSymbol{\{}\AgdaArgument{β}\AgdaSpace{}%
\AgdaSymbol{=}\AgdaSpace{}%
\AgdaBound{α}\AgdaSymbol{\}\{}\AgdaBound{ρᵃ}\AgdaSymbol{\}}\AgdaBound{ℓ}\AgdaSpace{}%
\AgdaBound{ι}\AgdaSpace{}%
\AgdaBound{𝒦}\AgdaSymbol{)}\<%
\\
\>[3]\AgdaFunction{spA}\AgdaSpace{}%
\AgdaSymbol{=}\AgdaSpace{}%
\AgdaBound{⨅A}\AgdaSpace{}%
\AgdaOperator{\AgdaInductiveConstructor{,}}\AgdaSpace{}%
\AgdaSymbol{(}\AgdaBound{p⨅A}\AgdaSpace{}%
\AgdaOperator{\AgdaInductiveConstructor{,}}\AgdaSpace{}%
\AgdaBound{A≤⨅A}\AgdaSymbol{)}\<%
\\
\>[3]\AgdaFunction{spK⊧pq}\AgdaSpace{}%
\AgdaSymbol{:}\AgdaSpace{}%
\AgdaFunction{S}\AgdaSpace{}%
\AgdaFunction{aℓι}\AgdaSpace{}%
\AgdaSymbol{(}\AgdaFunction{P}\AgdaSpace{}%
\AgdaBound{ℓ}\AgdaSpace{}%
\AgdaBound{ι}\AgdaSpace{}%
\AgdaBound{𝒦}\AgdaSymbol{)}\AgdaSpace{}%
\AgdaOperator{\AgdaFunction{⊫}}\AgdaSpace{}%
\AgdaBound{p}\AgdaSpace{}%
\AgdaOperator{\AgdaFunction{≈}}\AgdaSpace{}%
\AgdaBound{q}\<%
\\
\>[3]\AgdaFunction{spK⊧pq}\AgdaSpace{}%
\AgdaSymbol{=}\AgdaSpace{}%
\AgdaFunction{S-id1}\AgdaSymbol{\{}\AgdaArgument{ℓ}\AgdaSpace{}%
\AgdaSymbol{=}\AgdaSpace{}%
\AgdaFunction{aℓι}\AgdaSymbol{\}\{}\AgdaArgument{p}\AgdaSpace{}%
\AgdaSymbol{=}\AgdaSpace{}%
\AgdaBound{p}\AgdaSymbol{\}\{}\AgdaBound{q}\AgdaSymbol{\}}\AgdaSpace{}%
\AgdaSymbol{(}\AgdaFunction{P-id1}\AgdaSymbol{\{}\AgdaArgument{ℓ}\AgdaSpace{}%
\AgdaSymbol{=}\AgdaSpace{}%
\AgdaBound{ℓ}\AgdaSymbol{\}}\AgdaSpace{}%
\AgdaSymbol{\{}\AgdaArgument{𝒦}\AgdaSpace{}%
\AgdaSymbol{=}\AgdaSpace{}%
\AgdaBound{𝒦}\AgdaSymbol{\}\{}\AgdaArgument{p}\AgdaSpace{}%
\AgdaSymbol{=}\AgdaSpace{}%
\AgdaBound{p}\AgdaSymbol{\}\{}\AgdaBound{q}\AgdaSymbol{\}}\AgdaSpace{}%
\AgdaBound{σ}\AgdaSymbol{)}\<%
\end{code}
\fi
\end{AgdaAlign}

%% -------------------------------------------------------------------------------------

\section{Free Algebras}
\label{free-algebras}
\paragraph*{The absolutely free algebra}
The term algebra \af{𝑻} \ab X is \emph{absolutely free} (or \emph{universal}, or
\emph{initial}) for algebras in the signature \ab{𝑆}. That is, for every
\ab{𝑆}-algebra \ab{𝑨}, the following hold.

\begin{itemize}
\item Every function from \ab{X} to \af{𝕌[ \ab{𝑨} ]} lifts to a homomorphism from \af{𝑻} \ab{X} to \ab{𝑨}.
\item The homomorphism that exists by the previous item is unique.
\end{itemize}

We now prove the first of these facts in Agda which we call \af{free-lift}.\footnote{The
 \agdaalgebras library also defines
 \af{free-lift-func} \as{:} \aof{𝔻[~\af{𝑻}~\ab X~]}~\aor{⟶}~\aof{𝔻[~\ab{𝑨}~]}
 for constructing the analogous setoid function.}$^,$\footnote{For the proof of uniqueness,
see the \ualmodule{Terms.Func.Properties} module of the \agdaalgebras library.}

\begin{code}%
\>[0]\<%
\\
\>[0]\AgdaKeyword{module}\AgdaSpace{}%
\AgdaModule{\AgdaUnderscore{}}\AgdaSpace{}%
\AgdaSymbol{\{}\AgdaBound{X}\AgdaSpace{}%
\AgdaSymbol{:}\AgdaSpace{}%
\AgdaPrimitive{Type}\AgdaSpace{}%
\AgdaGeneralizable{χ}\AgdaSymbol{\}\{}\AgdaBound{𝑨}\AgdaSpace{}%
\AgdaSymbol{:}\AgdaSpace{}%
\AgdaRecord{Algebra}\AgdaSpace{}%
\AgdaGeneralizable{α}\AgdaSpace{}%
\AgdaGeneralizable{ρᵃ}\AgdaSymbol{\}(}\AgdaBound{h}\AgdaSpace{}%
\AgdaSymbol{:}\AgdaSpace{}%
\AgdaBound{X}\AgdaSpace{}%
\AgdaSymbol{→}\AgdaSpace{}%
\AgdaOperator{\AgdaFunction{𝕌[}}\AgdaSpace{}%
\AgdaBound{𝑨}\AgdaSpace{}%
\AgdaOperator{\AgdaFunction{]}}\AgdaSymbol{)}\AgdaSpace{}%
\AgdaKeyword{where}\<%
\\
\>[0][@{}l@{\AgdaIndent{0}}]%
\>[1]\AgdaFunction{free-lift}\AgdaSpace{}%
\AgdaSymbol{:}\AgdaSpace{}%
\AgdaOperator{\AgdaFunction{𝕌[}}\AgdaSpace{}%
\AgdaFunction{𝑻}\AgdaSpace{}%
\AgdaBound{X}\AgdaSpace{}%
\AgdaOperator{\AgdaFunction{]}}\AgdaSpace{}%
\AgdaSymbol{→}\AgdaSpace{}%
\AgdaOperator{\AgdaFunction{𝕌[}}\AgdaSpace{}%
\AgdaBound{𝑨}\AgdaSpace{}%
\AgdaOperator{\AgdaFunction{]}}\<%
\\
\>[1]\AgdaFunction{free-lift}\AgdaSpace{}%
\AgdaSymbol{(}\AgdaInductiveConstructor{ℊ}\AgdaSpace{}%
\AgdaBound{x}\AgdaSymbol{)}\AgdaSpace{}%
\AgdaSymbol{=}\AgdaSpace{}%
\AgdaBound{h}\AgdaSpace{}%
\AgdaBound{x}\<%
\\
\>[1]\AgdaFunction{free-lift}\AgdaSpace{}%
\AgdaSymbol{(}\AgdaInductiveConstructor{node}\AgdaSpace{}%
\AgdaBound{f}\AgdaSpace{}%
\AgdaBound{t}\AgdaSymbol{)}\AgdaSpace{}%
\AgdaSymbol{=}\AgdaSpace{}%
\AgdaSymbol{(}\AgdaBound{f}\AgdaSpace{}%
\AgdaOperator{\AgdaFunction{̂}}\AgdaSpace{}%
\AgdaBound{𝑨}\AgdaSymbol{)}\AgdaSpace{}%
\AgdaSymbol{(λ}\AgdaSpace{}%
\AgdaBound{i}\AgdaSpace{}%
\AgdaSymbol{→}\AgdaSpace{}%
\AgdaFunction{free-lift}\AgdaSpace{}%
\AgdaSymbol{(}\AgdaBound{t}\AgdaSpace{}%
\AgdaBound{i}\AgdaSymbol{))}\<%
\\
\>[0]\<%
\end{code}
\ifshort\else
\begin{code}%
\>[0][@{}l@{\AgdaIndent{1}}]%
\>[1]\AgdaFunction{free-lift-func}\AgdaSpace{}%
\AgdaSymbol{:}\AgdaSpace{}%
\AgdaOperator{\AgdaFunction{𝔻[}}\AgdaSpace{}%
\AgdaFunction{𝑻}\AgdaSpace{}%
\AgdaBound{X}\AgdaSpace{}%
\AgdaOperator{\AgdaFunction{]}}\AgdaSpace{}%
\AgdaOperator{\AgdaRecord{⟶}}\AgdaSpace{}%
\AgdaOperator{\AgdaFunction{𝔻[}}\AgdaSpace{}%
\AgdaBound{𝑨}\AgdaSpace{}%
\AgdaOperator{\AgdaFunction{]}}\<%
\\
\>[1]\AgdaFunction{free-lift-func}\AgdaSpace{}%
\AgdaOperator{\AgdaField{⟨\$⟩}}\AgdaSpace{}%
\AgdaBound{x}\AgdaSpace{}%
\AgdaSymbol{=}\AgdaSpace{}%
\AgdaFunction{free-lift}\AgdaSpace{}%
\AgdaBound{x}\<%
\\
\>[1]\AgdaField{cong}\AgdaSpace{}%
\AgdaFunction{free-lift-func}\AgdaSpace{}%
\AgdaSymbol{=}\AgdaSpace{}%
\AgdaFunction{flcong}\<%
\\
\>[1][@{}l@{\AgdaIndent{0}}]%
\>[2]\AgdaKeyword{where}\<%
\\
\>[2]\AgdaKeyword{open}\AgdaSpace{}%
\AgdaModule{Setoid}\AgdaSpace{}%
\AgdaOperator{\AgdaFunction{𝔻[}}\AgdaSpace{}%
\AgdaBound{𝑨}\AgdaSpace{}%
\AgdaOperator{\AgdaFunction{]}}\AgdaSpace{}%
\AgdaKeyword{using}\AgdaSpace{}%
\AgdaSymbol{(}\AgdaSpace{}%
\AgdaOperator{\AgdaField{\AgdaUnderscore{}≈\AgdaUnderscore{}}}\AgdaSpace{}%
\AgdaSymbol{)}\AgdaSpace{}%
\AgdaKeyword{renaming}\AgdaSpace{}%
\AgdaSymbol{(}\AgdaSpace{}%
\AgdaFunction{reflexive}\AgdaSpace{}%
\AgdaSymbol{to}\AgdaSpace{}%
\AgdaFunction{reflexiveᴬ}\AgdaSpace{}%
\AgdaSymbol{)}\<%
\\
\>[2]\AgdaFunction{flcong}\AgdaSpace{}%
\AgdaSymbol{:}\AgdaSpace{}%
\AgdaSymbol{∀}\AgdaSpace{}%
\AgdaSymbol{\{}\AgdaBound{s}\AgdaSpace{}%
\AgdaBound{t}\AgdaSymbol{\}}\AgdaSpace{}%
\AgdaSymbol{→}\AgdaSpace{}%
\AgdaBound{s}\AgdaSpace{}%
\AgdaOperator{\AgdaDatatype{≃}}\AgdaSpace{}%
\AgdaBound{t}\AgdaSpace{}%
\AgdaSymbol{→}\AgdaSpace{}%
\AgdaFunction{free-lift}\AgdaSpace{}%
\AgdaBound{s}\AgdaSpace{}%
\AgdaOperator{\AgdaFunction{≈}}\AgdaSpace{}%
\AgdaFunction{free-lift}\AgdaSpace{}%
\AgdaBound{t}\<%
\\
\>[2]\AgdaFunction{flcong}\AgdaSpace{}%
\AgdaSymbol{(}\AgdaInductiveConstructor{\AgdaUnderscore{}≃\AgdaUnderscore{}.rfl}\AgdaSpace{}%
\AgdaBound{x}\AgdaSymbol{)}\AgdaSpace{}%
\AgdaSymbol{=}\AgdaSpace{}%
\AgdaFunction{reflexiveᴬ}\AgdaSpace{}%
\AgdaSymbol{(}\AgdaFunction{≡.cong}\AgdaSpace{}%
\AgdaBound{h}\AgdaSpace{}%
\AgdaBound{x}\AgdaSymbol{)}\<%
\\
\>[2]\AgdaFunction{flcong}\AgdaSpace{}%
\AgdaSymbol{(}\AgdaInductiveConstructor{\AgdaUnderscore{}≃\AgdaUnderscore{}.gnl}\AgdaSpace{}%
\AgdaBound{x}\AgdaSymbol{)}\AgdaSpace{}%
\AgdaSymbol{=}\AgdaSpace{}%
\AgdaField{cong}\AgdaSpace{}%
\AgdaSymbol{(}\AgdaField{Interp}\AgdaSpace{}%
\AgdaBound{𝑨}\AgdaSymbol{)}\AgdaSpace{}%
\AgdaSymbol{(}\AgdaInductiveConstructor{≡.refl}\AgdaSpace{}%
\AgdaOperator{\AgdaInductiveConstructor{,}}\AgdaSpace{}%
\AgdaSymbol{(λ}\AgdaSpace{}%
\AgdaBound{i}\AgdaSpace{}%
\AgdaSymbol{→}\AgdaSpace{}%
\AgdaFunction{flcong}\AgdaSpace{}%
\AgdaSymbol{(}\AgdaBound{x}\AgdaSpace{}%
\AgdaBound{i}\AgdaSymbol{)))}\<%
\\
\>[0]\<%
\end{code}
\fi
Evidently, the proof is a straightforward structural induction argument.
\ifshort\else
At the base step, when the term has the form \aic{ℊ}
\ab x, the free lift of \ab h agrees with \ab h; at the inductive step, when the
term has the form \aic{node} \ab f \ab t, we assume (the induction hypothesis)
that the image of each subterm \ab t \ab i under the free lift of \ab h is known
and the free lift is defined by applying \ab f \aof{̂} \ab{𝑨} to these images.
\fi
Moreover, the free lift so defined is a homomorphism by construction;
\ifshort
for the proof---which is called \af{lift-hom} in the \agdaalgebras library---\seeshort.
\else
indeed, here is the trivial proof.

\begin{code}%
\>[0]\<%
\\
\>[0][@{}l@{\AgdaIndent{1}}]%
\>[1]\AgdaFunction{lift-hom}\AgdaSpace{}%
\AgdaSymbol{:}\AgdaSpace{}%
\AgdaFunction{hom}\AgdaSpace{}%
\AgdaSymbol{(}\AgdaFunction{𝑻}\AgdaSpace{}%
\AgdaBound{X}\AgdaSymbol{)}\AgdaSpace{}%
\AgdaBound{𝑨}\<%
\\
\>[1]\AgdaFunction{lift-hom}\AgdaSpace{}%
\AgdaSymbol{=}\AgdaSpace{}%
\AgdaFunction{free-lift-func}\AgdaSpace{}%
\AgdaOperator{\AgdaInductiveConstructor{,}}\AgdaSpace{}%
\AgdaFunction{hhom}\<%
\\
\>[1][@{}l@{\AgdaIndent{0}}]%
\>[2]\AgdaKeyword{where}\<%
\\
\>[2]\AgdaFunction{hfunc}\AgdaSpace{}%
\AgdaSymbol{:}\AgdaSpace{}%
\AgdaOperator{\AgdaFunction{𝔻[}}\AgdaSpace{}%
\AgdaFunction{𝑻}\AgdaSpace{}%
\AgdaBound{X}\AgdaSpace{}%
\AgdaOperator{\AgdaFunction{]}}\AgdaSpace{}%
\AgdaOperator{\AgdaRecord{⟶}}\AgdaSpace{}%
\AgdaOperator{\AgdaFunction{𝔻[}}\AgdaSpace{}%
\AgdaBound{𝑨}\AgdaSpace{}%
\AgdaOperator{\AgdaFunction{]}}\<%
\\
\>[2]\AgdaFunction{hfunc}\AgdaSpace{}%
\AgdaSymbol{=}\AgdaSpace{}%
\AgdaFunction{free-lift-func}\<%
\\
\\[\AgdaEmptyExtraSkip]%
\>[2]\AgdaFunction{hcomp}\AgdaSpace{}%
\AgdaSymbol{:}\AgdaSpace{}%
\AgdaFunction{compatible-map}\AgdaSpace{}%
\AgdaSymbol{(}\AgdaFunction{𝑻}\AgdaSpace{}%
\AgdaBound{X}\AgdaSymbol{)}\AgdaSpace{}%
\AgdaBound{𝑨}\AgdaSpace{}%
\AgdaFunction{free-lift-func}\<%
\\
\>[2]\AgdaFunction{hcomp}\AgdaSpace{}%
\AgdaSymbol{\{}\AgdaBound{f}\AgdaSymbol{\}\{}\AgdaBound{a}\AgdaSymbol{\}}\AgdaSpace{}%
\AgdaSymbol{=}\AgdaSpace{}%
\AgdaField{cong}\AgdaSpace{}%
\AgdaSymbol{(}\AgdaField{Interp}\AgdaSpace{}%
\AgdaBound{𝑨}\AgdaSymbol{)}\AgdaSpace{}%
\AgdaSymbol{(}\AgdaInductiveConstructor{≡.refl}\AgdaSpace{}%
\AgdaOperator{\AgdaInductiveConstructor{,}}\AgdaSpace{}%
\AgdaSymbol{(λ}\AgdaSpace{}%
\AgdaBound{i}\AgdaSpace{}%
\AgdaSymbol{→}\AgdaSpace{}%
\AgdaSymbol{(}\AgdaField{cong}\AgdaSpace{}%
\AgdaFunction{free-lift-func}\AgdaSymbol{)\{}\AgdaBound{a}\AgdaSpace{}%
\AgdaBound{i}\AgdaSymbol{\}}\AgdaSpace{}%
\AgdaFunction{≃-isRefl}\AgdaSymbol{))}\<%
\\
\\[\AgdaEmptyExtraSkip]%
\>[2]\AgdaFunction{hhom}\AgdaSpace{}%
\AgdaSymbol{:}\AgdaSpace{}%
\AgdaRecord{IsHom}\AgdaSpace{}%
\AgdaSymbol{(}\AgdaFunction{𝑻}\AgdaSpace{}%
\AgdaBound{X}\AgdaSymbol{)}\AgdaSpace{}%
\AgdaBound{𝑨}\AgdaSpace{}%
\AgdaFunction{hfunc}\<%
\\
\>[2]\AgdaFunction{hhom}\AgdaSpace{}%
\AgdaSymbol{=}\AgdaSpace{}%
\AgdaInductiveConstructor{mkhom}\AgdaSpace{}%
\AgdaSymbol{(λ\{}\AgdaBound{f}\AgdaSymbol{\}\{}\AgdaBound{a}\AgdaSymbol{\}}\AgdaSpace{}%
\AgdaSymbol{→}\AgdaSpace{}%
\AgdaFunction{hcomp}\AgdaSymbol{\{}\AgdaBound{f}\AgdaSymbol{\}\{}\AgdaBound{a}\AgdaSymbol{\})}\<%
\\
\>[0]\<%
\end{code}
\fi

It turns out that the interpretation of a term \ab p in an environment \ab{η} is the same
as the free lift of \ab{η} evaluated at \ab p.

\ifshort\else
\begin{code}%
\>[0]\<%
\\
\>[0]\AgdaKeyword{module}\AgdaSpace{}%
\AgdaModule{\AgdaUnderscore{}}\AgdaSpace{}%
\AgdaSymbol{\{}\AgdaBound{X}\AgdaSpace{}%
\AgdaSymbol{:}\AgdaSpace{}%
\AgdaPrimitive{Type}\AgdaSpace{}%
\AgdaGeneralizable{χ}\AgdaSymbol{\}\{}\AgdaBound{𝑨}\AgdaSpace{}%
\AgdaSymbol{:}\AgdaSpace{}%
\AgdaRecord{Algebra}\AgdaSpace{}%
\AgdaGeneralizable{α}\AgdaSpace{}%
\AgdaGeneralizable{ρᵃ}\AgdaSymbol{\}}\AgdaSpace{}%
\AgdaKeyword{where}\<%
\\
\>[0][@{}l@{\AgdaIndent{0}}]%
\>[1]\AgdaKeyword{open}\AgdaSpace{}%
\AgdaModule{Setoid}\AgdaSpace{}%
\AgdaOperator{\AgdaFunction{𝔻[}}\AgdaSpace{}%
\AgdaBound{𝑨}\AgdaSpace{}%
\AgdaOperator{\AgdaFunction{]}}%
\>[21]\AgdaKeyword{using}\AgdaSpace{}%
\AgdaSymbol{(}\AgdaSpace{}%
\AgdaOperator{\AgdaField{\AgdaUnderscore{}≈\AgdaUnderscore{}}}\AgdaSpace{}%
\AgdaSymbol{;}\AgdaSpace{}%
\AgdaFunction{refl}\AgdaSpace{}%
\AgdaSymbol{)}\<%
\\
\>[1]\AgdaKeyword{open}\AgdaSpace{}%
\AgdaModule{Environment}\AgdaSpace{}%
\AgdaBound{𝑨}%
\>[21]\AgdaKeyword{using}\AgdaSpace{}%
\AgdaSymbol{(}\AgdaSpace{}%
\AgdaOperator{\AgdaFunction{⟦\AgdaUnderscore{}⟧}}\AgdaSpace{}%
\AgdaSymbol{)}\<%
\end{code}
\fi
\begin{code}%
\>[0]\<%
\\
\>[1]\AgdaFunction{free-lift-interp}\AgdaSpace{}%
\AgdaSymbol{:}\AgdaSpace{}%
\AgdaSymbol{(}\AgdaBound{η}\AgdaSpace{}%
\AgdaSymbol{:}\AgdaSpace{}%
\AgdaBound{X}\AgdaSpace{}%
\AgdaSymbol{→}\AgdaSpace{}%
\AgdaOperator{\AgdaFunction{𝕌[}}\AgdaSpace{}%
\AgdaBound{𝑨}\AgdaSpace{}%
\AgdaOperator{\AgdaFunction{]}}\AgdaSymbol{)(}\AgdaBound{p}\AgdaSpace{}%
\AgdaSymbol{:}\AgdaSpace{}%
\AgdaDatatype{Term}\AgdaSpace{}%
\AgdaBound{X}\AgdaSymbol{)}\AgdaSpace{}%
\AgdaSymbol{→}\AgdaSpace{}%
\AgdaOperator{\AgdaFunction{⟦}}\AgdaSpace{}%
\AgdaBound{p}\AgdaSpace{}%
\AgdaOperator{\AgdaFunction{⟧}}\AgdaSpace{}%
\AgdaOperator{\AgdaField{⟨\$⟩}}\AgdaSpace{}%
\AgdaBound{η}\AgdaSpace{}%
\AgdaOperator{\AgdaFunction{≈}}\AgdaSpace{}%
\AgdaSymbol{(}\AgdaFunction{free-lift}\AgdaSymbol{\{}\AgdaArgument{𝑨}\AgdaSpace{}%
\AgdaSymbol{=}\AgdaSpace{}%
\AgdaBound{𝑨}\AgdaSymbol{\}}\AgdaSpace{}%
\AgdaBound{η}\AgdaSymbol{)}\AgdaSpace{}%
\AgdaBound{p}\<%
\\
\>[1]\AgdaFunction{free-lift-interp}\AgdaSpace{}%
\AgdaBound{η}\AgdaSpace{}%
\AgdaSymbol{(}\AgdaInductiveConstructor{ℊ}\AgdaSpace{}%
\AgdaBound{x}\AgdaSymbol{)}%
\>[32]\AgdaSymbol{=}\AgdaSpace{}%
\AgdaFunction{refl}\<%
\\
\>[1]\AgdaFunction{free-lift-interp}\AgdaSpace{}%
\AgdaBound{η}\AgdaSpace{}%
\AgdaSymbol{(}\AgdaInductiveConstructor{node}\AgdaSpace{}%
\AgdaBound{f}\AgdaSpace{}%
\AgdaBound{t}\AgdaSymbol{)}%
\>[32]\AgdaSymbol{=}\AgdaSpace{}%
\AgdaField{cong}\AgdaSpace{}%
\AgdaSymbol{(}\AgdaField{Interp}\AgdaSpace{}%
\AgdaBound{𝑨}\AgdaSymbol{)}\AgdaSpace{}%
\AgdaSymbol{(}\AgdaInductiveConstructor{≡.refl}\AgdaSpace{}%
\AgdaOperator{\AgdaInductiveConstructor{,}}\AgdaSpace{}%
\AgdaSymbol{(}\AgdaFunction{free-lift-interp}\AgdaSpace{}%
\AgdaBound{η}\AgdaSymbol{)}\AgdaSpace{}%
\AgdaOperator{\AgdaFunction{∘}}\AgdaSpace{}%
\AgdaBound{t}\AgdaSymbol{)}\<%
\end{code}

\paragraph*{The relatively free algebra in theory}
In this subsection, we describe, for a given class \ab{𝒦} of \ab{𝑆}-algebras, the
\emph{relatively free algebra} in \af{S} (\af{P} \ab{𝒦}) over \ab X, using the informal
language that is typical of mathematics literature. In the next section we will present
the relatively free algebra in Agda using the formal language of type theory.

Above we defined the term algebra \T{X}, which is free in the class of all
\ab{𝑆}-algebras; that is, \T{X} has the universal property and belongs to the class of
\ab{𝑆}-algebras.  Given an arbitrary class \ab{𝒦} of \ab{𝑆}-algebras, we can't expect that
\T{X} belongs to \ab{𝒦}, so, in general, we say that \T{X} is free \emph{for} \ab{𝒦}.
\ifshort\else
Indeed, it might not be possible to find a free algebra that belongs to \ab{𝒦}.
\fi
However, for any class \ab{𝒦} we can construct an algebra that is free for \ab{𝒦}
and belongs to the class \af{S} (\af{P} \ab{𝒦}), and for most applications this suffices.

The informal construction of the free algebra in \af{S} (\af{P} \ab{𝒦}), for an arbitrary
class \ab{𝒦} of \ab{𝑆}-algebras, proceeds by taking the quotient of \T{X} modulo a congruence relation
that we will denote by \afld{≈}.  One approach is to let
\afld{≈} := \af{⋂}\{\ab{θ} \af{∈} \af{Con} (\T{X}) : \T{X} \af{/} \ab{θ} \af{∈} \af{S}
\ab{𝒦}\}.\footnote{\af{Con} (\T{X}) is the set of congruences of \T{X}.}
Alternatively we could let \ab{ℰ} = \af{Th} \ab{𝒦} and take \afld{≈} to be the least equivalence relation
on the domain of \T{X} such that
\begin{enumerate}
\item for every equation (\ab p , \ab q) \af{∈} \af{Th} \ab{𝒦} and every
environment \ab{ρ} : \ab X \as{→} \Term{X}, we have\\
\af{⟦~\ab p~⟧} \afld{⟨\$⟩} \ab{ρ} \afld{≈} \af{⟦~\ab q~⟧} \afld{⟨\$⟩} \ab{ρ}, and
\item \afld{≈} is a congruence of \T{X}; that is, for every operation symbol \ab
f : \af{∣~\ab{𝑆}~∣}, and for all tuples \ab{s} \ab{t} : \af{∥~\ab{𝑆}~∥} \ab f
→ \Term{X}, the following implication holds:\footnote{Here all
interpretations, denoted by \af{⟦\au{}⟧}, are with respect to \T{X}.}\\[-8pt]

(∀ i → \af{⟦~\ab{s}~\ab i~⟧}~\afld{⟨\$⟩}~\ab{ρ}~\afld{≈}~\af{⟦~\ab{t}~\ab
i~⟧}~\afld{⟨\$⟩}~\ab{ρ})
\as{→} \af{⟦~\ab f~\ab s~⟧}~\afld{⟨\$⟩}~\ab{ρ}~\afld{≈}~\af{⟦~\ab f~\ab
t~⟧}~\afld{⟨\$⟩}~\ab{ρ}\\[-8pt]
\end{enumerate}
Whichever approach we choose, the \defn{relatively free algebra over} \ab{X} (relative to
\ab{𝒦}) is defined to be the quotient \Free{X} := \T{X}~\af{/}~\afld{≈}.

Evidently \Free{X} is a subdirect product of the algebras in \{\T{X}~\af{/}~\ab{θ}\!\},
where \ab{θ} ranges over congruences modulo which \T{X} belongs to \af{S}~\ab{𝒦}.
Thus, \Free{X} \af{∈} \af{P}(\af{S}~\ab{𝒦}) ⊆ \af{S}(\af{P}~\ab{𝒦}), and it follows
that \Free{X} satisfies the identities in \af{Th} \ab{𝒦} (those modeled by all members of
\ab{𝒦}).  Indeed, for each pair \ab p \ab q : \Term{X}, if \ab{𝒦} \af{⊫} \ab p \af{≈} \ab
q, then \ab p and \ab q must belong to the same \afld{≈}-class, so \ab p and \ab q are
identified in \Free{X}. \ifshort\else (Notice that \afld{≈} may be empty, in which case
\T{X}~\af{/}~\afld{≈} is trivial.) \fi

\paragraph*{The relatively free algebra in Agda}
We now define the relatively free algebra in Agda using the language of type theory.
Our approach will be different from the informal one described above, but the end result
will be the same.
We start with a type \ab{ℰ} representing a collection of identities and, instead of
forming a quotient, we take the domain of the free algebra to be a setoid whose
\afld{Carrier} is the type \Term{X} of {𝑆}-terms in \ab X and whose equivalence relation
includes all pairs (\ab p , \ab q) \af{∈} \Term{X} \af{×} \Term{X} such that \ab p \aod{≈}
\ab q is derivable from \ab{ℰ}; that is, \ab{ℰ} \aod{⊢} \ab X \aod{▹} \ab p \aod{≈} \ab q.
Observe that elements of this setoid are equal iff they belong to the same equivalence
class of the congruence \afld{≈} defined above.  Therefore, the setoid so defined represents
the quotient \T{X}~\af{/}~\afld{≈}.
Finally, the interpretation of an operation in the free algebra is simply the operation
itself, which works since \ab{ℰ} \aod{⊢} \ab X \aod{▹\au{}≈\au{}} is a congruence relation.

\begin{code}%
\>[0]\<%
\\
\>[0]\AgdaKeyword{module}\AgdaSpace{}%
\AgdaModule{FreeAlgebra}\AgdaSpace{}%
\AgdaSymbol{\{}\AgdaBound{χ}\AgdaSpace{}%
\AgdaSymbol{:}\AgdaSpace{}%
\AgdaPostulate{Level}\AgdaSymbol{\}(}\AgdaBound{ℰ}\AgdaSpace{}%
\AgdaSymbol{:}\AgdaSpace{}%
\AgdaSymbol{\{}\AgdaBound{Y}\AgdaSpace{}%
\AgdaSymbol{:}\AgdaSpace{}%
\AgdaPrimitive{Type}\AgdaSpace{}%
\AgdaBound{χ}\AgdaSymbol{\}}\AgdaSpace{}%
\AgdaSymbol{→}\AgdaSpace{}%
\AgdaFunction{Pred}\AgdaSpace{}%
\AgdaSymbol{(}\AgdaDatatype{Term}\AgdaSpace{}%
\AgdaBound{Y}\AgdaSpace{}%
\AgdaOperator{\AgdaFunction{×}}\AgdaSpace{}%
\AgdaDatatype{Term}\AgdaSpace{}%
\AgdaBound{Y}\AgdaSymbol{)}\AgdaSpace{}%
\AgdaSymbol{(}\AgdaFunction{ov}\AgdaSpace{}%
\AgdaBound{χ}\AgdaSymbol{))}\AgdaSpace{}%
\AgdaKeyword{where}\<%
\\
\\[\AgdaEmptyExtraSkip]%
\>[0][@{}l@{\AgdaIndent{0}}]%
\>[1]\AgdaFunction{FreeDomain}\AgdaSpace{}%
\AgdaSymbol{:}\AgdaSpace{}%
\AgdaPrimitive{Type}\AgdaSpace{}%
\AgdaBound{χ}\AgdaSpace{}%
\AgdaSymbol{→}\AgdaSpace{}%
\AgdaRecord{Setoid}\AgdaSpace{}%
\AgdaSymbol{\AgdaUnderscore{}}\AgdaSpace{}%
\AgdaSymbol{\AgdaUnderscore{}}\<%
\\
\>[1]\AgdaFunction{FreeDomain}\AgdaSpace{}%
\AgdaBound{X}\AgdaSpace{}%
\AgdaSymbol{=}\<%
\\
\>[1][@{}l@{\AgdaIndent{0}}]%
\>[2]\AgdaKeyword{record}%
\>[10]\AgdaSymbol{\{}\AgdaSpace{}%
\AgdaField{Carrier}%
\>[27]\AgdaSymbol{=}\AgdaSpace{}%
\AgdaDatatype{Term}\AgdaSpace{}%
\AgdaBound{X}\<%
\\
\>[10]\AgdaSymbol{;}\AgdaSpace{}%
\AgdaOperator{\AgdaField{\AgdaUnderscore{}≈\AgdaUnderscore{}}}%
\>[27]\AgdaSymbol{=}\AgdaSpace{}%
\AgdaBound{ℰ}\AgdaSpace{}%
\AgdaOperator{\AgdaDatatype{⊢}}\AgdaSpace{}%
\AgdaBound{X}\AgdaSpace{}%
\AgdaOperator{\AgdaDatatype{▹\AgdaUnderscore{}≈\AgdaUnderscore{}}}\<%
\\
\>[10]\AgdaSymbol{;}\AgdaSpace{}%
\AgdaField{isEquivalence}%
\>[27]\AgdaSymbol{=}\AgdaSpace{}%
\AgdaKeyword{record}\AgdaSpace{}%
\AgdaSymbol{\{}\AgdaSpace{}%
\AgdaField{refl}\AgdaSpace{}%
\AgdaSymbol{=}\AgdaSpace{}%
\AgdaInductiveConstructor{reflexive}\AgdaSpace{}%
\AgdaSymbol{;}\AgdaSpace{}%
\AgdaField{sym}\AgdaSpace{}%
\AgdaSymbol{=}\AgdaSpace{}%
\AgdaInductiveConstructor{symmetric}\AgdaSpace{}%
\AgdaSymbol{;}\AgdaSpace{}%
\AgdaField{trans}\AgdaSpace{}%
\AgdaSymbol{=}\AgdaSpace{}%
\AgdaInductiveConstructor{transitive}\AgdaSpace{}%
\AgdaSymbol{\}}\AgdaSpace{}%
\AgdaSymbol{\}}\<%
\\
\\[\AgdaEmptyExtraSkip]%
\>[1]\AgdaOperator{\AgdaFunction{𝔽[\AgdaUnderscore{}]}}\AgdaSpace{}%
\AgdaSymbol{:}\AgdaSpace{}%
\AgdaPrimitive{Type}\AgdaSpace{}%
\AgdaBound{χ}\AgdaSpace{}%
\AgdaSymbol{→}\AgdaSpace{}%
\AgdaRecord{Algebra}\AgdaSpace{}%
\AgdaSymbol{(}\AgdaFunction{ov}\AgdaSpace{}%
\AgdaBound{χ}\AgdaSymbol{)}\AgdaSpace{}%
\AgdaSymbol{\AgdaUnderscore{}}\<%
\\
\>[1]\AgdaField{Domain}\AgdaSpace{}%
\AgdaOperator{\AgdaFunction{𝔽[}}\AgdaSpace{}%
\AgdaBound{X}\AgdaSpace{}%
\AgdaOperator{\AgdaFunction{]}}\AgdaSpace{}%
\AgdaSymbol{=}\AgdaSpace{}%
\AgdaFunction{FreeDomain}\AgdaSpace{}%
\AgdaBound{X}\<%
\\
\>[1]\AgdaField{Interp}\AgdaSpace{}%
\AgdaOperator{\AgdaFunction{𝔽[}}\AgdaSpace{}%
\AgdaBound{X}\AgdaSpace{}%
\AgdaOperator{\AgdaFunction{]}}\AgdaSpace{}%
\AgdaSymbol{=}\AgdaSpace{}%
\AgdaFunction{FreeInterp}\AgdaSpace{}%
\AgdaKeyword{where}\<%
\\
\>[1][@{}l@{\AgdaIndent{0}}]%
\>[2]\AgdaFunction{FreeInterp}\AgdaSpace{}%
\AgdaSymbol{:}\AgdaSpace{}%
\AgdaSymbol{∀}\AgdaSpace{}%
\AgdaSymbol{\{}\AgdaBound{X}\AgdaSymbol{\}}\AgdaSpace{}%
\AgdaSymbol{→}\AgdaSpace{}%
\AgdaOperator{\AgdaFunction{⟨}}\AgdaSpace{}%
\AgdaBound{𝑆}\AgdaSpace{}%
\AgdaOperator{\AgdaFunction{⟩}}\AgdaSpace{}%
\AgdaSymbol{(}\AgdaFunction{FreeDomain}\AgdaSpace{}%
\AgdaBound{X}\AgdaSymbol{)}\AgdaSpace{}%
\AgdaOperator{\AgdaRecord{⟶}}\AgdaSpace{}%
\AgdaFunction{FreeDomain}\AgdaSpace{}%
\AgdaBound{X}\<%
\\
\>[2]\AgdaFunction{FreeInterp}\AgdaSpace{}%
\AgdaOperator{\AgdaField{⟨\$⟩}}\AgdaSpace{}%
\AgdaSymbol{(}\AgdaBound{f}\AgdaSpace{}%
\AgdaOperator{\AgdaInductiveConstructor{,}}\AgdaSpace{}%
\AgdaBound{ts}\AgdaSymbol{)}%
\>[32]\AgdaSymbol{=}\AgdaSpace{}%
\AgdaInductiveConstructor{node}\AgdaSpace{}%
\AgdaBound{f}\AgdaSpace{}%
\AgdaBound{ts}\<%
\\
\>[2]\AgdaField{cong}\AgdaSpace{}%
\AgdaFunction{FreeInterp}\AgdaSpace{}%
\AgdaSymbol{(}\AgdaInductiveConstructor{≡.refl}\AgdaSpace{}%
\AgdaOperator{\AgdaInductiveConstructor{,}}\AgdaSpace{}%
\AgdaBound{h}\AgdaSymbol{)}%
\>[32]\AgdaSymbol{=}\AgdaSpace{}%
\AgdaInductiveConstructor{app}\AgdaSpace{}%
\AgdaBound{h}\<%
\end{code}

\paragraph*{The natural epimorphism} % from 𝑻 X to 𝔽[ X ]}
We now define the natural epimorphism from \T{X} onto the relatively free algebra \Free{X} and prove that
its kernel is the congruence of \T{X} defined by the identities modeled by (\af S \ab{𝒦}, hence by) \ab{𝒦}.

\ifshort\else
\begin{code}%
\>[0]\<%
\\
\>[0]\AgdaKeyword{module}\AgdaSpace{}%
\AgdaModule{FreeHom}\AgdaSpace{}%
\AgdaSymbol{\{}\AgdaBound{𝒦}\AgdaSpace{}%
\AgdaSymbol{:}\AgdaSpace{}%
\AgdaFunction{Pred}\AgdaSymbol{(}\AgdaRecord{Algebra}\AgdaSpace{}%
\AgdaGeneralizable{α}\AgdaSpace{}%
\AgdaGeneralizable{ρᵃ}\AgdaSymbol{)}\AgdaSpace{}%
\AgdaSymbol{(}\AgdaGeneralizable{α}\AgdaSpace{}%
\AgdaOperator{\AgdaPrimitive{⊔}}\AgdaSpace{}%
\AgdaGeneralizable{ρᵃ}\AgdaSpace{}%
\AgdaOperator{\AgdaPrimitive{⊔}}\AgdaSpace{}%
\AgdaFunction{ov}\AgdaSpace{}%
\AgdaGeneralizable{ℓ}\AgdaSymbol{)\}}\AgdaSpace{}%
\AgdaKeyword{where}\<%
\\
\>[0][@{}l@{\AgdaIndent{0}}]%
\>[1]\AgdaKeyword{private}\AgdaSpace{}%
\AgdaFunction{c}\AgdaSpace{}%
\AgdaSymbol{=}\AgdaSpace{}%
\AgdaBound{α}\AgdaSpace{}%
\AgdaOperator{\AgdaPrimitive{⊔}}\AgdaSpace{}%
\AgdaBound{ρᵃ}\AgdaSpace{}%
\AgdaOperator{\AgdaPrimitive{⊔}}\AgdaSpace{}%
\AgdaBound{ℓ}\AgdaSpace{}%
\AgdaSymbol{;}\AgdaSpace{}%
\AgdaFunction{ι}\AgdaSpace{}%
\AgdaSymbol{=}\AgdaSpace{}%
\AgdaFunction{ov}\AgdaSpace{}%
\AgdaFunction{c}\<%
\\
\>[1]\AgdaKeyword{open}\AgdaSpace{}%
\AgdaModule{FreeAlgebra}\AgdaSpace{}%
\AgdaSymbol{\{}\AgdaArgument{χ}\AgdaSpace{}%
\AgdaSymbol{=}\AgdaSpace{}%
\AgdaFunction{c}\AgdaSymbol{\}}\AgdaSpace{}%
\AgdaSymbol{(}\AgdaFunction{Th}\AgdaSpace{}%
\AgdaBound{𝒦}\AgdaSymbol{)}\AgdaSpace{}%
\AgdaKeyword{using}\AgdaSpace{}%
\AgdaSymbol{(}\AgdaSpace{}%
\AgdaOperator{\AgdaFunction{𝔽[\AgdaUnderscore{}]}}\AgdaSpace{}%
\AgdaSymbol{)}\<%
\end{code}
\fi
\begin{code}%
\>[0]\<%
\\
\>[1]\AgdaOperator{\AgdaFunction{epiF[\AgdaUnderscore{}]}}\AgdaSpace{}%
\AgdaSymbol{:}\AgdaSpace{}%
\AgdaSymbol{(}\AgdaBound{X}\AgdaSpace{}%
\AgdaSymbol{:}\AgdaSpace{}%
\AgdaPrimitive{Type}\AgdaSpace{}%
\AgdaFunction{c}\AgdaSymbol{)}\AgdaSpace{}%
\AgdaSymbol{→}\AgdaSpace{}%
\AgdaFunction{epi}\AgdaSpace{}%
\AgdaSymbol{(}\AgdaFunction{𝑻}\AgdaSpace{}%
\AgdaBound{X}\AgdaSymbol{)}\AgdaSpace{}%
\AgdaOperator{\AgdaFunction{𝔽[}}\AgdaSpace{}%
\AgdaBound{X}\AgdaSpace{}%
\AgdaOperator{\AgdaFunction{]}}\<%
\\
\>[1]\AgdaOperator{\AgdaFunction{epiF[}}\AgdaSpace{}%
\AgdaBound{X}\AgdaSpace{}%
\AgdaOperator{\AgdaFunction{]}}\AgdaSpace{}%
\AgdaSymbol{=}\AgdaSpace{}%
\AgdaFunction{h}\AgdaSpace{}%
\AgdaOperator{\AgdaInductiveConstructor{,}}\AgdaSpace{}%
\AgdaFunction{hepi}\<%
\\
\>[1][@{}l@{\AgdaIndent{0}}]%
\>[2]\AgdaKeyword{where}\<%
\\
\>[2]\AgdaKeyword{open}\AgdaSpace{}%
\AgdaModule{Setoid}\AgdaSpace{}%
\AgdaOperator{\AgdaFunction{𝔻[}}\AgdaSpace{}%
\AgdaFunction{𝑻}\AgdaSpace{}%
\AgdaBound{X}\AgdaSpace{}%
\AgdaOperator{\AgdaFunction{]}}%
\>[27]\AgdaKeyword{using}\AgdaSpace{}%
\AgdaSymbol{()}%
\>[43]\AgdaKeyword{renaming}\AgdaSpace{}%
\AgdaSymbol{(}\AgdaSpace{}%
\AgdaOperator{\AgdaField{\AgdaUnderscore{}≈\AgdaUnderscore{}}}\AgdaSpace{}%
\AgdaSymbol{to}\AgdaSpace{}%
\AgdaOperator{\AgdaField{\AgdaUnderscore{}≈₀\AgdaUnderscore{}}}%
\>[67]\AgdaSymbol{;}\AgdaSpace{}%
\AgdaFunction{refl}\AgdaSpace{}%
\AgdaSymbol{to}\AgdaSpace{}%
\AgdaFunction{reflᵀ}\AgdaSpace{}%
\AgdaSymbol{)}\<%
\\
\>[2]\AgdaKeyword{open}\AgdaSpace{}%
\AgdaModule{Setoid}\AgdaSpace{}%
\AgdaOperator{\AgdaFunction{𝔻[}}\AgdaSpace{}%
\AgdaOperator{\AgdaFunction{𝔽[}}\AgdaSpace{}%
\AgdaBound{X}\AgdaSpace{}%
\AgdaOperator{\AgdaFunction{]}}\AgdaSpace{}%
\AgdaOperator{\AgdaFunction{]}}%
\>[27]\AgdaKeyword{using}\AgdaSpace{}%
\AgdaSymbol{(}\AgdaSpace{}%
\AgdaFunction{refl}\AgdaSpace{}%
\AgdaSymbol{)}%
\>[43]\AgdaKeyword{renaming}\AgdaSpace{}%
\AgdaSymbol{(}\AgdaSpace{}%
\AgdaOperator{\AgdaField{\AgdaUnderscore{}≈\AgdaUnderscore{}}}\AgdaSpace{}%
\AgdaSymbol{to}\AgdaSpace{}%
\AgdaOperator{\AgdaField{\AgdaUnderscore{}≈₁\AgdaUnderscore{}}}%
\>[67]\AgdaSymbol{)}\<%
\\
\\[\AgdaEmptyExtraSkip]%
\>[2]\AgdaFunction{con}\AgdaSpace{}%
\AgdaSymbol{:}\AgdaSpace{}%
\AgdaSymbol{∀}\AgdaSpace{}%
\AgdaSymbol{\{}\AgdaBound{x}\AgdaSpace{}%
\AgdaBound{y}\AgdaSymbol{\}}\AgdaSpace{}%
\AgdaSymbol{→}\AgdaSpace{}%
\AgdaBound{x}\AgdaSpace{}%
\AgdaOperator{\AgdaFunction{≈₀}}\AgdaSpace{}%
\AgdaBound{y}\AgdaSpace{}%
\AgdaSymbol{→}\AgdaSpace{}%
\AgdaBound{x}\AgdaSpace{}%
\AgdaOperator{\AgdaFunction{≈₁}}\AgdaSpace{}%
\AgdaBound{y}\<%
\\
\>[2]\AgdaFunction{con}\AgdaSpace{}%
\AgdaSymbol{(}\AgdaInductiveConstructor{rfl}\AgdaSpace{}%
\AgdaSymbol{\{}\AgdaBound{x}\AgdaSymbol{\}\{}\AgdaBound{y}\AgdaSymbol{\}}\AgdaSpace{}%
\AgdaInductiveConstructor{≡.refl}\AgdaSymbol{)}\AgdaSpace{}%
\AgdaSymbol{=}\AgdaSpace{}%
\AgdaFunction{refl}\<%
\\
\>[2]\AgdaFunction{con}\AgdaSpace{}%
\AgdaSymbol{(}\AgdaInductiveConstructor{gnl}\AgdaSpace{}%
\AgdaSymbol{\{}\AgdaBound{f}\AgdaSymbol{\}\{}\AgdaBound{s}\AgdaSymbol{\}\{}\AgdaBound{t}\AgdaSymbol{\}}\AgdaSpace{}%
\AgdaBound{x}\AgdaSymbol{)}\AgdaSpace{}%
\AgdaSymbol{=}\AgdaSpace{}%
\AgdaField{cong}\AgdaSpace{}%
\AgdaSymbol{(}\AgdaField{Interp}\AgdaSpace{}%
\AgdaOperator{\AgdaFunction{𝔽[}}\AgdaSpace{}%
\AgdaBound{X}\AgdaSpace{}%
\AgdaOperator{\AgdaFunction{]}}\AgdaSymbol{)}\AgdaSpace{}%
\AgdaSymbol{(}\AgdaInductiveConstructor{≡.refl}\AgdaSpace{}%
\AgdaOperator{\AgdaInductiveConstructor{,}}\AgdaSpace{}%
\AgdaFunction{con}\AgdaSpace{}%
\AgdaOperator{\AgdaFunction{∘}}\AgdaSpace{}%
\AgdaBound{x}\AgdaSymbol{)}\<%
\\
\\[\AgdaEmptyExtraSkip]%
\>[2]\AgdaFunction{h}\AgdaSpace{}%
\AgdaSymbol{:}\AgdaSpace{}%
\AgdaOperator{\AgdaFunction{𝔻[}}\AgdaSpace{}%
\AgdaFunction{𝑻}\AgdaSpace{}%
\AgdaBound{X}\AgdaSpace{}%
\AgdaOperator{\AgdaFunction{]}}\AgdaSpace{}%
\AgdaOperator{\AgdaRecord{⟶}}\AgdaSpace{}%
\AgdaOperator{\AgdaFunction{𝔻[}}\AgdaSpace{}%
\AgdaOperator{\AgdaFunction{𝔽[}}\AgdaSpace{}%
\AgdaBound{X}\AgdaSpace{}%
\AgdaOperator{\AgdaFunction{]}}\AgdaSpace{}%
\AgdaOperator{\AgdaFunction{]}}\<%
\\
\>[2]\AgdaFunction{h}\AgdaSpace{}%
\AgdaSymbol{=}\AgdaSpace{}%
\AgdaKeyword{record}\AgdaSpace{}%
\AgdaSymbol{\{}\AgdaSpace{}%
\AgdaField{f}\AgdaSpace{}%
\AgdaSymbol{=}\AgdaSpace{}%
\AgdaFunction{id}\AgdaSpace{}%
\AgdaSymbol{;}\AgdaSpace{}%
\AgdaField{cong}\AgdaSpace{}%
\AgdaSymbol{=}\AgdaSpace{}%
\AgdaFunction{con}\AgdaSpace{}%
\AgdaSymbol{\}}\<%
\\
\\[\AgdaEmptyExtraSkip]%
\>[2]\AgdaFunction{hepi}\AgdaSpace{}%
\AgdaSymbol{:}\AgdaSpace{}%
\AgdaRecord{IsEpi}\AgdaSpace{}%
\AgdaSymbol{(}\AgdaFunction{𝑻}\AgdaSpace{}%
\AgdaBound{X}\AgdaSymbol{)}\AgdaSpace{}%
\AgdaOperator{\AgdaFunction{𝔽[}}\AgdaSpace{}%
\AgdaBound{X}\AgdaSpace{}%
\AgdaOperator{\AgdaFunction{]}}\AgdaSpace{}%
\AgdaFunction{h}\<%
\\
\>[2]\AgdaField{compatible}\AgdaSpace{}%
\AgdaSymbol{(}\AgdaField{isHom}\AgdaSpace{}%
\AgdaFunction{hepi}\AgdaSymbol{)}\AgdaSpace{}%
\AgdaSymbol{=}\AgdaSpace{}%
\AgdaField{cong}\AgdaSpace{}%
\AgdaFunction{h}\AgdaSpace{}%
\AgdaFunction{reflᵀ}\<%
\\
\>[2]\AgdaField{isSurjective}\AgdaSpace{}%
\AgdaFunction{hepi}\AgdaSpace{}%
\AgdaSymbol{\{}\AgdaBound{y}\AgdaSymbol{\}}\AgdaSpace{}%
\AgdaSymbol{=}\AgdaSpace{}%
\AgdaInductiveConstructor{eq}\AgdaSpace{}%
\AgdaBound{y}\AgdaSpace{}%
\AgdaFunction{refl}\<%
\\
\\[\AgdaEmptyExtraSkip]%
\>[1]\AgdaOperator{\AgdaFunction{homF[\AgdaUnderscore{}]}}\AgdaSpace{}%
\AgdaSymbol{:}\AgdaSpace{}%
\AgdaSymbol{(}\AgdaBound{X}\AgdaSpace{}%
\AgdaSymbol{:}\AgdaSpace{}%
\AgdaPrimitive{Type}\AgdaSpace{}%
\AgdaFunction{c}\AgdaSymbol{)}\AgdaSpace{}%
\AgdaSymbol{→}\AgdaSpace{}%
\AgdaFunction{hom}\AgdaSpace{}%
\AgdaSymbol{(}\AgdaFunction{𝑻}\AgdaSpace{}%
\AgdaBound{X}\AgdaSymbol{)}\AgdaSpace{}%
\AgdaOperator{\AgdaFunction{𝔽[}}\AgdaSpace{}%
\AgdaBound{X}\AgdaSpace{}%
\AgdaOperator{\AgdaFunction{]}}\<%
\\
\>[1]\AgdaOperator{\AgdaFunction{homF[}}\AgdaSpace{}%
\AgdaBound{X}\AgdaSpace{}%
\AgdaOperator{\AgdaFunction{]}}\AgdaSpace{}%
\AgdaSymbol{=}\AgdaSpace{}%
\AgdaFunction{IsEpi.HomReduct}\AgdaSpace{}%
\AgdaOperator{\AgdaFunction{∥}}\AgdaSpace{}%
\AgdaOperator{\AgdaFunction{epiF[}}\AgdaSpace{}%
\AgdaBound{X}\AgdaSpace{}%
\AgdaOperator{\AgdaFunction{]}}\AgdaSpace{}%
\AgdaOperator{\AgdaFunction{∥}}\<%
\\
\\[\AgdaEmptyExtraSkip]%
\>[1]\AgdaFunction{kernel-in-theory}\AgdaSpace{}%
\AgdaSymbol{:}\AgdaSpace{}%
\AgdaSymbol{\{}\AgdaBound{X}\AgdaSpace{}%
\AgdaSymbol{:}\AgdaSpace{}%
\AgdaPrimitive{Type}\AgdaSpace{}%
\AgdaFunction{c}\AgdaSymbol{\}}\AgdaSpace{}%
\AgdaSymbol{→}\AgdaSpace{}%
\AgdaFunction{ker}\AgdaSpace{}%
\AgdaOperator{\AgdaFunction{∣}}\AgdaSpace{}%
\AgdaOperator{\AgdaFunction{homF[}}\AgdaSpace{}%
\AgdaBound{X}\AgdaSpace{}%
\AgdaOperator{\AgdaFunction{]}}\AgdaSpace{}%
\AgdaOperator{\AgdaFunction{∣}}\AgdaSpace{}%
\AgdaOperator{\AgdaFunction{⊆}}\AgdaSpace{}%
\AgdaFunction{Th}\AgdaSpace{}%
\AgdaSymbol{(}\AgdaFunction{V}\AgdaSpace{}%
\AgdaBound{ℓ}\AgdaSpace{}%
\AgdaFunction{ι}\AgdaSpace{}%
\AgdaBound{𝒦}\AgdaSymbol{)}\<%
\\
\>[1]\AgdaFunction{kernel-in-theory}\AgdaSpace{}%
\AgdaSymbol{\{}\AgdaArgument{X}\AgdaSpace{}%
\AgdaSymbol{=}\AgdaSpace{}%
\AgdaBound{X}\AgdaSymbol{\}}\AgdaSpace{}%
\AgdaSymbol{\{}\AgdaBound{p}\AgdaSpace{}%
\AgdaOperator{\AgdaInductiveConstructor{,}}\AgdaSpace{}%
\AgdaBound{q}\AgdaSymbol{\}}\AgdaSpace{}%
\AgdaBound{pKq}\AgdaSpace{}%
\AgdaBound{𝑨}\AgdaSpace{}%
\AgdaBound{vkA}\AgdaSpace{}%
\AgdaSymbol{=}\AgdaSpace{}%
\AgdaFunction{V-id1}\AgdaSymbol{\{}\AgdaArgument{ℓ}\AgdaSpace{}%
\AgdaSymbol{=}\AgdaSpace{}%
\AgdaBound{ℓ}\AgdaSymbol{\}\{}\AgdaArgument{p}\AgdaSpace{}%
\AgdaSymbol{=}\AgdaSpace{}%
\AgdaBound{p}\AgdaSymbol{\}\{}\AgdaBound{q}\AgdaSymbol{\}}\AgdaSpace{}%
\AgdaSymbol{(}\AgdaFunction{ζ}\AgdaSpace{}%
\AgdaBound{pKq}\AgdaSymbol{)}\AgdaSpace{}%
\AgdaBound{𝑨}\AgdaSpace{}%
\AgdaBound{vkA}\<%
\\
\>[1][@{}l@{\AgdaIndent{0}}]%
\>[2]\AgdaKeyword{where}\<%
\\
\>[2]\AgdaFunction{ζ}\AgdaSpace{}%
\AgdaSymbol{:}\AgdaSpace{}%
\AgdaSymbol{∀\{}\AgdaBound{p}\AgdaSpace{}%
\AgdaBound{q}\AgdaSymbol{\}}\AgdaSpace{}%
\AgdaSymbol{→}\AgdaSpace{}%
\AgdaSymbol{(}\AgdaFunction{Th}\AgdaSpace{}%
\AgdaBound{𝒦}\AgdaSymbol{)}\AgdaSpace{}%
\AgdaOperator{\AgdaDatatype{⊢}}\AgdaSpace{}%
\AgdaBound{X}\AgdaSpace{}%
\AgdaOperator{\AgdaDatatype{▹}}\AgdaSpace{}%
\AgdaBound{p}\AgdaSpace{}%
\AgdaOperator{\AgdaDatatype{≈}}\AgdaSpace{}%
\AgdaBound{q}\AgdaSpace{}%
\AgdaSymbol{→}\AgdaSpace{}%
\AgdaBound{𝒦}\AgdaSpace{}%
\AgdaOperator{\AgdaFunction{⊫}}\AgdaSpace{}%
\AgdaBound{p}\AgdaSpace{}%
\AgdaOperator{\AgdaFunction{≈}}\AgdaSpace{}%
\AgdaBound{q}\<%
\\
\>[2]\AgdaFunction{ζ}\AgdaSpace{}%
\AgdaBound{x}\AgdaSpace{}%
\AgdaBound{𝑨}\AgdaSpace{}%
\AgdaBound{kA}\AgdaSpace{}%
\AgdaSymbol{=}\AgdaSpace{}%
\AgdaFunction{sound}\AgdaSpace{}%
\AgdaSymbol{(λ}\AgdaSpace{}%
\AgdaBound{y}\AgdaSpace{}%
\AgdaBound{ρ}\AgdaSpace{}%
\AgdaSymbol{→}\AgdaSpace{}%
\AgdaBound{y}\AgdaSpace{}%
\AgdaBound{𝑨}\AgdaSpace{}%
\AgdaBound{kA}\AgdaSpace{}%
\AgdaBound{ρ}\AgdaSymbol{)}\AgdaSpace{}%
\AgdaBound{x}\AgdaSpace{}%
\AgdaKeyword{where}\AgdaSpace{}%
\AgdaKeyword{open}\AgdaSpace{}%
\AgdaModule{Soundness}\AgdaSpace{}%
\AgdaSymbol{(}\AgdaFunction{Th}\AgdaSpace{}%
\AgdaBound{𝒦}\AgdaSymbol{)}\AgdaSpace{}%
\AgdaBound{𝑨}\<%
\\
\>[0]\<%
\end{code}
Next we prove an important property of the relatively free algebra (relative to \ab{𝒦} and satisfying the identities in \af{Th} \ab{𝒦}), which will be used in the formalization of the HSP theorem; this is the assertion that for every algebra 𝑨, if \ab{𝑨} \af{⊨} \ab{Th} (\af{V} \ab{𝒦}), then there exists an epimorphism from \Free{A} onto \ab{𝑨}.

\ifshort\else
\begin{code}%
\>[0]\<%
\\
\>[0]\AgdaKeyword{module}\AgdaSpace{}%
\AgdaModule{\AgdaUnderscore{}}%
\>[10]\AgdaSymbol{\{}\AgdaBound{𝑨}\AgdaSpace{}%
\AgdaSymbol{:}\AgdaSpace{}%
\AgdaRecord{Algebra}\AgdaSpace{}%
\AgdaSymbol{(}\AgdaGeneralizable{α}\AgdaSpace{}%
\AgdaOperator{\AgdaPrimitive{⊔}}\AgdaSpace{}%
\AgdaGeneralizable{ρᵃ}\AgdaSpace{}%
\AgdaOperator{\AgdaPrimitive{⊔}}\AgdaSpace{}%
\AgdaGeneralizable{ℓ}\AgdaSymbol{)}\AgdaSpace{}%
\AgdaSymbol{(}\AgdaGeneralizable{α}\AgdaSpace{}%
\AgdaOperator{\AgdaPrimitive{⊔}}\AgdaSpace{}%
\AgdaGeneralizable{ρᵃ}\AgdaSpace{}%
\AgdaOperator{\AgdaPrimitive{⊔}}\AgdaSpace{}%
\AgdaGeneralizable{ℓ}\AgdaSymbol{)\}}\AgdaSpace{}%
\AgdaSymbol{\{}\AgdaBound{𝒦}\AgdaSpace{}%
\AgdaSymbol{:}\AgdaSpace{}%
\AgdaFunction{Pred}\AgdaSymbol{(}\AgdaRecord{Algebra}\AgdaSpace{}%
\AgdaGeneralizable{α}\AgdaSpace{}%
\AgdaGeneralizable{ρᵃ}\AgdaSymbol{)}\AgdaSpace{}%
\AgdaSymbol{(}\AgdaGeneralizable{α}\AgdaSpace{}%
\AgdaOperator{\AgdaPrimitive{⊔}}\AgdaSpace{}%
\AgdaGeneralizable{ρᵃ}\AgdaSpace{}%
\AgdaOperator{\AgdaPrimitive{⊔}}\AgdaSpace{}%
\AgdaFunction{ov}\AgdaSpace{}%
\AgdaGeneralizable{ℓ}\AgdaSymbol{)\}}\AgdaSpace{}%
\AgdaKeyword{where}\<%
\\
\>[0][@{}l@{\AgdaIndent{0}}]%
\>[1]\AgdaKeyword{private}\AgdaSpace{}%
\AgdaFunction{c}\AgdaSpace{}%
\AgdaSymbol{=}\AgdaSpace{}%
\AgdaBound{α}\AgdaSpace{}%
\AgdaOperator{\AgdaPrimitive{⊔}}\AgdaSpace{}%
\AgdaBound{ρᵃ}\AgdaSpace{}%
\AgdaOperator{\AgdaPrimitive{⊔}}\AgdaSpace{}%
\AgdaBound{ℓ}\AgdaSpace{}%
\AgdaSymbol{;}\AgdaSpace{}%
\AgdaFunction{ι}\AgdaSpace{}%
\AgdaSymbol{=}\AgdaSpace{}%
\AgdaFunction{ov}\AgdaSpace{}%
\AgdaFunction{c}\<%
\\
\>[1]\AgdaKeyword{open}\AgdaSpace{}%
\AgdaModule{FreeHom}\AgdaSpace{}%
\AgdaSymbol{\{}\AgdaArgument{ℓ}\AgdaSpace{}%
\AgdaSymbol{=}\AgdaSpace{}%
\AgdaBound{ℓ}\AgdaSymbol{\}}\AgdaSpace{}%
\AgdaSymbol{\{}\AgdaBound{𝒦}\AgdaSymbol{\}}\<%
\\
\>[1]\AgdaKeyword{open}\AgdaSpace{}%
\AgdaModule{FreeAlgebra}\AgdaSpace{}%
\AgdaSymbol{\{}\AgdaArgument{χ}\AgdaSpace{}%
\AgdaSymbol{=}\AgdaSpace{}%
\AgdaFunction{c}\AgdaSymbol{\}(}\AgdaFunction{Th}\AgdaSpace{}%
\AgdaBound{𝒦}\AgdaSymbol{)}%
\>[33]\AgdaKeyword{using}\AgdaSpace{}%
\AgdaSymbol{(}\AgdaSpace{}%
\AgdaOperator{\AgdaFunction{𝔽[\AgdaUnderscore{}]}}\AgdaSpace{}%
\AgdaSymbol{)}\<%
\\
\>[1]\AgdaKeyword{open}\AgdaSpace{}%
\AgdaModule{Setoid}\AgdaSpace{}%
\AgdaOperator{\AgdaFunction{𝔻[}}\AgdaSpace{}%
\AgdaBound{𝑨}\AgdaSpace{}%
\AgdaOperator{\AgdaFunction{]}}%
\>[33]\AgdaKeyword{using}\AgdaSpace{}%
\AgdaSymbol{(}\AgdaSpace{}%
\AgdaFunction{refl}\AgdaSpace{}%
\AgdaSymbol{;}\AgdaSpace{}%
\AgdaFunction{sym}\AgdaSpace{}%
\AgdaSymbol{;}\AgdaSpace{}%
\AgdaFunction{trans}\AgdaSpace{}%
\AgdaSymbol{)}\AgdaSpace{}%
\AgdaKeyword{renaming}%
\>[72]\AgdaSymbol{(}\AgdaSpace{}%
\AgdaField{Carrier}%
\>[83]\AgdaSymbol{to}\AgdaSpace{}%
\AgdaField{A}\AgdaSpace{}%
\AgdaSymbol{)}\<%
\end{code}
\fi
\begin{code}%
\>[0]\<%
\\
\>[1]\AgdaFunction{F-ModTh-epi}\AgdaSpace{}%
\AgdaSymbol{:}\AgdaSpace{}%
\AgdaBound{𝑨}\AgdaSpace{}%
\AgdaOperator{\AgdaFunction{∈}}\AgdaSpace{}%
\AgdaFunction{Mod}\AgdaSpace{}%
\AgdaSymbol{(}\AgdaFunction{Th}\AgdaSpace{}%
\AgdaSymbol{(}\AgdaFunction{V}\AgdaSpace{}%
\AgdaBound{ℓ}\AgdaSpace{}%
\AgdaFunction{ι}\AgdaSpace{}%
\AgdaBound{𝒦}\AgdaSymbol{))}\AgdaSpace{}%
\AgdaSymbol{→}\AgdaSpace{}%
\AgdaFunction{epi}\AgdaSpace{}%
\AgdaOperator{\AgdaFunction{𝔽[}}\AgdaSpace{}%
\AgdaFunction{A}\AgdaSpace{}%
\AgdaOperator{\AgdaFunction{]}}\AgdaSpace{}%
\AgdaBound{𝑨}\<%
\\
\>[1]\AgdaFunction{F-ModTh-epi}\AgdaSpace{}%
\AgdaBound{A∈ModThK}\AgdaSpace{}%
\AgdaSymbol{=}\AgdaSpace{}%
\AgdaFunction{φ}\AgdaSpace{}%
\AgdaOperator{\AgdaInductiveConstructor{,}}\AgdaSpace{}%
\AgdaFunction{isEpi}\<%
\\
\>[1][@{}l@{\AgdaIndent{0}}]%
\>[2]\AgdaKeyword{where}\<%
\\
\>[2]\AgdaFunction{φ}\AgdaSpace{}%
\AgdaSymbol{:}\AgdaSpace{}%
\AgdaOperator{\AgdaFunction{𝔻[}}\AgdaSpace{}%
\AgdaOperator{\AgdaFunction{𝔽[}}\AgdaSpace{}%
\AgdaFunction{A}\AgdaSpace{}%
\AgdaOperator{\AgdaFunction{]}}\AgdaSpace{}%
\AgdaOperator{\AgdaFunction{]}}\AgdaSpace{}%
\AgdaOperator{\AgdaRecord{⟶}}\AgdaSpace{}%
\AgdaOperator{\AgdaFunction{𝔻[}}\AgdaSpace{}%
\AgdaBound{𝑨}\AgdaSpace{}%
\AgdaOperator{\AgdaFunction{]}}\<%
\\
\>[2]\AgdaOperator{\AgdaField{\AgdaUnderscore{}⟨\$⟩\AgdaUnderscore{}}}\AgdaSpace{}%
\AgdaFunction{φ}\AgdaSpace{}%
\AgdaSymbol{=}\AgdaSpace{}%
\AgdaFunction{free-lift}\AgdaSymbol{\{}\AgdaArgument{𝑨}\AgdaSpace{}%
\AgdaSymbol{=}\AgdaSpace{}%
\AgdaBound{𝑨}\AgdaSymbol{\}}\AgdaSpace{}%
\AgdaFunction{id}\<%
\\
\>[2]\AgdaField{cong}\AgdaSpace{}%
\AgdaFunction{φ}\AgdaSpace{}%
\AgdaSymbol{\{}\AgdaBound{p}\AgdaSymbol{\}}\AgdaSpace{}%
\AgdaSymbol{\{}\AgdaBound{q}\AgdaSymbol{\}}\AgdaSpace{}%
\AgdaBound{pq}%
\>[21]\AgdaSymbol{=}%
\>[24]\AgdaFunction{trans}%
\>[31]\AgdaSymbol{(}\AgdaSpace{}%
\AgdaFunction{sym}\AgdaSpace{}%
\AgdaSymbol{(}\AgdaFunction{free-lift-interp}\AgdaSymbol{\{}\AgdaArgument{𝑨}\AgdaSpace{}%
\AgdaSymbol{=}\AgdaSpace{}%
\AgdaBound{𝑨}\AgdaSymbol{\}}\AgdaSpace{}%
\AgdaFunction{id}\AgdaSpace{}%
\AgdaBound{p}\AgdaSymbol{)}\AgdaSpace{}%
\AgdaSymbol{)}\<%
\\
\>[21]\AgdaSymbol{(}%
\>[24]\AgdaFunction{trans}%
\>[31]\AgdaSymbol{(}\AgdaSpace{}%
\AgdaBound{A∈ModThK}\AgdaSymbol{\{}\AgdaArgument{p}\AgdaSpace{}%
\AgdaSymbol{=}\AgdaSpace{}%
\AgdaBound{p}\AgdaSymbol{\}\{}\AgdaBound{q}\AgdaSymbol{\}}\AgdaSpace{}%
\AgdaSymbol{(}\AgdaFunction{kernel-in-theory}\AgdaSpace{}%
\AgdaBound{pq}\AgdaSymbol{)}\AgdaSpace{}%
\AgdaFunction{id}\AgdaSpace{}%
\AgdaSymbol{)}\<%
\\
\>[31]\AgdaSymbol{(}\AgdaSpace{}%
\AgdaFunction{free-lift-interp}\AgdaSymbol{\{}\AgdaArgument{𝑨}\AgdaSpace{}%
\AgdaSymbol{=}\AgdaSpace{}%
\AgdaBound{𝑨}\AgdaSymbol{\}}\AgdaSpace{}%
\AgdaFunction{id}\AgdaSpace{}%
\AgdaBound{q}\AgdaSpace{}%
\AgdaSymbol{)}\AgdaSpace{}%
\AgdaSymbol{)}\<%
\\
\>[2]\AgdaFunction{isEpi}\AgdaSpace{}%
\AgdaSymbol{:}\AgdaSpace{}%
\AgdaRecord{IsEpi}\AgdaSpace{}%
\AgdaOperator{\AgdaFunction{𝔽[}}\AgdaSpace{}%
\AgdaFunction{A}\AgdaSpace{}%
\AgdaOperator{\AgdaFunction{]}}\AgdaSpace{}%
\AgdaBound{𝑨}\AgdaSpace{}%
\AgdaFunction{φ}\<%
\\
\>[2]\AgdaField{compatible}\AgdaSpace{}%
\AgdaSymbol{(}\AgdaField{isHom}\AgdaSpace{}%
\AgdaFunction{isEpi}\AgdaSymbol{)}\AgdaSpace{}%
\AgdaSymbol{=}\AgdaSpace{}%
\AgdaField{cong}\AgdaSpace{}%
\AgdaSymbol{(}\AgdaField{Interp}\AgdaSpace{}%
\AgdaBound{𝑨}\AgdaSymbol{)}\AgdaSpace{}%
\AgdaSymbol{(}\AgdaInductiveConstructor{≡.refl}\AgdaSpace{}%
\AgdaOperator{\AgdaInductiveConstructor{,}}\AgdaSpace{}%
\AgdaSymbol{(λ}\AgdaSpace{}%
\AgdaBound{\AgdaUnderscore{}}\AgdaSpace{}%
\AgdaSymbol{→}\AgdaSpace{}%
\AgdaFunction{refl}\AgdaSymbol{))}\<%
\\
\>[2]\AgdaField{isSurjective}\AgdaSpace{}%
\AgdaFunction{isEpi}\AgdaSpace{}%
\AgdaSymbol{\{}\AgdaBound{y}\AgdaSymbol{\}}\AgdaSpace{}%
\AgdaSymbol{=}\AgdaSpace{}%
\AgdaInductiveConstructor{eq}\AgdaSpace{}%
\AgdaSymbol{(}\AgdaInductiveConstructor{ℊ}\AgdaSpace{}%
\AgdaBound{y}\AgdaSymbol{)}\AgdaSpace{}%
\AgdaFunction{refl}\<%
\end{code}
\ifshort\else

\medskip

\noindent Actually, we will need the following lifted version of this result.

\begin{code}%
\>[0]\<%
\\
\>[1]\AgdaFunction{F-ModTh-epi-lift}\AgdaSpace{}%
\AgdaSymbol{:}\AgdaSpace{}%
\AgdaBound{𝑨}\AgdaSpace{}%
\AgdaOperator{\AgdaFunction{∈}}\AgdaSpace{}%
\AgdaFunction{Mod}\AgdaSpace{}%
\AgdaSymbol{(}\AgdaFunction{Th}\AgdaSpace{}%
\AgdaSymbol{(}\AgdaFunction{V}\AgdaSpace{}%
\AgdaBound{ℓ}\AgdaSpace{}%
\AgdaFunction{ι}\AgdaSpace{}%
\AgdaBound{𝒦}\AgdaSymbol{))}\AgdaSpace{}%
\AgdaSymbol{→}\AgdaSpace{}%
\AgdaFunction{epi}\AgdaSpace{}%
\AgdaOperator{\AgdaFunction{𝔽[}}\AgdaSpace{}%
\AgdaFunction{A}\AgdaSpace{}%
\AgdaOperator{\AgdaFunction{]}}\AgdaSpace{}%
\AgdaSymbol{(}\AgdaFunction{Lift-Alg}\AgdaSpace{}%
\AgdaBound{𝑨}\AgdaSpace{}%
\AgdaFunction{ι}\AgdaSpace{}%
\AgdaFunction{ι}\AgdaSymbol{)}\<%
\\
\>[1]\AgdaFunction{F-ModTh-epi-lift}\AgdaSpace{}%
\AgdaBound{A∈ModThK}\AgdaSpace{}%
\AgdaSymbol{=}\AgdaSpace{}%
\AgdaFunction{∘-epi}\AgdaSpace{}%
\AgdaSymbol{(}\AgdaFunction{F-ModTh-epi}\AgdaSpace{}%
\AgdaSymbol{(λ}\AgdaSpace{}%
\AgdaSymbol{\{}\AgdaBound{p}\AgdaSpace{}%
\AgdaBound{q}\AgdaSymbol{\}}\AgdaSpace{}%
\AgdaSymbol{→}\AgdaSpace{}%
\AgdaBound{A∈ModThK}\AgdaSymbol{\{}\AgdaArgument{p}\AgdaSpace{}%
\AgdaSymbol{=}\AgdaSpace{}%
\AgdaBound{p}\AgdaSymbol{\}\{}\AgdaBound{q}\AgdaSymbol{\}))}\AgdaSpace{}%
\AgdaFunction{ToLift-epi}\<%
\end{code}
\fi

%% -------------------------------------------------------------------------------------

\section{Birkhoff's Variety Theorem}

Birkhoff's variety theorem, also known as the HSP theorem, asserts that a class of algebras
is a variety if and only if it is an equational class.  In this section, we present the
statement and proof of the HSP theorem---first in the familiar, informal style similar to
what one finds in standard textbooks (see, e.g.,~\cite[Theorem 4.41]{Bergman:2012}),
and then in the formal language of Martin-Löf type theory using Agda.

\subsection{Informal proof}
Let \ab{𝒦} be a class of algebras and recall that \ab{𝒦} is a \emph{variety} provided
\ifshort\else
it is closed under homomorphisms, subalgebras and products; equivalently,
\fi
\af{V} \ab{𝒦} ⊆ \ab{𝒦}.\footnote{Recall, \af{V} \ab{𝒦} := \af{H} (\af{S} (\af{P} \ab{𝒦})),
and observe that \ab{𝒦} ⊆ \af{V} \ab{𝒦} holds for all \ab{𝒦} since
\af{V} is a closure operator.}
We call \ab{𝒦} an \emph{equational class} if it is precisely the class of all models of some set of identities.

It is easy to prove that \emph{every equational class is a variety}.  Indeed, suppose \ab{𝒦} is an equational
class axiomatized by the set \ab{ℰ} of term identities; that is, \ab{𝑨} ∈ \ab{𝒦} iff
\ab{𝑨} \af{⊨} \ab{ℰ}. Since the classes \af H \ab{𝒦}, \af S \ab{𝒦}, \af P \ab{𝒦} and
\ab{𝒦} all satisfy the same set of equations, we have \af{V} \ab{𝒦} \af{⊫} \ab p
\af{≈} \ab q for all (\ab p , \ab q) \af{∈} \ab{ℰ}, so \af{V} \ab{𝒦} ⊆ \ab{𝒦}.

The converse assertion---that \emph{every variety is an equational class}---is less obvious.
Let \ab{𝒦} be an arbitrary variety.  We will describe a set of equations that axiomatizes
\ab{𝒦}.  A natural choice is the set
\af{Th} \ab{𝒦} of all equations that hold in \ab{𝒦}. Define \ab{𝒦⁺} = \af{Mod} (\af{Th}
\ab{𝒦}).  Clearly, \ab{𝒦} \aof{⊆} \ab{𝒦⁺}. We prove the reverse inclusion. Let \ab{𝑨}
\af{∈} \ab{𝒦⁺}; it suffices to find an algebra \ab{𝑭} \af{∈} \af{S} (\af{P} \ab{𝒦}) such
that \ab{𝑨} is a homomorphic image of \ab{𝑭}, as this will show that \ab{𝑨}
\af{∈} \af{H} (\af{S} (\af{P} \ab{𝒦})) = \ab{𝒦}.

Let \ab{X} be such that there exists a \emph{surjective} environment
\ab{ρ} : \ab{X} \as{→} \af{𝕌[~\ab{𝑨}~]}.
%\footnote{This is usually done by assuming \ab{X} has cardinality at least max(|~\af{𝕌[~\ab{𝑨}~]}~|, ω).}
By the \af{lift-hom} lemma, there is an epimorphism \ab{h} from \T{X} onto \af{𝕌[~\ab{𝑨}~]}
that extends \ab{ρ}.
Now, put \aof{𝔽[~\ab{X}~]}~:=~\T{X}/\ab{Θ}, and let \ab{g} : \T{X} \as{→} \aof{𝔽[~\ab{X}~]}
be the natural epimorphism with kernel \ab{Θ}. We claim that \af{ker} \ab g \af{⊆}
\af{ker} \ab h. If the claim is true, then there is a map \ab{f} : \aof{𝔽[~\ab{X}~]} \as{→} \ab{𝑨}
such that \ab f \af{∘} \ab g = \ab h. Since \ab h is surjective, so is \ab f. Hence \ab{𝑨}
\af{∈} \af{𝖧} (\af{𝔽} \ab X) \aof{⊆} \ab{𝒦⁺} completing the proof.
To prove the claim, let \ab u , \ab v \af{∈} \T{X} and assume that \ab g \ab u =
\ab g \ab v. Since \T{X} is generated by \ab X, there are terms \ab p, \ab q ∈
\T{X} such that \ab u = \af{⟦~\T{X}~⟧} \ab p and v = \af{⟦~\T{X}~⟧} \ab
q.\footnote{Recall, \af{⟦~\ab{𝑨}~⟧} \ab t denotes the interpretation of the term
\ab t in the algebra \ab{𝑨}.} Therefore,\\[-4pt]

\af{⟦~\Free{X}~⟧} \ab p = \ab g (\af{⟦~\T{X}~⟧} \ab p) = \ab g \ab u = \ab g \ab v =
\ab g (\af{⟦~\T{X}~⟧} \ab q) = \af{⟦~\Free{X}~⟧} \ab q,\\[8pt]
so \ab{𝒦} \af{⊫} \ab p \af{≈} \ab q, so (\ab p , \ab q) \af{∈} \af{Th}
\ab{𝒦}. Since \ab{𝑨} \af{∈} \ab{𝒦⁺} =
\af{Mod} (\af{Th} \ab{𝒦}), we obtain \ab{𝑨} \af{⊧} \ab p \af{≈} \ab q, so \ab h
\ab u = (\af{⟦~\ab{𝑨}~⟧} \ab p) \aofld{⟨\$⟩} \ab{ρ} = (\af{⟦~\ab{𝑨}~⟧} \ab q)
\aofld{⟨\$⟩} \ab{ρ} = \ab h \ab v, as desired.

\subsection{Formal proof}
We now show how to formally express and prove the twin assertions that
(i) every equational class is a variety and (ii) every variety is an equational class.

\paragraph*{Every equational class is a variety}
For (i), we need an arbitrary equational class. To obtain one, we start with an arbitrary
collection \ab{ℰ} of equations and let \ab{𝒦} = \af{Mod} \ab{ℰ}, the equational class
determined by \ab{ℰ}. We prove that \ab{𝒦} is a variety by showing that
\ab{𝒦} = \af{V} \ab{𝒦}. The inclusion \ab{𝒦} \aof{⊆} \af V \ab{𝒦}, which holds for all
classes \ab{𝒦}, is called the \defn{expansive} property of \af{V}. The converse inclusion
\af V \ab{𝒦} \aof{⊆} \ab{𝒦}, on the other hand, requires the hypothesis that \ab{𝒦} is an
equation class. We now formalize each of these inclusions.

\ifshort\else
\begin{code}%
\>[0]\<%
\\
\>[0]\AgdaKeyword{module}\AgdaSpace{}%
\AgdaModule{\AgdaUnderscore{}}\AgdaSpace{}%
\AgdaSymbol{(}\AgdaBound{𝒦}\AgdaSpace{}%
\AgdaSymbol{:}\AgdaSpace{}%
\AgdaFunction{Pred}\AgdaSymbol{(}\AgdaRecord{Algebra}\AgdaSpace{}%
\AgdaGeneralizable{α}\AgdaSpace{}%
\AgdaGeneralizable{ρᵃ}\AgdaSymbol{)}\AgdaSpace{}%
\AgdaSymbol{(}\AgdaGeneralizable{α}\AgdaSpace{}%
\AgdaOperator{\AgdaPrimitive{⊔}}\AgdaSpace{}%
\AgdaGeneralizable{ρᵃ}\AgdaSpace{}%
\AgdaOperator{\AgdaPrimitive{⊔}}\AgdaSpace{}%
\AgdaFunction{ov}\AgdaSpace{}%
\AgdaGeneralizable{ℓ}\AgdaSymbol{))\{}\AgdaBound{X}\AgdaSpace{}%
\AgdaSymbol{:}\AgdaSpace{}%
\AgdaPrimitive{Type}\AgdaSpace{}%
\AgdaSymbol{(}\AgdaGeneralizable{α}\AgdaSpace{}%
\AgdaOperator{\AgdaPrimitive{⊔}}\AgdaSpace{}%
\AgdaGeneralizable{ρᵃ}\AgdaSpace{}%
\AgdaOperator{\AgdaPrimitive{⊔}}\AgdaSpace{}%
\AgdaGeneralizable{ℓ}\AgdaSymbol{)\}}\AgdaSpace{}%
\AgdaKeyword{where}\<%
\\
\>[0][@{}l@{\AgdaIndent{0}}]%
\>[1]\AgdaKeyword{private}\AgdaSpace{}%
\AgdaFunction{ι}\AgdaSpace{}%
\AgdaSymbol{=}\AgdaSpace{}%
\AgdaFunction{ov}\AgdaSpace{}%
\AgdaSymbol{(}\AgdaBound{α}\AgdaSpace{}%
\AgdaOperator{\AgdaPrimitive{⊔}}\AgdaSpace{}%
\AgdaBound{ρᵃ}\AgdaSpace{}%
\AgdaOperator{\AgdaPrimitive{⊔}}\AgdaSpace{}%
\AgdaBound{ℓ}\AgdaSymbol{)}\<%
\end{code}
\fi
\begin{code}%
\>[0]\<%
\\
\>[1]\AgdaFunction{V-expa}\AgdaSpace{}%
\AgdaSymbol{:}\AgdaSpace{}%
\AgdaBound{𝒦}\AgdaSpace{}%
\AgdaOperator{\AgdaFunction{⊆}}\AgdaSpace{}%
\AgdaFunction{V}\AgdaSpace{}%
\AgdaBound{ℓ}\AgdaSpace{}%
\AgdaFunction{ι}\AgdaSpace{}%
\AgdaBound{𝒦}\<%
\\
\>[1]\AgdaFunction{V-expa}\AgdaSpace{}%
\AgdaSymbol{\{}\AgdaArgument{x}\AgdaSpace{}%
\AgdaSymbol{=}\AgdaSpace{}%
\AgdaBound{𝑨}\AgdaSymbol{\}}\AgdaSpace{}%
\AgdaBound{kA}\AgdaSpace{}%
\AgdaSymbol{=}\AgdaSpace{}%
\AgdaBound{𝑨}\AgdaSpace{}%
\AgdaOperator{\AgdaInductiveConstructor{,}}\AgdaSymbol{(}\AgdaBound{𝑨}\AgdaSpace{}%
\AgdaOperator{\AgdaInductiveConstructor{,}}\AgdaSymbol{(}\AgdaFunction{⊤}\AgdaSpace{}%
\AgdaOperator{\AgdaInductiveConstructor{,}}\AgdaSymbol{(λ}\AgdaSpace{}%
\AgdaBound{\AgdaUnderscore{}}\AgdaSpace{}%
\AgdaSymbol{→}\AgdaSpace{}%
\AgdaBound{𝑨}\AgdaSymbol{)}\AgdaOperator{\AgdaInductiveConstructor{,}}\AgdaSpace{}%
\AgdaSymbol{(λ}\AgdaSpace{}%
\AgdaBound{\AgdaUnderscore{}}\AgdaSpace{}%
\AgdaSymbol{→}\AgdaSpace{}%
\AgdaBound{kA}\AgdaSymbol{)}\AgdaOperator{\AgdaInductiveConstructor{,}}\AgdaSpace{}%
\AgdaFunction{Goal}\AgdaSymbol{)}\AgdaOperator{\AgdaInductiveConstructor{,}}\AgdaSpace{}%
\AgdaFunction{≤-reflexive}\AgdaSymbol{)}\AgdaOperator{\AgdaInductiveConstructor{,}}\AgdaSpace{}%
\AgdaFunction{IdHomImage}\<%
\\
\>[1][@{}l@{\AgdaIndent{0}}]%
\>[2]\AgdaKeyword{where}\<%
\\
\>[2]\AgdaKeyword{open}\AgdaSpace{}%
\AgdaModule{Setoid}\AgdaSpace{}%
\AgdaOperator{\AgdaFunction{𝔻[}}\AgdaSpace{}%
\AgdaBound{𝑨}\AgdaSpace{}%
\AgdaOperator{\AgdaFunction{]}}\AgdaSpace{}%
\AgdaKeyword{using}\AgdaSpace{}%
\AgdaSymbol{(}\AgdaSpace{}%
\AgdaFunction{refl}\AgdaSpace{}%
\AgdaSymbol{)}\<%
\\
\>[2]\AgdaKeyword{open}\AgdaSpace{}%
\AgdaModule{Setoid}\AgdaSpace{}%
\AgdaOperator{\AgdaFunction{𝔻[}}\AgdaSpace{}%
\AgdaFunction{⨅}\AgdaSpace{}%
\AgdaSymbol{(λ}\AgdaSpace{}%
\AgdaBound{\AgdaUnderscore{}}\AgdaSpace{}%
\AgdaSymbol{→}\AgdaSpace{}%
\AgdaBound{𝑨}\AgdaSymbol{)}\AgdaSpace{}%
\AgdaOperator{\AgdaFunction{]}}\AgdaSpace{}%
\AgdaKeyword{using}\AgdaSpace{}%
\AgdaSymbol{()}\AgdaSpace{}%
\AgdaKeyword{renaming}\AgdaSpace{}%
\AgdaSymbol{(}\AgdaSpace{}%
\AgdaFunction{refl}\AgdaSpace{}%
\AgdaSymbol{to}\AgdaSpace{}%
\AgdaFunction{refl⨅}\AgdaSpace{}%
\AgdaSymbol{)}\<%
\\
\\[\AgdaEmptyExtraSkip]%
\>[2]\AgdaFunction{to⨅}\AgdaSpace{}%
\AgdaSymbol{:}\AgdaSpace{}%
\AgdaOperator{\AgdaFunction{𝔻[}}\AgdaSpace{}%
\AgdaBound{𝑨}\AgdaSpace{}%
\AgdaOperator{\AgdaFunction{]}}\AgdaSpace{}%
\AgdaOperator{\AgdaRecord{⟶}}\AgdaSpace{}%
\AgdaOperator{\AgdaFunction{𝔻[}}\AgdaSpace{}%
\AgdaFunction{⨅}\AgdaSpace{}%
\AgdaSymbol{(λ}\AgdaSpace{}%
\AgdaBound{\AgdaUnderscore{}}\AgdaSpace{}%
\AgdaSymbol{→}\AgdaSpace{}%
\AgdaBound{𝑨}\AgdaSymbol{)}\AgdaSpace{}%
\AgdaOperator{\AgdaFunction{]}}\<%
\\
\>[2]\AgdaSymbol{(}\AgdaFunction{to⨅}\AgdaSpace{}%
\AgdaOperator{\AgdaField{⟨\$⟩}}\AgdaSpace{}%
\AgdaBound{x}\AgdaSymbol{)}\AgdaSpace{}%
\AgdaSymbol{=}\AgdaSpace{}%
\AgdaSymbol{λ}\AgdaSpace{}%
\AgdaBound{\AgdaUnderscore{}}\AgdaSpace{}%
\AgdaSymbol{→}\AgdaSpace{}%
\AgdaBound{x}\<%
\\
\>[2]\AgdaField{cong}\AgdaSpace{}%
\AgdaFunction{to⨅}\AgdaSpace{}%
\AgdaBound{xy}\AgdaSpace{}%
\AgdaSymbol{=}\AgdaSpace{}%
\AgdaSymbol{λ}\AgdaSpace{}%
\AgdaBound{\AgdaUnderscore{}}\AgdaSpace{}%
\AgdaSymbol{→}\AgdaSpace{}%
\AgdaBound{xy}\<%
\\
\\[\AgdaEmptyExtraSkip]%
\>[2]\AgdaFunction{from⨅}\AgdaSpace{}%
\AgdaSymbol{:}\AgdaSpace{}%
\AgdaOperator{\AgdaFunction{𝔻[}}\AgdaSpace{}%
\AgdaFunction{⨅}\AgdaSpace{}%
\AgdaSymbol{(λ}\AgdaSpace{}%
\AgdaBound{\AgdaUnderscore{}}\AgdaSpace{}%
\AgdaSymbol{→}\AgdaSpace{}%
\AgdaBound{𝑨}\AgdaSymbol{)}\AgdaSpace{}%
\AgdaOperator{\AgdaFunction{]}}\AgdaSpace{}%
\AgdaOperator{\AgdaRecord{⟶}}\AgdaSpace{}%
\AgdaOperator{\AgdaFunction{𝔻[}}\AgdaSpace{}%
\AgdaBound{𝑨}\AgdaSpace{}%
\AgdaOperator{\AgdaFunction{]}}\<%
\\
\>[2]\AgdaSymbol{(}\AgdaFunction{from⨅}\AgdaSpace{}%
\AgdaOperator{\AgdaField{⟨\$⟩}}\AgdaSpace{}%
\AgdaBound{x}\AgdaSymbol{)}\AgdaSpace{}%
\AgdaSymbol{=}\AgdaSpace{}%
\AgdaBound{x}\AgdaSpace{}%
\AgdaFunction{tt}\<%
\\
\>[2]\AgdaField{cong}\AgdaSpace{}%
\AgdaFunction{from⨅}\AgdaSpace{}%
\AgdaBound{xy}\AgdaSpace{}%
\AgdaSymbol{=}\AgdaSpace{}%
\AgdaBound{xy}\AgdaSpace{}%
\AgdaFunction{tt}\<%
\\
\\[\AgdaEmptyExtraSkip]%
\>[2]\AgdaFunction{Goal}\AgdaSpace{}%
\AgdaSymbol{:}\AgdaSpace{}%
\AgdaBound{𝑨}\AgdaSpace{}%
\AgdaOperator{\AgdaRecord{≅}}\AgdaSpace{}%
\AgdaFunction{⨅}\AgdaSpace{}%
\AgdaSymbol{(λ}\AgdaSpace{}%
\AgdaBound{x}\AgdaSpace{}%
\AgdaSymbol{→}\AgdaSpace{}%
\AgdaBound{𝑨}\AgdaSymbol{)}\<%
\\
\>[2]\AgdaFunction{Goal}\AgdaSpace{}%
\AgdaSymbol{=}\AgdaSpace{}%
\AgdaInductiveConstructor{mkiso}\AgdaSpace{}%
\AgdaSymbol{(}\AgdaFunction{to⨅}\AgdaSpace{}%
\AgdaOperator{\AgdaInductiveConstructor{,}}\AgdaSpace{}%
\AgdaInductiveConstructor{mkhom}\AgdaSpace{}%
\AgdaFunction{refl⨅}\AgdaSymbol{)}\AgdaSpace{}%
\AgdaSymbol{(}\AgdaFunction{from⨅}\AgdaSpace{}%
\AgdaOperator{\AgdaInductiveConstructor{,}}\AgdaSpace{}%
\AgdaInductiveConstructor{mkhom}\AgdaSpace{}%
\AgdaFunction{refl}\AgdaSymbol{)}\AgdaSpace{}%
\AgdaSymbol{(λ}\AgdaSpace{}%
\AgdaBound{\AgdaUnderscore{}}\AgdaSpace{}%
\AgdaBound{\AgdaUnderscore{}}\AgdaSpace{}%
\AgdaSymbol{→}\AgdaSpace{}%
\AgdaFunction{refl}\AgdaSymbol{)}\AgdaSpace{}%
\AgdaSymbol{(λ}\AgdaSpace{}%
\AgdaBound{\AgdaUnderscore{}}\AgdaSpace{}%
\AgdaSymbol{→}\AgdaSpace{}%
\AgdaFunction{refl}\AgdaSymbol{)}\<%
\\
\>[0]\<%
\end{code}
Earlier we proved the identity preservation lemma,
\af{V-id1} : \ab{𝒦} \aof{⊫} \ab p \aof{≈} \ab q \as{→} \af{V} \ab{ℓ} \ab{ι} \ab{𝒦} \aof{⊫} \ab p \aof{≈} \ab q.
Thus, if \ab{𝒦} is an equational class, then \af V \ab{𝒦} \aof{⊆} \ab{𝒦}, as we now confirm.

\begin{code}%
\>[0]\<%
\\
\>[0]\AgdaKeyword{module}\AgdaSpace{}%
\AgdaModule{\AgdaUnderscore{}}\AgdaSpace{}%
\AgdaSymbol{\{}\AgdaBound{ℓ}\AgdaSpace{}%
\AgdaSymbol{:}\AgdaSpace{}%
\AgdaPostulate{Level}\AgdaSymbol{\}\{}\AgdaBound{X}\AgdaSpace{}%
\AgdaSymbol{:}\AgdaSpace{}%
\AgdaPrimitive{Type}\AgdaSpace{}%
\AgdaBound{ℓ}\AgdaSymbol{\}\{}\AgdaBound{ℰ}\AgdaSpace{}%
\AgdaSymbol{:}\AgdaSpace{}%
\AgdaSymbol{\{}\AgdaBound{Y}\AgdaSpace{}%
\AgdaSymbol{:}\AgdaSpace{}%
\AgdaPrimitive{Type}\AgdaSpace{}%
\AgdaBound{ℓ}\AgdaSymbol{\}}\AgdaSpace{}%
\AgdaSymbol{→}\AgdaSpace{}%
\AgdaFunction{Pred}\AgdaSpace{}%
\AgdaSymbol{(}\AgdaDatatype{Term}\AgdaSpace{}%
\AgdaBound{Y}\AgdaSpace{}%
\AgdaOperator{\AgdaFunction{×}}\AgdaSpace{}%
\AgdaDatatype{Term}\AgdaSpace{}%
\AgdaBound{Y}\AgdaSymbol{)}\AgdaSpace{}%
\AgdaSymbol{(}\AgdaFunction{ov}\AgdaSpace{}%
\AgdaBound{ℓ}\AgdaSymbol{)\}}\AgdaSpace{}%
\AgdaKeyword{where}\<%
\\
\>[0][@{}l@{\AgdaIndent{0}}]%
\>[1]\AgdaKeyword{private}\AgdaSpace{}%
\AgdaFunction{𝒦}\AgdaSpace{}%
\AgdaSymbol{=}\AgdaSpace{}%
\AgdaFunction{Mod}\AgdaSymbol{\{}\AgdaArgument{α}\AgdaSpace{}%
\AgdaSymbol{=}\AgdaSpace{}%
\AgdaBound{ℓ}\AgdaSymbol{\}\{}\AgdaBound{ℓ}\AgdaSymbol{\}\{}\AgdaBound{X}\AgdaSymbol{\}}\AgdaSpace{}%
\AgdaBound{ℰ}%
\>[36]\AgdaComment{--\ an\ arbitrary\ equational\ class}\<%
\\
\>[1]\AgdaFunction{EqCl⇒Var}\AgdaSpace{}%
\AgdaSymbol{:}\AgdaSpace{}%
\AgdaFunction{V}\AgdaSpace{}%
\AgdaBound{ℓ}\AgdaSpace{}%
\AgdaSymbol{(}\AgdaFunction{ov}\AgdaSpace{}%
\AgdaBound{ℓ}\AgdaSymbol{)}\AgdaSpace{}%
\AgdaFunction{𝒦}\AgdaSpace{}%
\AgdaOperator{\AgdaFunction{⊆}}\AgdaSpace{}%
\AgdaFunction{𝒦}\<%
\\
\>[1]\AgdaFunction{EqCl⇒Var}\AgdaSpace{}%
\AgdaSymbol{\{}\AgdaBound{𝑨}\AgdaSymbol{\}}\AgdaBound{vA}\AgdaSymbol{\{}\AgdaBound{p}\AgdaSymbol{\}\{}\AgdaBound{q}\AgdaSymbol{\}}\AgdaSpace{}%
\AgdaBound{pℰq}\AgdaSpace{}%
\AgdaBound{ρ}\AgdaSpace{}%
\AgdaSymbol{=}\AgdaSpace{}%
\AgdaFunction{V-id1}\AgdaSymbol{\{}\AgdaArgument{ℓ}\AgdaSpace{}%
\AgdaSymbol{=}\AgdaSpace{}%
\AgdaBound{ℓ}\AgdaSymbol{\}\{}\AgdaArgument{𝒦}\AgdaSpace{}%
\AgdaSymbol{=}\AgdaSpace{}%
\AgdaFunction{𝒦}\AgdaSymbol{\}\{}\AgdaBound{p}\AgdaSymbol{\}\{}\AgdaBound{q}\AgdaSymbol{\}(λ}\AgdaSpace{}%
\AgdaBound{\AgdaUnderscore{}}\AgdaSpace{}%
\AgdaBound{x}\AgdaSpace{}%
\AgdaBound{τ}\AgdaSpace{}%
\AgdaSymbol{→}\AgdaSpace{}%
\AgdaBound{x}\AgdaSpace{}%
\AgdaBound{pℰq}\AgdaSpace{}%
\AgdaBound{τ}\AgdaSymbol{)}\AgdaBound{𝑨}\AgdaSpace{}%
\AgdaBound{vA}\AgdaSpace{}%
\AgdaBound{ρ}\<%
\\
\>[0]\<%
\end{code}
Together, \af{V-expa} and \af{Eqcl⇒Var} prove that every equational class is a variety.

\paragraph*{Every variety is an equational class}
To prove statement (ii), we need an arbitrary variety; to obtain one, we start with an arbitrary class
\ab{𝒦} of \ab{𝑆}-algebras and take its \emph{varietal closure}, \af{V} \ab{𝒦}.
We prove that \af{V} \ab{𝒦} is an equational class by showing it is precisely the collection of
algebras that model the equations in \af{Th} (\af{V} \ab{𝒦}); that is, we prove
\af{V} \ab{𝒦} = \af{Mod} (\af{Th} (\af{V} \ab{𝒦})).
The inclusion \af{V} \ab{𝒦} \aof{⊆} \af{Mod} (\af{Th} (\af{V} \ab{𝒦})) is a simple consequence of the fact that \af{Mod} \af{Th} is a closure operator. Nonetheless, completeness demands
that we formalize this fact, however trivial is its proof.

\ifshort\else
\begin{code}%
\>[0]\<%
\\
\>[0]\AgdaKeyword{module}\AgdaSpace{}%
\AgdaModule{\AgdaUnderscore{}}\AgdaSpace{}%
\AgdaSymbol{(}\AgdaBound{𝒦}\AgdaSpace{}%
\AgdaSymbol{:}\AgdaSpace{}%
\AgdaFunction{Pred}\AgdaSymbol{(}\AgdaRecord{Algebra}\AgdaSpace{}%
\AgdaGeneralizable{α}\AgdaSpace{}%
\AgdaGeneralizable{ρᵃ}\AgdaSymbol{)}\AgdaSpace{}%
\AgdaSymbol{(}\AgdaGeneralizable{α}\AgdaSpace{}%
\AgdaOperator{\AgdaPrimitive{⊔}}\AgdaSpace{}%
\AgdaGeneralizable{ρᵃ}\AgdaSpace{}%
\AgdaOperator{\AgdaPrimitive{⊔}}\AgdaSpace{}%
\AgdaFunction{ov}\AgdaSpace{}%
\AgdaGeneralizable{ℓ}\AgdaSymbol{))\{}\AgdaBound{X}\AgdaSpace{}%
\AgdaSymbol{:}\AgdaSpace{}%
\AgdaPrimitive{Type}\AgdaSpace{}%
\AgdaSymbol{(}\AgdaGeneralizable{α}\AgdaSpace{}%
\AgdaOperator{\AgdaPrimitive{⊔}}\AgdaSpace{}%
\AgdaGeneralizable{ρᵃ}\AgdaSpace{}%
\AgdaOperator{\AgdaPrimitive{⊔}}\AgdaSpace{}%
\AgdaGeneralizable{ℓ}\AgdaSymbol{)\}}\AgdaSpace{}%
\AgdaKeyword{where}\<%
\\
\>[0][@{}l@{\AgdaIndent{0}}]%
\>[1]\AgdaKeyword{private}\AgdaSpace{}%
\AgdaFunction{c}\AgdaSpace{}%
\AgdaSymbol{=}\AgdaSpace{}%
\AgdaBound{α}\AgdaSpace{}%
\AgdaOperator{\AgdaPrimitive{⊔}}\AgdaSpace{}%
\AgdaBound{ρᵃ}\AgdaSpace{}%
\AgdaOperator{\AgdaPrimitive{⊔}}\AgdaSpace{}%
\AgdaBound{ℓ}\AgdaSpace{}%
\AgdaSymbol{;}\AgdaSpace{}%
\AgdaFunction{ι}\AgdaSpace{}%
\AgdaSymbol{=}\AgdaSpace{}%
\AgdaFunction{ov}\AgdaSpace{}%
\AgdaFunction{c}\<%
\end{code}
\fi
\begin{code}%
\>[0]\<%
\\
\>[1]\AgdaFunction{ModTh-closure}\AgdaSpace{}%
\AgdaSymbol{:}\AgdaSpace{}%
\AgdaFunction{V}\AgdaSymbol{\{}\AgdaArgument{β}\AgdaSpace{}%
\AgdaSymbol{=}\AgdaSpace{}%
\AgdaGeneralizable{β}\AgdaSymbol{\}\{}\AgdaGeneralizable{ρᵇ}\AgdaSymbol{\}\{}\AgdaGeneralizable{γ}\AgdaSymbol{\}\{}\AgdaGeneralizable{ρᶜ}\AgdaSymbol{\}\{}\AgdaGeneralizable{δ}\AgdaSymbol{\}\{}\AgdaGeneralizable{ρᵈ}\AgdaSymbol{\}}\AgdaSpace{}%
\AgdaBound{ℓ}\AgdaSpace{}%
\AgdaFunction{ι}\AgdaSpace{}%
\AgdaBound{𝒦}\AgdaSpace{}%
\AgdaOperator{\AgdaFunction{⊆}}\AgdaSpace{}%
\AgdaFunction{Mod}\AgdaSymbol{\{}\AgdaArgument{X}\AgdaSpace{}%
\AgdaSymbol{=}\AgdaSpace{}%
\AgdaBound{X}\AgdaSymbol{\}}\AgdaSpace{}%
\AgdaSymbol{(}\AgdaFunction{Th}\AgdaSpace{}%
\AgdaSymbol{(}\AgdaFunction{V}\AgdaSpace{}%
\AgdaBound{ℓ}\AgdaSpace{}%
\AgdaFunction{ι}\AgdaSpace{}%
\AgdaBound{𝒦}\AgdaSymbol{))}\<%
\\
\>[1]\AgdaFunction{ModTh-closure}\AgdaSpace{}%
\AgdaSymbol{\{}\AgdaArgument{x}\AgdaSpace{}%
\AgdaSymbol{=}\AgdaSpace{}%
\AgdaBound{𝑨}\AgdaSymbol{\}}\AgdaSpace{}%
\AgdaBound{vA}\AgdaSpace{}%
\AgdaSymbol{\{}\AgdaBound{p}\AgdaSymbol{\}\{}\AgdaBound{q}\AgdaSymbol{\}}\AgdaSpace{}%
\AgdaBound{x}\AgdaSpace{}%
\AgdaBound{ρ}\AgdaSpace{}%
\AgdaSymbol{=}\AgdaSpace{}%
\AgdaBound{x}\AgdaSpace{}%
\AgdaBound{𝑨}\AgdaSpace{}%
\AgdaBound{vA}\AgdaSpace{}%
\AgdaBound{ρ}\<%
\\
\>[0]\<%
\end{code}

It remains to prove the converse inclusion, \af{Mod} (\af{Th} (V 𝒦)) \aof{⊆} \af{V} \ab{𝒦},
which is the main focus of the rest of the paper.  We proceed as follows:

\begin{enumerate}
\item \label{item:1} Let \ab{𝑪} be the product of all algebras in \af{S} \ab{𝒦}, so that \ab{𝑪} \af{∈} \af{P} (\af{S} \ab{𝒦}).
\item Prove \af{P} (\af{S} \ab{𝒦}) \af{⊆} \af{S} (\af{P} \ab{𝒦}), so \ab{𝑪} \af{∈} \af{S} (\af{P} \ab{𝒦}) by item~\ref{item:1}.
\item Prove \aof{𝔽[ \ab{X} ]} \af{≤} \ab{𝑪}, so that \aof{𝔽[ \ab{X} ]} \af{∈} \af{S} (\af{S} (\af{P} \ab{𝒦})) (= \af{S} (\af{P} \ab{𝒦})).
\item Prove that every algebra in \af{Mod} (\af{Th} (V 𝒦)) is a homomorphic image of
\aof{𝔽[ \ab{X} ]} and thus belongs to \af{H} (\af{S} (\af{P} \ab{𝒦})) (= \af{V} \ab{𝒦}).
\end{enumerate}

To define \ab{𝑪} as the product of all algebras in \af{S} \ab{𝒦}, we must first contrive
an index type for the class \af{S} \ab{𝒦}.  We do so by letting the indices be the algebras
belonging to \ab{𝒦}. Actually, each index will consist of a triple (\ab{𝑨} , \ab p ,
\ab{ρ}) where \ab{𝑨} is an algebra, \ab p : \ab{𝑨} \af{∈} \af{S} \ab{𝒦}) is a proof of membership in \ab{𝒦},
\ab{ρ} : \ab X \as{→} \aof{𝕌[ \ab{𝑨} ]} is an arbitrary environment.
Using this indexing scheme, we construct \ab{𝑪}, the product of all algebras in \ab{𝒦}
and all environments, as follows.

\ifshort\else
\begin{code}%
\>[0]\<%
\\
\>[0][@{}l@{\AgdaIndent{1}}]%
\>[1]\AgdaKeyword{open}\AgdaSpace{}%
\AgdaModule{FreeHom}\AgdaSpace{}%
\AgdaSymbol{\{}\AgdaArgument{ℓ}\AgdaSpace{}%
\AgdaSymbol{=}\AgdaSpace{}%
\AgdaBound{ℓ}\AgdaSymbol{\}}\AgdaSpace{}%
\AgdaSymbol{\{}\AgdaBound{𝒦}\AgdaSymbol{\}}\<%
\\
\>[1]\AgdaKeyword{open}\AgdaSpace{}%
\AgdaModule{FreeAlgebra}\AgdaSpace{}%
\AgdaSymbol{\{}\AgdaArgument{χ}\AgdaSpace{}%
\AgdaSymbol{=}\AgdaSpace{}%
\AgdaFunction{c}\AgdaSymbol{\}(}\AgdaFunction{Th}\AgdaSpace{}%
\AgdaBound{𝒦}\AgdaSymbol{)}%
\>[33]\AgdaKeyword{using}\AgdaSpace{}%
\AgdaSymbol{(}\AgdaSpace{}%
\AgdaOperator{\AgdaFunction{𝔽[\AgdaUnderscore{}]}}\AgdaSpace{}%
\AgdaSymbol{)}\<%
\\
\>[1]\AgdaKeyword{open}\AgdaSpace{}%
\AgdaModule{Environment}%
\>[33]\AgdaKeyword{using}\AgdaSpace{}%
\AgdaSymbol{(}\AgdaSpace{}%
\AgdaFunction{Env}\AgdaSpace{}%
\AgdaSymbol{)}\<%
\end{code}
\fi
\begin{code}%
\>[0]\<%
\\
\>[1]\AgdaFunction{ℑ⁺}\AgdaSpace{}%
\AgdaSymbol{:}\AgdaSpace{}%
\AgdaPrimitive{Type}\AgdaSpace{}%
\AgdaFunction{ι}\<%
\\
\>[1]\AgdaFunction{ℑ⁺}\AgdaSpace{}%
\AgdaSymbol{=}\AgdaSpace{}%
\AgdaFunction{Σ[}\AgdaSpace{}%
\AgdaBound{𝑨}\AgdaSpace{}%
\AgdaFunction{∈}\AgdaSpace{}%
\AgdaSymbol{(}\AgdaRecord{Algebra}\AgdaSpace{}%
\AgdaBound{α}\AgdaSpace{}%
\AgdaBound{ρᵃ}\AgdaSymbol{)}\AgdaSpace{}%
\AgdaFunction{]}\AgdaSpace{}%
\AgdaSymbol{(}\AgdaBound{𝑨}\AgdaSpace{}%
\AgdaOperator{\AgdaFunction{∈}}\AgdaSpace{}%
\AgdaFunction{S}\AgdaSpace{}%
\AgdaBound{ℓ}\AgdaSpace{}%
\AgdaBound{𝒦}\AgdaSymbol{)}\AgdaSpace{}%
\AgdaOperator{\AgdaFunction{×}}\AgdaSpace{}%
\AgdaSymbol{(}\AgdaField{Carrier}\AgdaSpace{}%
\AgdaSymbol{(}\AgdaFunction{Env}\AgdaSpace{}%
\AgdaBound{𝑨}\AgdaSpace{}%
\AgdaBound{X}\AgdaSymbol{))}\<%
\\
\\[\AgdaEmptyExtraSkip]%
\>[1]\AgdaFunction{𝔄⁺}\AgdaSpace{}%
\AgdaSymbol{:}\AgdaSpace{}%
\AgdaFunction{ℑ⁺}\AgdaSpace{}%
\AgdaSymbol{→}\AgdaSpace{}%
\AgdaRecord{Algebra}\AgdaSpace{}%
\AgdaBound{α}\AgdaSpace{}%
\AgdaBound{ρᵃ}\<%
\\
\>[1]\AgdaFunction{𝔄⁺}\AgdaSpace{}%
\AgdaBound{i}\AgdaSpace{}%
\AgdaSymbol{=}\AgdaSpace{}%
\AgdaOperator{\AgdaFunction{∣}}\AgdaSpace{}%
\AgdaBound{i}\AgdaSpace{}%
\AgdaOperator{\AgdaFunction{∣}}\<%
\\
\\[\AgdaEmptyExtraSkip]%
\>[1]\AgdaFunction{𝑪}\AgdaSpace{}%
\AgdaSymbol{:}\AgdaSpace{}%
\AgdaRecord{Algebra}\AgdaSpace{}%
\AgdaFunction{ι}\AgdaSpace{}%
\AgdaFunction{ι}\<%
\\
\>[1]\AgdaFunction{𝑪}\AgdaSpace{}%
\AgdaSymbol{=}\AgdaSpace{}%
\AgdaFunction{⨅}\AgdaSpace{}%
\AgdaFunction{𝔄⁺}\<%
\end{code}

\ifshort\else
\begin{code}%
\>[0]\<%
\\
\>[1]\AgdaFunction{skEqual}\AgdaSpace{}%
\AgdaSymbol{:}\AgdaSpace{}%
\AgdaSymbol{(}\AgdaBound{i}\AgdaSpace{}%
\AgdaSymbol{:}\AgdaSpace{}%
\AgdaFunction{ℑ⁺}\AgdaSymbol{)}\AgdaSpace{}%
\AgdaSymbol{→}\AgdaSpace{}%
\AgdaSymbol{∀\{}\AgdaBound{p}\AgdaSpace{}%
\AgdaBound{q}\AgdaSymbol{\}}\AgdaSpace{}%
\AgdaSymbol{→}\AgdaSpace{}%
\AgdaPrimitive{Type}\AgdaSpace{}%
\AgdaBound{ρᵃ}\<%
\\
\>[1]\AgdaFunction{skEqual}\AgdaSpace{}%
\AgdaBound{i}\AgdaSpace{}%
\AgdaSymbol{\{}\AgdaBound{p}\AgdaSymbol{\}\{}\AgdaBound{q}\AgdaSymbol{\}}\AgdaSpace{}%
\AgdaSymbol{=}\AgdaSpace{}%
\AgdaOperator{\AgdaFunction{⟦}}\AgdaSpace{}%
\AgdaBound{p}\AgdaSpace{}%
\AgdaOperator{\AgdaFunction{⟧}}\AgdaSpace{}%
\AgdaOperator{\AgdaField{⟨\$⟩}}\AgdaSpace{}%
\AgdaField{snd}\AgdaSpace{}%
\AgdaOperator{\AgdaFunction{∥}}\AgdaSpace{}%
\AgdaBound{i}\AgdaSpace{}%
\AgdaOperator{\AgdaFunction{∥}}\AgdaSpace{}%
\AgdaOperator{\AgdaFunction{≈}}\AgdaSpace{}%
\AgdaOperator{\AgdaFunction{⟦}}\AgdaSpace{}%
\AgdaBound{q}\AgdaSpace{}%
\AgdaOperator{\AgdaFunction{⟧}}\AgdaSpace{}%
\AgdaOperator{\AgdaField{⟨\$⟩}}\AgdaSpace{}%
\AgdaField{snd}\AgdaSpace{}%
\AgdaOperator{\AgdaFunction{∥}}\AgdaSpace{}%
\AgdaBound{i}\AgdaSpace{}%
\AgdaOperator{\AgdaFunction{∥}}\<%
\\
\>[1][@{}l@{\AgdaIndent{0}}]%
\>[2]\AgdaKeyword{where}\AgdaSpace{}%
\AgdaKeyword{open}\AgdaSpace{}%
\AgdaModule{Setoid}\AgdaSpace{}%
\AgdaOperator{\AgdaFunction{𝔻[}}\AgdaSpace{}%
\AgdaFunction{𝔄⁺}\AgdaSpace{}%
\AgdaBound{i}\AgdaSpace{}%
\AgdaOperator{\AgdaFunction{]}}\AgdaSpace{}%
\AgdaKeyword{using}\AgdaSpace{}%
\AgdaSymbol{(}\AgdaSpace{}%
\AgdaOperator{\AgdaField{\AgdaUnderscore{}≈\AgdaUnderscore{}}}\AgdaSpace{}%
\AgdaSymbol{)}\AgdaSpace{}%
\AgdaSymbol{;}\AgdaSpace{}%
\AgdaKeyword{open}\AgdaSpace{}%
\AgdaModule{Environment}\AgdaSpace{}%
\AgdaSymbol{(}\AgdaFunction{𝔄⁺}\AgdaSpace{}%
\AgdaBound{i}\AgdaSymbol{)}\AgdaSpace{}%
\AgdaKeyword{using}\AgdaSpace{}%
\AgdaSymbol{(}\AgdaSpace{}%
\AgdaOperator{\AgdaFunction{⟦\AgdaUnderscore{}⟧}}\AgdaSpace{}%
\AgdaSymbol{)}\<%
\\
\>[0]\<%
\end{code}
The type \af{skEqual} provides a term identity \ab p \af{≈} \ab q for each index \ab i = (\ab{𝑨} , \ab{p} , \ab{ρ}) of the product.
%(here, as above, \ab{𝑨} is an algebra, \ab{sA} is a proof that \ab{𝑨} belongs to \af{S} \ab{𝒦}, and \ab{ρ} is an environment).
%map assigning values in the domain of \ab{𝑨} to variable symbols in \ab X).
Later we prove that if the identity \ab{p} \af{≈} \ab q holds in all \ab{𝑨} \aof{∈} \af S \ab{𝒦} (for all environments), then \ab p \af{≈} \ab q
holds in the relatively free algebra \Free{X}; equivalently, the pair (\ab p , \ab q) belongs to the
kernel of the natural homomorphism from \T{X} onto \Free{X}. We will use that fact to prove
that the kernel of the natural hom from \T{X} to \ab{𝑪} is contained in the kernel of the natural hom from \T{X} onto \Free{X},
whence we construct a monomorphism from \Free{X} into \ab{𝑪}, and thus \Free{X} is a subalgebra of \ab{𝑪},
so belongs to \af S (\af P \ab{𝒦}).
\fi

\begin{code}%
\>[0]\<%
\\
\>[0][@{}l@{\AgdaIndent{1}}]%
\>[1]\AgdaFunction{homC}\AgdaSpace{}%
\AgdaSymbol{:}\AgdaSpace{}%
\AgdaFunction{hom}\AgdaSpace{}%
\AgdaSymbol{(}\AgdaFunction{𝑻}\AgdaSpace{}%
\AgdaBound{X}\AgdaSymbol{)}\AgdaSpace{}%
\AgdaFunction{𝑪}\<%
\\
\>[1]\AgdaFunction{homC}\AgdaSpace{}%
\AgdaSymbol{=}\AgdaSpace{}%
\AgdaFunction{⨅-hom-co}\AgdaSpace{}%
\AgdaFunction{𝔄⁺}\AgdaSpace{}%
\AgdaSymbol{(λ}\AgdaSpace{}%
\AgdaBound{i}\AgdaSpace{}%
\AgdaSymbol{→}\AgdaSpace{}%
\AgdaFunction{lift-hom}\AgdaSpace{}%
\AgdaSymbol{(}\AgdaField{snd}\AgdaSpace{}%
\AgdaOperator{\AgdaFunction{∥}}\AgdaSpace{}%
\AgdaBound{i}\AgdaSpace{}%
\AgdaOperator{\AgdaFunction{∥}}\AgdaSymbol{))}\<%
\end{code}
\ifshort\else
\begin{code}%
\>[0]\<%
\\
\>[1]\AgdaFunction{kerF⊆kerC}\AgdaSpace{}%
\AgdaSymbol{:}\AgdaSpace{}%
\AgdaFunction{ker}\AgdaSpace{}%
\AgdaOperator{\AgdaFunction{∣}}\AgdaSpace{}%
\AgdaOperator{\AgdaFunction{homF[}}\AgdaSpace{}%
\AgdaBound{X}\AgdaSpace{}%
\AgdaOperator{\AgdaFunction{]}}\AgdaSpace{}%
\AgdaOperator{\AgdaFunction{∣}}\AgdaSpace{}%
\AgdaOperator{\AgdaFunction{⊆}}\AgdaSpace{}%
\AgdaFunction{ker}\AgdaSpace{}%
\AgdaOperator{\AgdaFunction{∣}}\AgdaSpace{}%
\AgdaFunction{homC}\AgdaSpace{}%
\AgdaOperator{\AgdaFunction{∣}}\<%
\\
\>[1]\AgdaFunction{kerF⊆kerC}\AgdaSpace{}%
\AgdaSymbol{\{}\AgdaBound{p}\AgdaSpace{}%
\AgdaOperator{\AgdaInductiveConstructor{,}}\AgdaSpace{}%
\AgdaBound{q}\AgdaSymbol{\}}\AgdaSpace{}%
\AgdaBound{pKq}\AgdaSpace{}%
\AgdaSymbol{(}\AgdaBound{𝑨}\AgdaSpace{}%
\AgdaOperator{\AgdaInductiveConstructor{,}}\AgdaSpace{}%
\AgdaBound{sA}\AgdaSpace{}%
\AgdaOperator{\AgdaInductiveConstructor{,}}\AgdaSpace{}%
\AgdaBound{ρ}\AgdaSymbol{)}\AgdaSpace{}%
\AgdaSymbol{=}\AgdaSpace{}%
\AgdaFunction{Goal}\<%
\\
\>[1][@{}l@{\AgdaIndent{0}}]%
\>[2]\AgdaKeyword{where}\<%
\\
\>[2]\AgdaKeyword{open}\AgdaSpace{}%
\AgdaModule{Setoid}\AgdaSpace{}%
\AgdaOperator{\AgdaFunction{𝔻[}}\AgdaSpace{}%
\AgdaBound{𝑨}\AgdaSpace{}%
\AgdaOperator{\AgdaFunction{]}}%
\>[22]\AgdaKeyword{using}\AgdaSpace{}%
\AgdaSymbol{(}\AgdaSpace{}%
\AgdaOperator{\AgdaField{\AgdaUnderscore{}≈\AgdaUnderscore{}}}\AgdaSpace{}%
\AgdaSymbol{;}\AgdaSpace{}%
\AgdaFunction{sym}\AgdaSpace{}%
\AgdaSymbol{;}\AgdaSpace{}%
\AgdaFunction{trans}\AgdaSpace{}%
\AgdaSymbol{)}\<%
\\
\>[2]\AgdaKeyword{open}\AgdaSpace{}%
\AgdaModule{Environment}\AgdaSpace{}%
\AgdaBound{𝑨}%
\>[22]\AgdaKeyword{using}\AgdaSpace{}%
\AgdaSymbol{(}\AgdaSpace{}%
\AgdaOperator{\AgdaFunction{⟦\AgdaUnderscore{}⟧}}\AgdaSpace{}%
\AgdaSymbol{)}\<%
\\
\>[2]\AgdaFunction{fl}\AgdaSpace{}%
\AgdaSymbol{:}\AgdaSpace{}%
\AgdaSymbol{∀}\AgdaSpace{}%
\AgdaBound{t}\AgdaSpace{}%
\AgdaSymbol{→}\AgdaSpace{}%
\AgdaOperator{\AgdaFunction{⟦}}\AgdaSpace{}%
\AgdaBound{t}\AgdaSpace{}%
\AgdaOperator{\AgdaFunction{⟧}}\AgdaSpace{}%
\AgdaOperator{\AgdaField{⟨\$⟩}}\AgdaSpace{}%
\AgdaBound{ρ}\AgdaSpace{}%
\AgdaOperator{\AgdaFunction{≈}}\AgdaSpace{}%
\AgdaFunction{free-lift}\AgdaSpace{}%
\AgdaBound{ρ}\AgdaSpace{}%
\AgdaBound{t}\<%
\\
\>[2]\AgdaFunction{fl}\AgdaSpace{}%
\AgdaBound{t}\AgdaSpace{}%
\AgdaSymbol{=}\AgdaSpace{}%
\AgdaFunction{free-lift-interp}\AgdaSpace{}%
\AgdaSymbol{\{}\AgdaArgument{𝑨}\AgdaSpace{}%
\AgdaSymbol{=}\AgdaSpace{}%
\AgdaBound{𝑨}\AgdaSymbol{\}}\AgdaSpace{}%
\AgdaBound{ρ}\AgdaSpace{}%
\AgdaBound{t}\<%
\\
\\[\AgdaEmptyExtraSkip]%
\>[2]\AgdaFunction{ζ}\AgdaSpace{}%
\AgdaSymbol{:}\AgdaSpace{}%
\AgdaSymbol{∀\{}\AgdaBound{p}\AgdaSpace{}%
\AgdaBound{q}\AgdaSymbol{\}}\AgdaSpace{}%
\AgdaSymbol{→}\AgdaSpace{}%
\AgdaSymbol{(}\AgdaFunction{Th}\AgdaSpace{}%
\AgdaBound{𝒦}\AgdaSymbol{)}\AgdaSpace{}%
\AgdaOperator{\AgdaDatatype{⊢}}\AgdaSpace{}%
\AgdaBound{X}\AgdaSpace{}%
\AgdaOperator{\AgdaDatatype{▹}}\AgdaSpace{}%
\AgdaBound{p}\AgdaSpace{}%
\AgdaOperator{\AgdaDatatype{≈}}\AgdaSpace{}%
\AgdaBound{q}\AgdaSpace{}%
\AgdaSymbol{→}\AgdaSpace{}%
\AgdaBound{𝒦}\AgdaSpace{}%
\AgdaOperator{\AgdaFunction{⊫}}\AgdaSpace{}%
\AgdaBound{p}\AgdaSpace{}%
\AgdaOperator{\AgdaFunction{≈}}\AgdaSpace{}%
\AgdaBound{q}\<%
\\
\>[2]\AgdaFunction{ζ}\AgdaSpace{}%
\AgdaBound{x}\AgdaSpace{}%
\AgdaBound{𝑨}\AgdaSpace{}%
\AgdaBound{kA}\AgdaSpace{}%
\AgdaSymbol{=}\AgdaSpace{}%
\AgdaFunction{sound}\AgdaSpace{}%
\AgdaSymbol{(λ}\AgdaSpace{}%
\AgdaBound{y}\AgdaSpace{}%
\AgdaBound{ρ}\AgdaSpace{}%
\AgdaSymbol{→}\AgdaSpace{}%
\AgdaBound{y}\AgdaSpace{}%
\AgdaBound{𝑨}\AgdaSpace{}%
\AgdaBound{kA}\AgdaSpace{}%
\AgdaBound{ρ}\AgdaSymbol{)}\AgdaSpace{}%
\AgdaBound{x}\AgdaSpace{}%
\AgdaKeyword{where}\AgdaSpace{}%
\AgdaKeyword{open}\AgdaSpace{}%
\AgdaModule{Soundness}\AgdaSpace{}%
\AgdaSymbol{(}\AgdaFunction{Th}\AgdaSpace{}%
\AgdaBound{𝒦}\AgdaSymbol{)}\AgdaSpace{}%
\AgdaBound{𝑨}\<%
\\
\\[\AgdaEmptyExtraSkip]%
\>[2]\AgdaFunction{subgoal}\AgdaSpace{}%
\AgdaSymbol{:}\AgdaSpace{}%
\AgdaOperator{\AgdaFunction{⟦}}\AgdaSpace{}%
\AgdaBound{p}\AgdaSpace{}%
\AgdaOperator{\AgdaFunction{⟧}}\AgdaSpace{}%
\AgdaOperator{\AgdaField{⟨\$⟩}}\AgdaSpace{}%
\AgdaBound{ρ}\AgdaSpace{}%
\AgdaOperator{\AgdaFunction{≈}}\AgdaSpace{}%
\AgdaOperator{\AgdaFunction{⟦}}\AgdaSpace{}%
\AgdaBound{q}\AgdaSpace{}%
\AgdaOperator{\AgdaFunction{⟧}}\AgdaSpace{}%
\AgdaOperator{\AgdaField{⟨\$⟩}}\AgdaSpace{}%
\AgdaBound{ρ}\<%
\\
\>[2]\AgdaFunction{subgoal}\AgdaSpace{}%
\AgdaSymbol{=}\AgdaSpace{}%
\AgdaFunction{S-id1}\AgdaSymbol{\{}\AgdaArgument{ℓ}\AgdaSpace{}%
\AgdaSymbol{=}\AgdaSpace{}%
\AgdaBound{ℓ}\AgdaSymbol{\}\{}\AgdaArgument{p}\AgdaSpace{}%
\AgdaSymbol{=}\AgdaSpace{}%
\AgdaBound{p}\AgdaSymbol{\}\{}\AgdaBound{q}\AgdaSymbol{\}}\AgdaSpace{}%
\AgdaSymbol{(}\AgdaFunction{ζ}\AgdaSpace{}%
\AgdaBound{pKq}\AgdaSymbol{)}\AgdaSpace{}%
\AgdaBound{𝑨}\AgdaSpace{}%
\AgdaBound{sA}\AgdaSpace{}%
\AgdaBound{ρ}\<%
\\
\>[2]\AgdaFunction{Goal}\AgdaSpace{}%
\AgdaSymbol{:}\AgdaSpace{}%
\AgdaSymbol{(}\AgdaFunction{free-lift}\AgdaSymbol{\{}\AgdaArgument{𝑨}\AgdaSpace{}%
\AgdaSymbol{=}\AgdaSpace{}%
\AgdaBound{𝑨}\AgdaSymbol{\}}\AgdaSpace{}%
\AgdaBound{ρ}\AgdaSpace{}%
\AgdaBound{p}\AgdaSymbol{)}\AgdaSpace{}%
\AgdaOperator{\AgdaFunction{≈}}\AgdaSpace{}%
\AgdaSymbol{(}\AgdaFunction{free-lift}\AgdaSymbol{\{}\AgdaArgument{𝑨}\AgdaSpace{}%
\AgdaSymbol{=}\AgdaSpace{}%
\AgdaBound{𝑨}\AgdaSymbol{\}}\AgdaSpace{}%
\AgdaBound{ρ}\AgdaSpace{}%
\AgdaBound{q}\AgdaSymbol{)}\<%
\\
\>[2]\AgdaFunction{Goal}\AgdaSpace{}%
\AgdaSymbol{=}\AgdaSpace{}%
\AgdaFunction{trans}\AgdaSpace{}%
\AgdaSymbol{(}\AgdaFunction{sym}\AgdaSpace{}%
\AgdaSymbol{(}\AgdaFunction{fl}\AgdaSpace{}%
\AgdaBound{p}\AgdaSymbol{))}\AgdaSpace{}%
\AgdaSymbol{(}\AgdaFunction{trans}\AgdaSpace{}%
\AgdaFunction{subgoal}\AgdaSpace{}%
\AgdaSymbol{(}\AgdaFunction{fl}\AgdaSpace{}%
\AgdaBound{q}\AgdaSymbol{))}\<%
\end{code}
\fi
\begin{code}%
\>[0]\<%
\\
\>[1]\AgdaFunction{homFC}\AgdaSpace{}%
\AgdaSymbol{:}\AgdaSpace{}%
\AgdaFunction{hom}\AgdaSpace{}%
\AgdaOperator{\AgdaFunction{𝔽[}}\AgdaSpace{}%
\AgdaBound{X}\AgdaSpace{}%
\AgdaOperator{\AgdaFunction{]}}\AgdaSpace{}%
\AgdaFunction{𝑪}\<%
\\
\>[1]\AgdaFunction{homFC}\AgdaSpace{}%
\AgdaSymbol{=}\AgdaSpace{}%
\AgdaOperator{\AgdaFunction{∣}}\AgdaSpace{}%
\AgdaFunction{HomFactor}\AgdaSpace{}%
\AgdaFunction{𝑪}\AgdaSpace{}%
\AgdaFunction{homC}\AgdaSpace{}%
\AgdaOperator{\AgdaFunction{homF[}}\AgdaSpace{}%
\AgdaBound{X}\AgdaSpace{}%
\AgdaOperator{\AgdaFunction{]}}\AgdaSpace{}%
\AgdaFunction{kerF⊆kerC}\AgdaSpace{}%
\AgdaSymbol{(}\AgdaField{isSurjective}\AgdaSpace{}%
\AgdaOperator{\AgdaFunction{∥}}\AgdaSpace{}%
\AgdaOperator{\AgdaFunction{epiF[}}\AgdaSpace{}%
\AgdaBound{X}\AgdaSpace{}%
\AgdaOperator{\AgdaFunction{]}}\AgdaSpace{}%
\AgdaOperator{\AgdaFunction{∥}}\AgdaSymbol{)}\AgdaSpace{}%
\AgdaOperator{\AgdaFunction{∣}}\<%
\\
\>[0]\<%
\end{code}
If \AgdaPair{p}{q} belongs to the kernel of \af{homC}, then
\af{Th} \ab{𝒦} includes the identity \ab{p} \af{≈} \ab{q}---that is,
\af{Th} \ab{𝒦} \af{⊢} \ab X \af{▹} \ab{p} \af{≈} \ab{q}. Equivalently,
if the kernel of \af{homC} is contained in that of \af{homF[ X ]}.
\ifshort
We omit the proof of this lemma and merely display its formal statement.
\else
We formalize this fact as follows.
\fi

\begin{code}%
\>[0]\<%
\\
\>[0][@{}l@{\AgdaIndent{1}}]%
\>[1]\AgdaFunction{kerC⊆kerF}\AgdaSpace{}%
\AgdaSymbol{:}\AgdaSpace{}%
\AgdaSymbol{∀\{}\AgdaBound{p}\AgdaSpace{}%
\AgdaBound{q}\AgdaSymbol{\}}\AgdaSpace{}%
\AgdaSymbol{→}\AgdaSpace{}%
\AgdaSymbol{(}\AgdaBound{p}\AgdaSpace{}%
\AgdaOperator{\AgdaInductiveConstructor{,}}\AgdaSpace{}%
\AgdaBound{q}\AgdaSymbol{)}\AgdaSpace{}%
\AgdaOperator{\AgdaFunction{∈}}\AgdaSpace{}%
\AgdaFunction{ker}\AgdaSpace{}%
\AgdaOperator{\AgdaFunction{∣}}\AgdaSpace{}%
\AgdaFunction{homC}\AgdaSpace{}%
\AgdaOperator{\AgdaFunction{∣}}\AgdaSpace{}%
\AgdaSymbol{→}\AgdaSpace{}%
\AgdaSymbol{(}\AgdaBound{p}\AgdaSpace{}%
\AgdaOperator{\AgdaInductiveConstructor{,}}\AgdaSpace{}%
\AgdaBound{q}\AgdaSymbol{)}\AgdaSpace{}%
\AgdaOperator{\AgdaFunction{∈}}\AgdaSpace{}%
\AgdaFunction{ker}\AgdaSpace{}%
\AgdaOperator{\AgdaFunction{∣}}\AgdaSpace{}%
\AgdaOperator{\AgdaFunction{homF[}}\AgdaSpace{}%
\AgdaBound{X}\AgdaSpace{}%
\AgdaOperator{\AgdaFunction{]}}\AgdaSpace{}%
\AgdaOperator{\AgdaFunction{∣}}\<%
\end{code}
\ifshort
\vskip2mm
\else
\begin{code}%
\>[1]\AgdaFunction{kerC⊆kerF}\AgdaSpace{}%
\AgdaSymbol{\{}\AgdaBound{p}\AgdaSymbol{\}\{}\AgdaBound{q}\AgdaSymbol{\}}\AgdaSpace{}%
\AgdaBound{pKq}\AgdaSpace{}%
\AgdaSymbol{=}\AgdaSpace{}%
\AgdaFunction{S𝒦⊫→ker𝔽}\AgdaSpace{}%
\AgdaSymbol{(}\AgdaFunction{S𝒦⊫}\AgdaSpace{}%
\AgdaFunction{pqEqual}\AgdaSymbol{)}\<%
\\
\>[1][@{}l@{\AgdaIndent{0}}]%
\>[2]\AgdaKeyword{where}\<%
\\
\>[2]\AgdaFunction{S𝒦⊫}\AgdaSpace{}%
\AgdaSymbol{:}\AgdaSpace{}%
\AgdaSymbol{(∀}\AgdaSpace{}%
\AgdaBound{i}\AgdaSpace{}%
\AgdaSymbol{→}\AgdaSpace{}%
\AgdaFunction{skEqual}\AgdaSpace{}%
\AgdaBound{i}\AgdaSpace{}%
\AgdaSymbol{\{}\AgdaBound{p}\AgdaSymbol{\}\{}\AgdaBound{q}\AgdaSymbol{\})}\AgdaSpace{}%
\AgdaSymbol{→}\AgdaSpace{}%
\AgdaFunction{S}\AgdaSymbol{\{}\AgdaArgument{β}\AgdaSpace{}%
\AgdaSymbol{=}\AgdaSpace{}%
\AgdaBound{α}\AgdaSymbol{\}\{}\AgdaBound{ρᵃ}\AgdaSymbol{\}}\AgdaSpace{}%
\AgdaBound{ℓ}\AgdaSpace{}%
\AgdaBound{𝒦}\AgdaSpace{}%
\AgdaOperator{\AgdaFunction{⊫}}\AgdaSpace{}%
\AgdaBound{p}\AgdaSpace{}%
\AgdaOperator{\AgdaFunction{≈}}\AgdaSpace{}%
\AgdaBound{q}\<%
\\
\>[2]\AgdaFunction{S𝒦⊫}\AgdaSpace{}%
\AgdaBound{x}\AgdaSpace{}%
\AgdaBound{𝑨}\AgdaSpace{}%
\AgdaBound{sA}\AgdaSpace{}%
\AgdaBound{ρ}\AgdaSpace{}%
\AgdaSymbol{=}\AgdaSpace{}%
\AgdaBound{x}\AgdaSpace{}%
\AgdaSymbol{(}\AgdaBound{𝑨}\AgdaSpace{}%
\AgdaOperator{\AgdaInductiveConstructor{,}}\AgdaSpace{}%
\AgdaBound{sA}\AgdaSpace{}%
\AgdaOperator{\AgdaInductiveConstructor{,}}\AgdaSpace{}%
\AgdaBound{ρ}\AgdaSymbol{)}\<%
\\
\>[2]\AgdaFunction{S𝒦⊫→ker𝔽}\AgdaSpace{}%
\AgdaSymbol{:}\AgdaSpace{}%
\AgdaFunction{S}\AgdaSymbol{\{}\AgdaArgument{β}\AgdaSpace{}%
\AgdaSymbol{=}\AgdaSpace{}%
\AgdaBound{α}\AgdaSymbol{\}\{}\AgdaBound{ρᵃ}\AgdaSymbol{\}}\AgdaSpace{}%
\AgdaBound{ℓ}\AgdaSpace{}%
\AgdaBound{𝒦}\AgdaSpace{}%
\AgdaOperator{\AgdaFunction{⊫}}\AgdaSpace{}%
\AgdaBound{p}\AgdaSpace{}%
\AgdaOperator{\AgdaFunction{≈}}\AgdaSpace{}%
\AgdaBound{q}\AgdaSpace{}%
\AgdaSymbol{→}\AgdaSpace{}%
\AgdaSymbol{(}\AgdaBound{p}\AgdaSpace{}%
\AgdaOperator{\AgdaInductiveConstructor{,}}\AgdaSpace{}%
\AgdaBound{q}\AgdaSymbol{)}\AgdaSpace{}%
\AgdaOperator{\AgdaFunction{∈}}\AgdaSpace{}%
\AgdaFunction{ker}\AgdaSpace{}%
\AgdaOperator{\AgdaFunction{∣}}\AgdaSpace{}%
\AgdaOperator{\AgdaFunction{homF[}}\AgdaSpace{}%
\AgdaBound{X}\AgdaSpace{}%
\AgdaOperator{\AgdaFunction{]}}\AgdaSpace{}%
\AgdaOperator{\AgdaFunction{∣}}\<%
\\
\>[2]\AgdaFunction{S𝒦⊫→ker𝔽}\AgdaSpace{}%
\AgdaBound{x}\AgdaSpace{}%
\AgdaSymbol{=}\AgdaSpace{}%
\AgdaInductiveConstructor{hyp}\AgdaSpace{}%
\AgdaSymbol{(}\AgdaFunction{S-id2}\AgdaSymbol{\{}\AgdaArgument{ℓ}\AgdaSpace{}%
\AgdaSymbol{=}\AgdaSpace{}%
\AgdaBound{ℓ}\AgdaSymbol{\}\{}\AgdaArgument{p}\AgdaSpace{}%
\AgdaSymbol{=}\AgdaSpace{}%
\AgdaBound{p}\AgdaSymbol{\}\{}\AgdaBound{q}\AgdaSymbol{\}}\AgdaSpace{}%
\AgdaBound{x}\AgdaSymbol{)}\<%
\\
\\[\AgdaEmptyExtraSkip]%
\>[2]\AgdaFunction{pqEqual}\AgdaSpace{}%
\AgdaSymbol{:}\AgdaSpace{}%
\AgdaSymbol{∀}\AgdaSpace{}%
\AgdaBound{i}\AgdaSpace{}%
\AgdaSymbol{→}\AgdaSpace{}%
\AgdaFunction{skEqual}\AgdaSpace{}%
\AgdaBound{i}\AgdaSpace{}%
\AgdaSymbol{\{}\AgdaBound{p}\AgdaSymbol{\}\{}\AgdaBound{q}\AgdaSymbol{\}}\<%
\\
\>[2]\AgdaFunction{pqEqual}\AgdaSpace{}%
\AgdaBound{i}\AgdaSpace{}%
\AgdaSymbol{=}\AgdaSpace{}%
\AgdaFunction{goal}\<%
\\
\>[2][@{}l@{\AgdaIndent{0}}]%
\>[3]\AgdaKeyword{where}\<%
\\
\>[3]\AgdaKeyword{open}\AgdaSpace{}%
\AgdaModule{Environment}\AgdaSpace{}%
\AgdaSymbol{(}\AgdaFunction{𝔄⁺}\AgdaSpace{}%
\AgdaBound{i}\AgdaSymbol{)}%
\>[28]\AgdaKeyword{using}\AgdaSpace{}%
\AgdaSymbol{(}\AgdaSpace{}%
\AgdaOperator{\AgdaFunction{⟦\AgdaUnderscore{}⟧}}\AgdaSpace{}%
\AgdaSymbol{)}\<%
\\
\>[3]\AgdaKeyword{open}\AgdaSpace{}%
\AgdaModule{Setoid}\AgdaSpace{}%
\AgdaOperator{\AgdaFunction{𝔻[}}\AgdaSpace{}%
\AgdaFunction{𝔄⁺}\AgdaSpace{}%
\AgdaBound{i}\AgdaSpace{}%
\AgdaOperator{\AgdaFunction{]}}%
\>[28]\AgdaKeyword{using}\AgdaSpace{}%
\AgdaSymbol{(}\AgdaSpace{}%
\AgdaOperator{\AgdaField{\AgdaUnderscore{}≈\AgdaUnderscore{}}}\AgdaSpace{}%
\AgdaSymbol{;}\AgdaSpace{}%
\AgdaFunction{sym}\AgdaSpace{}%
\AgdaSymbol{;}\AgdaSpace{}%
\AgdaFunction{trans}\AgdaSpace{}%
\AgdaSymbol{)}\<%
\\
\>[3]\AgdaFunction{goal}\AgdaSpace{}%
\AgdaSymbol{:}\AgdaSpace{}%
\AgdaOperator{\AgdaFunction{⟦}}\AgdaSpace{}%
\AgdaBound{p}\AgdaSpace{}%
\AgdaOperator{\AgdaFunction{⟧}}\AgdaSpace{}%
\AgdaOperator{\AgdaField{⟨\$⟩}}\AgdaSpace{}%
\AgdaField{snd}\AgdaSpace{}%
\AgdaOperator{\AgdaFunction{∥}}\AgdaSpace{}%
\AgdaBound{i}\AgdaSpace{}%
\AgdaOperator{\AgdaFunction{∥}}\AgdaSpace{}%
\AgdaOperator{\AgdaFunction{≈}}\AgdaSpace{}%
\AgdaOperator{\AgdaFunction{⟦}}\AgdaSpace{}%
\AgdaBound{q}\AgdaSpace{}%
\AgdaOperator{\AgdaFunction{⟧}}\AgdaSpace{}%
\AgdaOperator{\AgdaField{⟨\$⟩}}\AgdaSpace{}%
\AgdaField{snd}\AgdaSpace{}%
\AgdaOperator{\AgdaFunction{∥}}\AgdaSpace{}%
\AgdaBound{i}\AgdaSpace{}%
\AgdaOperator{\AgdaFunction{∥}}\<%
\\
\>[3]\AgdaFunction{goal}%
\>[9]\AgdaSymbol{=}\AgdaSpace{}%
\AgdaFunction{trans}\AgdaSpace{}%
\AgdaSymbol{(}\AgdaFunction{free-lift-interp}\AgdaSymbol{\{}\AgdaArgument{𝑨}\AgdaSpace{}%
\AgdaSymbol{=}\AgdaSpace{}%
\AgdaOperator{\AgdaFunction{∣}}\AgdaSpace{}%
\AgdaBound{i}\AgdaSpace{}%
\AgdaOperator{\AgdaFunction{∣}}\AgdaSymbol{\}(}\AgdaField{snd}\AgdaSpace{}%
\AgdaOperator{\AgdaFunction{∥}}\AgdaSpace{}%
\AgdaBound{i}\AgdaSpace{}%
\AgdaOperator{\AgdaFunction{∥}}\AgdaSymbol{)}\AgdaSpace{}%
\AgdaBound{p}\AgdaSymbol{)}\<%
\\
\>[9]\AgdaSymbol{(}\AgdaSpace{}%
\AgdaFunction{trans}\AgdaSpace{}%
\AgdaSymbol{(}\AgdaBound{pKq}\AgdaSpace{}%
\AgdaBound{i}\AgdaSymbol{)(}\AgdaFunction{sym}\AgdaSpace{}%
\AgdaSymbol{(}\AgdaFunction{free-lift-interp}\AgdaSymbol{\{}\AgdaArgument{𝑨}\AgdaSpace{}%
\AgdaSymbol{=}\AgdaSpace{}%
\AgdaOperator{\AgdaFunction{∣}}\AgdaSpace{}%
\AgdaBound{i}\AgdaSpace{}%
\AgdaOperator{\AgdaFunction{∣}}\AgdaSymbol{\}}\AgdaSpace{}%
\AgdaSymbol{(}\AgdaField{snd}\AgdaSpace{}%
\AgdaOperator{\AgdaFunction{∥}}\AgdaSpace{}%
\AgdaBound{i}\AgdaSpace{}%
\AgdaOperator{\AgdaFunction{∥}}\AgdaSymbol{)}\AgdaSpace{}%
\AgdaBound{q}\AgdaSymbol{)))}\<%
\\
\>[0]\<%
\end{code}
\fi
\noindent We conclude that the homomorphism from \Free{X} to \af{𝑪} is injective, whence \Free{X} is (isomorphic to) a subalgebra of \af{𝑪}.

\begin{code}%
\>[0]\<%
\\
\>[0][@{}l@{\AgdaIndent{1}}]%
\>[1]\AgdaFunction{monFC}\AgdaSpace{}%
\AgdaSymbol{:}\AgdaSpace{}%
\AgdaFunction{mon}\AgdaSpace{}%
\AgdaOperator{\AgdaFunction{𝔽[}}\AgdaSpace{}%
\AgdaBound{X}\AgdaSpace{}%
\AgdaOperator{\AgdaFunction{]}}\AgdaSpace{}%
\AgdaFunction{𝑪}\<%
\\
\>[1]\AgdaFunction{monFC}\AgdaSpace{}%
\AgdaSymbol{=}\AgdaSpace{}%
\AgdaOperator{\AgdaFunction{∣}}\AgdaSpace{}%
\AgdaFunction{homFC}\AgdaSpace{}%
\AgdaOperator{\AgdaFunction{∣}}\AgdaSpace{}%
\AgdaOperator{\AgdaInductiveConstructor{,}}\AgdaSpace{}%
\AgdaFunction{isMon}\<%
\\
\>[1][@{}l@{\AgdaIndent{0}}]%
\>[2]\AgdaKeyword{where}\<%
\\
\>[2]\AgdaFunction{isMon}\AgdaSpace{}%
\AgdaSymbol{:}\AgdaSpace{}%
\AgdaRecord{IsMon}\AgdaSpace{}%
\AgdaOperator{\AgdaFunction{𝔽[}}\AgdaSpace{}%
\AgdaBound{X}\AgdaSpace{}%
\AgdaOperator{\AgdaFunction{]}}\AgdaSpace{}%
\AgdaFunction{𝑪}\AgdaSpace{}%
\AgdaOperator{\AgdaFunction{∣}}\AgdaSpace{}%
\AgdaFunction{homFC}\AgdaSpace{}%
\AgdaOperator{\AgdaFunction{∣}}\<%
\\
\>[2]\AgdaField{isHom}\AgdaSpace{}%
\AgdaFunction{isMon}\AgdaSpace{}%
\AgdaSymbol{=}\AgdaSpace{}%
\AgdaOperator{\AgdaFunction{∥}}\AgdaSpace{}%
\AgdaFunction{homFC}\AgdaSpace{}%
\AgdaOperator{\AgdaFunction{∥}}\<%
\\
\>[2]\AgdaField{isInjective}\AgdaSpace{}%
\AgdaFunction{isMon}\AgdaSpace{}%
\AgdaSymbol{\{}\AgdaBound{p}\AgdaSymbol{\}\{}\AgdaBound{q}\AgdaSymbol{\}}\AgdaSpace{}%
\AgdaBound{φpq}\AgdaSpace{}%
\AgdaSymbol{=}\AgdaSpace{}%
\AgdaFunction{kerC⊆kerF}\AgdaSpace{}%
\AgdaBound{φpq}\<%
\\
\\[\AgdaEmptyExtraSkip]%
\>[1]\AgdaFunction{F≤C}\AgdaSpace{}%
\AgdaSymbol{:}\AgdaSpace{}%
\AgdaOperator{\AgdaFunction{𝔽[}}\AgdaSpace{}%
\AgdaBound{X}\AgdaSpace{}%
\AgdaOperator{\AgdaFunction{]}}\AgdaSpace{}%
\AgdaOperator{\AgdaFunction{≤}}\AgdaSpace{}%
\AgdaFunction{𝑪}\<%
\\
\>[1]\AgdaFunction{F≤C}\AgdaSpace{}%
\AgdaSymbol{=}\AgdaSpace{}%
\AgdaFunction{mon→≤}\AgdaSpace{}%
\AgdaFunction{monFC}\<%
\\
\>[0]\<%
\end{code}
Using the last result we prove that \Free{X} belongs to \af{S} (\af{P} \ab{𝒦}). This
requires one more technical lemma concerning the classes \af{S} and \af{P}; specifically,
\ifshort
\af{P} (\af{S} \ab{𝒦}) \aof{⊆} \af{S} (\af{P} \ab{𝒦}) holds for every class \ab{𝒦}.
The \agdaalgebras library contains the formal statement and proof of this result, where
it is called \af{PS⊆SP} (\seeshort).
\else
a product of subalgebras of algebras in a class is a subalgebra of a product of algebras in the class;
in other terms, \af{P} (\af{S} \ab{𝒦}) \aof{⊆} \af{S} (\af{P} \ab{𝒦}), for every class \ab{𝒦}.
We state and prove this in Agda as follows.

\begin{code}%
\>[0]\<%
\\
\>[0][@{}l@{\AgdaIndent{1}}]%
\>[1]\AgdaKeyword{private}\AgdaSpace{}%
\AgdaFunction{a}\AgdaSpace{}%
\AgdaSymbol{=}\AgdaSpace{}%
\AgdaBound{α}\AgdaSpace{}%
\AgdaOperator{\AgdaPrimitive{⊔}}\AgdaSpace{}%
\AgdaBound{ρᵃ}\AgdaSpace{}%
\AgdaSymbol{;}\AgdaSpace{}%
\AgdaFunction{oaℓ}\AgdaSpace{}%
\AgdaSymbol{=}\AgdaSpace{}%
\AgdaFunction{ov}\AgdaSpace{}%
\AgdaSymbol{(}\AgdaFunction{a}\AgdaSpace{}%
\AgdaOperator{\AgdaPrimitive{⊔}}\AgdaSpace{}%
\AgdaBound{ℓ}\AgdaSymbol{)}\<%
\\
\\[\AgdaEmptyExtraSkip]%
\>[1]\AgdaFunction{PS⊆SP}\AgdaSpace{}%
\AgdaSymbol{:}\AgdaSpace{}%
\AgdaFunction{P}\AgdaSpace{}%
\AgdaSymbol{(}\AgdaFunction{a}\AgdaSpace{}%
\AgdaOperator{\AgdaPrimitive{⊔}}\AgdaSpace{}%
\AgdaBound{ℓ}\AgdaSymbol{)}\AgdaSpace{}%
\AgdaFunction{oaℓ}\AgdaSpace{}%
\AgdaSymbol{(}\AgdaFunction{S}\AgdaSymbol{\{}\AgdaArgument{β}\AgdaSpace{}%
\AgdaSymbol{=}\AgdaSpace{}%
\AgdaBound{α}\AgdaSymbol{\}\{}\AgdaBound{ρᵃ}\AgdaSymbol{\}}\AgdaSpace{}%
\AgdaBound{ℓ}\AgdaSpace{}%
\AgdaBound{𝒦}\AgdaSymbol{)}\AgdaSpace{}%
\AgdaOperator{\AgdaFunction{⊆}}\AgdaSpace{}%
\AgdaFunction{S}\AgdaSpace{}%
\AgdaFunction{oaℓ}\AgdaSpace{}%
\AgdaSymbol{(}\AgdaFunction{P}\AgdaSpace{}%
\AgdaBound{ℓ}\AgdaSpace{}%
\AgdaFunction{oaℓ}\AgdaSpace{}%
\AgdaBound{𝒦}\AgdaSymbol{)}\<%
\\
\>[1]\AgdaFunction{PS⊆SP}\AgdaSpace{}%
\AgdaSymbol{\{}\AgdaBound{𝑩}\AgdaSymbol{\}}\AgdaSpace{}%
\AgdaSymbol{(}\AgdaBound{I}\AgdaSpace{}%
\AgdaOperator{\AgdaInductiveConstructor{,}}\AgdaSpace{}%
\AgdaSymbol{(}\AgdaSpace{}%
\AgdaBound{𝒜}\AgdaSpace{}%
\AgdaOperator{\AgdaInductiveConstructor{,}}\AgdaSpace{}%
\AgdaBound{sA}\AgdaSpace{}%
\AgdaOperator{\AgdaInductiveConstructor{,}}\AgdaSpace{}%
\AgdaBound{B≅⨅A}\AgdaSpace{}%
\AgdaSymbol{))}\AgdaSpace{}%
\AgdaSymbol{=}\AgdaSpace{}%
\AgdaFunction{Goal}\<%
\\
\>[1][@{}l@{\AgdaIndent{0}}]%
\>[2]\AgdaKeyword{where}\<%
\\
\>[2]\AgdaFunction{ℬ}\AgdaSpace{}%
\AgdaSymbol{:}\AgdaSpace{}%
\AgdaBound{I}\AgdaSpace{}%
\AgdaSymbol{→}\AgdaSpace{}%
\AgdaRecord{Algebra}\AgdaSpace{}%
\AgdaBound{α}\AgdaSpace{}%
\AgdaBound{ρᵃ}\<%
\\
\>[2]\AgdaFunction{ℬ}\AgdaSpace{}%
\AgdaBound{i}\AgdaSpace{}%
\AgdaSymbol{=}\AgdaSpace{}%
\AgdaOperator{\AgdaFunction{∣}}\AgdaSpace{}%
\AgdaBound{sA}\AgdaSpace{}%
\AgdaBound{i}\AgdaSpace{}%
\AgdaOperator{\AgdaFunction{∣}}\<%
\\
\>[2]\AgdaFunction{kB}\AgdaSpace{}%
\AgdaSymbol{:}\AgdaSpace{}%
\AgdaSymbol{(}\AgdaBound{i}\AgdaSpace{}%
\AgdaSymbol{:}\AgdaSpace{}%
\AgdaBound{I}\AgdaSymbol{)}\AgdaSpace{}%
\AgdaSymbol{→}\AgdaSpace{}%
\AgdaFunction{ℬ}\AgdaSpace{}%
\AgdaBound{i}\AgdaSpace{}%
\AgdaOperator{\AgdaFunction{∈}}\AgdaSpace{}%
\AgdaBound{𝒦}\<%
\\
\>[2]\AgdaFunction{kB}\AgdaSpace{}%
\AgdaBound{i}\AgdaSpace{}%
\AgdaSymbol{=}%
\>[10]\AgdaField{fst}\AgdaSpace{}%
\AgdaOperator{\AgdaFunction{∥}}\AgdaSpace{}%
\AgdaBound{sA}\AgdaSpace{}%
\AgdaBound{i}\AgdaSpace{}%
\AgdaOperator{\AgdaFunction{∥}}\<%
\\
\>[2]\AgdaFunction{⨅A≤⨅B}\AgdaSpace{}%
\AgdaSymbol{:}\AgdaSpace{}%
\AgdaFunction{⨅}\AgdaSpace{}%
\AgdaBound{𝒜}\AgdaSpace{}%
\AgdaOperator{\AgdaFunction{≤}}\AgdaSpace{}%
\AgdaFunction{⨅}\AgdaSpace{}%
\AgdaFunction{ℬ}\<%
\\
\>[2]\AgdaFunction{⨅A≤⨅B}\AgdaSpace{}%
\AgdaSymbol{=}\AgdaSpace{}%
\AgdaFunction{⨅-≤}\AgdaSpace{}%
\AgdaSymbol{λ}\AgdaSpace{}%
\AgdaBound{i}\AgdaSpace{}%
\AgdaSymbol{→}\AgdaSpace{}%
\AgdaField{snd}\AgdaSpace{}%
\AgdaOperator{\AgdaFunction{∥}}\AgdaSpace{}%
\AgdaBound{sA}\AgdaSpace{}%
\AgdaBound{i}\AgdaSpace{}%
\AgdaOperator{\AgdaFunction{∥}}\<%
\\
\>[2]\AgdaFunction{Goal}\AgdaSpace{}%
\AgdaSymbol{:}\AgdaSpace{}%
\AgdaBound{𝑩}\AgdaSpace{}%
\AgdaOperator{\AgdaFunction{∈}}\AgdaSpace{}%
\AgdaFunction{S}\AgdaSymbol{\{}\AgdaArgument{β}\AgdaSpace{}%
\AgdaSymbol{=}\AgdaSpace{}%
\AgdaFunction{oaℓ}\AgdaSymbol{\}\{}\AgdaFunction{oaℓ}\AgdaSymbol{\}}\AgdaFunction{oaℓ}\AgdaSpace{}%
\AgdaSymbol{(}\AgdaFunction{P}\AgdaSpace{}%
\AgdaSymbol{\{}\AgdaArgument{β}\AgdaSpace{}%
\AgdaSymbol{=}\AgdaSpace{}%
\AgdaFunction{oaℓ}\AgdaSymbol{\}\{}\AgdaFunction{oaℓ}\AgdaSymbol{\}}\AgdaSpace{}%
\AgdaBound{ℓ}\AgdaSpace{}%
\AgdaFunction{oaℓ}\AgdaSpace{}%
\AgdaBound{𝒦}\AgdaSymbol{)}\<%
\\
\>[2]\AgdaFunction{Goal}\AgdaSpace{}%
\AgdaSymbol{=}\AgdaSpace{}%
\AgdaFunction{⨅}\AgdaSpace{}%
\AgdaFunction{ℬ}\AgdaSpace{}%
\AgdaOperator{\AgdaInductiveConstructor{,}}\AgdaSpace{}%
\AgdaSymbol{(}\AgdaBound{I}\AgdaSpace{}%
\AgdaOperator{\AgdaInductiveConstructor{,}}\AgdaSpace{}%
\AgdaSymbol{(}\AgdaFunction{ℬ}\AgdaSpace{}%
\AgdaOperator{\AgdaInductiveConstructor{,}}\AgdaSpace{}%
\AgdaSymbol{(}\AgdaFunction{kB}\AgdaSpace{}%
\AgdaOperator{\AgdaInductiveConstructor{,}}\AgdaSpace{}%
\AgdaFunction{≅-refl}\AgdaSymbol{)))}\AgdaSpace{}%
\AgdaOperator{\AgdaInductiveConstructor{,}}\AgdaSpace{}%
\AgdaSymbol{(}\AgdaFunction{≅-trans-≤}\AgdaSpace{}%
\AgdaBound{B≅⨅A}\AgdaSpace{}%
\AgdaFunction{⨅A≤⨅B}\AgdaSymbol{)}\<%
\\
\>[0]\<%
\end{code}
With this we can prove that \Free{X} belongs to \af{S} (\af{P} \ab{𝒦}).
\fi

\begin{code}%
\>[0]\<%
\\
\>[0][@{}l@{\AgdaIndent{1}}]%
\>[1]\AgdaFunction{SPF}\AgdaSpace{}%
\AgdaSymbol{:}\AgdaSpace{}%
\AgdaOperator{\AgdaFunction{𝔽[}}\AgdaSpace{}%
\AgdaBound{X}\AgdaSpace{}%
\AgdaOperator{\AgdaFunction{]}}\AgdaSpace{}%
\AgdaOperator{\AgdaFunction{∈}}\AgdaSpace{}%
\AgdaFunction{S}\AgdaSpace{}%
\AgdaFunction{ι}\AgdaSpace{}%
\AgdaSymbol{(}\AgdaFunction{P}\AgdaSpace{}%
\AgdaBound{ℓ}\AgdaSpace{}%
\AgdaFunction{ι}\AgdaSpace{}%
\AgdaBound{𝒦}\AgdaSymbol{)}\<%
\\
\>[1]\AgdaFunction{SPF}\AgdaSpace{}%
\AgdaSymbol{=}\AgdaSpace{}%
\AgdaOperator{\AgdaFunction{∣}}\AgdaSpace{}%
\AgdaFunction{spC}\AgdaSpace{}%
\AgdaOperator{\AgdaFunction{∣}}\AgdaSpace{}%
\AgdaOperator{\AgdaInductiveConstructor{,}}\AgdaSpace{}%
\AgdaSymbol{(}\AgdaField{fst}\AgdaSpace{}%
\AgdaOperator{\AgdaFunction{∥}}\AgdaSpace{}%
\AgdaFunction{spC}\AgdaSpace{}%
\AgdaOperator{\AgdaFunction{∥}}\AgdaSymbol{)}\AgdaSpace{}%
\AgdaOperator{\AgdaInductiveConstructor{,}}\AgdaSpace{}%
\AgdaSymbol{(}\AgdaFunction{≤-transitive}\AgdaSpace{}%
\AgdaFunction{F≤C}\AgdaSpace{}%
\AgdaSymbol{(}\AgdaField{snd}\AgdaSpace{}%
\AgdaOperator{\AgdaFunction{∥}}\AgdaSpace{}%
\AgdaFunction{spC}\AgdaSpace{}%
\AgdaOperator{\AgdaFunction{∥}}\AgdaSymbol{))}\<%
\\
\>[1][@{}l@{\AgdaIndent{0}}]%
\>[2]\AgdaKeyword{where}\<%
\\
\>[2]\AgdaFunction{psC}\AgdaSpace{}%
\AgdaSymbol{:}\AgdaSpace{}%
\AgdaFunction{𝑪}\AgdaSpace{}%
\AgdaOperator{\AgdaFunction{∈}}\AgdaSpace{}%
\AgdaFunction{P}\AgdaSpace{}%
\AgdaSymbol{(}\AgdaBound{α}\AgdaSpace{}%
\AgdaOperator{\AgdaPrimitive{⊔}}\AgdaSpace{}%
\AgdaBound{ρᵃ}\AgdaSpace{}%
\AgdaOperator{\AgdaPrimitive{⊔}}\AgdaSpace{}%
\AgdaBound{ℓ}\AgdaSymbol{)}\AgdaSpace{}%
\AgdaFunction{ι}\AgdaSpace{}%
\AgdaSymbol{(}\AgdaFunction{S}\AgdaSpace{}%
\AgdaBound{ℓ}\AgdaSpace{}%
\AgdaBound{𝒦}\AgdaSymbol{)}\<%
\\
\>[2]\AgdaFunction{psC}\AgdaSpace{}%
\AgdaSymbol{=}\AgdaSpace{}%
\AgdaFunction{ℑ⁺}\AgdaSpace{}%
\AgdaOperator{\AgdaInductiveConstructor{,}}\AgdaSpace{}%
\AgdaSymbol{(}\AgdaFunction{𝔄⁺}\AgdaSpace{}%
\AgdaOperator{\AgdaInductiveConstructor{,}}\AgdaSpace{}%
\AgdaSymbol{((λ}\AgdaSpace{}%
\AgdaBound{i}\AgdaSpace{}%
\AgdaSymbol{→}\AgdaSpace{}%
\AgdaField{fst}\AgdaSpace{}%
\AgdaOperator{\AgdaFunction{∥}}\AgdaSpace{}%
\AgdaBound{i}\AgdaSpace{}%
\AgdaOperator{\AgdaFunction{∥}}\AgdaSymbol{)}\AgdaSpace{}%
\AgdaOperator{\AgdaInductiveConstructor{,}}\AgdaSpace{}%
\AgdaFunction{≅-refl}\AgdaSymbol{))}\<%
\\
\>[2]\AgdaFunction{spC}\AgdaSpace{}%
\AgdaSymbol{:}\AgdaSpace{}%
\AgdaFunction{𝑪}\AgdaSpace{}%
\AgdaOperator{\AgdaFunction{∈}}\AgdaSpace{}%
\AgdaFunction{S}\AgdaSpace{}%
\AgdaFunction{ι}\AgdaSpace{}%
\AgdaSymbol{(}\AgdaFunction{P}\AgdaSpace{}%
\AgdaBound{ℓ}\AgdaSpace{}%
\AgdaFunction{ι}\AgdaSpace{}%
\AgdaBound{𝒦}\AgdaSymbol{)}\<%
\\
\>[2]\AgdaFunction{spC}\AgdaSpace{}%
\AgdaSymbol{=}\AgdaSpace{}%
\AgdaFunction{PS⊆SP}\AgdaSpace{}%
\AgdaFunction{psC}\<%
\\
\>[0]\<%
\end{code}
Finally, we prove that every algebra in \af{Mod} (\af{Th} (\af{V} \ab{𝒦})) is a homomorphic image of \af{𝔽[~\ab{X}~]}.

\ifshort\else
\begin{code}%
\>[0]\<%
\\
\>[0]\AgdaKeyword{module}\AgdaSpace{}%
\AgdaModule{\AgdaUnderscore{}}\AgdaSpace{}%
\AgdaSymbol{\{}\AgdaBound{𝒦}\AgdaSpace{}%
\AgdaSymbol{:}\AgdaSpace{}%
\AgdaFunction{Pred}\AgdaSymbol{(}\AgdaRecord{Algebra}\AgdaSpace{}%
\AgdaGeneralizable{α}\AgdaSpace{}%
\AgdaGeneralizable{ρᵃ}\AgdaSymbol{)}\AgdaSpace{}%
\AgdaSymbol{(}\AgdaGeneralizable{α}\AgdaSpace{}%
\AgdaOperator{\AgdaPrimitive{⊔}}\AgdaSpace{}%
\AgdaGeneralizable{ρᵃ}\AgdaSpace{}%
\AgdaOperator{\AgdaPrimitive{⊔}}\AgdaSpace{}%
\AgdaFunction{ov}\AgdaSpace{}%
\AgdaGeneralizable{ℓ}\AgdaSymbol{)\}}\AgdaSpace{}%
\AgdaKeyword{where}\<%
\\
\>[0][@{}l@{\AgdaIndent{0}}]%
\>[1]\AgdaKeyword{private}\AgdaSpace{}%
\AgdaFunction{c}\AgdaSpace{}%
\AgdaSymbol{=}\AgdaSpace{}%
\AgdaBound{α}\AgdaSpace{}%
\AgdaOperator{\AgdaPrimitive{⊔}}\AgdaSpace{}%
\AgdaBound{ρᵃ}\AgdaSpace{}%
\AgdaOperator{\AgdaPrimitive{⊔}}\AgdaSpace{}%
\AgdaBound{ℓ}\AgdaSpace{}%
\AgdaSymbol{;}\AgdaSpace{}%
\AgdaFunction{ι}\AgdaSpace{}%
\AgdaSymbol{=}\AgdaSpace{}%
\AgdaFunction{ov}\AgdaSpace{}%
\AgdaFunction{c}\<%
\\
\>[1]\AgdaKeyword{open}\AgdaSpace{}%
\AgdaModule{FreeAlgebra}\AgdaSpace{}%
\AgdaSymbol{\{}\AgdaArgument{χ}\AgdaSpace{}%
\AgdaSymbol{=}\AgdaSpace{}%
\AgdaFunction{c}\AgdaSymbol{\}(}\AgdaFunction{Th}\AgdaSpace{}%
\AgdaBound{𝒦}\AgdaSymbol{)}\AgdaSpace{}%
\AgdaKeyword{using}\AgdaSpace{}%
\AgdaSymbol{(}\AgdaSpace{}%
\AgdaOperator{\AgdaFunction{𝔽[\AgdaUnderscore{}]}}\AgdaSpace{}%
\AgdaSymbol{)}\<%
\end{code}
\fi
\begin{code}%
\>[0]\<%
\\
\>[1]\AgdaFunction{Var⇒EqCl}\AgdaSpace{}%
\AgdaSymbol{:}\AgdaSpace{}%
\AgdaSymbol{∀}\AgdaSpace{}%
\AgdaBound{𝑨}\AgdaSpace{}%
\AgdaSymbol{→}\AgdaSpace{}%
\AgdaBound{𝑨}\AgdaSpace{}%
\AgdaOperator{\AgdaFunction{∈}}\AgdaSpace{}%
\AgdaFunction{Mod}\AgdaSpace{}%
\AgdaSymbol{(}\AgdaFunction{Th}\AgdaSpace{}%
\AgdaSymbol{(}\AgdaFunction{V}\AgdaSpace{}%
\AgdaBound{ℓ}\AgdaSpace{}%
\AgdaFunction{ι}\AgdaSpace{}%
\AgdaBound{𝒦}\AgdaSymbol{))}\AgdaSpace{}%
\AgdaSymbol{→}\AgdaSpace{}%
\AgdaBound{𝑨}\AgdaSpace{}%
\AgdaOperator{\AgdaFunction{∈}}\AgdaSpace{}%
\AgdaFunction{V}\AgdaSpace{}%
\AgdaBound{ℓ}\AgdaSpace{}%
\AgdaFunction{ι}\AgdaSpace{}%
\AgdaBound{𝒦}\<%
\\
\>[1]\AgdaFunction{Var⇒EqCl}\AgdaSpace{}%
\AgdaBound{𝑨}\AgdaSpace{}%
\AgdaBound{ModThA}\AgdaSpace{}%
\AgdaSymbol{=}\AgdaSpace{}%
\AgdaOperator{\AgdaFunction{𝔽[}}\AgdaSpace{}%
\AgdaOperator{\AgdaFunction{𝕌[}}\AgdaSpace{}%
\AgdaBound{𝑨}\AgdaSpace{}%
\AgdaOperator{\AgdaFunction{]}}\AgdaSpace{}%
\AgdaOperator{\AgdaFunction{]}}\AgdaSpace{}%
\AgdaOperator{\AgdaInductiveConstructor{,}}\AgdaSpace{}%
\AgdaSymbol{(}\AgdaFunction{spFA}\AgdaSpace{}%
\AgdaOperator{\AgdaInductiveConstructor{,}}\AgdaSpace{}%
\AgdaFunction{AimgF}\AgdaSymbol{)}\<%
\\
\>[1][@{}l@{\AgdaIndent{0}}]%
\>[2]\AgdaKeyword{where}\<%
\\
\>[2]\AgdaFunction{spFA}\AgdaSpace{}%
\AgdaSymbol{:}\AgdaSpace{}%
\AgdaOperator{\AgdaFunction{𝔽[}}\AgdaSpace{}%
\AgdaOperator{\AgdaFunction{𝕌[}}\AgdaSpace{}%
\AgdaBound{𝑨}\AgdaSpace{}%
\AgdaOperator{\AgdaFunction{]}}\AgdaSpace{}%
\AgdaOperator{\AgdaFunction{]}}\AgdaSpace{}%
\AgdaOperator{\AgdaFunction{∈}}\AgdaSpace{}%
\AgdaFunction{S}\AgdaSymbol{\{}\AgdaFunction{ι}\AgdaSymbol{\}}\AgdaSpace{}%
\AgdaFunction{ι}\AgdaSpace{}%
\AgdaSymbol{(}\AgdaFunction{P}\AgdaSpace{}%
\AgdaBound{ℓ}\AgdaSpace{}%
\AgdaFunction{ι}\AgdaSpace{}%
\AgdaBound{𝒦}\AgdaSymbol{)}\<%
\\
\>[2]\AgdaFunction{spFA}\AgdaSpace{}%
\AgdaSymbol{=}\AgdaSpace{}%
\AgdaFunction{SPF}\AgdaSymbol{\{}\AgdaArgument{ℓ}\AgdaSpace{}%
\AgdaSymbol{=}\AgdaSpace{}%
\AgdaBound{ℓ}\AgdaSymbol{\}}\AgdaSpace{}%
\AgdaBound{𝒦}\<%
\\
\>[2]\AgdaFunction{epiFlA}\AgdaSpace{}%
\AgdaSymbol{:}\AgdaSpace{}%
\AgdaFunction{epi}\AgdaSpace{}%
\AgdaOperator{\AgdaFunction{𝔽[}}\AgdaSpace{}%
\AgdaOperator{\AgdaFunction{𝕌[}}\AgdaSpace{}%
\AgdaBound{𝑨}\AgdaSpace{}%
\AgdaOperator{\AgdaFunction{]}}\AgdaSpace{}%
\AgdaOperator{\AgdaFunction{]}}\AgdaSpace{}%
\AgdaSymbol{(}\AgdaFunction{Lift-Alg}\AgdaSpace{}%
\AgdaBound{𝑨}\AgdaSpace{}%
\AgdaFunction{ι}\AgdaSpace{}%
\AgdaFunction{ι}\AgdaSymbol{)}\<%
\\
\>[2]\AgdaFunction{epiFlA}\AgdaSpace{}%
\AgdaSymbol{=}\AgdaSpace{}%
\AgdaFunction{F-ModTh-epi-lift}\AgdaSymbol{\{}\AgdaArgument{ℓ}\AgdaSpace{}%
\AgdaSymbol{=}\AgdaSpace{}%
\AgdaBound{ℓ}\AgdaSymbol{\}}\AgdaSpace{}%
\AgdaSymbol{(λ}\AgdaSpace{}%
\AgdaSymbol{\{}\AgdaBound{p}\AgdaSpace{}%
\AgdaBound{q}\AgdaSymbol{\}}\AgdaSpace{}%
\AgdaSymbol{→}\AgdaSpace{}%
\AgdaBound{ModThA}\AgdaSymbol{\{}\AgdaArgument{p}\AgdaSpace{}%
\AgdaSymbol{=}\AgdaSpace{}%
\AgdaBound{p}\AgdaSymbol{\}\{}\AgdaBound{q}\AgdaSymbol{\})}\<%
\\
\>[2]\AgdaFunction{φ}\AgdaSpace{}%
\AgdaSymbol{:}\AgdaSpace{}%
\AgdaFunction{Lift-Alg}\AgdaSpace{}%
\AgdaBound{𝑨}\AgdaSpace{}%
\AgdaFunction{ι}\AgdaSpace{}%
\AgdaFunction{ι}\AgdaSpace{}%
\AgdaOperator{\AgdaFunction{IsHomImageOf}}\AgdaSpace{}%
\AgdaOperator{\AgdaFunction{𝔽[}}\AgdaSpace{}%
\AgdaOperator{\AgdaFunction{𝕌[}}\AgdaSpace{}%
\AgdaBound{𝑨}\AgdaSpace{}%
\AgdaOperator{\AgdaFunction{]}}\AgdaSpace{}%
\AgdaOperator{\AgdaFunction{]}}\<%
\\
\>[2]\AgdaFunction{φ}\AgdaSpace{}%
\AgdaSymbol{=}\AgdaSpace{}%
\AgdaFunction{epi→ontohom}\AgdaSpace{}%
\AgdaOperator{\AgdaFunction{𝔽[}}\AgdaSpace{}%
\AgdaOperator{\AgdaFunction{𝕌[}}\AgdaSpace{}%
\AgdaBound{𝑨}\AgdaSpace{}%
\AgdaOperator{\AgdaFunction{]}}\AgdaSpace{}%
\AgdaOperator{\AgdaFunction{]}}\AgdaSpace{}%
\AgdaSymbol{(}\AgdaFunction{Lift-Alg}\AgdaSpace{}%
\AgdaBound{𝑨}\AgdaSpace{}%
\AgdaFunction{ι}\AgdaSpace{}%
\AgdaFunction{ι}\AgdaSymbol{)}\AgdaSpace{}%
\AgdaFunction{epiFlA}\<%
\\
\>[2]\AgdaFunction{AimgF}\AgdaSpace{}%
\AgdaSymbol{:}\AgdaSpace{}%
\AgdaBound{𝑨}\AgdaSpace{}%
\AgdaOperator{\AgdaFunction{IsHomImageOf}}\AgdaSpace{}%
\AgdaOperator{\AgdaFunction{𝔽[}}\AgdaSpace{}%
\AgdaOperator{\AgdaFunction{𝕌[}}\AgdaSpace{}%
\AgdaBound{𝑨}\AgdaSpace{}%
\AgdaOperator{\AgdaFunction{]}}\AgdaSpace{}%
\AgdaOperator{\AgdaFunction{]}}\<%
\\
\>[2]\AgdaFunction{AimgF}\AgdaSpace{}%
\AgdaSymbol{=}\AgdaSpace{}%
\AgdaFunction{∘-hom}\AgdaSpace{}%
\AgdaOperator{\AgdaFunction{∣}}\AgdaSpace{}%
\AgdaFunction{φ}\AgdaSpace{}%
\AgdaOperator{\AgdaFunction{∣}}\AgdaSpace{}%
\AgdaSymbol{(}\AgdaField{from}\AgdaSpace{}%
\AgdaFunction{Lift-≅}\AgdaSymbol{)}\AgdaOperator{\AgdaInductiveConstructor{,}}\AgdaSpace{}%
\AgdaFunction{∘-IsSurjective}\AgdaSpace{}%
\AgdaSymbol{\AgdaUnderscore{}}\AgdaSpace{}%
\AgdaSymbol{\AgdaUnderscore{}}\AgdaSpace{}%
\AgdaOperator{\AgdaFunction{∥}}\AgdaSpace{}%
\AgdaFunction{φ}\AgdaSpace{}%
\AgdaOperator{\AgdaFunction{∥}}\AgdaSymbol{(}\AgdaFunction{fromIsSurjective}\AgdaSpace{}%
\AgdaSymbol{(}\AgdaFunction{Lift-≅}\AgdaSymbol{\{}\AgdaArgument{𝑨}\AgdaSpace{}%
\AgdaSymbol{=}\AgdaSpace{}%
\AgdaBound{𝑨}\AgdaSymbol{\}))}\<%
\\
\>[0]\<%
\end{code}
\af{ModTh-closure} and \af{Var⇒EqCl} show that
\af{V} \ab{𝒦} = \af{Mod} (\af{Th} (\af{V} \ab{𝒦})) holds for every class \ab{𝒦} of \ab{𝑆}-algebras.
Thus, every variety is an equational class. This completes the formal proof of Birkhoff's variety theorem.

%% -----------------------------------------------------------------------------
\section{Related work}
There have been a number of efforts to formalize parts of universal algebra in
type theory besides ours, most notably

\begin{enumerate}
\item
In~\cite{Capretta:1999}, Capretta formalized the basics of universal algebra in the
   Calculus of Inductive Constructions using the Coq proof assistant;
\item In~\cite{Spitters:2011}, Spitters and van der Weegen formalized the basics of universal algebra
   and some classical algebraic structures, also in the Calculus of Inductive Constructions using
   the Coq proof assistant and promoting the use of type classes;
\item In~\cite{Gunther:2018} Gunther et al developed what seemed (prior to the \agdaalgebras library) to be
   the most extensive library of formalized universal algebra to date; like \agdaalgebras, that work is based on dependent type theory, is programmed in Agda, and goes beyond the Noether isomorphism theorems to include some basic equational logic; although the coverage is less extensive than that of \agdaalgebras, Gunther et al do treat \emph{multi-sorted} algebras, whereas \agdaalgebras is currently limited to single sorted structures.
   \item In~\cite{Amato:2021}, ``Amato et al formalize multi-sorted algebras with finitary operators in UniMath. Limiting to finitary operators is due to the restrictions of the UniMath type theory, which does not have W-types nor user-defined inductive types. These restrictions also prompt the authors to code terms as lists of stack machine instructions rather than trees'' (quoting~\cite{Abel:2021}).
\item In~\cite{Lynge:2019}, ``Lynge and Spitters formalize multi-sorted algebras in HoTT, also restricting to finitary operators. Using HoTT they can define quotients as types, obsoleting setoids. They prove three isomorphism theorems concerning sub- and quotient algebras. A universal algebra or varieties are not formalized'' (quoting~\cite{Abel:2021}).
\item In~\cite{Abel:2021}, Abel gives a new formal proof of the soundness theorem and Birkhoff's completeness theorem for multi-sorted algebraic structures.
\end{enumerate}

%Some other projects aimed at formalizing mathematics generally, and algebra in particular, have developed into very extensive libraries that include definitions, theorems, and proofs about algebraic structures, such as groups, rings, modules, etc.  However, the goals of these efforts seem to be the formalization of special classical algebraic structures, as opposed to the general theory of (universal) algebras. Moreover, the part of universal algebra and equational logic formalized in the \agdaalgebras library extends beyond the scope of prior efforts.

%Prior to our work, a constructive version of Birkhoff's theorem was published by Carlstr\"om in~\cite{Carlstrom:2008}.  However, the logical foundations of that work is constructive set theory and, as far as we know, there was no attempt to formalized it in type theory and verify it with a proof assistant.

 %% for arXiv version (where no subdirectories are allowed)

% \input{1Introduction}
% \input{2Overture}
% \input{3Relations}
% \input{4Algebras}
% \input{5Conclusion}
% \include{1Introduction}
% \include{2Overture}
% \include{3Relations}
% \include{4Algebras}
% \include{5Conclusion}

\paragraph*{Acknowledgments}
This work would not have been possible without the wonderful \agda language and the
\agdastdlib, developed and
maintained by The Agda Team~\cite{agdastdlib}.
% (Andreas Abel, Guillaume Allais, Liang-Ting Chen, Jesper Cockx, Nils Anders Danielsson,
% Víctor López Juan, Ulf Norell, Andrés Sicard-Ramírez,Andrea Vezzosi, and Tesla Ice Zhang)
% , maintained by Matthew Daggitt and Guillaume Allais.
Most of the content of this paper was generated from the literate \agda file \HSPlagda and the \LaTeXe file \agdahsp (processed with the commands
\texttt{agda --latex} and \texttt{pdflatex}), which are available in the \agdaalgebras GitHub repository~\cite{ualib_v2.0.0}.
The first author was supported by the CoCoSym Project under the ERC Consolidator Grant (ERC CoG), No. 771005.

%%%%%%%%%%%%%%%%%%%%%%%%%%%%%%%%%%%%%%%%%%%%%%%%%%%%%%%%%%%%%%%%%%%%%%%%%%%%%%%%%%%%%%%%%%
%%%%%%%%%%%%%%%%%%%%%%%%%%%%%%%%%%%%%%%%%%%%%%%%%%%%%%%%%%%%%%%%%%%%%%%%%%%%%%%%%%%%%%%%%%

%% Bibliography
\bibliographystyle{plainurl}

\bibliography{ualib_refs}

% \appendix
% \newpage
% \section{Dependency Graph}

% ~\hskip-1.5cm\includegraphics[scale=0.65]{ualib-graph-top.png}

% \newpage

% ~\hskip-2.5cm\includegraphics[scale=0.55]{ualib-graph-bot.png}

%% \section{Some Components of the Type Topology Library}
%% Here we collect some of the components from the \typetopology library that we used above but did not have space to discuss.  They are collected here for the reader's convenience and to keep the paper somewhat self-contained.

%% \input{aux/typetopology.tex}

%%%%%%%%%%%%%%%%%%%%%%%%%%%%%%%%%%%%%%%%%%%%%%%%%%%%%%%%%%%%%%%%%%%%%%%%%%%
\end{document}